# PHYSICS. TASKS WITH SOLUTIONS

The study guide(textbook) is a component of a set of working materials created to ensure high-quality practical training of specialists in the course of physics.

The guide contains a set of tasks necessary for organizing comprehensive in-class and independent work for learners. It can serve as a foundation for further in-depth study of physics-technical training disciplines and specialized courses of the chosen specialty and is aligned with current educational training programs.

This textbook is a collection of physics problems with solutions, compiled in accordance with the curriculum for first-year full-time bachelor's students.

It contains 120 problems from six sections of the physics course, each provided with detailed solutions that take into account the most common questions raised by modern students.

Problem-solving is a crucial and indispensable part of effectively studying physics. It's not merely a test of acquired knowledge but a key tool that transforms theoretical concepts into practical skills. Applying laws and formulas to real-world situations not only deepens understanding of lecture material, but also develops logical and analytical thinking. It is through solving problems that students gain true insight into physical phenomena, form strategies for analysis, and develop solutions to tasks—making the learning process truly comprehensive.

This manual is designed primarily as a supplemental and holistic resource for students aiming not just for academic success, but for a deep and lasting grasp of the physics course. It will become a reliable companion in transitioning from passive absorption of information to active engagement with it, ensuring practical reinforcement of knowledge and fostering the skills needed to tackle both standard and nonstandard physics problems. We hope this guide helps you not only pass your exams successfully but also fall in love with physics by discovering its beauty and logic through the lens of practical tasks.

**МІНІСТЕРСТВО ОСВІТИ І НАУКИ УКРАЇНИ**

**ОДЕСЬКИЙ НАЦІОНАЛЬНИЙ МОРСЬКИЙ УНІВЕРСИТЕТ**

**Кафедра «Математика, фізика та астрономія»**

**Чінарова Л.Л., Андронов І.Л., Савчук Н.В.,**

**Іовчев С.І., Акопян Г.М.**

# ФІЗИКА

**Задачі з розв'язками**

Навчальний посібник для розв'язання задач

з курсу «Фізика»

Навчальний посібник складова комплексу робочих матеріалів, створених для забезпечення якісної практичної підготовки фахівців за освітньо-професійними програмами: «*Комп'ютерні науки*» за спеціальністю F3 Комп'ютерні науки; «*Кібербезпека та захист інформації*» за спеціальністю F5 Кібербезпека та захист інформації; «*Судноводіння*», «*Експлуатація суднових енергетичних установок*», «*Експлуатація суднового електрообладнання і засобів автоматики*» за спеціальністю J5 Морський та внутрішній водний транспорт.

Посібник містить набір задач необхідних для організації повноцінної аудиторної та самостійної роботи здобувачів освіти, може бути базовим для подальшого поглибленого вивчення навчальних дисциплін фізико-технічної підготовки та спецкурсів спеціальності, відповідає чинним навчальним програмам підготовки.

# ЗМІСТ

# ПЕРЕДМОВА

Посібник являє собою збірник задач з розв'язками укладений, відповідно до програми з фізики для студентів бакалаврів першого року денної форми навчання.

Посібник містить 120 задач з шости розділів курсу фізики, наданих з детальними розв'язками з урахуванням найпоширеніших запитань від сучасних студентів.

Розв'язання задач є суттєвим і невід'ємним елементом ефективного вивчення курсу фізики. Це не просто перевірка засвоєних знань, а ключовий інструмент, що трансформує теоретичні концепції в практичні навички. Застосування законів та формул до конкретних ситуацій є корисним не лише тому, що поглиблює розуміння лекційного матеріалу, але й тому, що розвиває логічне та аналітичне мислення. Саме через розв'язання задач відбувається справжнє усвідомлення фізичних явищ, формуються алгоритми аналізу та вирішення поставлених завдань, що робить навчання по-справжньому повним та цілісним.

Цей посібник призначений, в першу чергу, як додатковий та всебічний ресурс для студентів, які прагнуть досягти не просто успішного навчання, а й глибокого та міцного засвоєння курсу фізики. Він стане вашим надійним помічником у процесі переходу від пасивного сприйняття інформації до активного оперування нею, забезпечуючи практичне закріплення знань та розвиток навичок, необхідних для розв'язання як типових, так і нестандартних фізичних проблем. Сподіваємося, що цей посібник допоможе вам не лише успішно скласти іспити, а й по-справжньому полюбити фізику, розкривши її красу та логіку через призму практичних завдань.

# Розділ 1. МЕХАНІКА, КІНЕМАТИКА, ОСНОВИ ДИНАМІКИ, ЗАКОНИ ЗБЕРЕЖЕННЯ ЕНЕРГІЇ.

## 1.1. Основні кінематичні характеристики руху

**1. Траєкторія** – це уявна лінія, якиу описує кінець радіус-вектора в процесі руху або інакше – лінія, вздовж якої рухається точка. Форма траєкторії залежить від вибору системи відліку. В залежності від форми траєкторії рух поділяють на прямолінійний і криволінійний та рух по колу.

**2. Переміщення** – $\Delta\vec{r}$ – напрямлений відрізок прямої (вектор), який з'єднує початкове положення матеріальної точки на траєкторії з подальшими її положеннями. Це векторна фізична велиабона, що дорівнює різниці радіус-векторів в кінцевий і початковий моменти часу.

Модуль переміщення дорівнює довжині шляху при прямолінійному русі, але менший за довжину шляху при криволінійному русі.

**3. Шлях** – $s$ (Δs) – довжина траєкторії, тобто відстань, яку матеріальна точка проходить вздовж траєкторії. Це завжди додатна велиабона і визначається:

$$S = v \cdot t,$$

де $v$ – швидкість, $t$ – час руху.

**4. Швидкість.**

Середня швидкість описує рух в цілому на всій траєкторії руху, також можна розглядати швидкість на окремих частинах траєкторії і взагалі можна говорити про швидкість в кожній точці траєкторії.

Рух тіла може бути зі сталою швидкістю – це рівномірний і прямолінійний; із швидкістю, що змінюється за велиабоною і напрямком – прискорений, криволінійний рух.

Таким абоном, важливою характеристикою механічного руху являється швидкість зміни координат тіла з часом.

**Середня шляхова швидкість** нерівномірного руху визначається відношенням шляху ΔS, пройденого тілом за час Δt уздовж траєкторії:

$$\langle v \rangle = \frac{\Delta \mathrm{S}}{\Delta \mathrm{t}}$$

Середня швидкість переміщення – це векторна фізична велиабона, яка дорівнює відношенню вектора переміщення до проміжку часу, за який відбувається це переміщення:

$$\langle \vec{v} \rangle = \frac{\Delta \vec{r}}{\Delta t}$$

Напрямок середньої швидкості співпадає з напрямком $\Delta r$ . При необмеженому зменшені $\Delta t$ середня швидкість прагне до граничного значення, яке називається миттєвою швидкістю:

$$\vec{v} = \lim_{\Delta t \to 0} \frac{\Delta \vec{r}}{\Delta t} = \frac{d\vec{r}}{dt}$$

**Миттєва швидкість** це векторна велиабона, яка дорівнює похідній радіус-вектора матеріальної точки за часом:

$$\vec{v} = \frac{d\vec{r}}{dt}$$

Миттєва швидкість завжди напрямлена по дотичній до траєкторії.

**5. Прискорення** – характеризує швидкість зміни швидкості. Прискорення — це фізична велиабона, що показує, наскільки швидко змінюється швидкість об'єкта.

**Середнє прискорення** – це векторна велиабона, яка дорівнює відношенню вектора зміни (приросту) швидкості $\Delta \vec{v}$ до інтервалу часу, за який відбувається ця зміна:

$$\langle \vec{a} \rangle = \frac{\Delta \vec{v}}{\Delta t}$$

**Миттєве прискорення** – це векторна фізична велиабона, що дорівнює другій похідній від радіус-вектора за часом і, відповідно, першій похідній від миттєвої швидкості за часом:

$$\vec{a} = \frac{d\vec{v}}{dt} = \frac{d^2\vec{r}}{dt^2}$$

**Миттєве прискорення при криволінійному русі.**

Для визначення миттєвого прискорення $\vec{a}$ при криволінійному русі його зручно розкласти на дві складові: тангенціальне прискорення $\vec{a}_\tau$ та нормальне прискорення $\vec{a}_n$.

**Тангенціальне прискорення**

$$\vec{a}_\tau = \frac{dv}{dt}\vec{\tau}$$

характеризує зміну швидкості за велиабоною, спрямоване по дотичній до траєкторії.

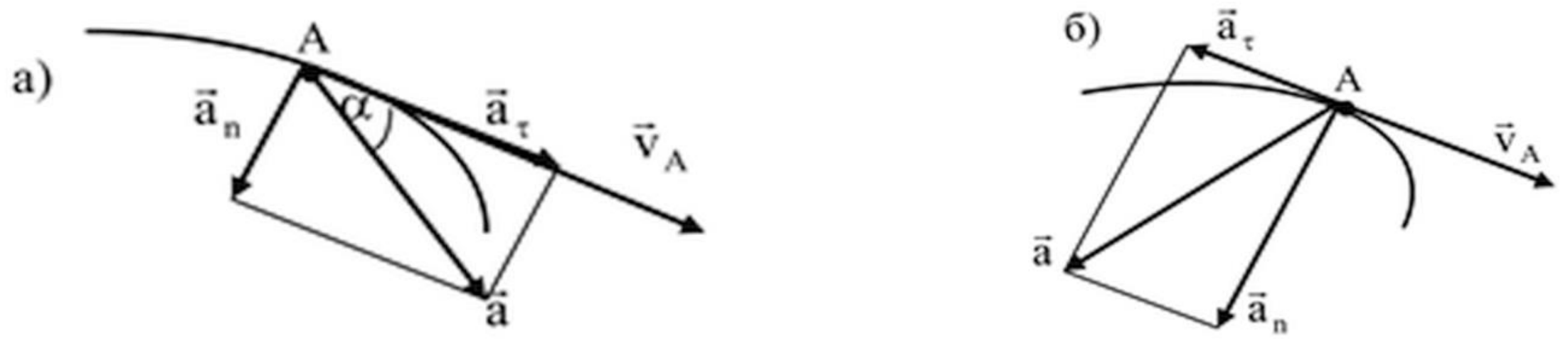

Рис.1.1.а) прискорений рух при криволінійному русі точки; б) уповільнений рух при криволінійному русі точки.

**Нормальне прискорення**

$$\vec{a}_n = \frac{v^2}{R}\vec{n}$$

характеризує зміну швидкості за напрямком, спрямоване до центра кривизни траєкторії ($R$ – радіус кривизни траєкторії).

Повне прискорення, виражене через компоненти $\vec{a}_\tau$ та $\vec{a}_n$:

$$\vec{a} = \vec{a}_\tau + \vec{a}_n = \frac{dv}{dt}\vec{\tau} + \frac{v^2}{R}\vec{n}$$

де $\vec{\tau}$ і $\vec{n}$ - орти, тобто одиничні вектори, дотичної та нормалі до траєкторії. Велиабона прискорення обабослюється за формулою

$$a = \sqrt{a_\tau^2 + a_n^2} = \sqrt{\left(\frac{dv}{dt}\right)^2 + \left(\frac{v^2}{R}\right)^2}$$

Таким абоном, прискорення криволінійного руху визначає зміну швидкості за напрямком і велиабоною.

**Рівномірний рух по колу**

Точка обертається по колу радіуса *r*. Швидкість точки постійна за модулем і дорівнює $v$. Швидкість $v$ називається лінійною швидкістю точки.

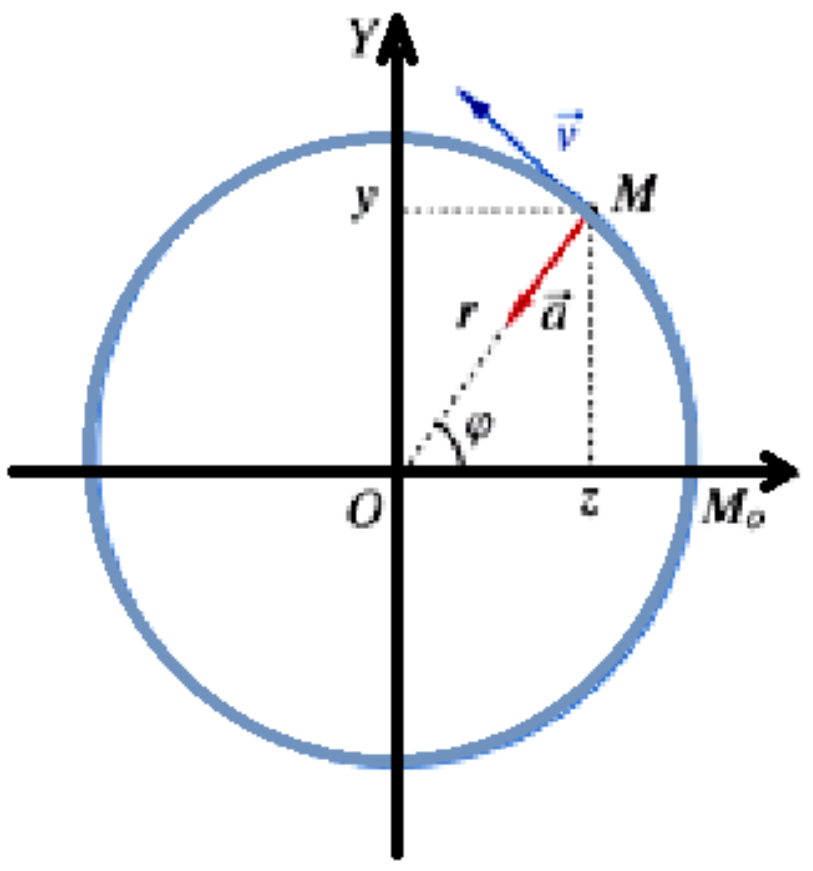

а)

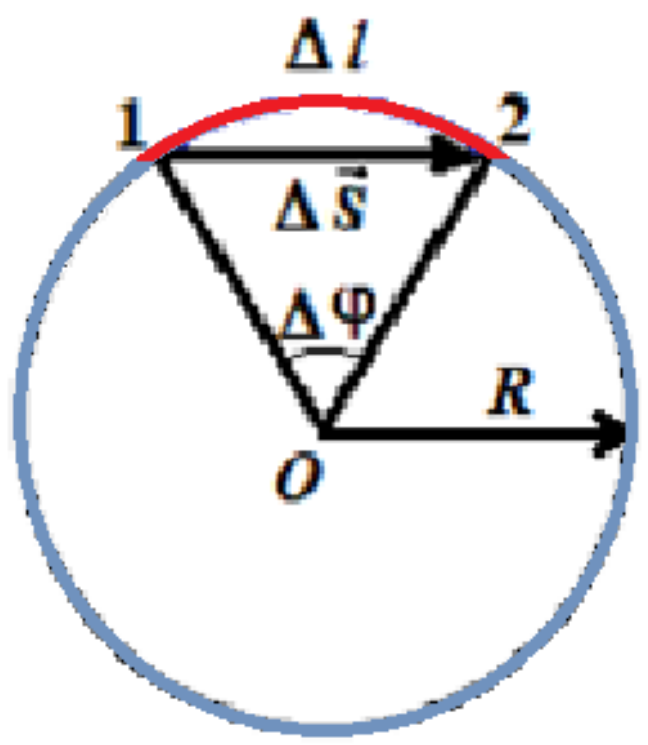

б)

Рис. 1.2. а) Обертання точки по колу радіуса *r*; б) вимірювання кута в радіанах.

*Маємо наступні характеристики такого руху*:

– **період обертання** – це час за який тіло здійснює повний оберт:

$$T = \frac{2\pi r \quad (\text{довжина кола})}{v \quad (\text{швидкість})}$$

– **частота обертання** – це велиабона, зворотна періоду:

$$\nu = \frac{1}{T}$$

частота показує, скільки повних обертів *N* точка здійснює за одиницю часу

$$\nu = \frac{N}{t} \qquad [\nu] = \frac{1}{c} = c^{-1}$$

– **кутова швидкість** це відношення кута повороту до часу:

$$\omega = \frac{\varphi}{\mathrm{t}}$$

Кут $\varphi = 2\pi n$, де $n$ число повних оборотів, як правило, вимірюється в радіанах, тому кутова швидкість вимірюється в рад/с. Радіан: кут, у якого довжина дуги дорівнює радіусу. Тобто, радіан - це «природна» одиниця вимірювання кута, яка пов'язана з геометричними властивостями кола.

$$\omega = \frac{2\pi}{T} \quad \text{і} \quad T = \frac{2\pi r}{v}$$

Підставивши $T$ отримаємо зв’язок лінійної та кутової швидкості:

$$v = \omega r.$$

Миттєви велиабони кутової швидкості і кутового прискорення визначаються аналогічно лінійним:

$$\omega = \frac{\mathrm{d}\varphi}{\mathrm{d}t}; \qquad \varepsilon = \frac{\mathrm{d}\omega}{\mathrm{dt}}$$

Підставивши $v = \omega r$ у формули:

$$\vec{a}_n = \frac{v^2}{R}\vec{n} \quad \text{і} \quad \vec{a}_\tau = \frac{dv}{dt}\vec{\tau}$$

отримаємо, що тангенціальне і нормальне прискорення точки тіла, що обертається виражаються формулами:

$$a_\tau = \varepsilon R; \;\; a_n = \omega^2 R,$$

де $\varepsilon$ – кутове прискорення тіла; $R$ – відстань точки від осі обертання; $\omega$ – кутова швидкість тіла. Таким абоном повне прискорення визначається за формулою:

$$a = \sqrt{\varepsilon^2 R^2 + \omega^4 R^2} = R\sqrt{\varepsilon^2 + \omega^4}$$

## 1.2. Обчислення швидкості та прискорення при наданому законі руху (приклади розв'язання задач)

**Задача 1.1.** Визнааботи швидкість $v_x$ та прискорення $a_x$ , якщо координата тіла $x$ змінюється з часом за формулою: $x(t) = 1 + 12t - 3t^2$.

**Розв'язання:**

За умовою тіло рухається вздовж осі $X$, а швидкість $v$ і прискорення $a$ визначається за формулами:

$$\vec{v} = \frac{d\vec{r}}{dt} \quad \text{та} \quad \vec{a} = \frac{d\vec{v}}{dt} = \frac{d^2\vec{r}}{dt^2}$$

Тобто щоб знайти швидкість диференціюємо

$$x(t) = 1 + 12t - 3t^2.$$

Існують декілька варіантів запису знаходження похідної від виразу:

$$\dot{x}(t) = \frac{dx}{dt} = \frac{d}{dt}(1 + 12t - 3t^2) = (1 + 12t - 3t^2)'$$

Символ $\frac{d}{dt}$ перед скобкою є теж саме, що штрих за скобкою, тобто похідна координати по часу $\frac{dx}{dt}$, це є дріб в абосельнику якого знаходиться достатньо мала зміна координати $dx$, а в знаменнику достатньо малий проміжок часу $dt$, на протязі якого ця зміна координати відбулась.

$$\frac{dx}{dt} = \lim_{\Delta t \to 0} \frac{\Delta x}{\Delta t} = \lim_{\Delta t \to 0} \frac{x(t + \Delta t) - x(t)}{\Delta t}$$

За основними формулами для обабослення похідних знайдемо:

$$(1 + 12t - 3t^2)' =$$

Основні формули обабослення похідних:

| | |
|---|---|
| $(C)' = 0$; $(x)' = 1$ | $(x^a)' = ax^{a-1}$ |
| $(e^x)' = e^x$ | $(\ln x)' = \frac{1}{x}$ |
| $(\sin x)' = \cos x$ | $(\cos x)' = -\sin x$ |
| $(\operatorname{tg} x)' = \dfrac{1}{\cos^2 x}$ | $(\operatorname{ctg} x)' = -\dfrac{1}{\sin^2 x}$ |
| $(\arcsin x)' = \dfrac{1}{\sqrt{1-x^2}}$ | $(\operatorname{arctg} x)' = \dfrac{1}{1+x^2}$ |
| $(u \pm v)' = u' \pm v'$ | $(Cu)' = Cu'$ |
| $\left(\dfrac{u}{v}\right)' = \dfrac{u'v - uv'}{v^2}$ | $(uv)' = u'v + uv'$ |
| $y' = y(\ln y)'$ | $(u^v)' = vu^{v-1}u' + u^v v' \ln u$ |

За формулою: $(u \pm v)' = u' \pm v'$ отримаємо суму

$$= (1)' + (12t)' - (3t^2)' =$$

і застосуємо формули: $(C)' = 0$; $(Cu)' = Cu'$; $(x)' = 1$;$(x^a)' = ax^{a-1}$

і отримаємо $= 12 - 6t$

Таким абоном, похідна координати є швидкість

$$\dot{x}(t) = v_x = 12 - 6t$$

Знайдемо $\ddot{x}(t)$, тобто візьмемо похідну вже від $\dot{x}(t)$

$$\ddot{x}(t) = (12 - 6t)' = 6$$

Таким абоном, друга похідна координати є прискорення

$$\ddot{x}(t) = a_x = 6$$

**Задача 1.2.** Нехай тіло рухається вздовж осі *X* за законом: $x = 5 \sin 2t$. Визначити $v_x$ та $a_x$.

**Розв'язання:**

Рух тіла за таким законом являється прикладом гармонічних коливань, коли координата змінюється за законом синуса. Диференціюємо перший раз і отримаємо $v_x$, диференціюємо другий раз і отримаємо $a_x$.

$$v_x = \dot{x}(t) = (5 \sin 2t)' =$$

Тут ми маємо справу зі складною функцією тому, що під знаком sin маємо $2t$, а не $t$. Застосовуваємо відповідну формулу для складної функції:

$\left(f\left(u(x)\right)\right)' = f_u'(u)\cdot u_x'(x)$ та $(\sin x)' = \cos x$

Але спочатку винесемо постійний знак за скобки, тобто застосуємо формулу

$$(Cu)' = Cu'$$

$$= 5(\sin 2t)' = 5(\sin 2t)'(2t)' = 5\cdot 2\cdot \cos 2t = 10\cos 2t$$

$$a_x = \ddot{x}(t) = (10\cos 2t)' = 10(\cos 2t)'(2t)' =$$

$(\cos x)' = -\sin x$

$$= 10\cdot 2\cdot(-\sin 2t) = -20\sin 2t$$

**Задача 1.3.** Матеріальна точка рухається вздовж осі О$x$ за законом $x=A+Bt+Ct^3$, де $A$=3 м, $B$=2 м/с, $C$=0,05 м/с$^3$. Визнааботи координату $x$, швидкість $v$ й прискорення $a$ у моменти часу $t_1 = 0$; $t_2 = 4$ с, а також шлях $S$, середні значення швидкості $\langle v\rangle$ й прискорення $\langle a\rangle$ за перші 4 с руху.

**Дано:**

$x=A+Bt+Ct^3$

$A$=3 м,

$B$=2 м/с,

$C$=0,05 м/с$^3$

$t_1$=0, $t_2$=4 с

**Знайти:**

$x_1$, $x_2$, $v_1$, $v_2$, $a_1$, $a_2$, $S$, $\langle v\rangle$, $\langle a\rangle$ - ?

**Розв'язання:**

Підставимо цифрові значення коефіцієнтів в рівняння $x = A + Bt + Ct^3$ і отримаємо рівняння:

$$x = 3 + 2t + 0{,}05t^3$$

Координату $x_1$ знаходимо підстановкою $t_1 = 0$ с в отримане рівняння руху

$$x = 3 + 2t + 0{,}05t^3$$

а $x_2$ підстановкю $t_2$=4 с в теж саме рівняння руху і отримаємо, що $x_1$=3 м і $x_2$=14,2 м.

При прямолінійному русі шлях $S$ дорівнюється велиабоні переміщення $\Delta x$:

$$S = x_2 - x_1 = 14{,}2 - 3 = 11{,}2 \text{ м}$$

Середня швидкість визначається за формулою: $\langle v\rangle = \frac{\Delta S}{\Delta t}$ або просто

$$v = \frac{S}{t}$$

$$\langle v\rangle = \frac{x_2 - x_1}{t_2 - t_1} = \frac{14{,}2 - 3}{4} = 2{,}8 \text{ м/с}$$

Миттєва швидкість це перша похідна від координати за часом:

$$v = v_x = \frac{dx}{dt} = (3 + 2t + 0{,}05t^3)' = 2 + 3 \cdot 0{,}05t^2$$

в отримане рівняння підставимо $t_1$=0 с і отримаємо $v_1 = 2\frac{\text{м}}{\text{с}}$, потім підставимо $t_2$=4 с і отримаємо $v_2 = 4{,}4\frac{\text{м}}{\text{с}}$ .

Середнє прискорення знаходимо як:

$$\langle a\rangle = \frac{v_2 - v_1}{t_2 - t_1} = \frac{14{,}4 - 2}{4} = 0{,}6 \text{ м}/c^2$$

Миттєве прискорення:

$$a = a_x = \frac{dv_x}{dt} = (2 + 3 \cdot 0{,}05t^2)' = 2 \cdot 3 \cdot 0{,}05t$$

в отриманий вираз підставимо $t_2$=4 с і отримаємо $a_2 = 1{,}2 \text{ м/с}^2$.

**Відповідь:** $\langle v\rangle = 2{,}8\frac{\text{м}}{\text{с}}$; $\langle a\rangle = 0{,}6 \text{ м}/c^2$.

**Задача 1.4.** Шлях першого тіла визначається за формулою $S_1 = A_1t + B_1t^2 + Ct^3$ ($A_1 = 6$ м/с; $B_1 = -1$ м/с$^2$; $C=\frac{1}{3}$ м/с$^3$), а другого $S_2 = A_2t + B_2t^2$ ($A_2 = -2$ м/с; $B_2 = 2$ м/с$^2$). Через який час швидкості тіл будуть однаковими? Визнааботи швидкості тіл у цей момент.

**Розв'язання:**

Підставимо надані абослові значення в формули:

$$S_1 = 6t - t^2 + \frac{1}{3}t^3$$

$$S_2 = -2t + 2t^2$$

Треба визнаботи через який час швидкості двох тіл будуть однаковими, тобто $v_1 = v_2$. Для цього визнаабомо вирази для швидкостей на якийсь момент часу, тобто візьмемо похідну від шляху для першого і для другого тіла.

$$v_1 = \frac{dS_1}{dt} = \left(6t - t^2 + \frac{1}{3}t^3\right)' = 6 - 2t + t^2$$

$$v_2 = \frac{dS_2}{dt} = (-2t + 2t^2)' = -2 + 4t$$

В момент часу коли $v_1 = v_2$ можна прирівняти і праві частини цих виразів:

$$6 - 2t + t^2 = -2 + 4t$$

Отримали квадратне рівняння:

$$t^2 - 6t + 8 = 0$$

Корні рівняння знайдемо за формулою:

$$t_{1,2} = \frac{-b \pm \sqrt{b^2 - 4ac}}{2a} = \frac{6 \pm \sqrt{4}}{2}$$

Отриманий перший корінь $t_1 = 4$ підставимо в один із двох виразів, або в $v_1 = 6 - 2t + t^2$, або в $v_2 = -2 + 4t$, не має значення в який, результат буде однаковий, тобто отримаємо $v(t_1) = 14$ м/с; потім так саме зробимо підставивши $t_2 = 2$ в один із виразів для $v_1$ або $v_2$ і отриаємо $v(t_2) = 6$ м/с .

**Відповідь:** $v(t_1) = 14$ м/с і $v(t_2) = 6$ м/с.

**Задача 1.5.** Залежність кута повороту тіла від часу дається рівнянням: $\varphi = A + Bt + Ct^2 + Dt^3$ , де $A = 1$ рад, $B = 0{,}1$ рад/с, $C = 0{,}02$ рад/с$^2$ , $D = 0{,}01$ рад/с$^3$. Знайти: кутовий шлях, якій тіло пройде за 3 с від початку відліку часу, середню кутову швидкість та середнє кутове прискорення за цей час.

**Розв'язання:**

Підставимо надані абослові значення в формулу руху і отримаємо:

$$\varphi(\mathrm{t}) = 1 + 0{,}1t + 0{,}02t^2 + 0{,}01t^3$$

Кутовий шлях, якій тіло пройде за 3 с визначимо за формулою:

$$\varphi = \varphi_2 - \varphi_1$$

де $\varphi_2$ – кутовий шлях, що тіло пройшло за 3 с ($t_2$ = 3 с); $\varphi_1$ – кутовий шлях на моменту часу $t_1$ = 0 с. Із залежності кутового шляху від часу $\varphi$(t) знайдемо $\varphi_1$ підставивши в вираз $\varphi(t)$ значення $t_1$ = 0 с і $\varphi_2$ підставивши в цей же вираз значення $t_2$ = 3 с.

$\varphi_1 = 1$ рад;

$\varphi_2 = 1 + 0{,}1 \cdot 3 + 0{,}02 \cdot 3^2 + 0{,}01 \cdot 3^3 = 1{,}75$ рад;

$\varphi = \varphi_2 - \varphi_1 = 1{,}75 - 1 = 0{,}75$ рад.

Середня кутова швидкість за 3 с від початку обертання знаходиться за формулою:

$$\langle\omega\rangle = \frac{\varphi_2 - \varphi_1}{t_2 - t_1},$$

аналогічно знаходженню середньої лінійної швидкості: $\langle v\rangle = \frac{x_2 - x_1}{t_2 - t_1}$.

$$\langle\omega\rangle = \frac{\varphi_2 - \varphi_1}{t_2 - t_1} = \frac{1{,}75 - 1}{3 - 0} = 0{,}25 \text{ рад/с}.$$

Середнє кутове прискорення за 3 с від початку обертання визначається як:

$$\langle\varepsilon\rangle = \frac{\omega_2 - \omega_1}{t_2 - t_1},$$

де $\omega_2$ – кутова швидкість в момент часу $t_2$ = 3 с; $\omega_1$ – кутова швидкість в момент часу $t_1$ = 0 с. Тобто $\omega_1$ і $\omega_2$ – це є миттєві кутові швидкості.

Миттєву кутову швидкість знайдемо за визначенням, як:

$$\omega = \frac{d\varphi}{dt} = (1 + 0{,}1t + 0{,}02t^2 + 0{,}01t^3)' = 0{,}1 + 0{,}02 \cdot 2 \cdot t + 0{,}01 \cdot 3 \cdot t^2$$

Підставимо числові данні: $t_1$ = 0 с. – отримаємо $\omega_1$= 0,1 рад/с, та $t_2$ = 3 с. – отримаємо

$$\omega_2 = 0{,}1 + 0{,}02 \cdot 2 \cdot 3 + 0{,}01 \cdot 3 \cdot 3^2 = 0{,}49 \text{ рад/с}.$$

Отримані значення миттєвих кутових швидкостей підставимо в формулу для середнього кутового прискорення і отримаємо:

$$\langle\varepsilon\rangle = \frac{0{,}49 - 0{,}1}{3 - 0} = 0{,}13 \text{ рад/с}^2$$

**Відповідь:** $\varphi = 0{,}75$ рад; $\langle\omega\rangle = 0{,}25$ рад/с; $\langle\varepsilon\rangle = 0{,}13$ рад/с$^2$

**Задача 1.6.** Тіло обертається навколо нерухомої осі за законом $\varphi = 10 + 20t - 2t^2$. Знайти повне прискорення точки, що знаходиться на відстані 0,1 м від осі обертання, для моменту часу $t = 4$ с.

**Розв'язання:**

Кожна точка тіла, що обертається, описує коло. Повне прискорення точки, що рухається по кривій лінії, може бути знайдено як геометрична сума тангенціального прискорення $\vec{a}_\tau$, спрямованого по дотичній до траєкторії, і нормального прискорення $\vec{a}_n$, спрямованого до центру кривизни траєкторії:

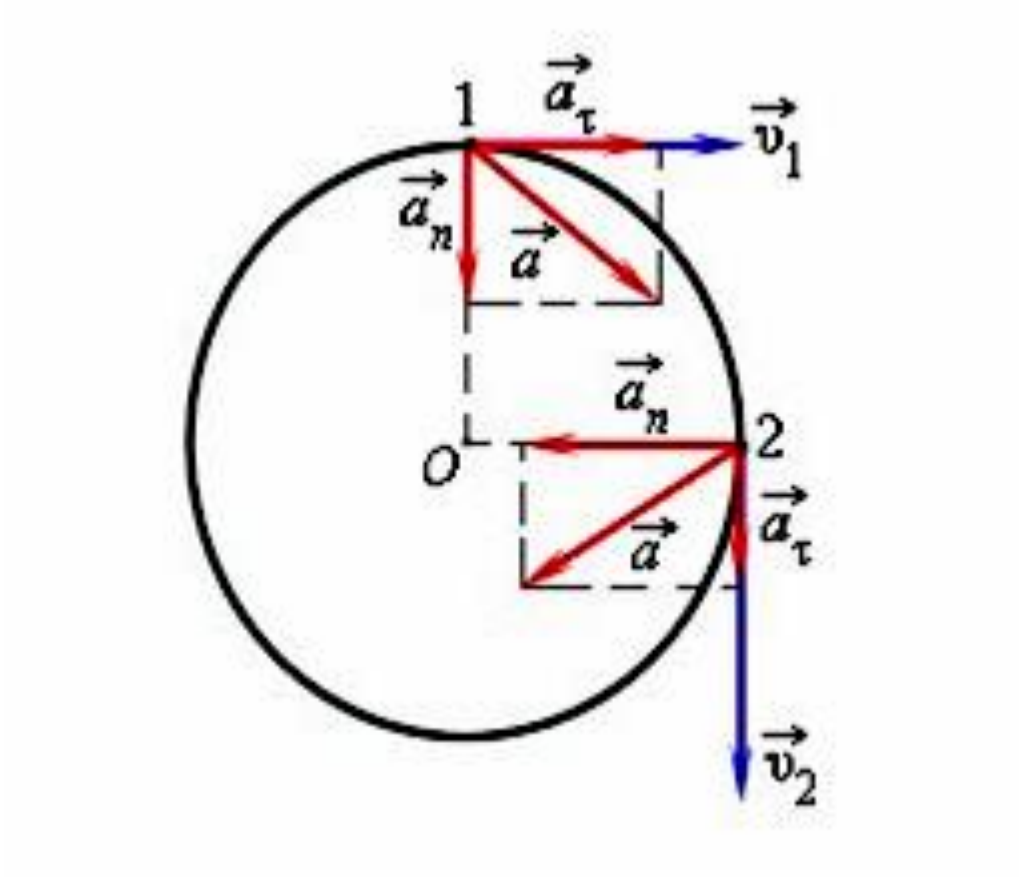

$$a = \sqrt{a_\tau^2 + a_n^2}$$

Тангенціальне і нормальне прискорення точки тіла, що обертається виражаються формулами:

$$a_\tau = \varepsilon R, \qquad a_n = \omega^2 R,$$

де ε – кутове прискорення тіла; $R$ – відстань точки від осі обертання; ω – кутова швидкість тіла.

Підставимо ці формули в вираз для повного прискорення:

$$a = \sqrt{a_\tau^2 + a_n^2}$$

і отримаємо формулу за якою зможемо знайти повне прискорення:

$$a = \sqrt{\varepsilon^2 R^2 + \omega^4 R^2} = R\sqrt{\varepsilon^2 + \omega^4}$$

Кутова швидкість тіла, що обертається, дорівнює першій похідній від кута повороту по часу:

$$\omega = \frac{d\varphi}{dt} = (10\ +\ 20t - 2t^2)' = 20 - 4t$$

Підставимо в отриманий вираз замість $t$ значення 4 с і отримаємо кутову швидкість на момент часу $t$ = 4 с:

$$\omega = 20 - 4t = (20 - 4\ \cdot\ 4) = 4\ \text{рад/с}.$$

Кутове прискорення тіла, що обертається дорівнюється першій похідній від кутової швидкості по часу:

$$\varepsilon = \frac{d\omega}{dt} = (20 - 4t)' = -4\ \text{рад}/c^2\ .$$

Цей вираз не має залежності від аргументу часу *t*, тому, кутове прискорення має стале значення, яке не залежить від часу. Підставимо значення $\omega$ и $\varepsilon$ в формулу:

$$a = R\sqrt{\varepsilon^2 + \omega^4} = 0{,}1 \cdot \sqrt{(-4)^2 + 4^4} = 1{,}65\ \text{м/с}^2$$

**Задача 1.7.** Колесо обертається рівноприскорено. Через 2 с. від початку руху вектор повного прискорення точки, що лежить на ободі колеса, утворює кут 60° з вектором її лінійної швидкості. Визначте кутове прискорення в цей час.

| **Дано:** | **Розв’язання:** |
|---|---|
| $t$=2 с<br>α=60° | Відомо, що тангенціальне прискорення як і швидкість однаково спрямовані по дотичній до траєкторії, тому кут $\alpha = 60^{o}$ це є кут між |
| **Знайти:**<br>ε - ? | повним прискоренням $a$ і його тангенційною складовою $a_\tau$ . В прямокутному трикутнику за означенням: |

$$\cos\alpha = \frac{a_\tau}{a} \;;\; \text{тому} \quad a_\tau = a\cos\alpha.$$

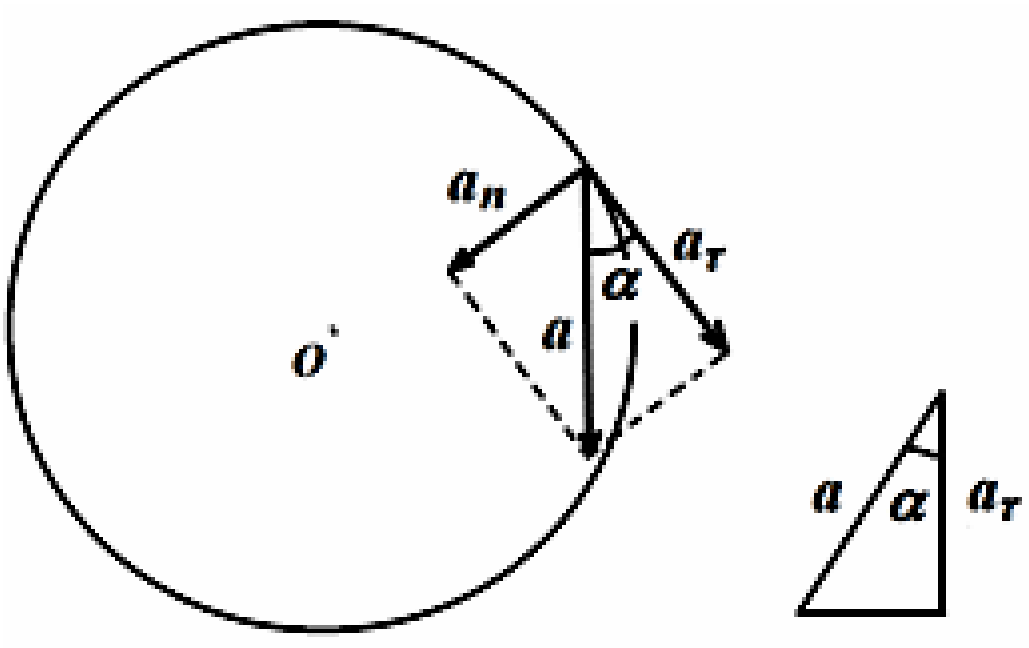

Повне прискорення:

$$a = \sqrt{a_\tau^2 + a_n^2}$$

Тангенціальне і нормальне прискорення точки тіла, що обертається виражаються формулами:

$$a_\tau = \varepsilon R, \quad a_n = \omega^2 R.$$

Кутова швидкість це $\omega = \omega_0 + \varepsilon t$, якщо на початку руху кутова швидкість $\omega_0 = 0$, то кутова швидкість на момент часу $t$ буде $\omega = \varepsilon t$ підставимо в вираз для $a_n$ і отримаємо:

$$a_n = \omega^2 R = \varepsilon^2 t^2 R = \varepsilon R(\varepsilon t^2) = a_\tau \varepsilon t^2$$

де ε – кутове прискорення тіла; $R$ – відстань точки, що рухається, від осі обертання.

Підставив замість $a_n$ в формулу для повного прискорення вираз $a_\tau \varepsilon t^2$ отримаємо:

$$a = \sqrt{a_\tau^2 + a_n^2} = \sqrt{(a_\tau)^2 + (\varepsilon^2 t^2 R)^2} = \sqrt{(a_\tau)^2 + (a_\tau)^2 \varepsilon^2 t^4} = a_\tau \sqrt{1 + \varepsilon^2 t^4}$$

$$\text{тобто} \quad a = a_\tau \sqrt{1 + \varepsilon^2 t^4}$$

Із формули $a_\tau = a\cos\alpha$ виділимо $a = \frac{a_\tau}{\cos\alpha}$ і після підстановки отримаємо:

$$\frac{a_\tau}{\cos\alpha} = a_\tau\sqrt{1+\varepsilon^2 t^4} \text{ і скоротшуємо } a_\tau$$

В результаті отримаємо:

$$\frac{1}{\cos\alpha} = \sqrt{1+\varepsilon^2 t^4}$$

Щоб позбавитися квадратного кореня піднесемо обидві частини рівняння до квадрата:

$$\frac{1}{(\cos\alpha)^2} = 1+\varepsilon^2 t^4$$

За тригонометричною формулою:

$$1+tg^2\alpha = \frac{1}{\cos^2\alpha}$$

тобто $1+tg^2\alpha = 1+\varepsilon^2 t^4 \quad \Longrightarrow \quad \varepsilon^2 t^4 = tg^2\alpha; \quad \varepsilon^2 = \frac{tg^2\alpha}{t^4};$

$$\sqrt{\varepsilon^2} = \sqrt{\frac{tg^2\alpha}{t^4}}; \qquad \varepsilon = \frac{tg\alpha}{t^2};$$

$$tg60^o = \sqrt{3} \quad \Longrightarrow \quad \varepsilon = \frac{\sqrt{3}}{2^2} = \frac{\sqrt{3}}{4}$$

Відповідь отримана в град/с$^2$, яку можна перевести в рад/с$^2$. Відомо, що повний оберт 360° це 6,28 рад., тобто 1° градусу відповідає 0,017 радіан.

**Відповідь:** через 2 с від початку руху $\varepsilon = 7{,}35 \cdot 10^{-3}$рад/с$^2$

## 1.3. Траєкторії руху у декартової системі координат.

Щоб мати змогу абосельно розглядати переміщення вводиться система відліку – сукупність системи координат та часу. Тобто положення тіла в просторі завжди визначають відносно якого-небудь іншого тіла, яке називається тілом відліку. З тілом відліку пов’язують систему координат. Як правило користуються декартовою системою координат, в якій положення точки на даний момент часу характеризується трьома координатами *x*, *y*, *z* або радіус-вектором *r* (на рис. позначено, як вектор А), проведеним із початку системи координат в дану точку.

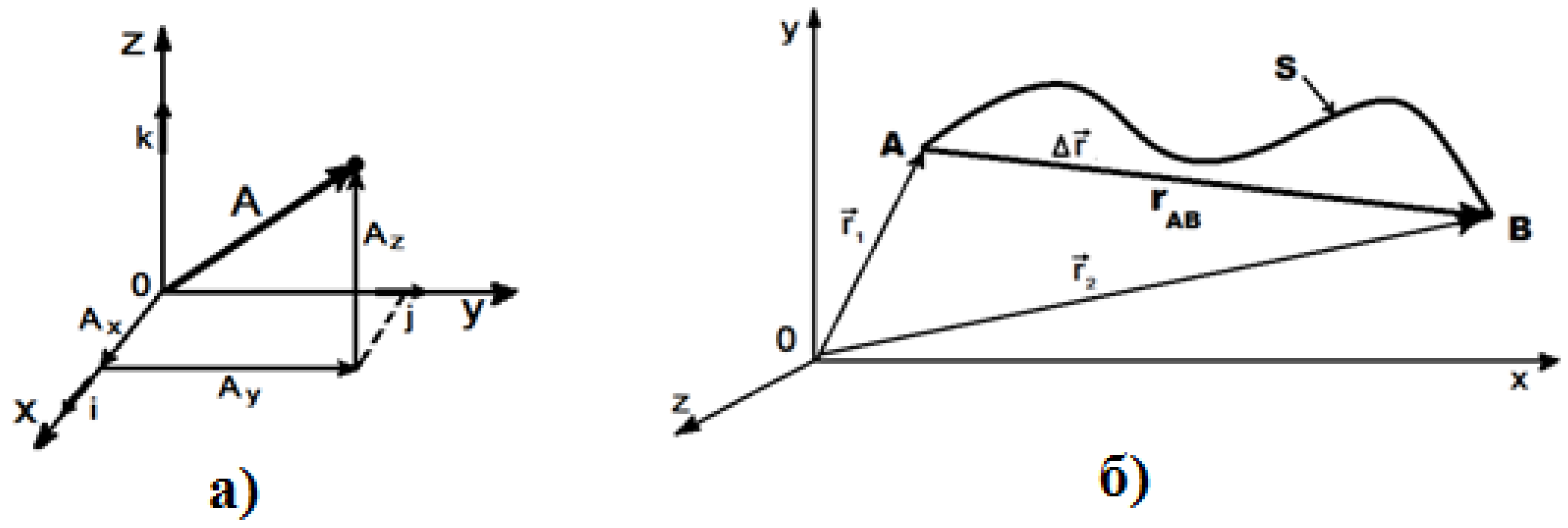

Рис. 1.3. а) декартова система координат; б) визначення зміни положення точки у просторі за допомогою радіус вектора.

Радіус-вектор задає положення матеріальної точки (центра мас твердого тіла) у просторі, а його проекції на координатні осі $A_X$, $A_Y$, $A_Z$, дорівнюють декартовим координатам точки, де $\vec{\imath}, \vec{\jmath}, \overrightarrow{k}$ - одиничні вектори напрямів (орти); *x*, *y*, *z* – координати точки, тобто:

$$\vec{r} = x\vec{\imath} + y\vec{\jmath} + z\vec{k}$$

Переміщення матеріальної точки:

$$\Delta\vec{r} = \vec{r}_2 - \vec{r}_1$$

Основна задача механіки – це визнаaботи залежність координат тіла від часу. Якщо тіло рухається рівномірно та прямолінійно зі швидкістю $\vec{v}$, за час $t$ воно переміститься із точки $M_0$ в точку M.

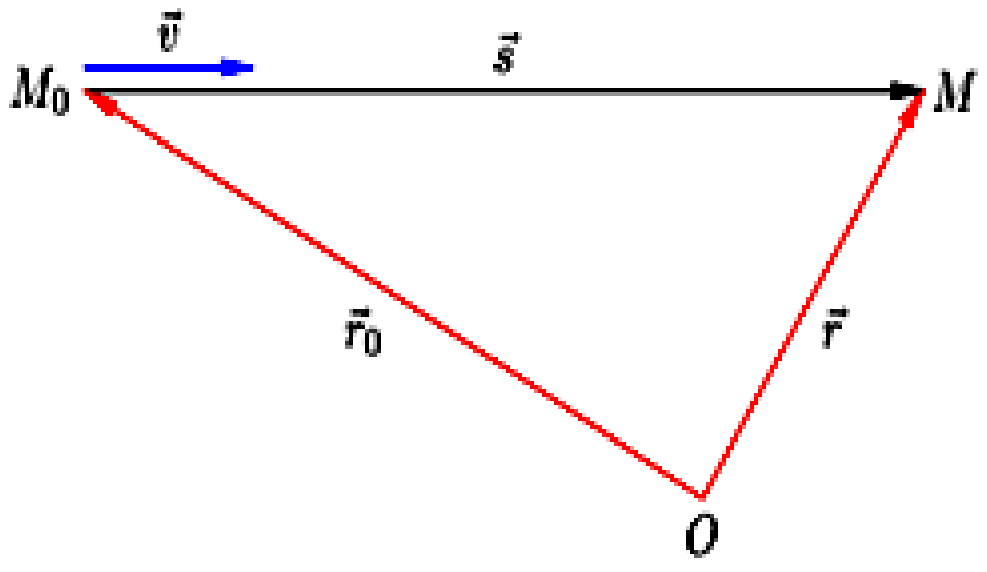

$$|\Delta\vec{r}| = \vec{r} - \vec{r}_0 \text{ тобто } \vec{s} = \vec{r} - \vec{r}_0, \qquad \text{також } \vec{s} = \vec{v}t$$

Таким абоном маємо:

$$\vec{r} - \vec{r}_0 = \vec{v}t \text{ звідсі } \qquad \vec{r} = \vec{r}_0 + \vec{v}t \quad - \text{ закон руху.}$$

Для рівноприскоренного руху:

$$\vec{r} = \vec{r}_0 + \vec{v}_0 t + \frac{\vec{a}t^2}{2}$$

При переході від цієї рівності до проекцій на координатні осі отримаємо:

$$x = x_0 + v_{0x}t + \frac{\vec{a}_x t^2}{2},$$
$$y = y_0 + v_{0y}t + \frac{\vec{a}_y t^2}{2},$$
$$z = z_0 + v_{0z}t + \frac{\vec{a}_z t^2}{2}.$$

Ці формули, представляють залежність координат тіла від часу і являються розв'язанням основної задачі механіки для рівноприскореного руху.

Маюабо на увазі, що переміщення тіла $\vec{r} - \vec{r}_0 = \vec{s}$, отримаємо:

$$\vec{s} = \vec{v}_0 t + \frac{\vec{a}t^2}{2}$$

У випадку падаючого тіла, рівняння зміни висоти з часом прийме вигляд:

$$h = h_0 + v_{0y}t + \frac{\vec{g}_y t^2}{2}.$$

**Задача 1.8.** Повітряна куля за час $t = 4$ хв піднялася на висоту $h = 800$ м, при цьому була віднесена вітром вбік на відстань $\ell = 600$ м. Визнаaботи: модуль переміщення $\Delta\vec{r}$ кулі відносно точки запуску та швидкість $v_в$ вітру, вважаюабо її постійною.

**Дано:**

$t$ =4 хв=240 с

$h$ =800 м

$\ell$ =600 м

**Знайти:**

$\Delta\vec{r}$ - ?

$v_в$ – ?

**Розв'язання.**

В цієї задачі розглядаємо окремо дві складові руху – вздовж осі О$X$ це $\ell$ =600 м і вздовж осі О$Y$ це $h$ =800 м, результуючою руху, тобто переміщенням $\Delta r$, знайдемо як квадрат гіпотенузи прямокутного трикутника з катетами $\ell$ і $h$:

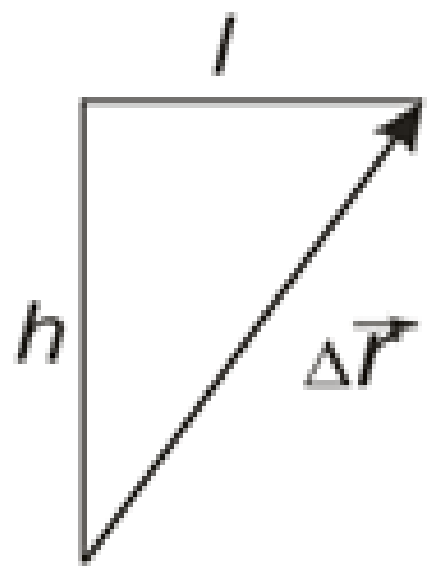

$$\Delta r = \sqrt{h^2 + \ell^2} = \sqrt{800^2 + 600^2} = \sqrt{1000000} = 1000\,\text{м}$$

Відповідаючи на друге питання – знайти швидкість вітру, звернемо увагу на те, що вітер віє вздовж осі О$X$, тому за пройдений шлях, підставимо лише $\ell$ і отримаємо:

$$v_в = \frac{l}{t} = \frac{600}{240} = 2{,}5\ \text{м/с}$$

**Відповідь:** $\Delta r = 1000$ м; $v_в = 2{,}5$ м/с.

**Задача 1.9.** Частинка, що рухається зі швидкістю $v$ = 1 м/с, перемістилася за час $t$ = 4 с із точки $M$ з координатами ($x_1$ = 2,6 м, $y_1$ = 1 м) у точку $N$ з координатами ($x_2$ = 0,6 м, $y_2$ = 2,4 м). Визнааботи модуль переміщення $\Delta r$ частинки і шлях, який вона пройшла.

**Дано:**

$x_1$ = 2,6 м, $y_1$ = 1 м

$x_2$ = 0,6 м, $y_2$ = 2,4 м

$t$ =4 с

$v$ = 1 м/с

**Знайти:**

$\Delta r$ - ?

**Розв'язання:**

Довжина вектору - це є модуль вектору і визначається за формулою:

$$|\vec{a}| = \sqrt{a_1^2 + a_2^2}$$

(за теоремою Піфагора квадрат гіпотенузи = сумі квадратів катетів),

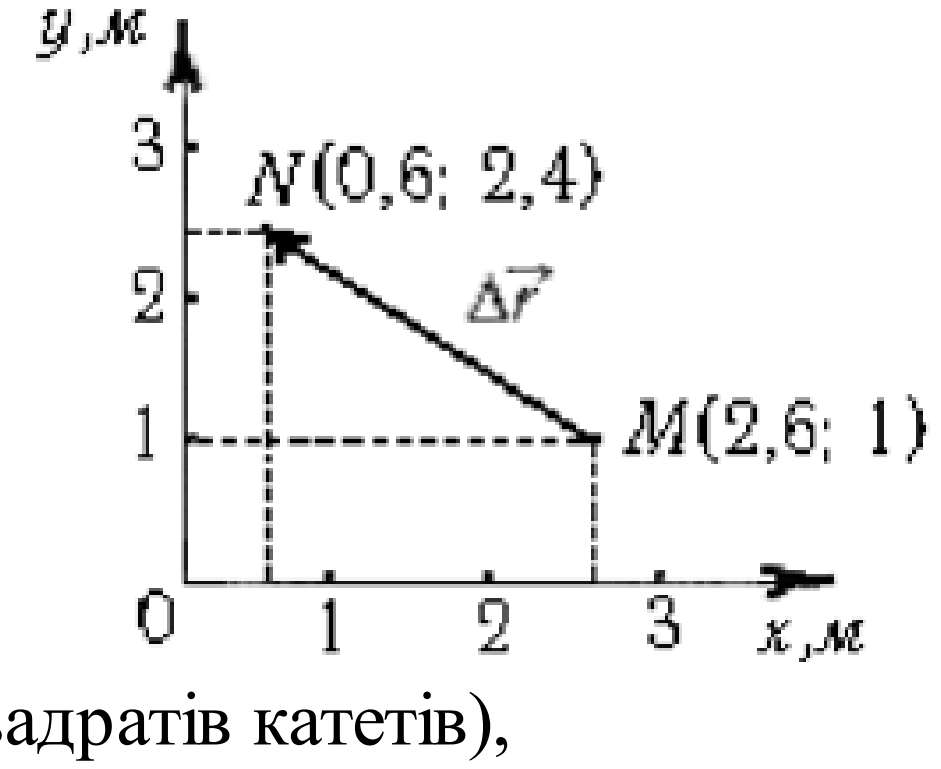

де $a_1 = x_2 - x_1$ і $a_2 = y_2 - y_1$

В нашій задачі вектор позначений як переміщення $\Delta r$ , застосуємо формули до нього і отримаємо вираз:

$$(\Delta r)^2 = (x_2 - x_1)^2 + (y_2 - y_1)^2 = (0{,}6 - 2{,}6)^2 + (2{,}4 - 1)^2$$

$$\Delta r = \sqrt{(0{,}6 - 2{,}6)^2 + (2{,}4 - 1)^2} = 2{,}44 \text{ м}$$

Щоб знайти шлях застосуємо формулу: $S = vt$

Всі необхідні дані, тобто час і швидкість, були надані в умовах задачі, тому просто підставимо абослові дані у формулу:

$$S = vt = 1 \text{ м/с} \cdot 4 \text{ с} = 4 \text{ м}$$

**Відповідь**: $\Delta r = 2{,}44$ м; $S = 4$ м.

**Задача 1.10.** Якою є лінійна швидкість точок земної поверхні на широті $\varphi$ = 60° під час добового обертання Землі? Радіус Землі прийняти рівним 6400 км.

| **Дано:** |
|---|
| $R$ = 6400 км=$6{,}4\cdot10^6$ м |
| $\varphi$ = 60° |
| Т =24 год =86400 с |
| **Знайти:** |
| $\upsilon$ - ? |

**Розв’язання.**

На рисунку точка $B$ – це точка земної поверхні, що знаходиться на широті $\varphi$ = 60°. Звичайна швидкість (лінійна) визначається за формулою:

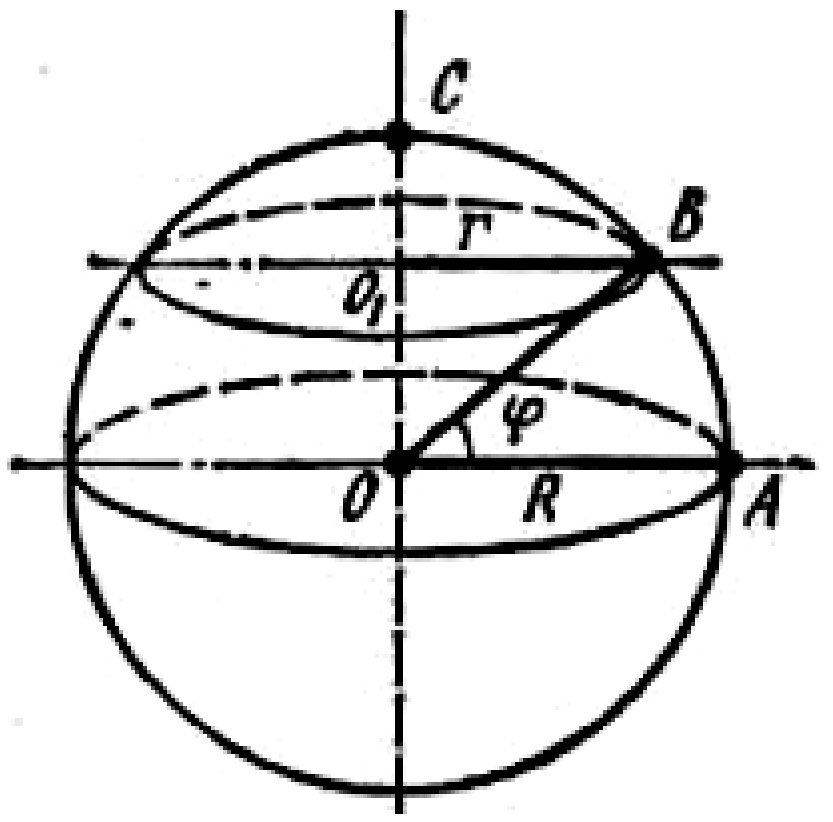

$$v = \frac{S - \text{шлях}}{t - \text{час}}$$

Але ми можемо записати, що тіло В робить повний оборот на широті $\varphi$ = 60° за час, що називається періодом, тобто шлях в цьому випадку буде довжина кола, яка визначається за формулою:

$$\ell = 2\pi r,$$

де $r$ – це радіус кола, що описує точка $B$.

Лінійну швидкість $v = \frac{S}{t}$ ми можемо знайти, підставивши замість $S$ формулу довжини кола $\ell$:

$$v = \frac{2\pi r}{\mathrm{T}}.$$

На цю формулу ми також можемо вийти якщо розглянути зв'язок між лінійною швидкістю та кутовою:

$$v = \omega \cdot r$$

Кутова швидкість це повній оберт = $2\pi = 360^{\circ}$ по колу за час $T$ (період), тобто

$$\omega = \frac{2\pi}{\mathrm{T}}$$

підставимо в формулу і отримаємо

$$v = \frac{2\pi r}{\mathrm{T}}$$

де $r$ – це радіус кола на широті $\varphi = 60^{\circ}$ , а $R$ це радіус Землі.

Розглянемо детальніше рисунок, що наданий до задачі, із прямокутного трикутника $O_1BO$ виразимо $r$, згадавши що:

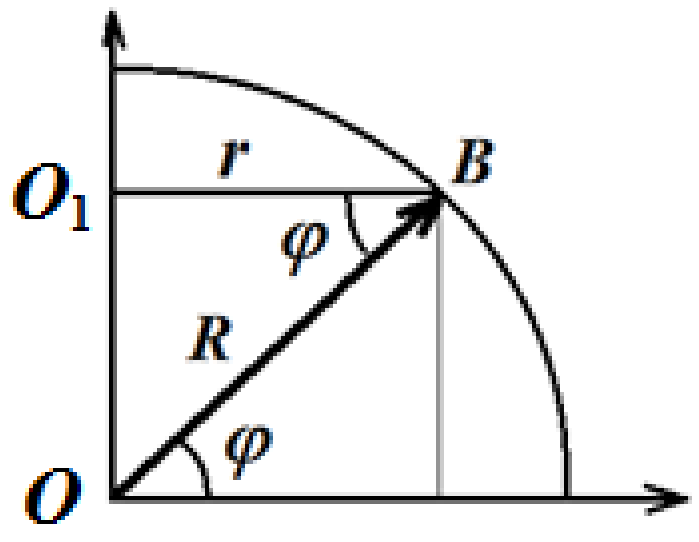

$$\cos\varphi = \frac{r}{R}$$

тобто $r = R\cos\varphi$, а $\cos 60^{\circ} = \frac{1}{2}$.

Підставимо в $v = \frac{2\pi r}{\mathrm{T}}$ отриманий вираз для $r = R\cos\varphi$ і отримаємо:

$$v = \frac{2\pi R\cos\varphi}{\mathrm{T}}$$

Після підстановки абословихданих: $\cos\varphi = \cos 60^{\circ} = \frac{1}{2} = 0{,}5$,

$R = 6{,}4 \cdot 10^6$м, $T$ = 24 год = 24· 60 хв = 24 · 60 · 60 с.= 86400 с

отримаємо:

$$v = \frac{2\pi R\cos\varphi}{\mathrm{T}} = \frac{2 \cdot 3{,}14 \cdot 6{,}4 \cdot 10^6 \cdot 0{,}5}{86400} = 232{,}5 \text{ м/с}$$

**Відповідь:** 232,5 м/с.

**Задача 1.11.** Тіло кинуто горизонтально зі швидкістю $v_0$ = 25 м/с та з висоти $h$ = 60 м. Необхідно знайти час $t$ та дальність польоту $L$.

| **Дано:** |
|---|
| $v_0$ = 25 м/с |
| $h$ = 60 м |
| **Знайти:** |
| $t$ - ? |
| $L$ - ? |

**Розв'язання.**

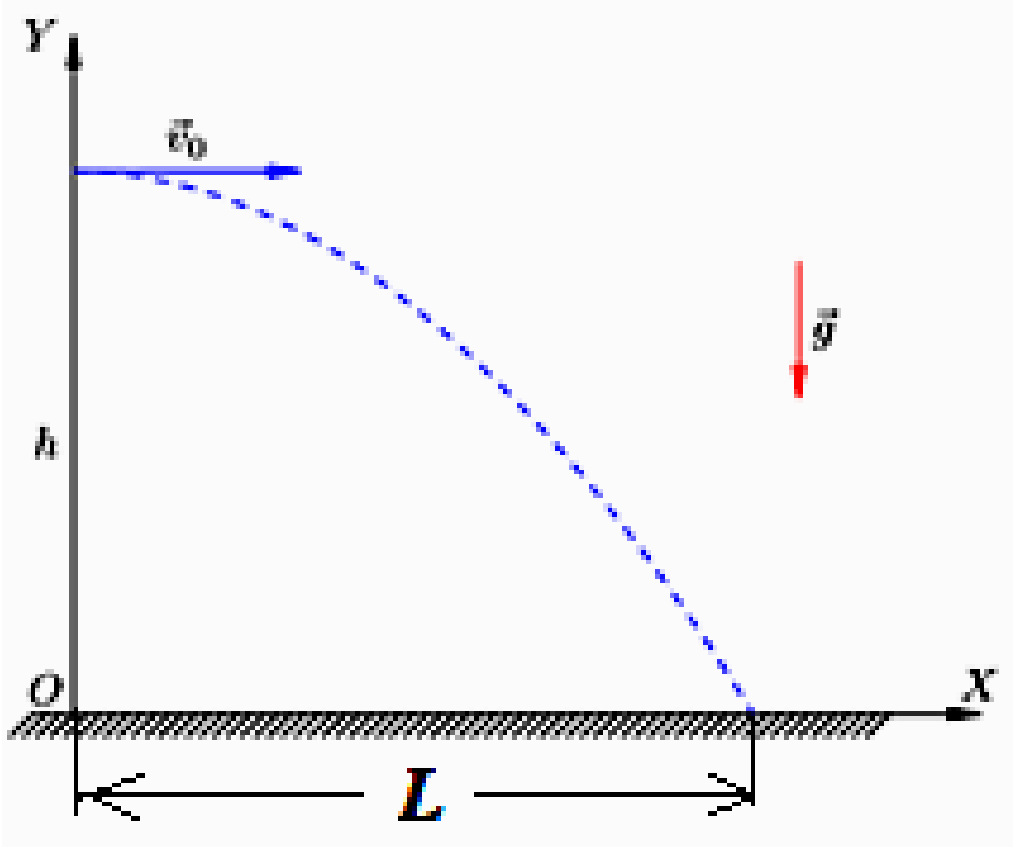

Запишемо формулу рівноприскоренного руху як проєкції на координатні осі O$X$ і O$Y$:

$$x = x_0 + v_{0x}t + \frac{a_x t^2}{2}, \qquad y = y_0 + v_{0y}t + \frac{a_y t^2}{2}.$$

Дивлячись на рисунок, де траєкторія руху зображена блакитним кольором, визначаємо значення кожного члена рівняння відповідно до умови задачі:

– для проекції на ось O$X$ маємо, що на початок руху координата по осі абсцис $x_0 = 0$, за умовою задачі тіло кинуто горизонтально зі швидкістю $v_0$ , тобто вздовж осі O$X$ і тому початкова швидкість $v_{0x} = v_0$, і маюабо на увазі визначення прискорення «це є швидкість зміни швидкості тіла» на початок руху така зміна ще відсутня, тому $a_x = 0$;

– для проекції на ось O$Y$ маємо, що на початок руху, за умовою задачі, тіло знаходиться на висоті $h$, тобто координата $y_0 = h$, проекція на ось ординат початкової швидкості, для будь якого тіла, що кинуто горизонтально, буде $v_{0y} = 0$, а проекція прискорення, на будь якій висоті, буде дорівнюватись $g = 9{,}8$ м/с$^2$ , маюабо на увазі, що напрямок прискорення протилежний напрямку осі O$Y$ запишемо $a_y = -g$ .

Підставимо у формули визначенні данні і отримаємо:

$$x = x_0 + v_{0x}t + \frac{a_x t^2}{2} = 0 + v_0 t + \frac{0 \cdot t^2}{2} = v_0 t$$

$$y = y_0 + v_{0y}t + \frac{a_y t^2}{2} = h + 0 + \frac{-gt^2}{2} = h - \frac{gt^2}{2}$$

За умовою задачі необхідно дізнатися час $t$ і дальність польоту $L$. Час польоту визначається моментом удару о поверхню, коли $h = 0$, тобто в момент падіння координата тіла $y$ перетворюється в нуль $y = 0$.

$$y = h - \frac{gt^2}{2}$$

$$h - \frac{gt^2}{2} = 0 \quad \Rightarrow \quad \frac{gt^2}{2} = h;\ t^2 = \frac{2h}{g} \quad \Longrightarrow$$

$$t = \sqrt{\frac{2h}{g}} = \sqrt{\frac{2 \cdot 60}{9{,}8}} = \sqrt{12{,}24} = 3{,}5\, c$$

Дальність польоту $L$ це значення координати $\boldsymbol{x}$ в момент часу $t$.

$$L = x(t) = v_0 t = v_0 \sqrt{\frac{2h}{g}} = 25 \cdot 3{,}5 = 87{,}5 \text{ м}$$

**Відповідь:** $t = 3{,}5\, c$; $L = 87{,}5$ м.

**Задача 1.12.** Камінь, що кинуто горизонтально з висоти $h = 5$ м, впав на відстані $S = 10$ м від місця кидання. Визнаaботи його кінцеву швидкість.

**Розв'язання.**

Запишемо координати тіла від часу, як проєкції на координатні осі O$X$ і O$Y$:

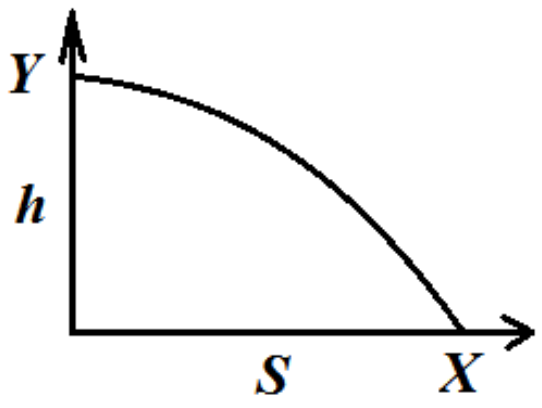

$$x = x_0 + v_{0x}t + \frac{gt^2}{2}; \quad y = y_0 + v_{0y}t + \frac{gt^2}{2}$$

Розглянемо складові швидкості за проекціями на координатні осі, тобто $v_{0y}$ – вертикальна складова або проекція на вісь O$Y$, а $v_{0x}$ – горизонтальна складова або проекція на вісь O$X$.

Визначимо всі значення відповідно до умов задачі (аналогічно як у 1.11):

$$v_{0x} = v_0; \; x_0 = 0; \quad a_x = 0; \; y_0 = h; \; v_{0y} = 0; \; a_y = -g;$$

Підставимо їх у наведені формули:

– в $x = x_0 + v_{0x}t + \frac{a_x t^2}{2}$ отримаємо шлях:

$$\text{S} = x - x_0 = v_0 t + \frac{0 \cdot t^2}{2} = v_0 t,$$

тобто

$$\text{S} = v_0 t,$$

– в $y = y_0 + v_{0y}t - \frac{gt^2}{2}$ врахуємо, що в момент падіння: $h = 0$ тобто $y = 0$, $y_0 = h$ і $v_{0y} = 0$, після підстановки отримаємо:

$$0 = h + 0 \cdot t - \frac{gt^2}{2}; \qquad h = \frac{gt^2}{2}$$

із отриманого рівняння виразимо час

$$t = \sqrt{\frac{2h}{g}}$$

Кінцеву швидкість знайдемо через складові швидкості:

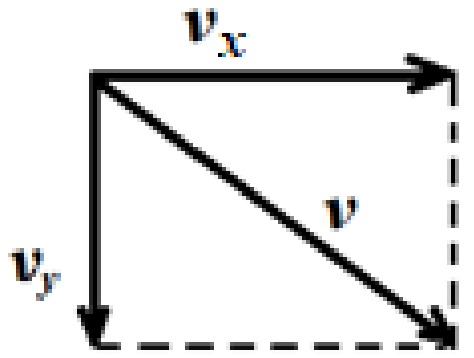

запишемо скалярні рівняння для складових швидкості:

$$v_x = v_{0x} + a_x t\,,$$

де $v_{0x} = v_0$, а $a_x = 0 \;\Rightarrow\; v_x = v_0$

$$v_y = v_{0y} + a_y t,$$

де $v_{0y} = 0,\ a_y = -g \ \Longrightarrow\ v_y = a_y t$

Кінцеву швидкість визначимо за формулою

$$v^2 = v_x^2 + v_y^2$$

або

$$v = \sqrt{v_0^2 + v_y^2}.$$

Із формули, яку отримали раніше $S = v_0 t$ виразимо $v_0$:

$$v_0 = \frac{S}{t}$$

і підставимо

$$t = \sqrt{\frac{2h}{g}}$$

тоді

$$v_0 = \frac{S}{\sqrt{\frac{2h}{g}}}$$

Обидві частини отриманого рівняння підведемо до квадрата

$$v_0^2 = \frac{S^2}{\frac{2h}{g}} = \frac{gS^2}{2h}$$

це є горизонтальна складові швидкості.

Вертикальна складова швидкості

$$v_y = a_y t = gt = g\sqrt{\frac{2h}{g}},$$

тоді

$$v_y^2 = g^2\frac{2h}{g} = 2gh$$

Тепер маємо визнааботи кінцеву швидкість за формулою:

$$v = \sqrt{v_0^2 + v_y^2} = \sqrt{\frac{gS^2}{2h} + 2gh} = 14\ \text{м/с}$$

**Відповідь:** $v = 14$ м/с.

**Задачі для самостійного розв'язання**

1. Пароплав йде по річці від пункту $A$ до пункту $B$ зі швидкістю $\upsilon_1$=10 км/год, а назад зі швидкістю $\upsilon_2$=16 км/год. Швидкість пароплава $\upsilon$ відносно берега є постійною. Знайти середню швидкість $\bar{\upsilon}$ пароплава та швидкість $u$ течії ріки. [0,8 м/с$^2$]

2. Тіло падає з висоти $h$=19,6 м з початковою швидкістю $\upsilon_0$=0. Який шлях пройде тіло за першу та за останню 0,1 с свого руху. [0,049*м*, 1,9*м* ]

3. Камінь, кинутий горизонтально, через час $t$=0,5 с після початку руху мав швидкість $\upsilon$, в 1,5 разів більше швидкості $\upsilon_x$ в момент кидання. З якою швидкістю $\upsilon_x$ кинуто камінь?[$\upsilon_x$=4,47 м/с].

4. Точка рухається по колу радіусом $R$=2 см. Залежність шляху від часу дається рівнянням $s=Ct^3$, де $C$=0,1 см/с$^3$. Знайти нормальне $a_n$ і тангенціальне $a_\tau$ прискорення точки в момент, коли лінійна швидкість точки $\upsilon$=0,3 м/с. [$a_n$=4,5 м/с$^2$, $a_\tau$=0,06 м/с$^2$].

5. Залежність пройденого тілом шляху від часу задається рівнянням $s = A + Ct^2 + Dt^3$ ($C$=0,1м/с$^2$, $D$=0,03м/с$^3$). Визначити час після початку руху, за який прискорення $a$ тіла дорівнюватиме 2 м/с$^2$. [10 с]

## 1.4. Основні закони динаміки.

**Динаміка** – це розділ механіки, який вивчає рух тіла або системи тіл у взаємозв'язку з приабонами виникнення руху. До таких приабон відносять дію одного тіла або системи тіл на інші тіла, а мірою взаємодії тіл є сила $\vec{F}$.

Вимірювання сили здійснюється за зміною швидкості або деформації. Дію декількох сил, прикладених у деякій точці тіла, можна замінити однією силою, яку називають рівнодійною; вона дорівнює:

$$\vec{F} = \vec{F}_1 + \vec{F}_2 + \cdots + \vec{F}_n = \sum_{i=1}^{i=n} \vec{F}_i$$

Зокрема, коли дві сили діють на одній прямій, то рівнодійна буде спрямована щодо напрямку більшої з них. За величиною рівнодійна буде дорівнюватиме $F=F_1+F_2$, коли напрямки сил співпадають і $F = |F_1 - F_2|$, коли сили мають протилежні напрямки. Якщо дві сили прикладені в одній точці тіла $O$, а кут між ними $\alpha$, то рівнодійною буде сила, спрямована по діагоналі паралелограма побудованого на них, і за велиабоною може бути обабослена таким абоном:

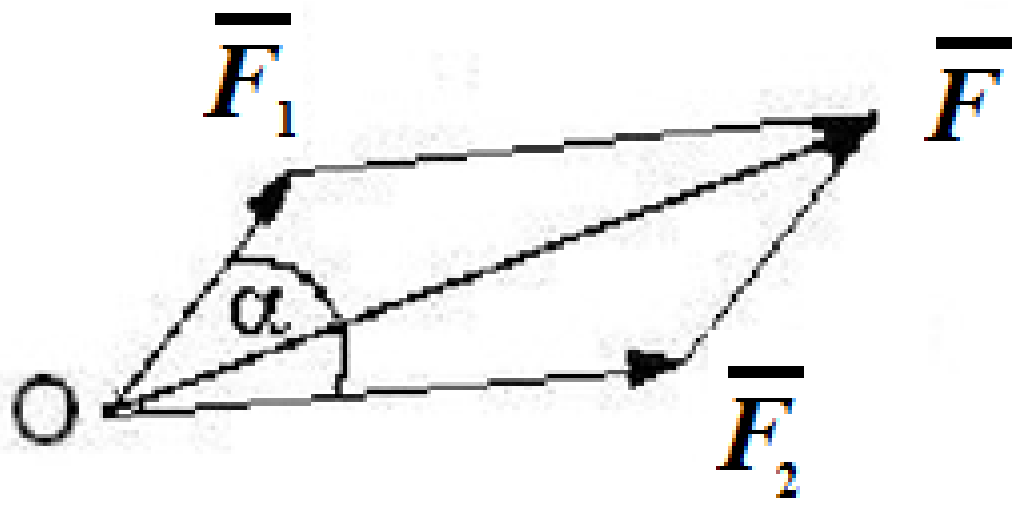

$$|\vec{F}| = \sqrt{\left(\vec{F}_1 + \vec{F}_2\right)^2} = \sqrt{F_1^2 + F_2^2 + 2(F_1F_2)} = \sqrt{F_1^2 + F_2^2 + 2(F_1F_2)\cos\alpha}$$

**2 закон Ньютона**:

$$\vec{a} = \frac{\vec{F}}{m}$$

Чим більша сума прикладених до тіла зовнішніх сил, тим більше тим більшого прискорення набуває тіло. Одночасно, абом масивніше тіло, до якого прикладена сума зовнішніх сил, тим меншого прискорення воно набуває.

$$F = ma$$

де $F$ – сила, $m$ – маса, $a$ – прискорення.

**Густина однорідного тіла** називається відношення маси тіла до його об'єму:

$$\rho = \frac{m}{V}$$

Густина не залежить від геометричних властивостей тіла (форми, об'єму) і є характеристикою речовини тіла.

**3 закон Ньютона**: сили, з якими взаємодіють тіла, рівні за велиабоною і спрямовані протилежно.

$$\vec{\boldsymbol{F}}_{\mathbf{1,2}} = -\vec{\boldsymbol{F}}_{\mathbf{2,1}}$$

**Сила тертя ковзання**

$$F = \mu N$$

де $\mu$ - коефіцієнт тертя ковзання, $N$ - нормальна складова сили реакції опори. Усі величини тут скалярні, бо вектори $\vec{F}$ і $\vec{N}$ взаємно перпендикулярні.

**Сила пружності**

$$F = -kx$$

де $k$ - коефіцієнт пружності (жорсткість пружини), $x$ - абсолютна деформація

**Сила гравітаційної взаємодії** двох точкових мас $m$ і $m_0$ визначається законом всесвітнього тяжіння між тілами:

$$F = G\frac{m \cdot m_0}{r^2},$$

де $G$ = 6.67 $10^{-11}$Н $м^2/кг^2$ – гравітаційна стала; $r$ – відстань між тілами.

**Сила тяжіння Землі** визначається гравітаційною взаємодією Землі та тіла, що знаходиться на поверхні Землі. За формулою взаємодії двох тіл маємо:

$$F = G\frac{M_з \cdot m}{(R_з + h)^2}$$

де $h$ – висота тіла над поверхнею Землі; $R_З$ = 6400 км – радіус земної сфери (радіус Землі); $M_З = 5.96 \cdot 10^{24}$ кг – маса Землі.

Якщо в цієї формулі замість $F$ підставити $mg$ ($F = mg$ за 2 законом Ньютона), то отримаємо, що:

$$g = G\frac{M_з}{(R_з + h)^2}$$

де $g$ – прискорення вільного падіння тіла в полі тяжіння Землі. На поверхні Землі ($h$ = 0) прискорення сили тяжіння:

$$g = G\frac{M_з}{(R_з)^2} \text{ має значення } g = 9{,}8 \text{ м/с}^2.$$

**Задача 1.13.** Автомобіль масою $m = 1{,}5$ т рухаюабось прямолінійно та рівноприскорено, за $t = 10$ с збільшив швидкість свого руху від 22 до 58 км/ч. Визначте модуль $F$ рівнодійної сил, які впливали на автомобіль.

**Дано:**
$m$ = 1,5 т
$t$ =10 с
$\upsilon_1$ =22 км/ч
$\upsilon_2$=58 км/ч

**Знайти:**
$F$ - ?

**Розв'язання.**

Рівнодійна сила визначається за другим законом Ньютона

$$F = ma$$

Маса $m$ надана в умовах задачі, треба шукати прискорення $a$.

За час $t$=10 с автомобіль збільшив швидкість на $\Delta v =$ 36 км/ч. Звернемо увагу, що відповідь треба надати в Ньютонах, а сила в 1 Н це 1Н=$\frac{\text{м·кг}}{\text{с}^2}$, тому спочатку треба перевести км в метри, час у секунди, а тони у кг і лише потім визначати прискорення за формулою

$$a = \frac{\Delta v}{t}$$

Тобто маємо:

$$36\frac{\text{км}}{\text{ч}} = \frac{36 \cdot 1000 \text{ м}}{60 \text{ мин.} \cdot 60 \text{ с.}} = \frac{36000}{3600} = 10\frac{\text{м}}{\text{с}}; \quad 1 \text{ т} = 1000 \text{ кг} \Rightarrow 1{,}5 \text{ т} = 1500 \text{ кг}$$

тому $a = \dfrac{10 \text{ м/с}}{10 \text{ с}} = 1 \text{ м/с}^2 \quad \Longrightarrow \quad F = ma = 1500 \text{ кг} \cdot 1 \text{ м/с}^2 = 1500 \text{ Н}$

**Відповідь:** 1500 Н.

**Задача 1.14.** Визнааботь модуль прискорення тіла $a$ масою $m = 4$ кг, на яке діють сили вказані на рис. Модулі даних сил: $F_1 = 2{,}8$ Н; $F_2 = F_3 = 4{,}8$ Н.

| **Дано:** | **Розв'язання.** |
|---|---|
| $m = 4$ кг<br>$F_1 = 2{,}8$ Н;<br>$F_2 = F_3 = 4{,}8$ Н. | Згідно з другим законом Ньютона прискорення тіла |
| **Знайти:**<br>$a$ - ? | |

$$\vec{a} = \frac{\vec{F}}{m},$$

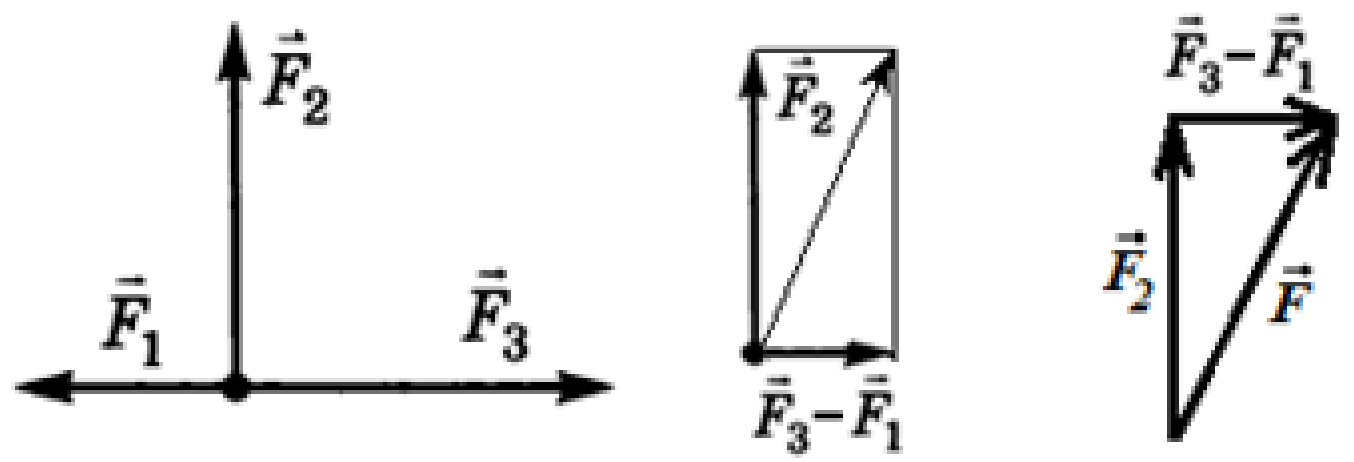

де

$$\vec{F} = \vec{F}_1 + \vec{F}_2 + \vec{F}_3.$$

рівнодійна, пракладених сил.

Виразимо модуль прискорення тіла через модуль рівнодійної сили

$$|a| = \frac{|F|}{m}$$

Зручніше спочатку визнааботи рівнодійну двох сил на однієї лінії, але протилежно спрямованих:

$$F_{1,3} = F_3 - F_1 = 4{,}8 - 2{,}8 = 2{,}0\ H.$$

Рівнодійну двох сил $F_{1,3}$ та $F_2$ знайдемо за теоремою Піфагора – квадрат гіпотенузи дорівнюється сумі квадратів катетів. Тому отримаємо:

$$F = \sqrt{F_{1,3}^2 + F_2^2} = \sqrt{2^2 + 4{,}8^2} = \sqrt{4 + 23{,}04} = \sqrt{27{,}04} = 5{,}2\ \text{Н}$$

$$a = \frac{F}{m} = \frac{5{,}2}{4} = 1{,}3\ \text{м/с}^2$$

**Відповідь:** $a = 1{,}3\ \text{м/с}^2$ .

**Задача 1.15.** Порівняйте прискорення вільного падіння тіл на Землі ($g_З$) і на Марсі ($g_М$). Радіус Марса ($R_М$) становить $n = 0{,}53$ радіуса Землі ($R_З$), маса Марса ($М_М$) становить $k = 0{,}107$ від маси Землі ($М_З$).

| **Дано:** | **Розв'язання.** |
|---|---|
| $M_М = k \cdot M_З$ | |
| $R_М = n \cdot R_З$ | |
| **Знайти:** | |
| $g_М$–? | |

Сила тяжіння визначається за формулою:

$$F = G\frac{M_з \cdot m}{(R_з + h)^2}$$

Сила гравітаційної взаємодії двох тіл визначається за формулою:

$$F = G\frac{m \cdot m_0}{r^2},$$

яка для взаємодії планети та тіла в полі тяжіння цієї планети приймає вигляд:

$$F = G\frac{M_з \cdot m}{(R_з + h)^2}$$

Одночасно ми маємо записати, що на тіло діє сила тяжіння $F = m \cdot g$. Прирівняємо їх і при $h$=0 отримаємо:

$$mg = G\frac{M_з \cdot m}{(R_з)^2} \quad \text{або} \quad g = G\frac{M_з}{(R_з)^2}$$

За цією формулою запишемо прискорення вільного падіння для Землі і для Марса:

$$\text{для Землі } g_З = G\frac{M_З}{(R_З)^2}, \qquad \text{для Марса } g_М = G\frac{M_М}{(R_М)^2}$$

Знайдемо відношення $g_М/g_З$:

$$\frac{g_М}{g_З} = \frac{G\frac{M_М}{(R_М)^2}}{G\frac{M_З}{(R_З)^2}} = \frac{M_М}{(R_М)^2} \cdot \frac{(R_З)^2}{M_З}$$

За умовою задачі $R_М = n \cdot R_З$ і $M_М = k \cdot M_З$, підставимо в отриманий вираз.

$$\frac{k \cdot M_З}{(n \cdot R_З)^2} \cdot \frac{(R_З)^2}{M_З} = \frac{k}{n^2} = \frac{0{,}107}{0{,}53^2} = 0{,}38$$

$$\frac{g_М}{g_З} = 0{,}38 \quad \text{або} \quad g_М = 0{,}38 \cdot g_З \quad \text{де} \quad g_З = 9{,}8\frac{м}{с^2}$$

$$g_М = 0{,}38 \cdot 9{,}8 = 3{,}72 \text{ м/с}^2$$

**Відповідь:** $g_М = 3{,}72\ м/с^2$ або в 0,38 разів менше ніж на Землі.

**Задача 1.16.**

Маюабо на увазі, що маса Землі невідома, визначте висоту $h$, на якій прискорення вільного падіння $g_1$, буде в $n = 3$ разів менше, ніж прискорення вільного падіння $g$ у поверхні Землі. Радіус Землі $R_З = 6{,}37 \cdot 10^6$ м.

**Дано:**

$$g_1 = \frac{1}{n} g$$

$$n = 3$$

$$R_З = 6{,}37 \cdot 10^6\ м$$

**Знайти:**

$h$–?

**Розв'язання.**

Сила тяжіння на висоті $h$:

$$F_т = m \cdot g_1$$

Сила гравітаційної взаємодії:

$$F = G\frac{M_з \cdot m}{(R_з + h)^2}$$

Запишемо сили вздаємодії, що діють на висоті $h$, це сила тяжіння $F_т = m \cdot g_1$ та сила гравітаційної взаємодії $F = G\frac{M_з \cdot m}{(R_з + h)^2}$. Прирівняємо їх і отримаємо:

$$m \cdot g_1 = G\frac{M_з \cdot m}{(R_з + h)^2}, \qquad g_1 = G\frac{M_з}{(R_з + h)^2}$$

Підставимо $g_1 = \frac{1}{n} g$ і отпимаємо:

$$\frac{1}{n} g = G\frac{M_з}{(R_з + h)^2}$$

Позбавимось квадрата у знаменнику:

$$\sqrt{\frac{g}{n}} = \sqrt{\frac{GM_з}{(R_з + h)^2}} \quad \text{або} \quad \sqrt{\frac{g}{n}} = \frac{\sqrt{GM_з}}{R_з + h}$$

Ми знаємо, що $g = G\frac{M_з}{(R_з)^2}$, підставимо це в отримане рівняння:

$$\sqrt{\frac{G\frac{M_з}{(R_з)^2}}{n}} = \frac{\sqrt{GM_з}}{R_з + h}; \qquad \sqrt{G\frac{M_з}{(R_з)^2}} = \frac{\sqrt{nGM_з}}{R_з + h}; \qquad \frac{1}{R_з} = \frac{\sqrt{n}}{R_з + h}$$

$$\sqrt{n}\,R_з = R_з + h\,;$$

$$h = R_з\left(\sqrt{n} - 1\right) = 6{,}37 \cdot 10^6\left(\sqrt{3} - 1\right) = 4{,}65 \cdot 10^6 \text{ м}$$

**Відповідь:** $h = 4{,}65 \cdot 10^6$ м.

**Задача 1.17.** Яку швидкість повинен мати штучний супутник, щоб обертатися по кругової орбіті на висоті 600 км над поверхнею Землі? Який період його обертання? Радіус Землі вважати рівним 6400 км, $M_З = 6 \cdot 10^{24}$ кг. Обертання Землі можна не враховувати.

**Дано:**

$R$ = 6400 км=$6{,}4 \cdot 10^6$ м

$h$ = 600 км=$6 \cdot 10^5$ м

$M_З$ =$6 \cdot 10^{24}$ кг

**Знайти:**

$v$ - ?

$T$ - ?

**Розв'язання.**

При рівномірному русі по колу вектор прискорення $a_д$ завжди є перпендикулярний вектору швидкості, та має напрям по радіусу в центр кола.

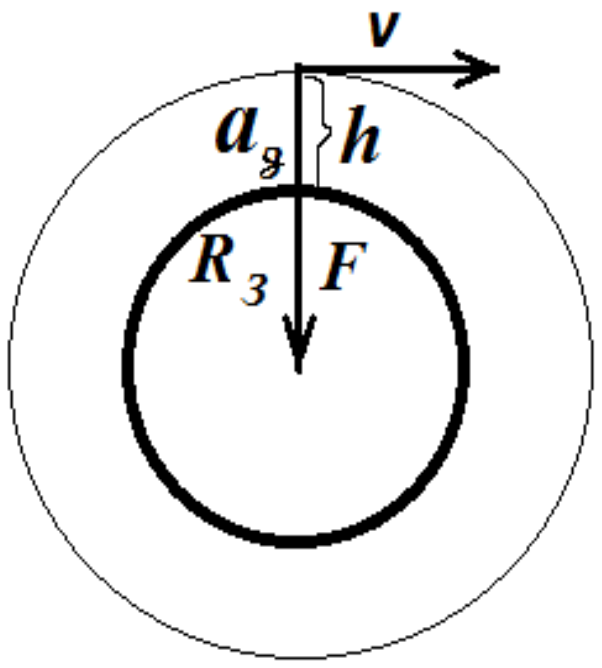

$$a_д = \frac{v^2}{R_{кола}}\,; \qquad R_{кола} = R_З + h;$$

По закону всесвітнього тяжіння

$$F_т = G\frac{M_З \cdot m}{(R_З + h)^2};$$

також $F_т = ma_д$ прирівняємо праві частини рівнянь:

$$ma_д = G\frac{M_З \cdot m}{(R_З + h)^2}; \qquad a_д = G\frac{M_З}{(R_З + h)^2}$$

Замість $a_д$ підставимо

$$a_д = \frac{v^2}{R_{кола}}$$

$$\frac{v^2}{R_З + h} = G\frac{M_З}{(R_З + h)^2}; \quad v^2 = G\frac{M_З}{R_З + h}; \qquad v = \sqrt{G\frac{M_З}{R_З + h}}$$

Пам'ятаємо, що $G = 6{,}67 \cdot 10^{-11}\ \frac{\text{н}\cdot\text{м}^2}{\text{кг}^2}$, а $1\ \text{Н} = 1\ \frac{\text{кг}\cdot\text{м}}{\text{с}^2}$, тому

$$G = 6{,}67 \cdot 10^{-11}\ \frac{\text{м}^3}{\text{кг}\cdot\text{с}^2}$$

Підставимо значення, що надані в умові задачі і отримаємо:

$$v = \sqrt{6{,}67 \cdot 10^{-11} \frac{\text{м}^3}{\text{кг}\cdot\text{с}^2} \cdot \frac{6 \cdot 10^{24}\text{кг}}{64 \cdot 10^5\text{м} + 6 \cdot 10^5\text{м}}} = \sqrt{0{,}572 \cdot 10^8} = 7{,}56\ \text{км/с}$$

Період обертання знайдемо за формулою:

$$\text{T} = \frac{2\pi R}{v},$$

де $R = R_3 + h$

$$\text{T} = \frac{2\pi(R_3 + h)}{v} = 5709\ \text{с} = 95\ \text{мин} = 1\ \text{ч}\ 35\ \text{мин}$$

**Відповідь:** $v = 7{,}7$ км/с; $\text{T} = 1$ ч 35 мин

**Задача 1.18.** Визначте масу Місяця, якщо відомо, що його штучний супутник обертається майже по коловій орбіті, радіус якої 1890 км, і має період обертання 2 год 3 хв 30 с.

**Дано:**

$R$ = 1890 км=1890·$10^3$ м

$T$ =2 год 3 хв 30 с=7410 с

$G = 6{,}67 \cdot 10^{-11} \frac{\text{Н}\cdot\text{м}^2}{\text{кг}^2}$

**Знайти:**

$M$ - ?

**Розв'язання.**

За 2 законом Ньютона сила всесвітнього тяжіння надає штучному супутнику Місяця доцентрове прискорення, яке за законом Ньютона:

$$\text{F} = m \cdot a, \quad \Rightarrow \quad a = \frac{\text{F}}{m}$$

також $F = G\frac{\text{M}\cdot m}{R^2}$ сила всесвітнього тяжіння і, якщо підставити його у попереднє рівняння, то отримаємо:

$$a = \frac{\text{F}}{m} = \frac{G\frac{\text{M}\cdot m}{R^2}}{m} = G\frac{\text{M}}{R^2}$$

Також доцентрове прискорення можна визнаaботи за формулою:

$$a_{д} = \frac{v^2}{R},$$

де $v$ - лінійна швидкість руху супутника, $R$ - радіус отбіти супутника. Прирівняємо праві сторони рівнянь для доцентрового прискорення:

$$\frac{v^2}{R} = G\frac{\text{М}}{R^2}$$

із отриманого рівняння виразимо

$$\text{М} = \frac{v^2 R^2}{GR},$$

тобто

$$\text{М} = \frac{v^2 R}{G}$$

Щоб визнааботи за цією формулою масу спочатку треба визнааботи лінійну швидкість руху супутника. Відомо, що тіло робить повний оборот за час, який має назву період, а шлях в цьому випадку буде довжина кола, яка визначається за формулою: $\ell = 2\pi R$, де $R -$ це радіус кола, що описує тіло. Лінійну швидкість $v = \frac{S}{t}$ ми можемо знайти, підставивши замість $S$ формулу довжини кола $\ell$:

$$v = \frac{2\pi R}{\text{T}}$$

Підставивши це рівняння в формулу для знаходження маси Місяця маємо:

$$M = \frac{v^2 R}{G} = \frac{\left(\frac{2\pi R}{\text{T}}\right)^2 \cdot R}{G} = \frac{4\pi^2 R^3}{GT^2}$$

$$M = \frac{4 \cdot 3{,}14^2 (1890 \cdot 10^3)^3 \text{м}^3}{6{,}67 \cdot 10^{-11} \frac{\text{Н} \cdot \text{м}^2}{\text{кг}^2} \cdot 7410^2 \text{с}^2} \approx 7{,}35 \cdot 10^{22} \text{кг}.$$

**Відповідь:** $M \approx 7{,}35 \cdot 10^{22}$ кг.

**Задача 1.19.** Супутник рухається навколо деякої планети по коловій орбіті радіуса $R = 4{,}7 \cdot 10^9$м зі швидкістю $v = 10$ км/с. Радіус планети $R_1 = 1{,}5 \cdot 10^8$м. Встановити середню густину $\rho$ речовини планети.

| Дано: | Розв'язання |
|---|---|
| $R = 4{,}7 \cdot 10^9$м | Сила тяжіння, яка діє на супутник: |
| $v = 10$ км/с $= 10^4$м/с | $F_т = ma_д$, |
| $R_1 = 1{,}5 \cdot 10^8$м | де $a_Д$ - доцентрове прискорення можна знайти за |
| **Знайти:** | формулою: |
| $\rho$ - ? | $$a_д = \frac{v^2}{R}.$$ |

Сила гравітаційної взаємодії

$$F = G\frac{mM}{R^2}$$

і сила тяжіння $F_т = ma_д$ врівноважені, тому маємо:

$$F_т = G\frac{М \cdot m}{R^2} \quad \text{тобто} \quad m\frac{v^2}{R} = G\frac{М \cdot m}{R^2} \quad \text{або} \quad \frac{v^2}{R} = G\frac{М}{R^2}$$

Треба знайти густину $\rho$ речовини планети. Густина визначається за формулою:

$$\rho = \frac{M}{V}, \qquad \Longrightarrow \; M = \rho V, \qquad \text{а об'єм кулі це} \;\; V_к = \frac{4}{3}\pi R^3.$$

Підставимо в формулу для маси об'єм кулі:

$$M = \rho \cdot \frac{4}{3}\pi R_1^3$$

і отриманий вираз

$$\frac{v^2}{R} = G\frac{М}{R^2}$$

перепишемо в наступному вигляді:

$$M = \frac{v^2 \cdot R}{G}$$

з врахуванням попередньої формули

$$\frac{v^2 \cdot R}{G} = \rho \cdot \frac{4}{3}\pi R_1^3$$

Із отриманого рівняння виразимо густину:

$$\rho = \frac{3v^2 \cdot R}{4\pi G R_1^3} = \frac{3 \cdot (10^4)^2 \cdot 4{,}7 \cdot 10^9}{4 \cdot 3{,}14 \cdot 6{,}673 \cdot 10^{-11} \cdot (1{,}5 \cdot 10^8)^3} = \frac{14{,}1 \cdot 10^8 \cdot 10^9}{83{,}8 \cdot 3{,}375 \cdot 10^{13}} =$$

$$= \frac{14{,}1 \cdot 10^4}{282{,}8} = 0{,}05 \cdot 10^4 = 500 \text{ кг/м}^3$$

**Відповідь:** $\rho = 500 \text{ кг/м}^3$.

**Задача 1.20.** Визнааботи прискорення вільного падіння $g_з$ на поверхні нейтронної зірки, якщо її радіус $R$=10 км, а середня густина речовини $\rho = 5 \cdot 10^{17}\text{кг/м}^3$.

| Дано: | Розв’язання. |
|---|---|
| $R$ =10 км | Аналогічно тому, як знаходили прискорення вільного падіння тіла в полі тяжіння Землі використаємо тіж формули для знаходження прискорення вільного падіння на поверхні нейтронної зірки $g_з$: |
| $\rho = 5 \cdot 10^{17}\text{кг/м}^3$ | Закон тяжіння |
| $G = 6{,}673 \cdot 10^{-11}\ \frac{\text{Н} \cdot \text{м}^2}{\text{кг}^2}$ | |
| **Знайти:** | |
| $g_з$ - ? | |

$$F_т = mg$$

та сила гравітаційної взаємодії

$$F = G\frac{mM}{R^2},$$

$$mg = G\frac{mM}{R^2}; \Longrightarrow \quad g = G\frac{M}{R^2};$$

Запишемо формулу для визначення густини речовини

$$\rho = \frac{M}{V}, \quad \text{де об}'\text{єм } V$$

де $V$ об’єм

Виражається як:

$$V = \frac{4}{3}\pi R^3,$$

тобто підставивши її маємо, що

$$M = \rho \cdot V = \frac{4}{3}\pi\rho R^3$$

тоді

$$g_з = \frac{G\frac{4}{3}\pi\rho R^3}{R^2} = \frac{4}{3}\pi G\rho R = \frac{4 \cdot 3{,}14 \cdot 6{,}673 \cdot 10^{-11} \cdot 5 \cdot 10^{17} \cdot 10^4}{3} =$$

$$= 140 \cdot 10^{10} = 1{,}4 \cdot 10^{12} \text{ м/с}^2$$

**Відповідь:** $g_з = 1{,}4 \cdot 10^{12}$ м/с$^2$.

**Задача 1.21.** Що сильніше притягує Місяць Сонце або Земля? Яке прискорення Місяцю надається цим надлишком сили?

**Дано:**

$M_{\odot} = 1{,}989 \cdot 10^{30}$ кг

$M_З = 5{,}97 \cdot 10^{24}$ кг

$M_Л = 7{,}35 \cdot 10^{22}$ кг

$G = 6{,}673 \cdot 10^{-11} \ \frac{Н \cdot м^2}{кг^2}$

$R_{ЗЛ} = 3{,}84 \cdot 10^8$м

$R_{СЗ} = 1{,}496 \cdot 10^{11}$м

**Знайти:**

$\frac{F_1}{F_2}$ – ?

Розв’язання.

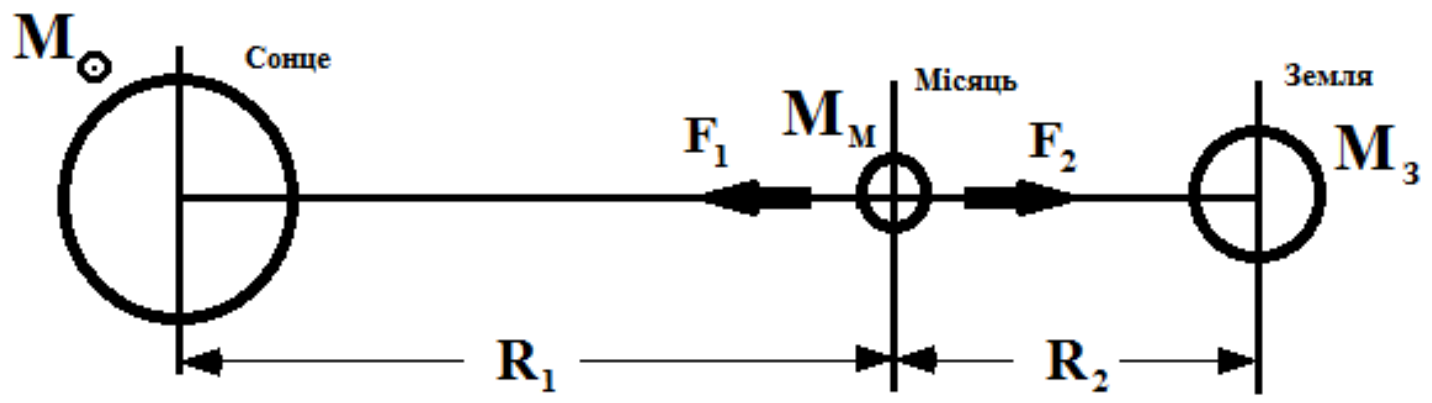

Приймаємо до уваги, що відстань між Місяцем та Землею набагато менша в порівнянні з відстанню між Землею та Сонцем і той факт, що відстань між Сонцем та Місяцем може змінюватись за рахунок обертання навколо Землі, тобто Місяц періодично займає положення між Сонцем та Землею – мінімальна відстань і, коли між Сонцем і Місяцем знаходиться Земля – максимальна відстань. Тому замість відстані між Сонцем та Місяцем можна взяти відстань між Сонцем та Землею.

Запишемо для сил, закон всесвітнього тяжіння:

$$F = G\frac{M_1 M_2}{R^2}$$

та другий закон Ньютона:

$$F = ma$$

що діють на Місяц закон всесвітнього тяжіння:

зі сторони Сонця

$$F_1 = G\frac{M_{\odot} M_{\text{М}}}{R_1^2},$$

замість $R_1$ підставимо $R_{\text{СЗ}}$, зі сторони Землі

$$F_2 = G\frac{M_{З} M_{\text{М}}}{R_2^2}$$

Щоб відповісти на питання, що сильніше притягує Місяць Сонце або Земля знайдемо відношення сил $F_1$ до $F_2$ :

$$\frac{F_1}{F_2} = G\frac{M_{\odot} M_M}{R_{\text{СЗ}}^2} \div G\frac{M_З M_M}{R_2^2} = \frac{M_{\odot}}{M_З} \cdot \left(\frac{R_2}{R_{\text{СЗ}}}\right)^2 = \frac{1{,}989 \cdot 10^{30}}{5{,}97 \cdot 10^{24}} \cdot \left(\frac{3{,}84 \cdot 10^8}{1{,}496 \cdot 10^{11}}\right)^2$$

$$= 3{,}33 \cdot 10^5 \cdot (2{,}57 \cdot 10^{-3})^2 \approx 22 \cdot 10^{-1} = 2{,}2$$

Таким абоном, Сонце притягує Місяць в 2,2 рази сильніше ніж Земля.

Обабослимо прискорення, яке надає Місяцю цій надлишок сили

$$a = \frac{F}{M_M},$$

де $F$ це різниця між діюабоми силами від Сонця $F_1$ і від Землі $F_2$:

$$F = F_1 - F_2 = GM_M\left(\frac{M_{\odot}}{R_{\text{СЗ}}^2} - \frac{M_З}{R_{\text{ЗЛ}}^2}\right)$$

Підставляємо отриманий вираз $F$ в формулу $a = \frac{F}{M_M}$ і отримаємо:

$$a = G\left(\frac{M_{\odot}}{R_{\text{СЗ}}^2} - \frac{M_З}{R_{\text{ЗЛ}}^2}\right) \approx 3{,}23 \cdot 10^{-3} \text{ м/с}^2$$

**Відповідь:** $\frac{F_1}{F_2} = 2{,}2,\ a \approx 3{,}23 \cdot 10^{-3}$ м/с$^2$.

**Задача 1.22.** Центр мас системи «Сонце+планети» не співпадає з центром Сонця, а змінюється в відповідності з положенням планет. Якщо маса Сонця $M_{\odot} = 6 \cdot 10^{30}$кг знайти радіуси кіл $R_C$, що описує Сонце під впливом:

а) Землі (маса $M_З = 6 \cdot 10^{24}$кг, відстань до Сонця $R = 150$ млн. км);

б) Юпітера (маса в 318 раз більша Земної, радіус орбіти в 5,2 раз більше Земного).

| **а) Дано:** | **Розв'язання.** |
|---|---|
| $M_{\odot} = 6 \cdot 10^{30}$кг<br>$M_З = 6 \cdot 10^{24}$кг<br>$R = 150$ млн. км | 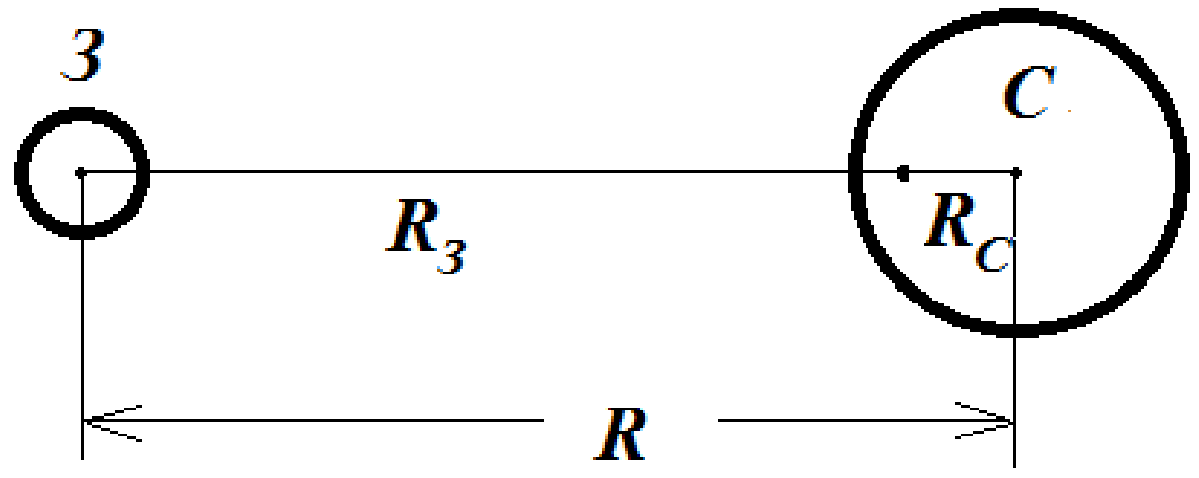 |
| **Знайти:**<br>$R_C$ - ? | Сила тяжіння |

$$F = G\frac{M_З \cdot M_С}{R^2}$$

врівноважена доцентровою силою $F_д = ma_д$, де доцентрове прискорення знаходимо за формулою $a_д = \frac{v^2}{R}$, тобто

$$F_д = m\frac{v^2}{R}.$$

Пам'ятаємо, що лінійна та кутова швидкості пов'язані формулою $v = \omega R$, тоді доцентрову силу виразимо через кутову швидкість:

$$F_д = \frac{m\omega^2 R^2}{R} = m\omega^2 R.$$

Запишимо для Землі

$$F_{дЗ} = M_З \cdot \omega^2 \cdot R_З;$$

і для Сонця

$$F_{дС} = M_С \cdot \omega^2 \cdot R_С.$$

Розглянемо повне прискорення, що складається із тангенціального та нормального прискорення:

$$a = \sqrt{a_\tau^2 + a_n^2}$$

Тангенціальне і нормальне прискорення точки тіла, що обертається виражаються формулами:

$$a_\tau = \varepsilon R, \qquad a_n = \omega^2 R,$$

де нормальне прискорення і є доцентровим. За умовою задачі сили, що діють зі сторони Сонця та Землі врівноважені, тому маємо записати, що $F_{дЗ} = F_{дС}$, тобто:

$$M_З \cdot \omega^2 \cdot R_З = M_С \cdot \omega^2 \cdot R_С; \quad M_З \cdot R_З = M_С \cdot R_С, \qquad \text{де} \qquad R = R_З + R_С$$

Склали і розв'язали систему рівнянь:

$$\begin{cases} R = R_З + R_С \\ M_З \cdot R_З = M_С \cdot R_С \end{cases} \qquad \begin{cases} R_З = R - R_С \\ R_З = \dfrac{M_С \cdot R_С}{M_З} \end{cases}$$

$$R - R_С = \frac{M_С \cdot R_С}{M_З}; \quad \frac{M_С}{M_З} = \frac{R - R_С}{R_С} = \frac{R}{R_С} - 1; \quad R_С = \frac{R \cdot M_З}{M_С + M_З} = 450 \text{ км}$$

**Відповідь:** під дією Землі $R_С = 450$ км.

**б) Дано:**

$М_\odot = 6 \cdot 10^{30}$кг

$М_Ю = 318 \cdot М_З$

$М_З = 6 \cdot 10^{24}$кг

$R_Ю = 5{,}2 \cdot R$

$R = 150$ млн.км

**Знайти:**

$R_C$–?

**Розв'язання**.

Радіус кола $R_C$, що описує Сонце під впливом Юпітера знаходиться за тією ж формулою, що ми вивели для знаходження $R_C$ під впливом Землі:

$$R_С = \frac{R \cdot M_З}{M_С + M_З}$$

Запишемо її підставивши дані для Юпітера:

$$R_С = \frac{R_Ю \cdot M_Ю}{M_С + M_Ю}$$

За умовою задачі:

$$R_Ю = 15 \cdot 10^7 \cdot 5{,}2 = 78 \cdot 10^7 \text{км і } M_Ю = 6 \cdot 10^{24} \cdot 318 = 1908 \cdot 10^{24} \text{кг}.$$

Підставивши ці значення отримаємо:

$$R_c = 744120 \text{ км}$$

Під впливом Юпітера зміщення центра мас системи 744120 км є набагато більшим ніж під впливом Землі 450 км.

**Відповідь:** під впливом Землі $R_c = 450$ км і Юпітера $R_c = 744120$ км.

**Задача 1.21.** Трамвай, рушаючи з місця, рухається з прискоренням $a_1$=0,5 м/с$^2$. Через час $t_1$=12 с після початку руху мотор вимикається і трамвай рухається до зупинки рівносповільнено. Коефіцієнт тертя на всьому шляху μ=0,01. Знайти найбільшу швидкість $\upsilon_1$ і час $t$ руху трамвая. Чому дорівнює його прискорення $a_2$ при рівносповільненому русі? Яку відстань $s$ пройде трамвай за час руху?

**Дано:**

$a_1$=0,5 м/с$^2$

$t_1$=12 с

μ=0,01

$g$=9,8 м/с$^2$

$\upsilon_0$=0

$\upsilon_2$=0

---

$\upsilon_1$ – ?

$t$ – ?

$s$ – ?

**Розв'язання:**

На першій ділянці шляху трамвай рухався рівноприскорено, на другій – рівносповільнено. Рівняння швидкості при рівноприскореному русі

$$\upsilon_1 = \upsilon_0 + a_1 t_1,$$

звідси

$$\upsilon_1 = a_1 t_1;$$

$$\upsilon_1 = 0{,}5 \cdot 12 = 6 \text{ м/с}.$$

Загальний час руху дорівнює

$$t = t_1 + t_2 \qquad (1)$$

Рівняння швидкості при рівносповільненому русі

$$\upsilon_2 = \upsilon_1 - a_2 t_2,$$

звідси

$$t_2 = \frac{\upsilon_1}{a_2} \qquad (2)$$

Закон Ньютона для руху на другої ділянці

$$\vec{F}_{\text{тер}} = m\vec{a}_2 .$$

В проекції на напрям протилежний напряму руху отримаємо

$$F_{\text{тер}} = ma_2 = \mu mg ,$$

звідси

$$a_2 = \mu g . \qquad (3)$$

Підставимо (3) в (2), а (2) в (1)

$$t = t_1 + \frac{\upsilon_1}{\mu g} ;$$

$$t = 12 + \frac{6}{0{,}01 \cdot 9{,}8} = 73{,}2\ \text{с} .$$

Довжину першої ділянки знайдемо з формули

$$s_1 = \frac{\upsilon_1^2 - \upsilon_0^2}{2a_1} = \frac{\upsilon_1^2}{2a_1} \qquad (4)$$

Таким чином довжина другої ділянки дорівнює

$$s_2 = \frac{\upsilon_2^2 - \upsilon_1^2}{-2a_2} = \frac{\upsilon_1^2}{2a_2} ,$$

звідси враховуючи (3) отримаємо

$$s_2 = \frac{\upsilon_1^2}{2\mu g} \qquad (5)$$

Згідно з (4) і (5) шлях, який пройшов трамвай дорівнює

$$s = s_1 + s_2 = \frac{\upsilon_1^2}{2a_1} + \frac{\upsilon_1^2}{2\mu g};$$

$$s = \frac{6^2}{2 \cdot 0{,}5} + \frac{6^2}{2 \cdot 0{,}01 \cdot 9.8} = 219{,}7\ \text{м}$$

**Відповідь:** $\upsilon_1 = 6\ \text{м/с}$, $t = 73{,}2\ \text{с}$, $s = 219{,}7\ \text{м}$

**Задача 1.22.** На автомобіль масою $m$=1 т під час руху діє сила тертя $F_{\text{тер}}$, яка дорівнює 0,1 діючої на нього сили тяжіння $mg$. Знайти силу тяги $F$, яку розвиває мотор автомобіля, якщо автомобіль рухається з прискоренням $a$=1 м/с$^2$ в гору з ухилом 1 м на кожні 25 м шляху.

| Дано: | Розв'язання: |
|---|---|
| $m$=1000 кг<br>$F_{\text{тер}}$=0,1$mg$<br>$g$=9,8 м/с$^2$<br>$a$=1 м/с$^2$<br>$h$=1 м<br>$l$=25 м | 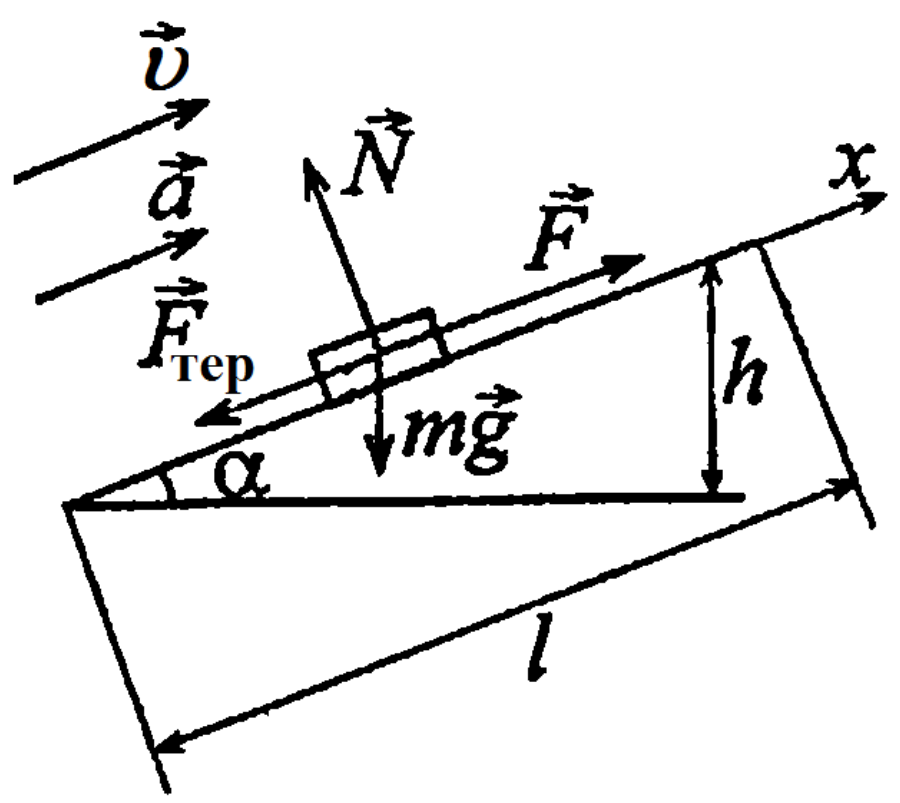 |
| $F$ – ? | |

Задамо напрям осі $x$ вздовж похилої за напрямом руху автомобіля і запишемо другий закон Ньютона в проекції на цю ось

$$F - mg\sin\alpha - F_{\text{тер}} = ma, \tag{1}$$

де

$$\sin\alpha = \frac{h}{l}. \tag{2}$$

З урахуванням рівнянь (1) та (2), сила тяги, яку розвиває мотор автомобіля дорівнює

$$F = ma + mg\frac{h}{l} + 0{,}1mg = m\left(a + g\frac{h}{l} + 0{,}1g\right)$$

$$F = 1000\left(1 + 9{,}8\frac{1}{25} + 0{,}1 \cdot 9{,}8\right) = 2372 \text{ Н}$$

**Відповідь:** $F$=2372 Н

### Задачі для самостійного розв’язання

1. Маса ліфта з пасажирами $m$=800 кг. З яким прискоренням $a$ і в якому напряму рухається ліфт, якщо відомо, що сила натягування троса, який тримає ліфт: а) $T$=12 кН; б) $T$=6 кН? [а) $a$=5,2 м/с$^2$, б) $a$=–2,3м/с$^2$].

2. Тіло ковзає по похилій площині, яка утворює із горизонтом кут $\alpha$=45°. Пройшовши шлях $s$=36,4 см, тіло набуває швидкості $\upsilon$=2 м/с. Знайти коефіцієнт тертя $\mu$ тіла о площину. [$\mu$=0,2].

3. Тіло ковзає по похилій площині, яка утворює із горизонтом кут $\alpha$=45°. Залежність пройденого тілом шляху $s$ від часу $t$ дається рівнянням $s=Ct^2$, де $C$=1,73м/с$^2$. Знайти коефіцієнт тертя $\mu$ тіла о площину. [$\mu$=0,5].

4. Дві гирі масами $m_1$=2 кг і $m_2$=1 кг з’єднані ниткою і перекинуті через невагомий блок. Знайти прискорення $a$, з яким рухаються гирі, і силу натягування нитки $T$. Тертям у блоці знехтувати. [ $T$=13 Н, $a$=3,27 м/с$^2$].

5. Невагомий блок закріплений на кінці стола (див.рис.). Гирі *1* і *2* однакової маси $m_1=m_2$=1 кг з’єднані ниткою і перекинуті через блок. Коефіцієнт тертя гирі *2* о стіл $\mu$=0,1. Знайти прискорення $a$, з яким рухаються гирі, і силу натягування нитки $T$. Тертям в блоці знехтувати. [$T_1=T_2$=5,4 Н].

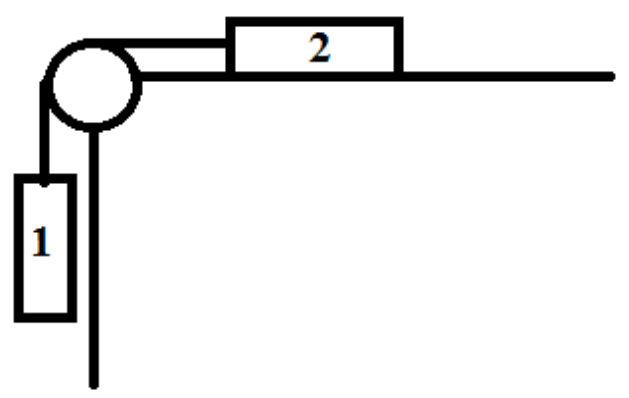

## 1.5. Енергія, робота, потужність. Закони збереження енергії.

Однією з характеристик дії сили є робота. У процесі роботи відбувається перетворення енергії одного тіла або системи тіл на енергію іншого тіла або системи тіл.

**Механічна робота** $A$ – це фізична велиабона, яка дорівнює добутку модуля сили F на модуль переміщення s, яке його здійснює тіло під дією цієї сили, і на косинус кута $\alpha$ між вектором сили і вектором переміщення:

$$A = F \cdot s \cdot \cos\alpha$$

**Потужність** $\boldsymbol{P}$ – це фізична велиабона, яка характеризує швидкість виконання роботи й дорівнює відношенню роботи $A$ до проміжку часу $t$, за який вона виконана:

$$P = \frac{A}{t}.$$

Якщо деяке тіло рухається з постійною швидкістю $v$, то його переміщення дорівнюватиме: $s = vt$, робота сили тяги становитиме:

$$\mathrm{A} = F_x \mathrm{s} = F_x vt,$$

отже потужність можна обабослити за формулою:

$$P = \frac{A}{t} = \frac{F_x vt}{t} = F_x v$$

або $A = F_x s$ при $F_x = F \cdot \cos\alpha$ отримаємо, що

$$P = \frac{A}{t} = \frac{F \cdot s \cdot \cos\alpha}{t} = F \cdot v \cdot \cos\alpha,$$

де $P$ – потужність в даний момент часу; $F_x$ – проекція сили в даний момент часу; $v$ – миттєва швидкість руху тіла.

Миттєва потужність характеризує швидкість виконання роботи в кожен момент часу, тобто є відношенням елементарної роботи до нескінченно малого проміжка часу, за який вона виконується:

$$P = \frac{dA}{dt}$$

**Механічна енергія *E*** – це фізична велиабона, яка характеризує здатність тіла (системи тіл або поля) виконати роботу. Під час виконання механічної роботи енергія тіла змінюється, тобто механічна робота є мірою зміни енергії тіла. У механіці розрізняють два види енергії – кінетичну та потенціальну.

Кінетична енергія $E_к$ – це фізична велиабона, яка характеризує тіло, що рухається, і дорівнює половині добутку маси m тіла на квадрат швидкості $v$ його руху:

$$E_k = \frac{mv^2}{2}$$

Робота рівнодійної всіх сил, які діють на тіло, дорівнює зміні кінетичної енергії тіла:

$$A = E_k - E_{k0} = \frac{mv^2}{2} - \frac{mv_0^2}{2} = \Delta E_k$$

де $m$ – маса тіла, $v_0$ – початкова швидкість тіла, $v$ – кінцева швидкість тіла.

**Потенціальна енергія $\boldsymbol{E_п}$** – це енергія, яку має тіло внаслідок взаємодії з іншими тілами або внаслідок взаємодії частин тіла між собою. Потенціальна енергія тіла масою m, піднятого на висоту $h$ поблизу поверхні Землі:

$$E_п = mgh\ .$$

Робота сили тяжіння дорівнює зміні потенціальної енергії тіла, взятій з протилежним знаком:

$$A = -(mgh - mgh_0) = -mg(h - h_0) = -(E_п - E_{п0}) = -\Delta E_п$$

Потенціальна енергія пружно деформованого тіла (пружини):

$$E_п = \frac{kx^2}{2},$$

де $x$ – видовження тіла (пружини), $k$ – жорсткість пружини (коефіцієнт пропорційності).

Робота сили пружності дорівнює зміні потенціальної енергії тіла, взятій з протилежним знаком:

$$A = -\left(\frac{kx^2}{2} - \frac{kx_0^2}{2}\right) = -(E_п - E_{п0}) = -\Delta E_п$$

Повна механічна енергія системи тіл – це сума кінетичної та потенціальної енергій системи: $E = E_к + E_п$ . Повна механічна енергія замкненої системи тіл за відсутністю в системі сил тертя не змінюється з часом, тобто зберігається:

$$E_{к0} + E_{п0} = E_к + E_п$$

Закон збереження повної механічної енергії може мати вигляд:

$$\frac{mv_1^2}{2} + mgh_1 + \frac{kx_1^2}{2} = \frac{mv_2^2}{2} + mgh_2 + \frac{kx_2^2}{2}$$

**Пружний і абсолютно непружний удари.**

Удар або зіткнення – це короткочасна взаємодія тіл, у ході якої вони безпосередньо торкаються одне одного. Оскільки систему тіл, що стикаються, можна вважати замкненою (під час удару внутрішні сили в системі в багато разів більші за зовнішні сили), то під час удару виконується закон збереження імпульсу: в ізольованій системі імпульс тіл, які входять до неї, залишається постійним, тобто

$$m_1 v_1 + m_2 v_2 + ... + m_n v_n = \text{const}$$

До і після удару потенціальні енергії тіл дорівнюють нулю. Повна механічна енергія $E_0$ тіл на початку удару й повна механічна енергія $E$ тіл наприкінці удару дорівнюють сумі кінетичних енергій цих тіл:

$$E_0 = \frac{m_1 v_{01}^2}{2} + \frac{m_2 v_{02}^2}{2}; \qquad E = \frac{m_1 v_1^2}{2} + \frac{m_2 v_2^2}{2}$$

**Абсолютно непружний удар** – зіткнення тіл, у результаті якого тіла рухаються як одне ціле. За законом збереження імпульсу:

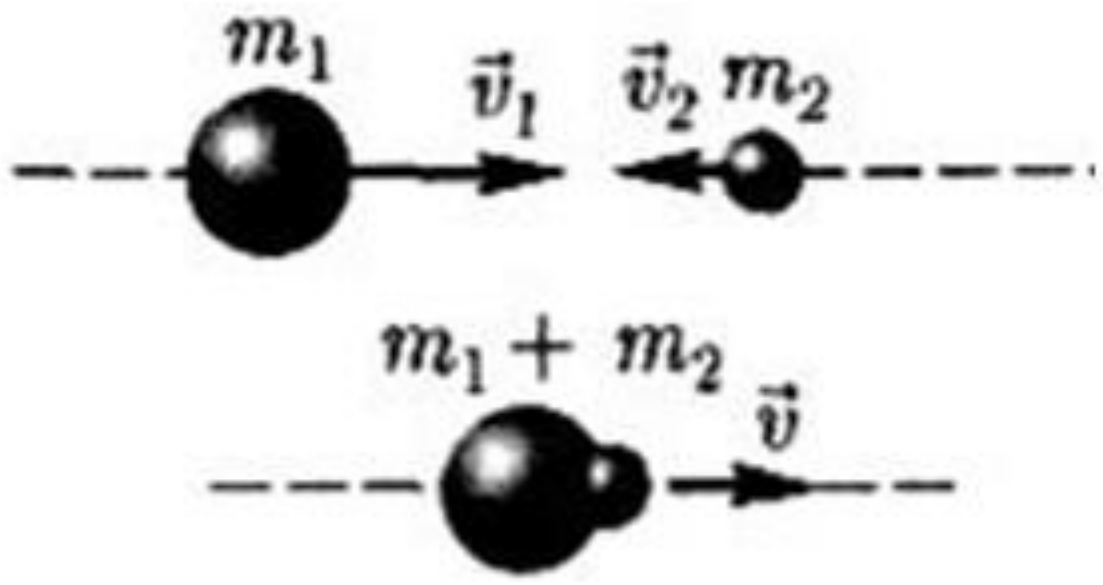

Рис. 1.4. Абсолютно непружний удар.

$$m_1\vec{v}_1 + m_2\vec{v}_2 = (m_1 + m_2)\vec{v}.$$

$$\vec{v} = \frac{m_1\vec{v}_1 + m_2\vec{v}_2}{m_1 + m_2}.$$

$$m_1 v_1 + m_2 v_2 = m_1 v_1' + m_2 v_2'.$$

**Пружний удар** – зіткнення тіл, за якого деформація тіл виявляється зворотною, тобто повністю зникає після припинення взаємодії.

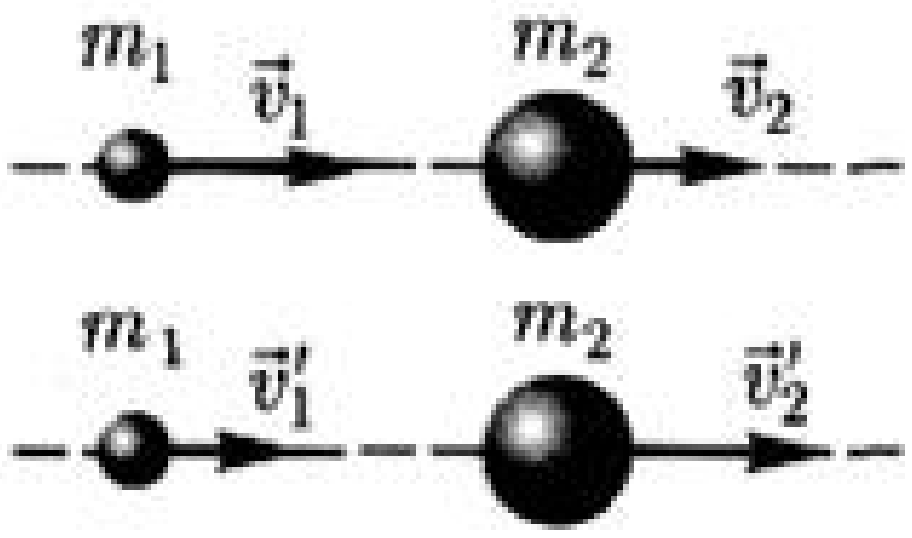

Рис. 1.5. Абсолютно пружний удар.

$$v_1 + v_1' = v_2 + v_2'.$$

В залежності від напрямку руху куль (назустріч або в одному напрямку), центральний удар або ні та їх мас існують різні варіанти руху після зіткнення, наприклад:

1) $m_1$=$m_2$ в випадку, що друга куля була нерухомою $v_2 = 0$, то після удару $v_1' = 0$, а друга куля буде рухатися з тією ж швидкістю і в тому ж напрямку, що й перша куля до удару $v_2' = v_1$ .

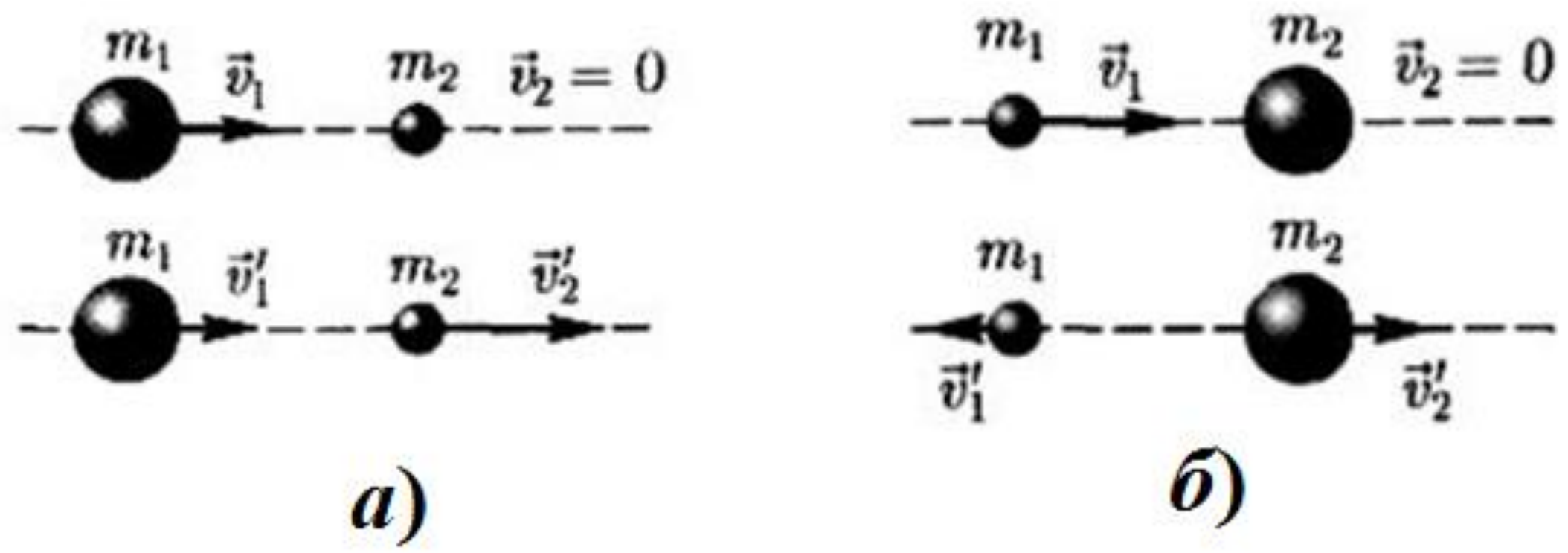

Рис. 1.6. Варіанти абсолютно пружного удару.

2) $m_1 > m_2$ ; $v_2 = 0$. Перша куля продовжує рухатися в тому ж напрямку як до удару, але з меншою швидкістю $v_1' < v_1$. Швидкість другої кулі після удару більша ніж швидкість першої після удару $v_2' > v_1'$ ( рис. 6*а*);

3) $m_1 < m_2$ ; $v_2 = 0$. Напрямок руху першої кулі при ударі змінюється - куля відскакує назад. Друга куля рухається в ту ж сторону, в яку рухалася перша куля до удару, але з меншою швидкістю, $v_2' < v_1'$ (рис. 6*б*).

4) $m_2 \gg m_1$; $v_2 = 0$. Випадок зіткнення зі стіною: $v_1' = -v_1$, $v_2' \approx 0$

5) $m_1 = m_2$ і обидві кулі рухались, то кулі рівної маси обмінюються швидкостями: $v_1' = v_2$ , $v_2' = v_1$.

При прямому, центральному і абсолютно пружному ударі зберігаються імпульс і кінетична енергія системи. Запишемо закон збереження механічної енергії і закон збереження імпульсу у систему і розв'яжемо її:

$$\begin{cases} \dfrac{m_1 v_1^2}{2} + \dfrac{m_2 v_2^2}{2} = \dfrac{m_1 v_1'^2}{2} + \dfrac{m_2 v_2'^2}{2} \\ m_1 v_1 + m_2 v_2 = m_1 v_1' + m_2 v_2' \end{cases}$$

перетворимо рівняння:

$$\begin{cases} m_1(v_1^2 - v_1'^2) = m_2(v_2'^2 - v_2^2) \\ m_1(v_1 - v_1') = m_2(v_2' - v_2) \end{cases}$$

Поділимо ліву частину першого рівняння на ліву частину другого, і відповідно праву частину першого на праву частину другого і отримаємо:

$$v_1 + v_1' = v_2' + v_2 \text{ звідси } v_2' = v_1 + v_1' - v_2$$

Підставимо отриманий вираз для $v_2'$ в рівняння закону збереження імпульсу

$$m_1 v_1 + m_2 v_2 = m_1 v_1' + m_2(v_1 + v_1' - v_2)$$

$$m_1 v_1 + m_2 v_2 = m_1 v_1' + m_2 v_1 + m_2 v_1' - m_2 v_2$$

$$v_1'(m_1 + m_2) = (m_1 - m_2)v_1 + 2m_2 v_2$$

Далі вирішивши цю систему рівнянь відносно $v_1'$ вірішуємо для $v_2'$, і отримаємо проекції швидкості тіл після удару на виділений напрямок, який збігається з напрямком руху одного з тіл до удару.

$$v_1' = \frac{2m_2 v_2 + (m_1 - m_2)v_1}{m_1 + m_2} \text{ і } v_2' = \frac{2m_1 v_1 + (m_2 - m_1)v_2}{m_1 + m_2}$$

**Задача 1.23.** Яку роботу $A$ виконує сила тяжіння під час одного оберта супутника навколо Землі по коловій орбіті? Опором середовища знехтувати.

| **Дано:** | **Розв'язання**. |
|---|---|
| $m$<br>$g$ =9,8 м/с$^2$ | Сила тяжіння для будь-якого тіла на коловій орбіті визначається за формулой: |
| **Знайти:**<br>$A$ - ? | $F = m \cdot g.$<br>Роботу, яку виконує сила тяжіння знайдемо за формулою: |

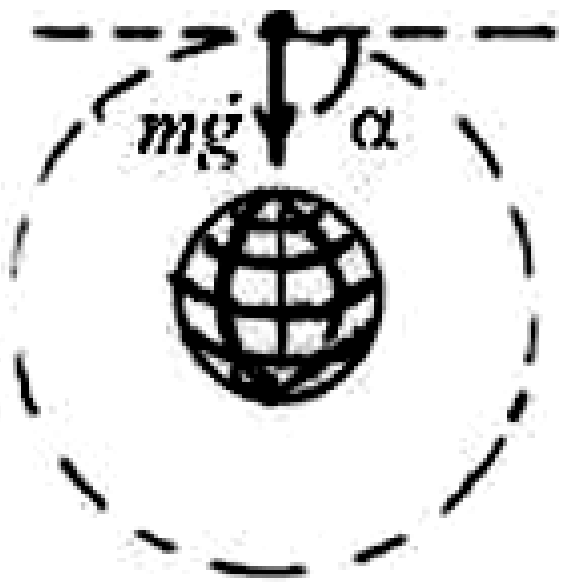

$$A = F \cdot s \cdot \cos\alpha$$

На кожному малому відрізку $\Delta s$ орбіти сила тяжіння спрямована до центру Землі, а вектор руху по дотичній до кожної точкі орбіти, тобто кут між напрямком руху і силою тяжіння $\alpha = 90^o$, тобто:

$$A = mg\Delta s \cdot \cos\alpha, \quad \alpha = 90^o, \quad \cos 90^\circ = 0 \;\Rightarrow\; A = 0$$

**Відповідь:** $A = 0$.

**Задача 1.24.** Тіло масою 1 кг вільно падає без початкової швидкості з деякої висоти. Знайдіть потужність сили тяжіння через 3 с. польоту. Вважайте, що $g$=10 м/с.

| **Дано:** | **Розв'язання**. |
|---|---|
| $m$ =1 кг<br>$g$ =10 м/с$^2$<br>$v_0 = 0$<br>$t$=3 с | Переміщення<br>$$S = h - h_0 = v_{0y}t + \frac{gt^2}{2}$$<br>де $v_{0y} = 0$ за умовою задачі, але, при підстановці цього значення ми отримаємо середню потужність, а нам треба потужність на момент через 3 с. польоту, тобто миттєву потужність на момент часу у 3 с. Тому нам треба знайти миттєву потужність: |
| **Знайти:**<br>$P$ - ? | |

$$P = \frac{dA}{dt}.$$

Складаємо рівняння для роботи:

$$A = F_y \cdot S$$

де

$$F_y = F_т = mg,$$

$$S = h - h_0 = v_{0y}t + \frac{gt^2}{2}$$

Підставимо $F_y$ і $S$ в формулу $A = F_y \cdot S$ та отримаємо:

$$A = mg\left(v_{0y}t + \frac{gt^2}{2}\right)$$

$$P = \frac{dA}{dt} = \left(mg\left(v_{0y}t + \frac{gt^2}{2}\right)\right)' = mg\left(v_{0y}t + \frac{gt^2}{2}\right)' = mgv_{0y} + mg \cdot gt =$$

за умовою задачі $v_{0y} = 0$ (вільно падає без початкової швидкості) тому

$$= mg^2t = 1 \cdot 10^2 \cdot 3 = 300\,кг\frac{м^2}{с^3}$$

**Відповідь:** $A = 300\,кг\frac{м^2}{с^3}$

**Задача 1.25.** Куля масою $m_1$=10 кг, що рухається зі швидкістю $v_1$=12 м/с, зіштовхується з кулею масою $m_2$=4 кг, швидкість $v_2$ якої дорівнює 5 м/с. Вважаюабо удар центральним і абсолютно непружним, визнааботи швидкість $v'$ куль після удару в випадку коли мала куля наздоганяє велику, що рухається в тому же напрямку.

**Дано:**

$m_1$ =10 кг

$\upsilon_1$ =12 м/с-

$m_2$ =4 кг

$\upsilon_2$ =5 м/с

**Знайти:**

$\upsilon$' - ?

**Розв'язання.**

Закон збереження імпульсу $m_1v_1 + m_2v_2 = m_1v_1' + m_2v_2'$

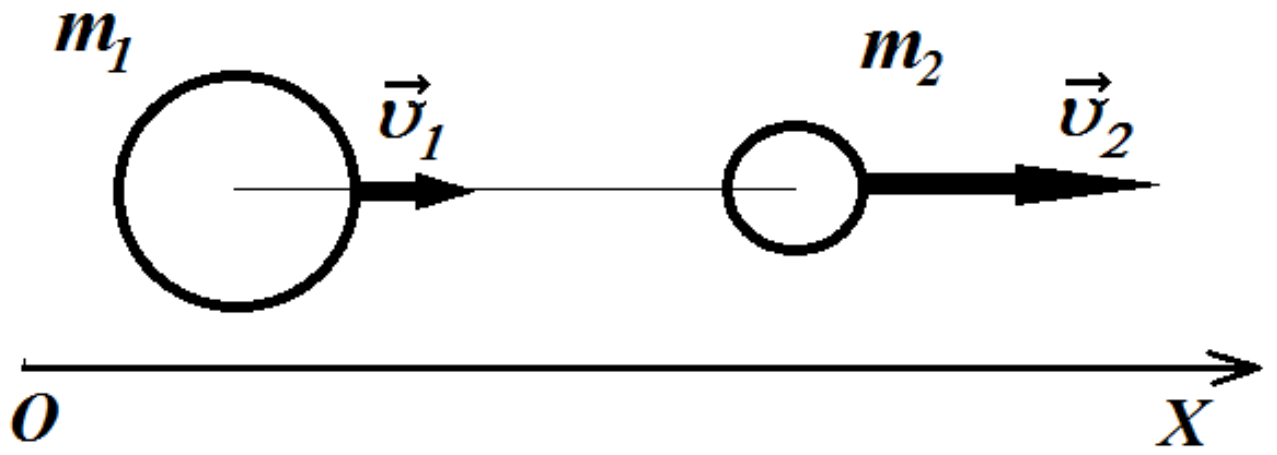

За умовою задачі удар є абсолютно непружним, тобто після зіткнення кулі продовжують рухатися як одне ціле зі швидкістю $v'$. Тому перепишемо рівняння збереження імпульсу в наступному вигляді:

$$m_1v_1 + m_2v_2 = v'(m_1 + m_2).$$

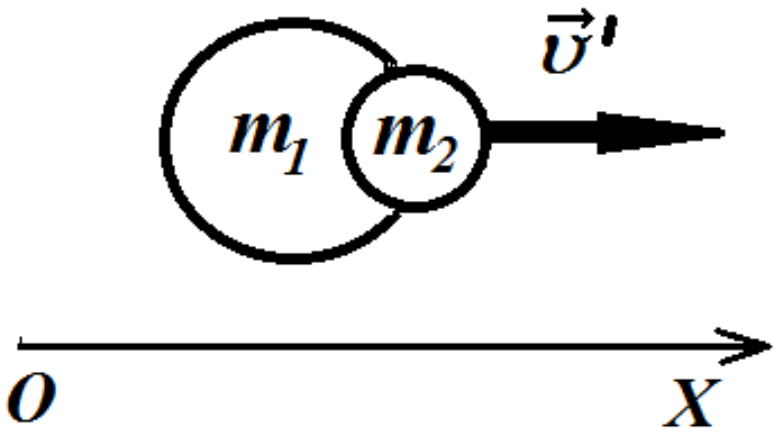

Звідси виразимо $v'$:

$$v' = \frac{m_1v_1 + m_2v_2}{m_1 + m_2}$$

Підставимо значення, що надані в умовах задачі і отримаємо:

$$v' = \frac{10 \cdot 12 + 4 \cdot 5}{10 + 4} = \frac{140}{14} = 10 \text{ м/с}$$

**Відповідь:** $v' = 10$ м/с.

**Задача 1.26.** Куля масою $m_1$=10 кг, що рухається зі швидкістю $v_1$=4 м/с, зіштовхується з кулею масою $m_2$=4 кг, швидкість якої дорівнює $v_2$ =12 м/с. Вважаюабо удар центральним і абсолютно непружним, визнааботи швидкість $v'$ куль після удару в випадку коли кулі рухаються назустріч одна одній.

| **Дано:** | **Розв'язання.** |
|---|---|
| $m_1$ =10 кг<br>$v_1$ =4 м/с-<br>$m_2$ =4 кг<br>$v_2$ =12 м/с | 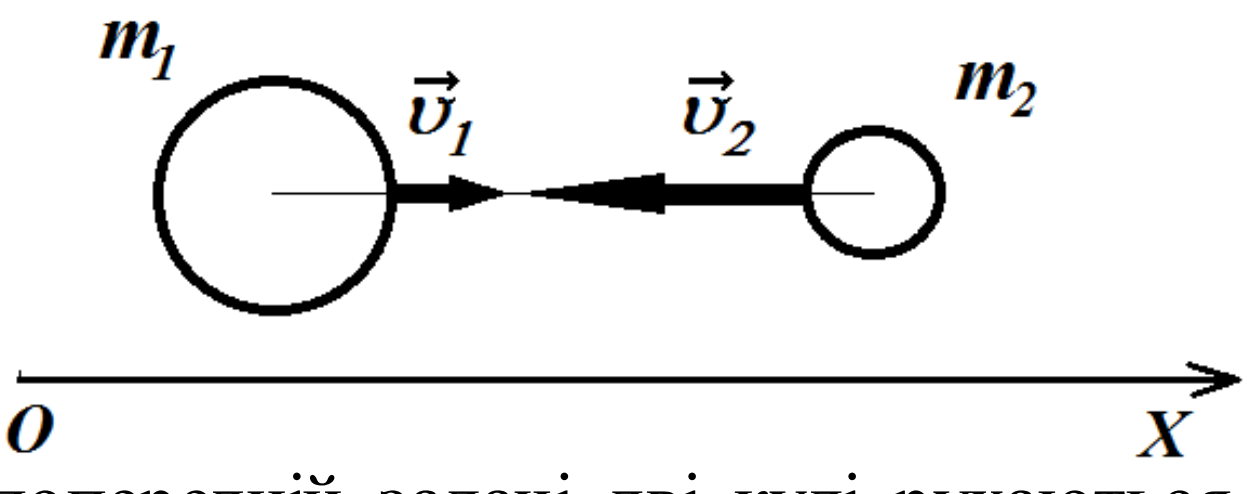 |
| **Знайти:**<br>$v$' - ? | В попередній задачі дві кулі рухаються в В цій задачі кулі рухалися назустріч одна одній і після удару продовжили рухатись |

як одне ціле. В цьому випадку швидкість $v_2$ відносно осі О$X$ має протилежний напрямок, тому і проекція $\vec{v}_2$ на вісь О$X$ буде записана зі знаком (-) і буде підставлена у формулу закону збереження імпульсу зі знаком (-):

$$m_1 v_1 + m_2(-v_2) = v'(m_1 + m_2)$$

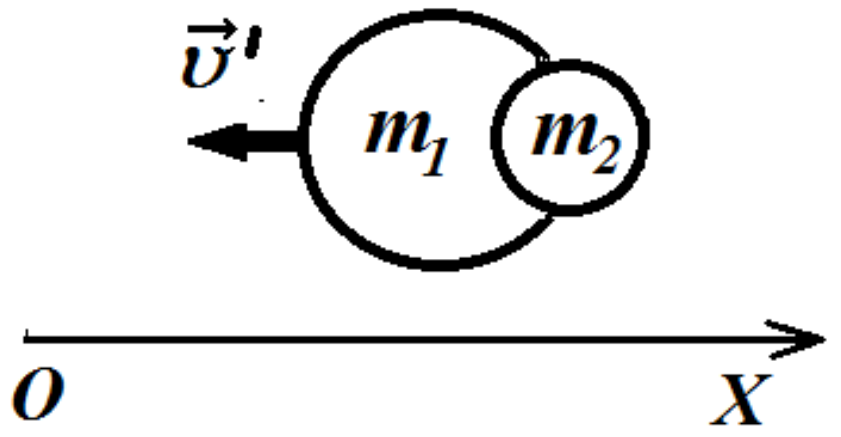

$$m_1 v_1 - m_2 v_2 = v'(m_1 + m_2)$$

$$v' = \frac{m_1 v_1 - m_2 v_2}{m_1 + m_2}$$

$$v' = \frac{10 \cdot 4 - 4 \cdot 12}{10 + 4} = \frac{40 - 48}{14} = -\frac{8}{14} = -0{,}57 \text{ м/с}$$

З отриманих результатів ми баабомо, що після зіткнення, кулі продовжили рухатися в протилежному напрямку відносно осі О$X$, про що свідаботь знак мінус. Тобто напрямок руху об'єднаних куль співпадає з напрямком швидкості $v_2$ .

**Відповідь:** $v' = -0{,}57\frac{м}{с}$.

**Задача 1.27.** Розглянемо аналогічну задачу, але у випадку коли $v_2$=0, тобто перше тіло налітає на нерухоме друге тіло (кулю), наприклад: куля, що мала швидкість $v_1$ =3 м/с, після пружного зіткнення з нерухомою кулею продовжила рух у тому самому напрямі зі швидкістю $v_1'$= 2 м/с. Визнааботи: відношення мас куль $m_1/m_2$ та швидкість другої кулі $v_2'$ після зіткнення.

| Дано: | Розв'язання. |
|---|---|
| $v_1$ =3 м/с- <br> $v_2$ =0 м/с <br> $v_1' = 2$ м/с | 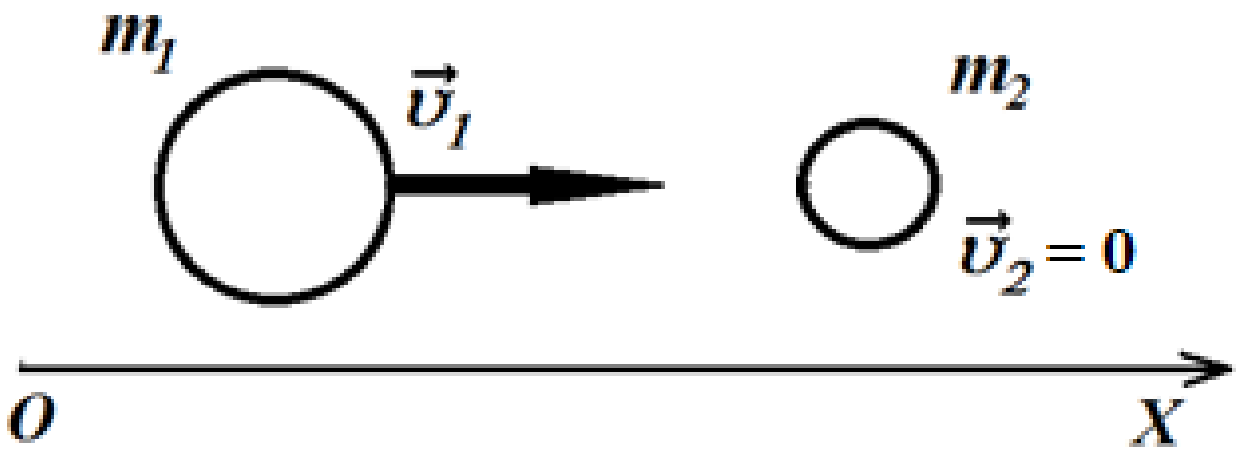 |
| **Знайти:** <br> $m_1/m_2$ - ? <br> $v_2'$ - ? | |

При центральному абсолютно пружному зіткненні двох куль можливі варіанти, наприклад: $v_2'$>0, тобто друга куля відскоабоить в напрямку руху першої, але напрям руху першої кулі після удару не є однозначним і залежить від співвідношення мас. Якщо перша куля є масивнішою ($m_1$>$m_2$), то $v_1'$>0, тобто вона продовжить рух у тому самому напрямі, проте з меншою швидкістю. Відповідно, при $m_1$<$m_2$ перша куля після удару відскоабоить у зворотньому напрямі ($v_1'$<0). При рівних масах куль ($m_1$=$m_2$) з формул випливає, що $v_1' = v_2$, $v_2' = v_1$. Таким абоном, швидкості тіл після абсолютно пружного удару не можуть бути однаковими по велиабоні і по напрямку і визначаються за формулою:

для першої кулі $$v_1' = \frac{2m_2v_2 + (m_1 - m_2)v_1}{m_1 + m_2}$$ і

для другої кулі $$v_2' = \frac{2m_1v_1 + (m_2 - m_1)v_2}{m_1 + m_2}$$

В випадку, коли друга куля до удару була нерухомою при підстановці у них $v_2$=0 приймуть вид:

$$v_1' = \frac{(m_1 - m_2)v_1}{m_1 + m_2} \quad \text{і} \quad v_2' = \frac{2m_1v_1}{m_1 + m_2}$$

Для нашого випадку напрям першої кулі після зіткнення не змінився, відповідно удар є центральним. Введемо позначення $m_1/m_2=k$ і зробимо заміну $m_1=km_2$ в формулі для швидкості першої кулі після зіткнення:

$$v_1' = \frac{(km_2 - m_2)v_1}{km_2 + m_2} = \frac{m_2(k-1)v_1}{m_2(k+1)} = \frac{(k-1)v_1}{(k+1)} \text{ виразимо } k$$

$$v_1'(k+1) = (k-1)v_1 \quad kv_1' + v_1' = kv_1 - v_1$$

$$kv_1' - kv_1 = -(v_1' + v_1)$$

$$k = \frac{v_1' + v_1}{v_1 - v_1'} = \frac{2+3}{3-2} = 5$$

З другої формули знайдемо $v_2'$. Спочатку підставимо $m_1=km_2$ і отримаємо

$$v_2' = \frac{2km_2v_1}{km_2 + m_2} = \frac{2km_2v_1}{m_2(k+1)} = \frac{2kv_1}{k+1}$$

Підставимо числові дані

$$v_2' = \frac{2 \cdot 5 \cdot 3}{5+1} = \frac{30}{6} = 5 \text{ м/с}$$

**Відповідь:** $v_2' = 5$ м/с.

**Задача 1.28.** У ядерному реакторі необхідно сповільнювати нейтрони, тобто зменшувати їх кінетичну енергію. Зіткнення нейтронів з ядрами речовини-сповільнювача можна розглядати як абсолютно пружний удар куль. Маса кулі, що рухається (нейтрона) $m_1$ , маса нерухомої кулі (ядра сповільнювача) $m_2 \approx 2m_1$. Визнааботи, яку частку $\varepsilon$ своєї кінетичної енергії нейтрон передає ядру під час зіткнення. Удар прямий, центральний, абсолютно пружний.

| **Дано:** | **Розв'язання.** |
|---|---|
| $m_2 \approx 2m_1$. | Позначимо через $\Delta E_1$ енергію яку втратила перша куля під час зіткнення, а E1 – це початкова кінетична енергія першої кулі, тоді відношення $\frac{\Delta E_1}{E_1}$ і буде тією |
| **Знайти:** | |
| $\varepsilon$ - ? | |

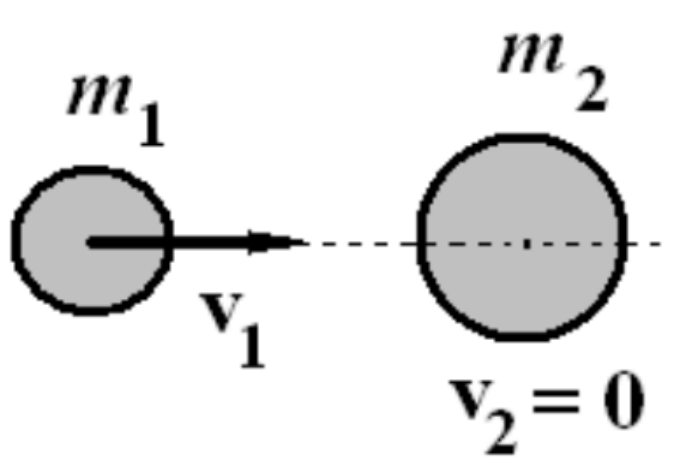

часткою кінетичної енергії $\varepsilon$, яку треба знайти.

$$\varepsilon = \frac{\Delta E_1}{E_1}$$

Швидкість куль до удару $v_1$ та $v_2$, після удару познаабомо як $u_1$ та $u_2$. Енергія, що втрачена першою кулею, дорівнює енергії, яку отримала друга куля, тобто це кінетична енергія другої кулі $u_2$ після удару, яку знайдемо за формулою:

$$E_k = \frac{mv^2}{2}$$

Запишемо в позначеннях нашої задаабо:

$$\Delta E_1 = E_2 = \frac{m_2 u_2^2}{2}$$

Аналогічно запишемо кінетичну енергію для першої кулі до зіткнення:

$$E_1 = \frac{m_1 v_1^2}{2}$$

Отримані вирази підставимо у вираз пошуку часткової кінетичної енергії $\varepsilon$:

$$\varepsilon = \frac{E_2}{E_1} = \frac{\frac{m_2 u_2^2}{2}}{\frac{m_1 v_1^2}{2}} = \frac{m_2 u_2^2}{2} \cdot \frac{2}{m_1 v_1^2} = \frac{m_2 u_2^2}{m_1 v_1^2}$$

Швидкість другої кулі після пружного зіткнення можна знайти за формулою:

$$u_2 = \frac{2m_1 v_1 + (m_2 - m_1) v_2}{m_1 + m_2}$$

Підставимо $v_2 = 0$ тому, що друга куля до удару не рухалась, тоді:

$$u_2 = \frac{2m_1 v_1}{m_1 + m_2}$$

Знайдений вираз підставимо в рівняння:

$$\varepsilon = \frac{m_2 u_2^2}{m_1 v_1^2}$$

$$\varepsilon = \frac{m_2\left(\frac{2m_1 v_1}{m_1 + m_2}\right)^2}{m_1 v_1^2} = \frac{\frac{m_2(2m_1 v_1)^2}{(m_1+m_2)^2}}{m_1 v_1^2} = \frac{m_2(2m_1 v_1)^2}{(m_1+m_2)^2}\frac{1}{m_1 v_1^2} = \frac{4m_1 m_2}{(m_1+m_2)^2}$$

Із отриманого рівняння видно, що частка переданої енергії залежить тільки від мас куль, що зіштовхуються. В отриманий вираз підставимо

$m_1 = \frac{m_2}{2}$ (за умовою задачі $m_2 \approx 2m_1$) і отримаємо:

$$\varepsilon = \frac{4m_1 m_2}{(m_1+m_2)^2} = \frac{4\frac{m_2}{2}m_2}{\left(\frac{m_2}{2} + m_2\right)^2} = \frac{2m_2^2}{\left(\frac{3m_2}{2}\right)^2}$$

$$\varepsilon = \frac{2m_2^2}{\left(\frac{3}{2}\right)^2 m_2^2} = \frac{2\cdot 4}{9} = \frac{8}{9} = 0{,}89 = 89\%$$

**Відповідь:** $\varepsilon = 89\%$.

**Задача 1.29.** Дві однакові кульки, що рухались із швидкостями $\vec{v}_1$та $\vec{v}_2$ під кутом $\alpha$ одна до одної, після пружного удару розлетілись із швидкостями $\vec{u}_1$ та $\vec{u}_2$. Визначити: кут розльоту кульок $\beta$.

| Дано: | Розв'язання. |
|---|---|
| $\vec{v}_1; \vec{v}_2; \vec{u}_1; \vec{u}_2;\ \alpha$ | Закон збереження імпульсу: |
| **Знайти:** | $m\vec{v}_1 + m\vec{v}_2 = m\vec{u}_1 + m\vec{u}_2$ |
| $\beta$–? | Закон збереження кінетичної енергії: |

$$\frac{m_1 v_1^2}{2} + \frac{m_1 v_2^2}{2} = \frac{m_1 u_1^2}{2} + \frac{m_1 u_2^2}{2}$$

Пружне зіткнення кульок відбувається із збереженням імпульсу та кинетичної енергії відповідно до рівнянь:

$$\begin{cases} m\vec{v}_1 + m\vec{v}_2 = m\vec{u}_1 + m\vec{u}_2 \\ \dfrac{m_1 v_1^2}{2} + \dfrac{m_1 v_2^2}{2} = \dfrac{m_1 u_1^2}{2} + \dfrac{m_1 u_2^2}{2} \end{cases} \Rightarrow \quad \begin{cases} \vec{v}_1 + \vec{v}_2 = \vec{u}_1 + \vec{u}_2 \\ v_1^2 + v_2^2 = u_1^2 + u_2^2 \end{cases}$$

Вектори $\vec{v}_1 + \vec{v}_2$ та $\vec{u}_1 + \vec{u}_2$ являють собою діагоналі паралелограмів швидкостей, причому $\vec{v}_1 + \vec{v}_2 = \vec{u}_1 + \vec{u}_2$. Отже, за теоремою косинусів маємо

$$v_1^2 + v_2^2 + 2v_1 v_2 \cos\alpha = u_1^2 + u_2^2 + 2u_1 u_2 \cos\beta$$

Врахувавши друге рівняння:

$$v_1^2 + v_2^2 = u_1^2 + u_2^2$$

Отримаємо:

$$v_1 v_2 \cos\alpha = u_1 u_2 \cos\beta \Rightarrow \cos\beta = \frac{v_1 v_2}{u_1 u_2}\cos\alpha.$$

В окремому випадку, коли одна з кульок, наприклад, друга перебуває в спокої , то $\cos\beta = 0$ і $\beta = 90°$, тобто кульки розлітаються під прямим кутом.

**Відповідь:** $\cos\beta = \frac{v_1 v_2}{u_1 u_2}\cos\alpha$.

**Задача 1.30.** Пластилінова кулька маси $m = 80$ г рухається зі швидкістю $v = 3\frac{\text{м}}{\text{с}}$ і зіштовхується з нерухомою пластиліновою кулькою тієї ж маси. Визнааботи кількість теплоти $Q$, що виділилась під час зіткнення.

| **Дано:** | **Розв'язання.** |
|---|---|
| $m_1 = m_2 = 80$ г $= 8 \cdot 10^{-2}$ кг | Закон збереження імпульсу: |
| $v_1 = 3\dfrac{\text{м}}{\text{с}}$ | $m\vec{v}_1 + m\vec{v}_2 = m\vec{u}_1 + m\vec{u}_2$ |
| $v_2 = 0$ м/с | Кінетична енергія кулі: |
| **Знайти:** | $E_k = \dfrac{mv^2}{2}$ |
| $Q$–? | Із закону збереження енергії виходить, що |

різниця енергії перетворюється на тепло, тому:

$$E_0 - E = Q$$

За законом збереження імпульсу:

$$m_1 v_1 + m_2 v_2 = m_1 u_1 + m_2 u_2$$

За умовою цієї задачі $v_2 = 0$, а швидкість після зіткнення двох кульок (після зіткнення вони рухаються разом) $u_1 = u_2 = u,$ тому закон збереження імпульсу запишемо як:

$$m_1 v_1 = u(m_1 + m_2)$$

крім того, за умовою задачі $m_1 = m_2$, тому:

$$m v_1 = 2mu; \qquad u = \frac{v_1}{2}$$

Кількість теплоти, що виділилася при зіткненні, дорівнюється різниці кінетичних енергій до і після зіткнення:

$$Q = (E_{k1} + E_{k2}) - E_k'$$

Запишемо кінетичну енергію тіл перед непружним ударом:

$$(E_{k1} + E_{k2}) = \frac{m_1 v_1^2}{2} + \frac{m_2 v_2^2}{2}$$

та кінетичну енергію цієї системи після зіткнення:

$$E_k' = \frac{(m_1 + m_2)u^2}{2} \quad \text{при } m_1 = m_2 \quad E_k = \frac{2mu^2}{2} = mu^2$$

$$E_{k1} + E_{k2} = \frac{m v_1^2}{2} \quad \text{при } E_{k2} = 0 \text{ (тому що } v_2 = 0)$$

Підставимо отримані рівняння в формулу і отримаємо:

$$Q = \frac{m v_1^2}{2} - mu^2 \quad \text{підставимо} \quad u = \frac{v_1}{2}$$

$$Q = \frac{m {v_1}^2}{2} - m\left(\frac{v_1}{2}\right)^2 = \frac{m {v_1}^2}{2} - \frac{m {v_1}^2}{4} = \frac{2m {v_1}^2 - m {v_1}^2}{4} = \frac{m {v_1}^2}{4}$$

$$Q = \frac{m {v_1}^2}{4} = \frac{0{,}08 \cdot 3^2}{4} = 0{,}04 \cdot 9 = 0{,}36 \text{ Дж}$$

**Відповідь:** $Q = 0{,}36$ Дж.

**Задача 1.31.** Визначте кількість теплоти, яка виділилась під час абсолютно не пружного зіткнення двох однакових кульок з масою 2 кг, якщо одна куля до зіткнення рухалась назустріч зі швидкістю $v_{01}$ =10 м/с, а друга $v_{02} = 2$ м/с.

**Дано:**

$m_1 = m_2 = 2$ кг

$v_{01} = 10 \frac{м}{с}$

$v_{02} = 2$ м/с

$E_0 > E$

**Знайти:**

$Q$–?

**Розв'язання.**

Маємо абсолютно непружний удар при зустпічнрму русі, тобто швидкість другої кулі буде записана зі знаком (-) і після зіткнення кулі стали одним цілим, тому можна записати:

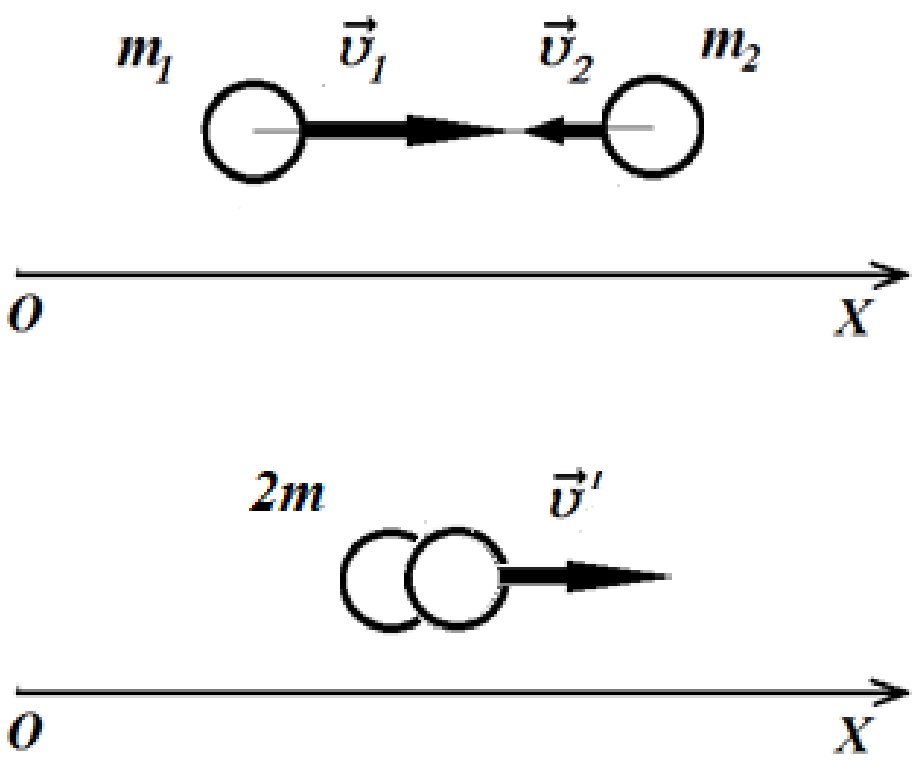

$$m_1\vec{v}_{01} - m_2\vec{v}_{02} = (m_1 + m_2)\vec{v}'$$

Із закону збереження енергії виходить, що різниця енергії перетворюється на тепло, тому:

$$E_0 - E = Q$$

Формула повної кінетичної енергії має вигляд:

до удару

$$E_0 = \frac{m_1 v_{01}^2}{2} + \frac{m_2 v_{02}^2}{2}$$

$((-\vec{v}_{02})^2$ – позитивне)

після удару

$$E = \frac{m_1 (v')^2}{2} + \frac{m_2 (v')^2}{2}$$

За умовою задачі кулі мають однакові маси і ми можемо писати просто $m$

$$m_1 = m_2 = m$$

З урахуванням вище сказаного маємо із

$$m_1\vec{v}_{01} - m_2\vec{v}_{02} = (m_1 + m_2)\vec{v}',$$

що після зіткнення швидкість кульок:

$$v' = \frac{m\vec{v}_{01} - m\vec{v}_{02}}{2m} = \frac{m(v_{01} - v_{02})}{2m} = \frac{2(10-2)}{2\cdot 2} = \frac{8}{2} = 4 \text{ м/с}$$

$$E_0 = \frac{m(v_{01}^2 + v_{02}^2)}{2} = \frac{2(10^2 + 2^2)}{2} = 104 \text{ Дж}$$

$$E = \frac{(m_1 + m_2)(v')^2}{2} = \frac{2m(v')^2}{2} = m(v')^2 = 2\cdot 4^2 = 32 \text{ Дж}$$

$$Q = E_0 - E = 104 - 32 = 72 \text{ Дж.}$$

**Відповідь:** $Q = 72$ Дж.

**Задача 1.32.** З якою швидкістю $\upsilon$ рухався вагон масою $m$=20 т, якщо при ударі о стінку буфер стиснувся на $l$=10 см? Жорсткість пружини кожного буфера $k$=1 МН/м.

| Дано: | Розв’язання: |
|---|---|
| $m=2\cdot 10^4$ кг | Жорсткість пружин з’єднаних паралельно |
| $l=0{,}1$ м | $k_{12} = k_1 + k_2 = 2k$. |
| $k=1\cdot 10^6$ Н/м | Потенційна енергія пружної взаємодії буферів зі стінкою |
| $\upsilon - ?$ | $W_п = \frac{k_{12}l^2}{2} = \frac{2kl^2}{2} = kl^2$. |

Кінетична енергія потяга, який рухався

$$W_к = \frac{m\upsilon^2}{2}.$$

По закону збереження енергії

$$W_п = W_к,\ kl^2 = \frac{m\upsilon^2}{2},$$

звідси

$$\upsilon = l\sqrt{\frac{2k}{m}} = 0{,}1\sqrt{\frac{2\cdot 10^6}{2\cdot 10^4}} = 1 \text{ м/с.}$$

**Відповідь:** $\upsilon$=1 м/с

## Задачі для самостійного розв’язання

1. Шофер автомобіля, що має масу $m$=1 т, починає гальмувати на відстані $s$=25 м від перешкоди на дорозі. Сила тертя в гальмівних колодках автомобіля $F_{тертя}$=3,84 кН. При якій граничній швидкості $\upsilon$ руху автомобіль встигне зупинитися перед перешкодою? Тертя коліс об дорогу знехтувати. [$\upsilon$=13,9 м/с].

2. З вежі висотою $h$=25 м горизонтально кинуто камінь зі швидкістю $\upsilon_0$=15 м/с. Знайти кінетичну $W_к$ і потенційну $W_п$ енергії каменю через час $t$=1 с після початку руху. Маса каменю $m$=0,2 кг. [$W_п$=39,4 Дж].

3. Тіло масою $m$=1 кг ковзає спочатку з похилої площини висотою $h$=1 м і довжиною схилу $l$=10 м, а потім по горизонтальній поверхні. Коефіцієнт тертя на всьому шляху $\mu$=0,05. Знайти: а) кінетичну енергію $W$ тіла біля основи площини; б) швидкість $\upsilon$ тіла біля основи площини; в) відстань $s$, яка пройдена тілом по горизонтальній поверхні до зупинки. [$W_к$=4,9Дж, $\upsilon$=3,1 м/с, $s$=10 м ].

4. Тіло масою $m_1$=1 кг, що рухається горизонтально зі швидкістю $\upsilon_1$=1 м/с, наздоганяє друге тіло масою $m_2$=0,5 кг і неупружньо стикається з ним. Яку швидкість $u$ отримають тіла, якщо: а) друге тіло стояло нерухомо; б) друге тіло рухалося зі швидкістю $\upsilon_2$=0,5 м/с в тому ж напрямку, що і перше тіло; в) друге тіло рухалося зі швидкістю $\upsilon_2$=0,5 м/с у напрямку, протилежному напрямку руху першого тіла. [$u_1$=0,67 м/с, $u_2$=0,87 м/с, $u_3$=0,5 м/с].

5. Куля, що летить горизонтально, потрапляє в шар, підвішений на невагомому жорсткому стержні, і застряє в ньому. Маса кулі в 1000 разів менше маси шару. Відстань від центру шару до точки підвісу стрижня $l$=1 м. Знайти швидкість кулі $\upsilon$, якщо відомо, що стрижень з шаром відхилився від удару кулі на кут $\alpha$=10$^o$. [$\upsilon$=550 м/с].

6. Вантаж масою $m$=1 кг падає на чашку терезів з висоти $H$=10 см. Які показання терезів $F$ в момент удару, якщо після заспокоєння хитань чашка терезів опускається на $h$=0,5 см? [$F$=72,5 Н].

# Розділ 2. МОЛЕКУЛЯРНА ФІЗИКА. ГАЗИ. ОСНОВИ ТЕРМОДИНАМІКИ.

## 2.1. Основи молекулярно-кінетичної теорії (МКТ).

Кількість речовини вимірюється у молях. Один моль – це кількість речовини, яка містить стільки частинок (атомів, молекул), скільки атомів міститься у масі 0,012 кг ізотопу вуглецю ${}^{12}_{6}\mathrm{C}$.

Відносною атомною масою $A_r$ хімічного елементу називають відношення маси атома $m_0$ цього елемента до 1/12 маси атома $m_{0C}$ ізотопу вуглецю ${}^{12}_{6}\mathrm{C}$, тобто це відношення маси атома до атомної одиниці маси (1одиниця=1,66$\cdot 10^{-24}$г).

$$A_r(X) = \frac{m_0}{\frac{m_{0c}}{12}} = \frac{m_a(X)}{\frac{m_a(C)}{12}} = \frac{m_a(X)}{1{,}66 \cdot 10^{-24}\,\text{г}}$$

Таким чином, відносна атомна маса $A_r$ — безрозмірна величина, яка показує, у скільки разів маса атома більше атомної одиниці маси. Відносні атомні маси хімічних елементів наведені у періодичній таблиці.

Кількість атомів або молекул в одному молі речовини, має назву числа (сталої) Авогадро $N_A$.

$$\mathrm{N_A} = 6{,}02 \cdot 10^{23}\,\text{моль}^{-1}.$$

Масу одного моля речовини називають молярною масою $M$. Молярна маса дорівнює добутку маси молекули $m_0$ на сталу Авогадро:

$$M = m_0 N_A$$

Для визначення молярної маси речовини можна також скористатися співвідношенням:

$$M = M_r \cdot 10^{-3}\text{кг/моль}$$

де $M_r$ – відносною молекулярною масою.

$M_r$ речовини називають відношення маси молекули цієї речовини (або формульної одиниці) до 1/12 маси атома ізотопу вуглецю $^{12}_{6}\text{C}$, тобто до атомної одиниці маси:

$$\text{M}_r = \frac{m_0}{\frac{m_{0c}}{12}} \quad \text{або} \quad \text{M}_r(X) = \frac{m_a(X)}{1{,}66 \cdot 10^{-24}\text{г}}$$

Відносна молекулярна маса показує, у скільки разів маса молекули або формульної одиниці більше атомної одиниці маси. Кількість речовини, тобто кількість молів ν у даній масі $m$, може бути визначена як відношення кількості молекул $N$ до числа Авогадро $N_A$:

$$\nu = \frac{N}{\text{N}_\text{A}}, \text{ або } \nu = \frac{m}{M}.$$

**Закон рівнорозподілу енергії**: середня енергія, яка припадає на одну ступінь вільності молекули, дорівнює $\frac{1}{2}kT$. Таким чином, для багатоатомної молекули середня енергія:

$$\langle \varepsilon_k \rangle = \frac{i}{2}kT$$

де $i$ - кількість ступенів свободи молекули.

**Кількість ступенів свободи молекули:**

для одноатомного газу $i = 3$;

для двохатомного $i = 5$ ($i_{\text{пост.}}=3$; $i_{\text{оберт.}}=2$);

для багатоатомного $i=6$ ($i_{\text{пост.}}=3$; $i_{\text{оберт.}}=3$).

Середня кінетична енергія поступального руху молекул ідеального газу прямо пропорційна абсолютній температурі:

$$\bar{E}_k = \frac{3}{2}kT$$

де $k$ – стала Больцмана – коефіцієнт пропорційності, який не залежить ані від температури, ані від складу та кількості газу:

$$k = 1{,}38 \cdot 10^{-23}\,\frac{\text{Дж}}{\text{К}}$$

Середня кінетична енергія обертального руху однієї молекули:

$$\langle \varepsilon_{\text{об}} \rangle = \frac{i_{\text{об}}}{2}kT$$

Середня сумарна кінетична енергія однієї молекули:

$$\langle \varepsilon \rangle = \frac{i}{2}kT,$$

де $i$ – число ступенів свободи молекули ($i = i_{\text{пост}} + i_{\text{об}}$).

**Задача 2.1.** Визначити кількість речовини і кількість молекул, що містяться в 1 кг вуглекислого газу.

| **Дано:** | **Розв'язання**. |
|---|---|
| $m$=1 кг<br>$M(CO_2)=44\cdot10^{-3}$кг/моль<br>$N_A=6{,}02\cdot10^{23}$моль$^{-1}$ | Знайдемо молекулярну масу $CO_2$ за формулою:<br>$M = M_r \cdot 10^{-3}\,\text{кг/моль}$<br>$M_r = A_r(C) + 2A_r(O) = 12 + 2\cdot16 = 44$<br>$\text{M} = M_r \cdot 10^{-3} = 44\cdot10^{-3}$ |
| **Знайти:**<br>$N$ - ? ν -? | $\nu = \frac{m}{M}$ |

кількість речовин,

$$N = \nu \cdot N_A = \frac{m}{M} \cdot N_A$$

кількість молекул,

$$\nu = \frac{1\,\text{кг}}{44\cdot10^{-3}\,\text{кг/моль}} = 23\,\text{моль}$$

$$N = 23\,\text{моль} \cdot 6{,}02 \cdot 10^{23}\text{моль}^{-1} = 1.4\ \cdot 10^{23}$$

**Відповідь:** $N = 1.4\ \cdot 10^{23}$

**Задача 2.2.** Обабослити середню кінетичну енергію $\langle\varepsilon_{об}\rangle$ обертального руху однієї молекули кисню за температури *T*=350 К і середню кінетичну енергію $\langle E\rangle$ обертального руху всіх молекул кисню, маса якого $m$ = 4 г.

**Дано:**
$m$=4 г = 4· $10^{-3}$ кг
*T*=350 К
$O_2$

**Знайти:**
$\langle\varepsilon_{об}\rangle$ –?;
$\langle E_{об}\rangle$–?

**Розв'язання.**

Формула енергії обертального руху однієї молекули:

$$\varepsilon = \frac{i}{2}kT$$

де $i$ – кількість ступенів вільності, $k = 1{,}38\cdot 10^{-23}\frac{Дж}{К}$ - стала Больцмана

При обертальному рухові двохатомної молекули маємо $i_{об} = 2$ , підставимо у формулу для енергії обертального руху однієї молекули:

$$\varepsilon = \frac{i}{2}kT$$

і отримаємо:

$$\langle\varepsilon_{об}\rangle = \frac{2}{2}kT = kT = 1{,}38\cdot 10^{-23}\cdot 350 = 483\cdot 10^{-23}\ Дж.$$

Щоб знайти енергію обертального руху всіх молекул помножимо кількість молекул *N*, що є в 4 г. на знайдену енергію обертального руху однієї молекули:

$$\langle E_{об}\rangle = \langle\varepsilon_{об}\rangle\cdot N$$

Знайдемо кількість молекул в речовині кисню масою $m$ = 4 г. за формулою:

$$N = \nu\cdot N_A$$

де $N_A$ – стала Авогадро ($N_A = 6{,}02\cdot 10^{23}$ моль$^{-1}$); $\nu$ (ню) – кількість речовини виражена в молях: $\nu = \frac{m}{M}$ , $M$ –молярна маса (маса 1 моля), яка дорівнюється:

$$M = кM_r; \quad к = 10^{-3}\ кг/моль$$

відносна молекулярна маса

$$M_r = \sum n_i A_{r,i}$$

де $n_i$ –кількість атомів; $A_{r,i}$ –атомний номер.

$$M_r(O_2) = 2 \cdot 16 = 32$$

$$M = \text{к}M_r = 32 \cdot 10^{-3} \text{ кг/моль}$$

Підставимо в формулу:

$$N = \nu \cdot N_A = \frac{m}{M} \cdot N_A = \frac{4 \cdot 10^{-3}}{32 \cdot 10^{-3}} \cdot 6{,}02 \cdot 10^{23} = 7{,}525 \cdot 10^{22} \text{ молекул}$$

Таким чином, підставивши знайдену кількість можемо знайти середню кінетичну енергію обертального руху всіх молекул кисню масою $m$=4 г.

$$\langle E_{об} \rangle = \langle \varepsilon_{об} \rangle \cdot N = 483 \cdot 10^{-23} \cdot 7{,}525 \cdot 10^{22} = 362{,}25 \text{ Дж.}$$

**Відповідь:** $\langle \varepsilon_{об} \rangle = 483 \cdot 10^{-23}$ Дж; $\langle E_{об} \rangle = 362{,}25$ Дж.

**Задача 2.3**. Чому дорівнюють середні кінетичні енергії поступального й обертального руху всіх молекул, що містяться в $m$ = 2 кг водню за температури $T$ = 400 К?

**Дано:**

$m$=2 кг

$T$=400 К

**Знайти:**

$\langle E_{пост} \rangle$–?;

$\langle E_{об} \rangle$–?

**Розв'язання.**

Середні кінетичні енергії однієї молекули:

$$\langle \varepsilon_{пост} \rangle = \frac{i_{пост}}{2} kT$$

$$\langle \varepsilon_{об} \rangle = \frac{i_{об}}{2} kT$$

Вважаємо водень ідеальним газом. Молекула водню – $H_2$ двохатомна (складається з двох атомів водню), зв'язок між атомами вважаємо жорстким, тобто коливальних ступенів вільності не враховуємо, тоді число ступенів вільності молекули водню $i$ = 5.

У середньому на одну ступінь вільності припадає енергія:

$$\langle \varepsilon_i \rangle = \frac{1}{2} kT, \qquad \text{та} \quad \langle \varepsilon_i \rangle = \frac{i}{2} kT$$

де $k$ – стала Больцмана, $T$ – термодинамічна температура.

Кількість ступенів свободи для поступального руху $i_{п}$ =3, а для обертального руху двохатомної молекули – $i_{об}$ =2 ступеням свободи. Тому, після підстановки $i$ в формулу енергії для однієї молекули маємо середні енергії:

$$\langle \varepsilon_{пост} \rangle = \frac{3}{2} kT; \qquad \langle \varepsilon_{об} \rangle = \frac{2}{2} kT$$

Щоб визначити середні енергії всіх молекул газу в 2 кг водню, треба помножити енергію однієї молекули на відповідну кількість молекул, що приходяться на 2 кг водню, тобто для 2 кг запишемо:

$$\langle \mathrm{E}_{пост} \rangle = \frac{3}{2} kT \cdot N; \quad \langle \mathrm{E}_{об} \rangle = \frac{2}{2} kT \cdot N$$

Число молекул $N$ в довільній масі газу, можна визначити за формулою:

$$N = \nu N_{\mathrm{A}}$$

де $\nu$ – число молів (кількість речовини $\nu$ в будь-якій фізичній системі визначається кількістю молей, яка міститься в ній.

Моль – кількість речовини системи, яка містить стільки ж структурних елементів, скільки атомів міститься в нукліді вуглецю-12 масою 0,012 кг), $N_{\mathrm{A}}$ – стала Авогадро ($N_{\mathrm{A}} = 6{,}02 \cdot 10^{23}$ моль$^{-1}$).

Таким чином, кількість речовини:

$$\nu = \frac{N}{N_A} \quad \text{або} \quad \nu = \frac{m}{M}.$$

Підставимо

$$\nu = \frac{m}{M}$$

в формулу:

$$N = \nu N_{\mathrm{A}} = \frac{m}{M} N_{\mathrm{A}}$$

Тоді середня кінетична енергія поступального руху всіх молекул водню буде дорівнювати:

$$\langle E_{\text{пост}} \rangle = \frac{3}{2} kT \cdot N = \frac{3}{2} kT \cdot \frac{m}{M(\text{H}_2)} N_{\text{A}};$$

де стала Больцмана $k = \frac{R}{N_A}$.

Підставимо її в формули енергії і отримаємо:

$$\langle E_{\text{пост}} \rangle = \frac{3}{2} \frac{R}{N_A} \cdot T \cdot \frac{m}{M(\text{H}_2)} N_{\text{A}} = \frac{3m}{2M(\text{H}_2)} RT$$

де $R$ – газова стала ($R = 8{,}31$ Дж / (моль · К)).

Аналогічно середня кінетична енергія обертального руху молекул водню:

$$\langle E_{\text{об}} \rangle = \frac{2}{2} kT \cdot N = kT \cdot \frac{m}{M(\text{H}_2)} \text{N}_{\text{A}}$$

$$\langle E_{\text{об}} \rangle = \frac{m}{M(\text{H}_2)} RT$$

В цих формулах невідома тільки молярна маса водно $M(\text{H}_2)$.

Молярна маса, виражена в грамах на моль, чисельно дорівнює відносній молекулярній масі

$$M = M_r k$$

де $k = 10^{-3}$ кг/моль.

Дивимось в таблицю Менделєєва знаходимо водень та його масове число дорівнює 1, але молекула водню складається з двох атомів, тому маємо $M = 2 \cdot 10^{-3}$ кг/моль. Підставляючи числові значення у ці формули, маємо:

$$\langle E_{\text{пост}} \rangle = \frac{3 \cdot 2\text{кг}}{2 \cdot 2 \cdot 10^{-3} \frac{\text{кг}}{\text{моль}}} 8{,}31 \frac{\text{Дж}}{\text{моль} \cdot \text{К}} 400\ \text{К} = 4{,}99 \cdot 10^6 \text{Дж}$$

$$\langle E_{\text{об}} \rangle = \frac{2\text{кг}}{2 \cdot 10^{-3} \frac{\text{кг}}{\text{моль}}} 8{,}31 \frac{\text{Дж}}{\text{моль} \cdot \text{К}} 400\ \text{К} = 3{,}32 \cdot 10^6 \text{Дж}$$

**Відповідь:** $\langle E_{\text{пост}} \rangle = 4{,}99 \cdot 10^6$ Дж; $\langle E_{\text{об}} \rangle = 3{,}32 \cdot 10^6$ Дж.

**Задача 2.4.** У посудині місткістю $V$ = 1 л знаходиться кисень масою $m$ = 1 г. Визначити концентрацію молекул кисню в посудині.

| **Дано:** | **Розв'язання:** |
|---|---|
| $m$=1 г = $1\cdot 10^{-3}$ кг | $N = \nu \cdot N_A$ |
| $V = 1$ л $= 0{,}001$ м$^3$ | $n = \frac{N}{V}$ |
| $O_2$ | Концентрація молекул визначається за формулою: |
| **Знайти:** | |
| $n$–? | $n = \frac{N - \text{абосло молекул}}{V - \text{об'єм}}$ |

Кількість молекул $N$ знайдемо за формулою:

$$N = \nu \cdot N_A \,,$$

де $\nu$ - кількість речовини, що виражена кількістю молей: $\nu = \frac{m}{M}$

Тобто кількість молекул визначається:

$$N = \nu \cdot N_A = \frac{m}{M} \cdot N_A,$$

де $N_A$ – стала Авогадро ($N_A = 6{,}02 \cdot 10^{23}$ моль$^{-1}$); $M$ –молярна маса (маса 1 моля), яка дорівнюється:

$$M = kM_r \,; \qquad k = 10^{-3} \text{ кг/моль}$$

відносна молекулярна маса

$$M_r = \sum n_i A_{r,i}$$

де $n_i$ –кількість атомів $i$-го хім. елемента, що входить до складу молекули даної речовини; $A_{r,i}$ – атомний номер (відносна атомна маса елемента).

$$M_r(O_2) = 2 \cdot 16 = 32$$

$$M(O_2) = kM_r(O_2) = 32 \cdot 10^{-3} \text{ кг/моль}$$

$$N = \frac{m}{M(O_2)} \cdot N_A = \frac{1 \cdot 10^{-3}}{32 \cdot 10^{-3}} \cdot 6{,}02 \cdot 10^{23} = 1{,}88 \cdot 10^{22} \text{ молекул}$$

Тепер ми можемо знайти концентрацію молекул кисню за формулою:

$$n = \frac{N}{V} = \frac{1{,}88 \cdot 10^{22}}{0{,}001} = 18{,}8 \cdot 10^{24}\ 1/\text{м}^3$$

**Відповідь:** $n = 18{,}8 \cdot 10^{24}$ $(1/\text{м}^3)$.

**Задача 2.5**. Визнааботи абосло $N$ молекул, що містяться в об'ємі $V = 1$ мм$^3$ води, і масу $m_0$ молекули води. Вважаюабо, що молекули води мають вигляд кульок, які стикаються одна з одною, визнааботи діаметр $d$ молекул.

| **Дано:** | **Розв'язання.** |
|---|---|
| $V = 1\text{мм}^3$ | $$\text{N} = \nu\text{N}_\text{A} = \frac{m}{M}\text{N}_\text{A};\ \ M = m_0 N_A$$ |
| $H_2O$ | $$d = \frac{L}{n} = \frac{L}{\sqrt[3]{N}}$$ |
| **Знайти:** | |
| $N, m_0, d$ –? | |

Число $N$ молекул, що містяться в деякій системі масою $m$ дорівнює добуткові сталої Авогадро $N_\text{A}$ на кількість речовини $\nu$:

$$\text{N} = \nu\text{N}_\text{A},$$

де $\nu = \frac{m}{M}$ підставимо і отримаємо, що

$$N = \frac{m}{M}\text{N}_\text{A}$$

де $M$ – молярна маса, $m$ – маса речовини.

По умовам в задачі наданий об'єм, тому нам треба виразити масу так, щоб в формулі був об'єм. Згадаємо, що за визначенням густина $\rho = \frac{m}{V}$.

Виразивши масу як добуток густини на об'єм $V$, одержимо $m = \rho\text{V}$ і підставивши отримаємо:

$$N = \frac{\rho\text{V}}{M}\text{N}_\text{A}$$

Знайдемо молярну масу $M(H_2O) = M(H_2) + M(O)$ , масове число визначимо по таблиці Менделєєва: для водню ($H$) це 1, а для кисню ($O$) це 16.

$$M_r(H_2O) = M_r(H_2) + M_r(O) = 2 \cdot 1 + 16 = 18$$

Зробимо обчислення числа молекул $N$ з урахуванням того, що

$M$=18·10$^{-3}$ кг/моль, а густина води $\rho = 997$ кг/м$^3 \approx 10^3$ кг/м$^3$,
$V = (1\text{мм})^3 = (10^{-3}\text{м})^3 = 10^{-9}\text{м}^3$ та число Авогадро $N_A$=6,02·$10^{23}$моль$^{-1}$:

$$N = \frac{\rho V}{M} N_A = \frac{10^3 \frac{\text{кг}}{\text{м}^3} \cdot 10^{-9}\text{м}^3}{18 \cdot 10^{-3}\text{кг/моль}} \cdot 6{,}02 \cdot 10^{23}\text{моль}^{-1} = 3{,}34 \cdot 10^{19} \text{ молекул.}$$

Маса $m_0$ однієї молекули , а молярна маса виражається через масу однієї молекули та число Авогадро.

$$M = m_0 N_A \text{ звідки } m_0 = \frac{M}{N_A} = \frac{18 \cdot 10^{-3}\text{кг/моль}}{6{,}02 \cdot 10^{23}\text{моль}^{-1}} = 2{,}99 \cdot 10^{-26}\text{кг}$$

Якщо молекули води щільно прилягають одна до одної, то можна вважати, що кожна молекула має об'єм (кубічна комірка) $V_0$=$d^3$, де $d$ – діаметр молекули, тобто діаметр молекул:

$$\text{d} = \sqrt[3]{V_0}$$

Об'єм $V_0$ знайдемо, розділивши молярний об'єм $V_m$ на число молекул у молі, тобто на $N_A$ :

$$V_0 = \frac{V_m}{N_A}$$

де $V_m$ - молярний об'єм, це відношення об'єму речовини до її кількості, який чисельно дорівнюється об'єму 1 моля речовини. Молярний об'єм також можна визначити як відношення молярної маси до густини:

$$V_m = \frac{M}{\rho}$$

де $\rho$ – густина води ($\rho$= 1000 кг/м$^3$), а молярна маса вже знайдена М=18·10$^{-3}$ кг/моль. Таким чином:

$$V_m = \frac{18 \cdot 10^{-3} \frac{\text{кг}}{\text{моль}}}{1 \cdot 10^3 \frac{\text{кг}}{\text{м}^3}} = 18 \cdot 10^{-6} \frac{\text{м}^3}{\text{моль}}$$

Підставимо отримане значення у формулу:

$$V_0 = \frac{V_m}{N_A} = \frac{18 \cdot 10^{-6} \frac{\text{м}^3}{\text{моль}}}{6{,}02 \cdot 10^{23} \text{моль}^{-1}} = 2{,}99 \cdot 10^{-29} \text{ м}^3, \quad \text{тоді}$$

$$d = \sqrt[3]{V_0} = \sqrt[3]{2{,}99 \cdot 10^{-29}} = \sqrt[3]{30 \cdot 10^{-30}} = 3{,}11 \cdot 10^{-10} \text{ м}$$

**Відповідь:** $N = 3{,}34 \cdot 10^{19}$ молекул; $m_0 = 2{,}99 \cdot 10^{-26}$кг; $d = 3{,}11 \cdot 10^{-10}$ м.

## 2.2. Газові закони, термодинамічні системи.

При вивченні зміни стану важливо встановити зв'язок між різними величинами, що характеризують властивості речовини і називаються параметрами стану. Найважливішими з них є густина або зв'язаний з нею питомий об'єм, що займає одиниця маси речовини, тиск і температура.

Основне рівняння кінетичної теорії газів має вигляд

$$p = \frac{2}{3} n \frac{m_0 \overline{\upsilon}^2}{2} = \frac{2}{3} n \overline{W}_0,$$

де $n$ – кількість молекул в одиниці об'єму, $m_0$ – маса однієї молекули, $\sqrt{\overline{\upsilon}^2}$ – середня квадратична швидкість молекул, $\overline{W}_0$ – середня кінетична енергія поступального руху однієї молекули.

Кількість молекул в одиниці об'єму

$$n = \frac{p}{kT},$$

Середня квадратична швидкість молекул

$$\sqrt{\overline{\upsilon}^2} = \sqrt{\frac{3kT}{m_0}} = \sqrt{\frac{3RT}{M}}$$

Рівняння, що зв'язує для певної маси речовини об'єм, тиск і температуру, називають термічним рівнянням стану речовини

Для з'ясування загальних властивостей, притаманних усім газам, використовують спрощену модель реальних газів, яка має назву ідеальний газ. Закони ідеального газу виведені для незмінної маси газу і розрізняють такі ізопроцеси:

$T$ = const – ізотермічний процес;

$V$ = const – ізохорний процес;

$p$ = const – ізобарний процес;

$Q$ = 0 – адіабатний процес (без теплообміну з зовнішніми тілами).

**Ізотермічний процес. Закон Бойля–Маріотта**:

Для даної маси газу $m$ = const, при $T$ = const тиск змінюється обернено пропорційно об'єму:

$$p \cdot V = \text{const}$$

Для однакової маси газу при тій самій температурі газ займає тим більший об'єм, чим нижчий тиск: $p_1 > p_2$ при $V_2 > V_1$.

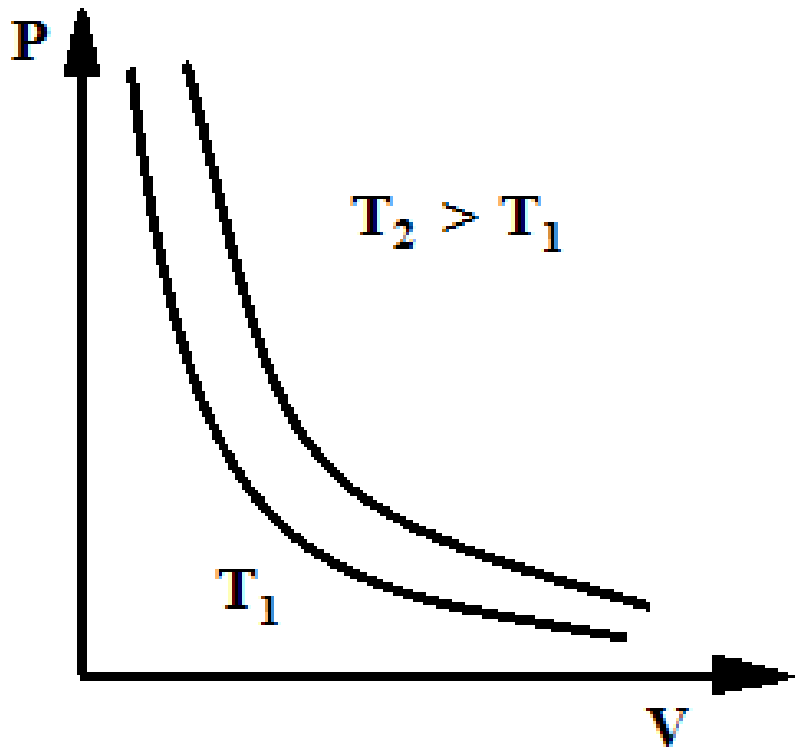

Рис. 2.1. Графік ізотермічного процесу.

**Ізобарний процес. Закон Гей–Люссака**.

Для даної маси газу $m$ = const, при $p$ = const об'єм зростає лінійно з температурою:

$$V = V_0(1 + \alpha t);\ \ V = \alpha V_0(273 + t) = \alpha V_0 T$$

$$\text{або} \quad \frac{V}{T} = \text{const}.$$

де $V_0$ – об'єм газу при $t$=0°C; α = (1/273) $град^{-1}$ – температурний коефіцієнт об'єму (температурний коефіцієнт теплового поширення).

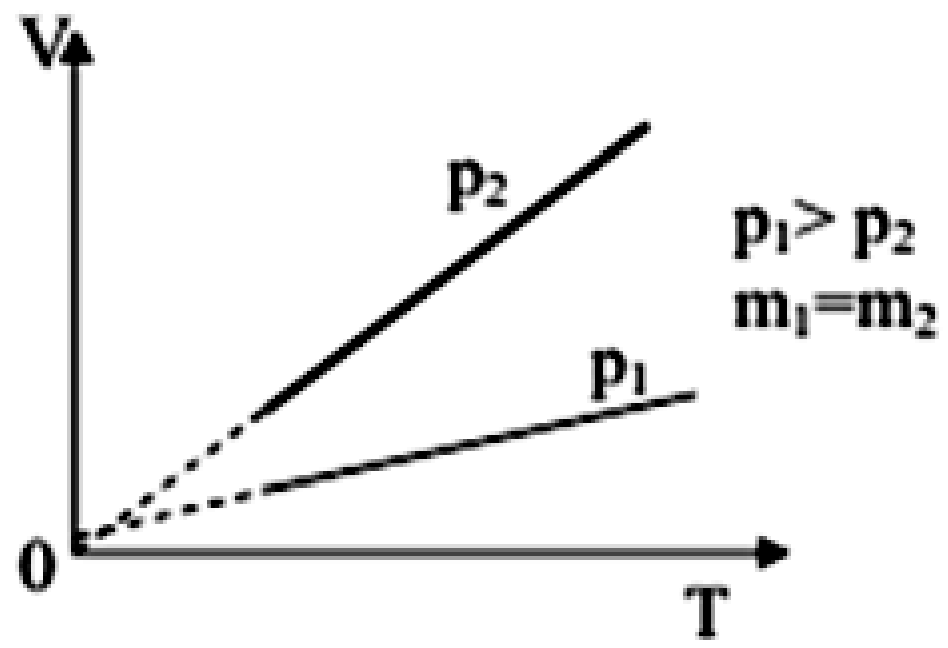

Рис. 2.2. Графік ізобарного процесу.

**Ізохорний процес: Закон Шарля.**

Для даної маси газу $m$ = const, при $V$ = const тиск зростає лінійно з температурою. При $m$ = const, $V$ = const маємо рівняння для цього процесу:

$$p = p_0(1 + \gamma t),$$

де $p_0$ – тиск газу при $t$=0°C; γ = (1/273) $град^{-1}$ – термічний коефіцієнт тиску.

Інакше:

$$p = \gamma p_0 T\,, \qquad \text{або} \quad \frac{p}{T} = const$$

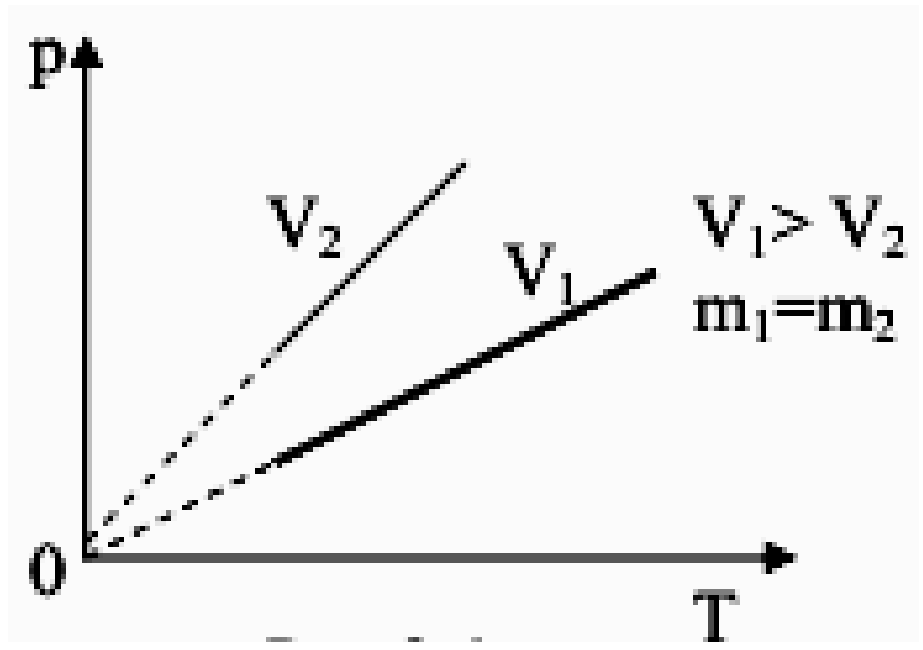

Рис. 2.3. Графік ізохорного процесу.

Об'єднання законів Бойля–Маріотта і Гей–Люссака приводить до рівняння стану ідеального газу (рівняння Клапейрона), яке зв'язує всі основні його параметри:

$$\frac{pV}{T} = \text{const при } m = \text{const}.$$

Для довільної маси рівняння стану ідеального газу - рівняння Мендєлєєва–Клапейрона:

$$pV = \nu RT,$$

де $\nu = \frac{m}{M}$ – кількість молів газу, $M$ – молярна маса.

Інакше:

$$p = \frac{\rho RT}{M}, \qquad \text{де } \rho - \text{густина газу} \quad \rho = \frac{m}{V}.$$

З рівняння Менделєєва – Клапейрона можна отримати кожний з газових законів. Рівняння Менделєєва – Клапейрона можна записати в іншій формі, якщо виразити $\rho$ і $M$ у вигляді

$$\rho = m_0 n; \;\; M = m_0 NA$$

де $m_0$ – маса молекули; $N_{\text{А}}$ – число Авогадро; $n$ – чисельна густина молекул газу) і використати сталу Больцмана:

$$k = \frac{R}{N_A} = 1{,}38 \cdot 10^{-23} \text{Дж/К},$$

отримаємо рівняння:

$$p = nkT.$$

Стала Больцмана є числовим еквівалентом, який зв'язує абсолютну температуру, виміряну в одиницях енергії, з температурою виміряною в кельвінах, де $R$ = 8,31 Дж/(моль·К) – універсальна газова стала, яка чисельно дорівнює роботі поширення, яку виконує один моль ідеального газу при нагріванні на 1 К в ізобарному процесі та $N_{\text{А}} = 6{,}02 \cdot 10^{23}$ моль$^{-1}$ – кількість атомів або молекул в одному молі речовини, має назву числа (сталої) Авогадро $N_{\text{А}}$.

Внутрішня енергія системи – сума кінетичної енергії хаотичного руху частинок, які входять до неї, і потенціальної енергії їх взаємодії. Тобто енергія теплового руху молекул:

$$U = \frac{i}{2}\frac{m}{M}RT$$

де $i -$ число ступенів вільності молекули, $m -$ маса газу, $M -$ молярна маса даної речовини, $T -$ абсолютна температура, $R -$ універсальна газова стала ($R = kN_A$).

Перший закон термодинаміки – це закон збереження енергії при її перетвореннях в системах, які складаються з великої кількості частинок. Перший закон термодинаміки може бути сформульований так: елементарна кількість теплоти, що надана системі йде на зміну її внутрішньої енергії і на виконання системою роботи проти зовнішніх сил. Для кінцевих змін параметрів системи, тобто для переходу системи з одного стану в інший стан маємо рівняння:

$$Q = \Delta U + A.$$

**Ізохорний процес**.

У ході цього процесу об'єм газу не змінюється ($\Delta V = 0$) і газ роботу не виконує ($A = 0$), тому рівняння першого закону термодинаміки має вигляд:

$$Q = \Delta U.$$

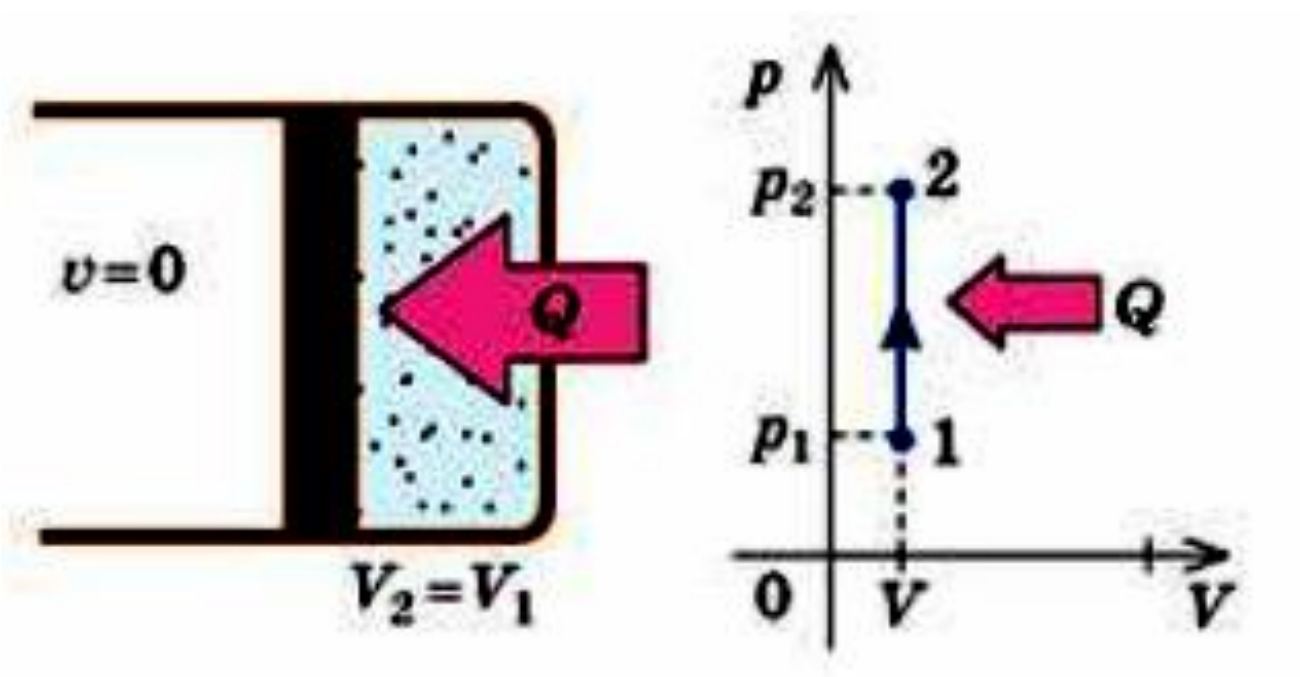

$$m = const; V = const; Q = \Delta U$$

Рис. 2.4. Ізохорний процес, з точки зору першого закону термодинаміки.

При ізохорному процесі вся передана газу кількість теплоти витрачається на збільшення внутрішньої енергії газу. Якщо газ ідеальний одноатомний, то кількість теплоти, передана газу, дорівнює:

$$Q = \Delta U = \frac{3}{2}\frac{m}{M}R\Delta T = \frac{3}{2}V\Delta p$$

**Ізотермічний процес**.

У ході цього процесу температура, а отже, і внутрішня енергія газу не змінюється ($\Delta U = 0$), тому рівняння першого закону термодинаміки має вигляд:

$$Q = A.$$

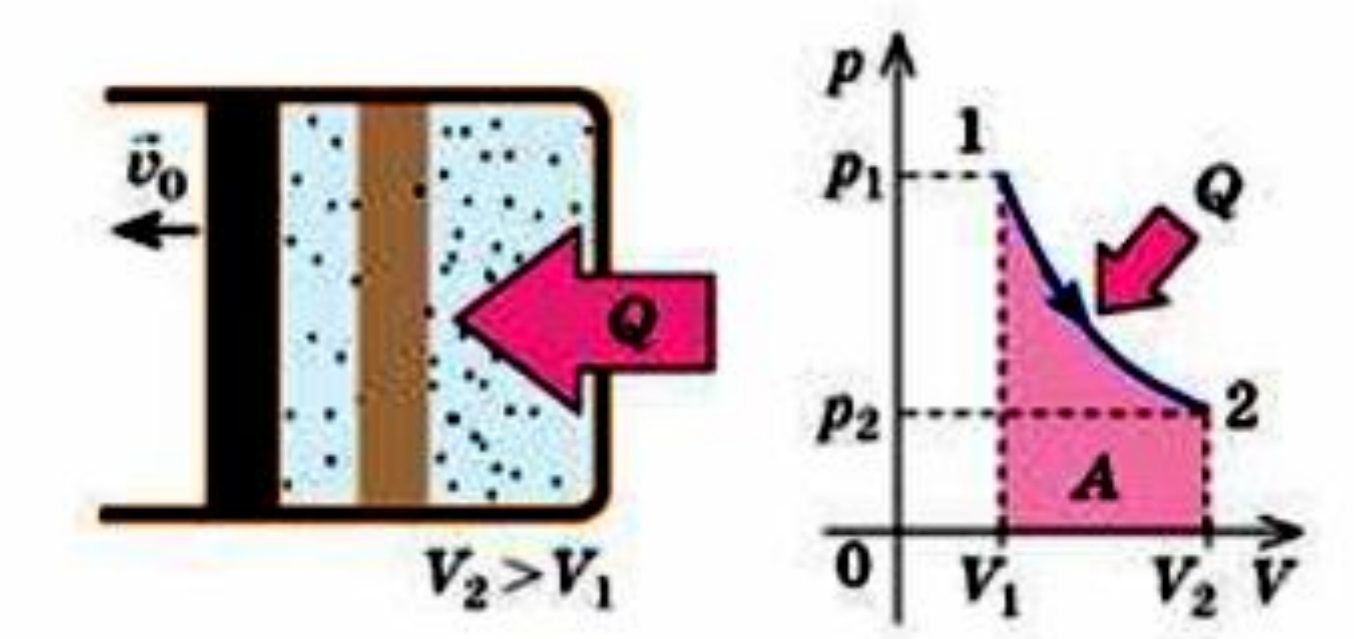

Рис.2.5. Ізотермічний процес, з точки зору першого закону термодинаміки.

При ізотермічному процесі вся передана газу кількість теплоти йде на виконання механічної роботи.

Робота яка виконується при ізотермічній зміні об'єму газу

$$A_{із} = RT\frac{m}{M}\ln\frac{V_2}{V_1}.$$

**Ізобарний процес**.

У ході цього процесу виконується робота і змінюється внутрішня енергія газу, тому рівняння першого закону термодинаміки має вигляд:

$$Q = \Delta U + A.$$

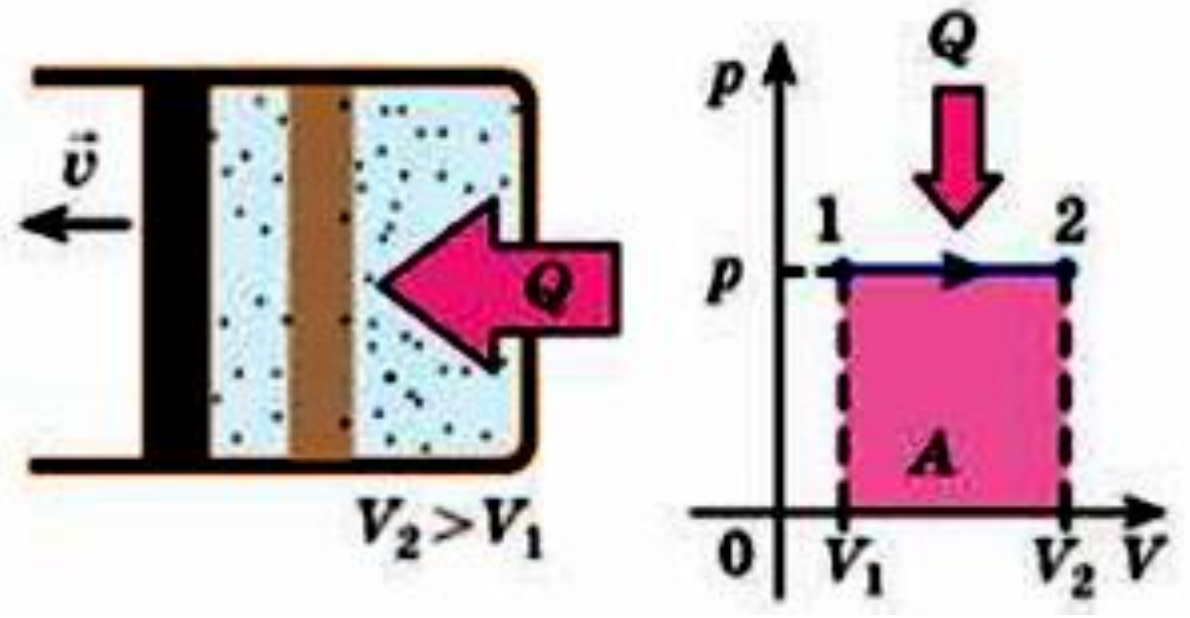

$$m = const; p = const; Q = \Delta U + A$$

Рис.2.6. Ізобарний процес, з точки зору першого закону термодинаміки.

При ізобарному процесі передана газу кількість теплоти йде і на збільшення внутрішньої енергії газу, і на виконання механічної роботи. При одноатомному ідеальному газі робота дорівнює:

$$\mathrm{A} = p\Delta V$$

зміна внутрішньої енергії

$$\Delta \mathrm{U} = \frac{3}{2}\mathrm{p}\Delta V.$$

Кількість теплоти, передана газу, дорівнює:

$$\mathrm{Q} = \Delta \mathrm{U} + \mathrm{A} = \frac{3}{2}\mathrm{p}\Delta V + \mathrm{p}\Delta V = \frac{5}{2}\mathrm{p}\Delta V, \quad \text{або} \quad \mathrm{Q} = \frac{5}{2}\frac{m}{M}R\Delta \mathrm{T}.$$

**Адіабатний процес** — це процес, який відбувається без теплообміну з навколишнім середовищем. При адіабатному процесі кількість теплоти $Q$, передана системі, дорівнює нулю, тому перший закон термодинаміки має вигляд: $\Delta U + A = 0$, або $\mathrm{A} = -\Delta U$, тобто збільшення внутрішньої енергії системи відбувається за рахунок виконання над газом роботи (адіабатне стиснення): $\Delta U = \mathrm{A}'$. Якщо ж газ сам виконує роботу (адіабатне розширення), то його внутрішня енергія зменшується: $A = -\Delta U$.

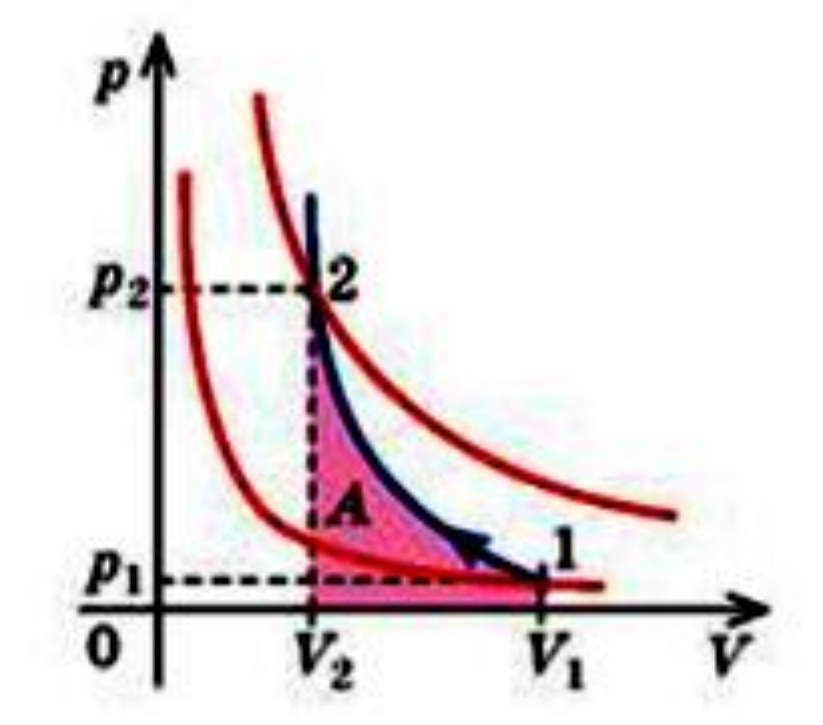

Рис. 2.7. Графіки ізопроцесів, де синій колір - адіабата, червоний - ізотерма.

Таким чином, у ході адіабатного розширення газ виконує додатну роботу за рахунок зменшення внутрішньої енергії, при цьому температура газу зменшується.

**Задача 2.6.** Якої довжини $\ell$ вийде ланцюжок із молекул кисню, які знаходяться за нормальних умов в об'ємі $V = 2$ см$^3$? Діаметр молекул кисню $d$ = 0,3 нм.

| Дано: | Розв'язання. |
|---|---|
| $V = 2\text{ см}^3 = 2 \cdot 10^{-6}\text{ м}^3$ | $\ell = d \cdot N$ |
| $T_0 = 273\text{ К}$ | $N = n \cdot V$ |
| $p_0 = 1{,}013 \cdot 10^5\text{ Па}$ | За умовою задачі кисень знаходиться за нормальних умов, що означає при температурі $T_0$ =0°С= $273\text{ К}$ і тиску $p_0$ =1 атмосфери= $1{,}013 \cdot 10^5\text{ Па}$. Щоб знайти довжину ланцюжка $\ell$ за формулою $\ell = d \cdot N$ треба визнааботи кількість молекул: |
| $d = 0{,}3\text{ нм} = 0{,}3 \cdot 10^{-9}\text{ м}$ | |
| $O_2$ | |
| **Знайти:** | |
| $\ell$ - ? | |

$$N = nV$$

де $n$ – концентрація молекул, яку можна визнааботи за рівнянням стану ідеального газу (рівняння Менделеева-Клайперона) $p = nkT$ і отримаємо:

$$n = \frac{p_0}{kT_0}$$

де $k$ – стала Больцмана: $k = \frac{R}{N_A} = 1{,}38 \cdot 10^{-23}\text{Дж/К}$.

Підставимо вираз для концентрації в формулу для визначення кількості і отримаємо, що:

$$\ell = d\frac{p_0}{kT_0}V = \frac{0{,}3 \cdot 10^{-9} \cdot 1{,}013 \cdot 10^5 \cdot 2 \cdot 10^{-6}}{1{,}38 \cdot 10^{-23} \cdot 273} = 1{,}6 \cdot 10^{10}\text{ м}$$

**Відповідь:** $\ell = 1{,}6 \cdot 10^{10}$ м

**Задача 2.7.** При температурі $T_1$= 280 К об'єм газу $V_1$= 49 л. Визначить, яким стане об'єм газу $V_2$ після нагріву до $T_2$= 360 К і збільшенні тиску в 3 рази.

| **Дано:** | **Розв'язання.** |
|---|---|
| $T_1$= 280 К<br>$T_2$= 360 К<br>$V_1$= 49 л = 0,049 м$^3$<br>$P_2 = 3P_1$ | Для довільної маси рівняння стану ідеального газу:<br>$$pV = \frac{m}{M}RT$$<br>(рівняння Мендєлєєва–Клапейрона)<br>Запишемо рівняння ідеального газу для першого стану, тобто при температурі $T_1$= 280 К: |
| **Знайти:**<br>$V_2$ - ? | |

$$P_1 V_1 = \frac{m}{M} R T_1$$

І для другого стану при температурі $T_2$= 360 К:

$$P_2 V_2 = \frac{m}{M} R T_2$$

Перетворимо ці рівняння так, щоб змінні велиабони – тиск, об'єм, температура були з однієї сторони рівняння, а ті, що є незмінні з другої:

$$\frac{P_1 V_1}{T_1} = \frac{m}{M} R \quad \text{і} \quad \frac{P_2 V_2}{T_2} = \frac{m}{M} R$$

$$\Rightarrow \quad \frac{P_1 V_1}{T_1} = \frac{P_2 V_2}{T_2} \text{ виразимо } V_2$$

$$V_2 = \frac{P_1 V_1 T_2}{P_2 T_1} \text{ за умовою задачі } P_2 = 3P_1$$

$$V_2 = \frac{P_1 V_1 T_2}{3P_1 T_1} = \frac{V_1 T_2}{3T_1} = \frac{0{,}049 \cdot 360}{3 \cdot 280} = \frac{176{,}4}{84} = 2{,}1 \text{ м}^3$$

**Відповідь:** $V_2 = 2{,}1$ м$^3$.

**Задача 2.8.** Якщо температуру газу, що знаходиться в балоні, збільшити у 3 рази, то як зміниться густина газу?

| **Дано:** | **Розв'язання.** |
|---|---|
| $T_1;\ T_2 = 3T_1$ | Густина |
| $m$ | $$\rho = \frac{m\ (\text{маса})}{V\ (\text{об'єм})}$$ |
| **Знайти:** $\rho - ?$ | |

Щоб зрозуміти, як буде поводитись густина газу в балоні з підвищенням температури в 3 рази, треба подивитись від чого залежить густина. За формулою для визначення густини:

$$\rho = \frac{m}{V}$$

В абосельнику знаходиться маса газу, а за умовою задачі вона не змінюється. В знаменнику знаходиться об'єм газу. Відомо, що підвищення температури газу призводить до його розширення, тобто до зміни об'єму. В нашому випадку газ знаходиться в балоні, тобто об'єм не може збільшитись, тому зміна температури призведе до підвищення тиску. Тиск це є характеристика середньої кінетичної енергії поступального руху молекул газу. Тиск газу на стінки посудини виникає внаслідок багаторазових зіткнень молекул газу зі стінками. Кожна молекула, вдаряюабось об стінку, передає їй імпульс. Сумарний імпульс, переданий молекулами за одиницю часу на одиницю площі, і є тиском.

В задачі запитання йде про густину або інакше щільність, а це зовсім інша характеристика газу. Густина газу показує, наскільки "щільно" упаковані молекули газу в певному об'ємі. густина газів сильно залежить від:

1. Температури: при підвищенні температури (за постійного тиску) молекули газу починають рухатися швидше, віддаляються одна від одної, і об'єм газу збільшується, що призводить до зменшення густини.

2. Тиску: при підвищенні тиску (за постійної температури) молекули газу "стискаються" у менший об'єм, що призводить до збільшення густини.

Тобто збільшення температури може призвести до зменшення густини внаслідок збільшення об'єму. З підвищенням тиску густина може збільшитись, але також при цьому повинен змінитись, а саме зменшитись, об'єм. Таким чином, щоб змінилась густина повинен змінитися об'єм.

В умовах задачі сказано «газ, що знаходиться в балоні», а об'єм балону це величина, яка є незмінною, тобто густина газу з підвищенням температури не зміниться. Зміниться тільки тиск.

**Відповідь:** $\rho$ не зміниться.

**Задача 2.9.** При випуску з балона частини газу його температура зменшилась у $n$ разів, тиск – у $k$ разів. Із балона була випущена частина газу $\frac{\Delta m}{m} = \frac{m-m_1}{m}$, чому вона буде дорівнюватись?

**Дано:**

$T_1;\ T_2 = \frac{1}{n}T_1$

$m$

$P_1;\ P_2 = \frac{1}{k}P_1$

**Знайти:**

$\Delta m-?$

**Розв'язання.**

По рівнянню Менделеева-Клайперона

$$PV = \nu RT$$

де $\nu$ – кількість молів $\nu = \frac{m}{M-\text{молярна маса}}$, тобто:

$$PV = \frac{m}{M}RT.$$

Наданий вираз частини газу, що був випущений, в умовах задачі представлений як: $\frac{\Delta m}{m} = \frac{m-m_1}{m}$, тобто масу для зміненого стану газу можна представити як:

$$\Delta m = m - m_1$$

Підставимо параметри, що зазнали змін у рівняння стану ідеального газу:

$$PV = \frac{m}{M}RT$$

і отримаємо новий вираз:

$$\frac{1}{k}PV = \frac{\Delta m}{M}R\frac{1}{n}T$$

де $\left(\frac{1}{k}\right)$ – зменшення тиску у $k$ разів по умовам задачі та $\left(\frac{1}{n}\right)$ – зменшення температури в $n$ разів. В задачі запитується чому буде дорівнюватись випущена частина газу $\frac{\Delta m}{m}$.

Для початкового стану газу, до зміни маси, можно записати вираз для маси $m$ через рівняння Менделеева-Клайперона:

$$PV = \frac{m}{M}RT; \quad MPV = mRT \quad \Rightarrow \; m = \frac{MPV}{RT}$$

Аналогічно знайдемо вираз для $\Delta m$ із:

$$\frac{1}{k}PV = \frac{\Delta m}{M}R\frac{1}{n}T; \quad nMPV = \Delta mRkT \quad \Rightarrow$$

$$\Delta m = \frac{nMPV}{kRT}$$

Знайдемо відношення $\frac{\Delta m}{m}$:

$$\frac{\Delta m}{m} = \frac{\frac{nMPV}{kRT}}{\frac{MPV}{RT}} = \frac{nMPV}{kRT} \cdot \frac{RT}{MPV} = \frac{n}{k}$$

**Відповідь:** $\frac{\Delta m}{m} = \frac{n}{k}$.

**Задача 2.10.** У посудині об’ємом 5 л міститься газ при тиску 200 кПа і температурі 17°С. Під час ізобарного розширення газом була виконана робота 196 Дж. На скільки градусів нагрівся газ.

| **Дано:** | **Розв’язання:** |
|---|---|
| $V$ = 5 л = 5· $0{,}001$ м$^3$<br>$p$=200 кПа= $200 \cdot 10^3 \frac{\text{Н}}{\text{м}^2}$<br>$t$ = 17°С=17+273=290 К<br>$A$ = 196 Дж | В умовах задачі сказано, що робота була виконана під час ізобарного процесу, тобто $p = const$. При ізобарному процесі для пошуку роботи треба використовувати наступну формулу: |
| **Знайти:**<br>$\Delta T$–? | $$A = p(V_2 - V_1)$$ |

Перетворимо формулу $A = p(V_2 - V_1)$ так, щоб виразити $V_2$

$$V_2 - V_1 = \frac{\text{A}}{p}; \qquad \text{де за умовою } V_1 = 5 \cdot 10^{-3} \text{ м}^3$$

$$V_2 = \frac{\text{A}}{p} + V_1 = \frac{196\,[\text{Н} \cdot \text{м}]}{200 \cdot 10^3 \left[\frac{\text{Н}}{\text{м}^2}\right]} + 5 \cdot 10^{-3}\ [\text{м}^3] = 5{,}96 \cdot 10^{-3}\ \text{м}^3$$

Щоб знайти температуру газу після нагріву розглянемо закон Гей-Люссака, тобто ізобарний процес $p = const$ при $m = const.$

$$\frac{V}{T} = \text{const} \ \text{ або } \ \frac{V_1}{T_1} = \frac{V_2}{T_2}$$

Запишемо цей закон для стану 1 і стану 2 та виразимо $T_2$:

$$\frac{V_2}{T_2} = \frac{V_1}{T_1} \ \text{ перетворимо на } V_2 T_1 = V_1 T_2$$

$$T_2 = \frac{V_2 T_1}{V_1} = \frac{5{,}96 \cdot 10^{-3} \cdot 290}{5 \cdot 10^{-3}} = 345{,}68\,\text{К} - 273 \approx 73^\text{o}\text{C}$$

**Задача 2.11.** Об'єм газу при температурі $t_1 = 55^\text{o}\text{C}$ становить $V_1$ = 300 дм$^3$. Якщо тиск зменшити в три рази, то газ займе об'єм $V_2$ = 750 дм$^3$ при температурі $t_2$. Визнаaботи цю температуру.

**Дано:**

$\text{T}_1 = 55^\text{o}\text{C} + 273 = 328\ \text{К}$

$\text{V}_1 = 300\ \text{дм}^3 = 0{,}3\ \text{м}^3$

$\text{V}_2 = 750\ \text{дм}^3 = 0{,}75\ \text{м}^3$

T(K) = 273 + $t^o C$

$\text{P}_1 = 3\text{P}_2$

**Знайти:**

$t_2$–?

**Розв'язання.**

Запишемо рівняння ідеального газу для першого стану, тобто при $\text{T}_1 = 328$ К і для другого стану при $\text{T}_2$:

$$P_1 \text{V}_1 = \frac{m}{M} \text{R} T_1 \quad \text{і} \quad P_2 \text{V}_2 = \frac{m}{M} \text{R} T_2$$

Перетворимо ці рівняння так, щоб змінні велиабони – тиск, об'єм, температура були з однієї сторони рівняння, а ті, що є незмінні з другої:

$$\frac{P_1 \text{V}_1}{T_1} = \frac{m}{M} \text{R} \quad \text{і} \quad \frac{P_2 \text{V}_2}{T_2} = \frac{m}{M} \text{R}$$

$$\Longrightarrow \frac{P_1 V_1}{T_1} = \frac{P_2 V_2}{T_2} \text{ виразимо } T_2, \text{пам}'\text{ятаюабо, що } P_1 = 3P_2$$

$$T_2 = \frac{P_2 V_2 T_1}{P_1 V_1} = \frac{P_2 V_2 T_1}{3P_2 V_1} = \frac{V_2 T_1}{3V_1} = \frac{0{,}75 \cdot 328}{3 \cdot 0{,}3} = 273{,}3\ \text{К}$$

$$t_2 = 273{,}3 - 273 = 0{,}3^o C$$

**Відповідь:** $t_2 = 0{,}3^o C$.

**Задача 2.12**. У балоні ємності $V$ = 40 л міститься азот при температурі $t_1$= 57°C і тиску $p_1$= 0,3·МПа. Частину газу з балона випустили. При цьому тиск не змінився, а температура підвищилася до $t_2$ = 87°C. Яку масу $\Delta m$ газу випустили?

**Дано:**

$V_1 = 40\ \text{л} = 4 \cdot 10^{-2}\ \text{м}^3$

$T_1 = 330\ \text{К}$

$p_1 = 0{,}3\ \text{Мпа} = 0{,}3 \cdot 10^6 \text{Па}$

$T_2 = 360\ \text{К}$

$p_2 = p_1$

**Знайти:**

$\Delta m$–?

**Розв'язання**

Рівняння стану ідеального газу:

$$pV = \frac{m}{M} RT$$

Запишемо рівняння ідеального газу для першого стану:

$$p_1 V_1 = \frac{m}{M} R T_1$$

У другому стані частину газу з балона випустили, тобто кількість газу стала на $\Delta m$ менше, тому для 2 стану маємо:

$$p_2 V_2 = \frac{m - \Delta m}{M} R T_2$$

За умовою задачі зміна стану здійснилась при $p$=const, крім того за умовою задачі $V_1 = V_2$ (балон ємністю $V$ = 40 л). Тобто $p_1 = p_2$ і $V_1 = V_2$ отже

$$p_1 V_1 = p_2 V_2$$

Поглянемо на ліву частину рівнянь стану ідеального газу, які є рівні, то і праві частині рівні:

$$\frac{m}{M}RT_1 = \frac{m - \Delta m}{M}RT_2 \quad \text{і виразимо } \Delta m$$

$$mT_1 = (m - \Delta m)T_2$$

$$\frac{mT_1}{T_2} = m - \Delta m; \qquad \Delta m = m - \frac{mT_1}{T_2};$$

$$\Delta m = m\left(1 - \frac{T_1}{T_2}\right) \quad \text{або} \quad \Delta m = m\left(\frac{T_2 - T_1}{T_2}\right)$$

В отриманому рівнянні для знаходження $\Delta m$ не відоме значення початкової $m$.

Так саме виразимо $m$ із рівняння для першого стану

$$\mathrm{p}_1\mathrm{V}_1 = \frac{m}{M}\mathrm{R}T_1$$

і отримаємо:

$$m = \frac{p_1 V_1 M}{RT_1},$$

де газова стала $R = 8{,}31$ Дж/моль · К

Отриманий вираз підставимо в виведене рівняння для $\Delta m$:

$$\Delta m = \frac{p_1 V_1 M}{RT_1}\left(\frac{T_2 - T_1}{T_2}\right) = \frac{p_1 V_1 M}{RT_1 T_2}(T_2 - T_1)$$

Молярну масу азоту $M(N_2)$ знайдемо за формулою: $\mathrm{M} = M_r \cdot 10^{-3}$

($M_r = nA_r$; $n -$ кількість атомів і молекулах; $A_r$ – атомна маса)

Підставимо дані для азоту ($n = 2,\ A_r\,(\mathrm{N}_2) = 14$)

$$\mathrm{M(N_2)} = 2 \cdot 14 \cdot 10^{-3} = 28 \cdot 10^{-3}\ \text{кг/моль}$$

$$\Delta m = \frac{p_1 V_1 M}{RT_1 T_2}(T_2 - T_1) = \frac{0{,}3 \cdot 10^6 \cdot 4 \cdot 10^{-2} \cdot 28 \cdot 10^{-3}}{8{,}31 \cdot 330 \cdot 360}(360 - 330) =$$

$$= \frac{336 \cdot 30}{987228} = \frac{10080}{987228} = 0{,}01\ \text{кг}$$

**Відповідь:** $\Delta m = 0{,}01$ кг.

**Задача 2.13**. Кисень нагрівається при незмінному тиску 80 кПа. Його об'єм збільшується від 1 м$^3$ до 3 м$^3$. Знайти: 1) зміну внутрішньої енергії кисню; 2) виконану при розширенні роботу; 3) надану газу кількість теплоти.

| **Дано:** | **Розв'язання.** |
|---|---|
| $p$=80 кПа=8$\cdot$ 10$^4$ Па<br>$V_1$= 1 м$^3$<br>$V_2$= 3 м$^3$ | При $p = \mathrm{const}$ перший закон термодинаміки матиме вигляд: $$Q = A + \Delta U$$ |
| **Знайти:**<br>$Q$ - ? | Виконана газом робота може бути знайдена за формулою: |

$$A = p(V_2 - V_1) = 8 \cdot 10^4\ \text{Па} \cdot (3\ \text{м}^3 - 1\ \text{м}^3)$$
$$= 160 \cdot 103\ \text{Дж}$$

Зміна внутрішньої енергії задається виразом:

$$\Delta U = \frac{i}{2}\frac{m}{M}R\Delta T$$

де $i$ — кількість ступенів вільності, у двохатомної молекули $i = 5$.

Розглянемо рівняння стану ідеального газу (рівняння Клайперона-Менделєєва) для довільної маси газу:

$$pV = \nu RT \quad \text{або} \quad pV = \frac{m}{M}RT$$

Запишемо це рівняння для двох станів:

$$pV_1 = \frac{m}{M}RT_1 \qquad pV_2 = \frac{m}{M}RT_2$$

знайдемо їх різницю:

$$pV_2 - pV_1 = \frac{m}{M}RT_2 - \frac{m}{M}RT_1$$
$$p(V_2 - V_1) = \frac{m}{M}R(T_2 - T_1)$$

Поглянемо на отриману формулу і порівняємо її з рівнянням

$$A = p(V_2 - V_1).$$

Бачимо, що ліві частини отриманих рівнянь є однакові, тобто дорівнюються $p(V_2 - V_1)$. Подивимось на рівняння:

$$\Delta U = \frac{i}{2}\frac{m}{M}R\Delta T$$

Замість лівої частини можемо записати:

$$A = \frac{m}{M}R(T_2 - T_1)$$

і порівнявши з формулою для зміни внутрішньої енергії, зробимо заміну $\frac{m}{M}R\Delta T$ на А:

$$\Delta U = \frac{i}{2}\frac{m}{M}R\Delta T = \frac{i}{2}A$$

$$\Delta U = \frac{5}{2}160 \cdot 10^3 \text{ Дж} = 400 \cdot 10^3 \text{Дж.}$$

$$Q = A + \Delta U = 160 \cdot 10^3 \text{ Дж} + 400 \cdot 10^3 \text{Дж.} = 560 \cdot 10^3 \text{Дж.}$$

**Відповідь**: $A$ = 160 $\cdot 10^3$ Дж.; $\Delta U = 400 \cdot 10^3$Дж.; $Q$ =$560 \cdot 10^3$Дж.

**Задача 2.14**. Під час нагрівання ідеального газу на $\Delta T = 1$ К при сталому тиску об'єм його збільшився на 1/350 первісного об'єму. Визначити початкову температуру $T_1$ газу.

| **Дано:** | **Розв'язання:** |
|---|---|
| $\Delta T = 1$ К<br>$\Delta V = \frac{1}{350}V_1$ | Запишемо закон Гей-Люссака для стану *1* і стану *2*:<br>$\frac{V_1}{T_1} = \frac{V_2}{T_2}$ |
| **Знайти:**<br>$T_1 - ?$ | За умовою задачі нам дано $\Delta T = 1$ К або інакше<br>$T_2 - T_1 = 1$ К |

$$T_2 = 1 + T_1$$

Також дано, що $\Delta V$ =1/350 $V_1$ або $V_2 - V_1 = \frac{1}{350} V_1$

$$V_2 = \frac{1}{350} V_1 + V_1 = \frac{351}{350}V_1$$

Підставимо в першу формулу перетворений запис наданих умов:

$$\frac{V_2}{T_2} = \frac{V_1}{T_1}; \qquad \frac{\frac{351}{350}V_1}{1 + T_1} = \frac{V_1}{T_1}$$

$$V_1(1 + T_1) = T_1 \frac{351}{350} V_1$$

$$1 + T_1 = T_1 \frac{351}{350}$$

$$\frac{351}{350} T_1 - T_1 = 1;$$

$$\frac{T_1}{350} = 1; \qquad T_1 = 350\,\text{К}$$

**Відповідь:** $T_1 = 350\,\text{К}$.

### 2.3. Кругові процеси (цикли). Теплові машини, коефіцієнт корисної дії (ККД) теплової машини.

Процес, який складається із неперервної черги рівноважних станів, називають рівноважним або квазістатичним. Рівноважним може бути тільки нескінченно повільний процес.

Термодинамічний процес, який здійснює система, називають **оборотним**, якщо після нього можливо повернути систему і все навколишнє середовище у початковий стан без будь-яких змін. Якщо процес не задовольняє цю умову, то він є необоротним. Необхідною умовою оборотності процесів є їх рівноважність (квазістатичність).

**Коловим** процесом або **циклом** називають сукупність декількох термодинамічних процесів, внаслідок яких система повертається в початковий стан. Колові процеси на діаграмах станів (у координатах *p-V, p-T, V-T*) зображується замкненими кривими.

З трьох агрегатних станів речовини найбільш простим є газоподібний стан, оскільки в цьому випадку сили, що діють між молекулами вельми малі і ними за певних умов можна знехтувати. Для пояснення властивостей речовини у газоподібному стані використовують фізичну модель – ідеальний газ. **Ідеальним газом** називають такий газ, в якому відсутні сили міжмолекулярної взаємодії.

Тепловою машиною називається будь-який періодично діюабой пристрій, який виробляє роботу за рахунок одержуваної ззовні теплоти. Прямим круговим процесом (циклом теплової машини) називається цикл, в якому отримана ззовні теплота перетворюється в корисну роботу. Зворотним круговим процесом (циклом холодильної машини) називається цикл, в якому отримана ззовні робота витрачається на перенесення теплоти від менш нагрітих тіл до більш нагрітих тіл.

Циклом Карно називається цикл теплової машини, який пов'язаний тільки з двома тепловими резервуарами: нагрівачем і холодильником. Цикл Карно складається з двох рівноважних ізотермічних процесів і двох рівноважних адіабатичних процесів. Як робоче тіло використовується ідеальний газ.

Термодинамічний коефіцієнт корисної дії (ККД) теплової машини

$$\eta = \frac{A}{Q_1} = \frac{Q_1 - Q_2}{Q_1}$$

де $Q_1$ – теплота, яку отримало робоче тіло машини за один цикл; $Q_2$ – теплота, яку робоче тіло віддало за цикл; $A = Q_1 - Q_2$ – корисна робота, яку виконало робоче тіло за цикл.

У реальних теплових машинах ККД обмежений нерівністю Карно

$$\eta \leq 1 - \frac{T_1}{T_2}$$

де $T_1$ – температура нагрівника, $T_2$ – температура холодильника, тобто максимальна і мінімальна температури робочої речовини за цикл.

Для циклу Карно:

$$\eta = \frac{T_1 - T_2}{T_1}$$

**Задача 2.15.** На рис. зображений в координатах $p$ і $T$ графік замкненого процесу, який відбувся з газом незмінної маси. Визначте, який вигляд має графік даного циклу в координатах $p$ і $V$ серед наданих варіантов відповіді.

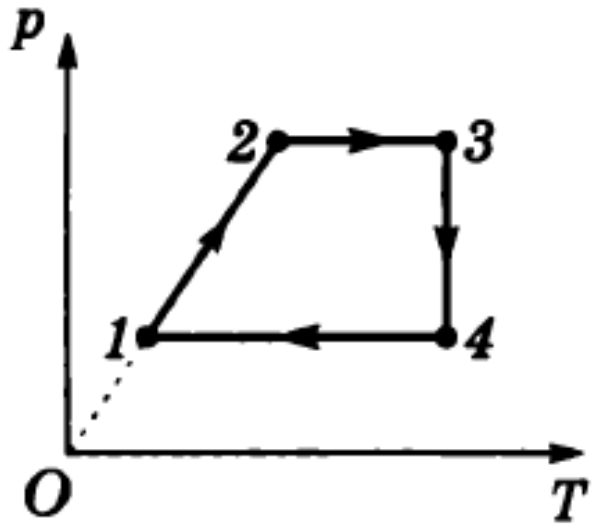

| А | Б | В | Г |
|---|---|---|---|
| 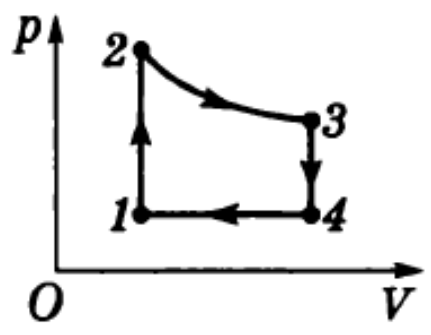 | 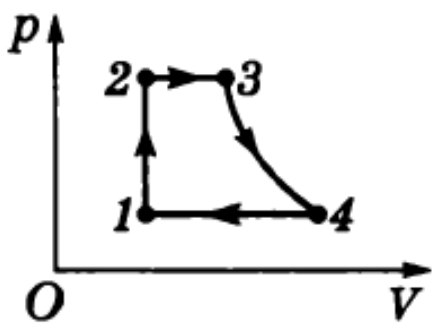 | 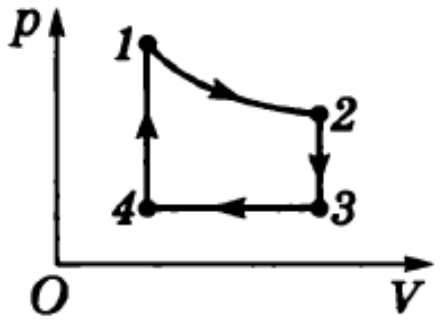 | 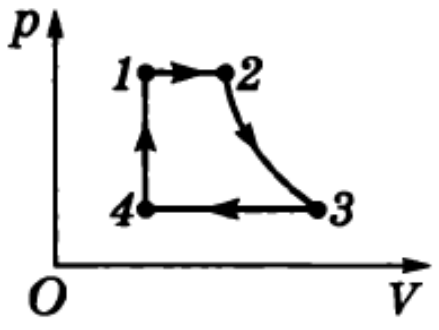 |

**Розв'язання.**

Такі задачі зустрічаються у тестах і розв'язуються почерговим розглядом ділянок графіку даного замкненого процесу і порівнянням їх з аналогічними ділянками графіків кожного варіанту.

Почнемо з ділянки *1-2* замкненого процесу наданого в координатах $p$—$T$, тобто залежності тиску від температури. Баабомо, що на цій ділянці *1-2* зростає тиск і зростає температура. Тепер дивимось на варіант А залежності тиску від об'єму, а саме аналізуємо поведінку показника тиску на ділянці *1-2*. Тиск зростає при незмінному об'ємі. Такаж поведінку тиску спостерігається і у варіанті Б. Варіант В – тиск на ділянці *1-2* зменшується, а варіант Г – тиск на ділянці *1-2* не змінюється. Таким абоном, вже при аналізі поведінки показника тиску на першій ділянці замкненого процесу, змогли прибрати із розгляду варіанти В і Г, як не відповідаючі умові, що тиск повинен зростати.

Подивимось поведінку тиску на ділянці *2-3* замкненого процесу наданого в координатах $p-T$ і баабомо, що тиск з ростом температури не збільшується. Дивимось поведінку тиску на ділянці *2-3* у залишившихся варіантах А і Б. Варіант А – тиск на ділянці *2-3* зменшується, а у варіанті Б – спостерігається незмінність тиску зі зростанням об'єму. Правильна відповідь відповідає варіанту Б, тому що тиск в цьому випадку не змінюється. Для впевненості проаналізуємо поведінку тиску в цьому варіанті для решти ділянок. На ділянці *3-4* графіку замкненого процесу залежності тиску від температури – тиск зменшується, порівнюємо з ділянкой *3-4* варіанту Б і баабомо, що тиск також зменшується. На останній ділянці *4-1* залежності тиску від температури спостерігаємо незмінність тиску, яка також спостерігається і на ділянці *4-1* варіанту Б. Тобто на всіх ділянках поведінка показника тиску замкненого процесу залежності тиску від температури співпала з поведінкою тиску замкненого процесу наданого в координатах $p-T$ у варіанту Б.

**Відповідь:** Варіант Б.

**Задача 2.16.** *Я*ка точка на *p-V* діаграмі зміни стану ідеального газу, що зображено на рисунку, відповідає мінімальній температурі?

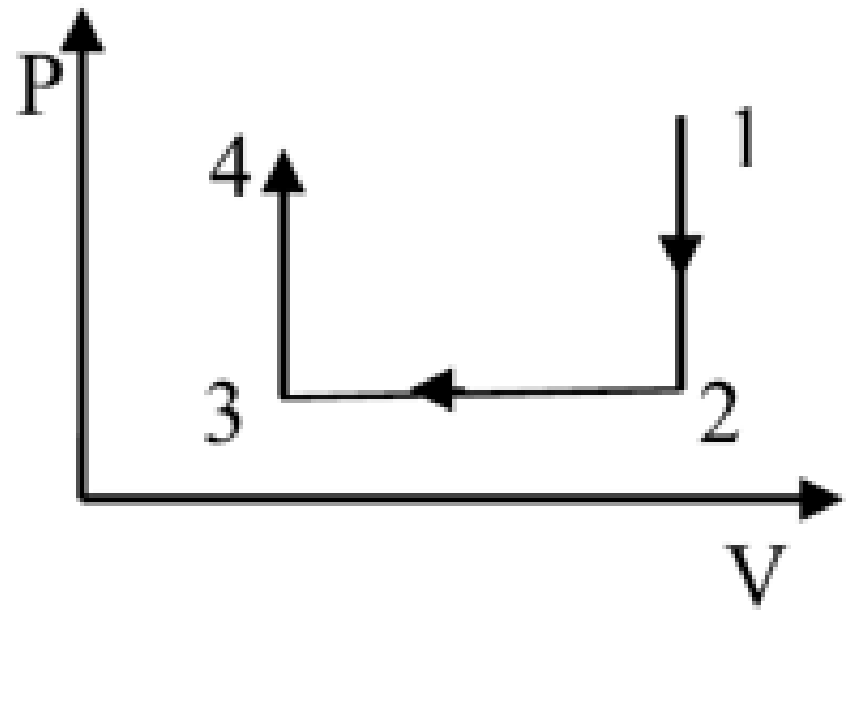

**Розв'язання.**

Щоб дати відповідь на запитання цієї задачі поглянемо на закон Бойля-Маріотта або ізотермічний процес ідеального газу, для незмінної маси газу при $T$ = const: $p \cdot V$=const, тобто для газу даної маси, при постійній температурі добуток тиску газу на його об’єм буде постійним. Ця залежність тиску газу від об’єму при постійній температурі графічно зображають кривою, яка має назву ізотерми.

Аналізуючи графік ізотермічного процесу бачимо, що різним постійним температурам відповідають різні ізотерми і для однакової маси газу при тій самій температурі газ займає тим більший об’єм, чим нижчий тиск.

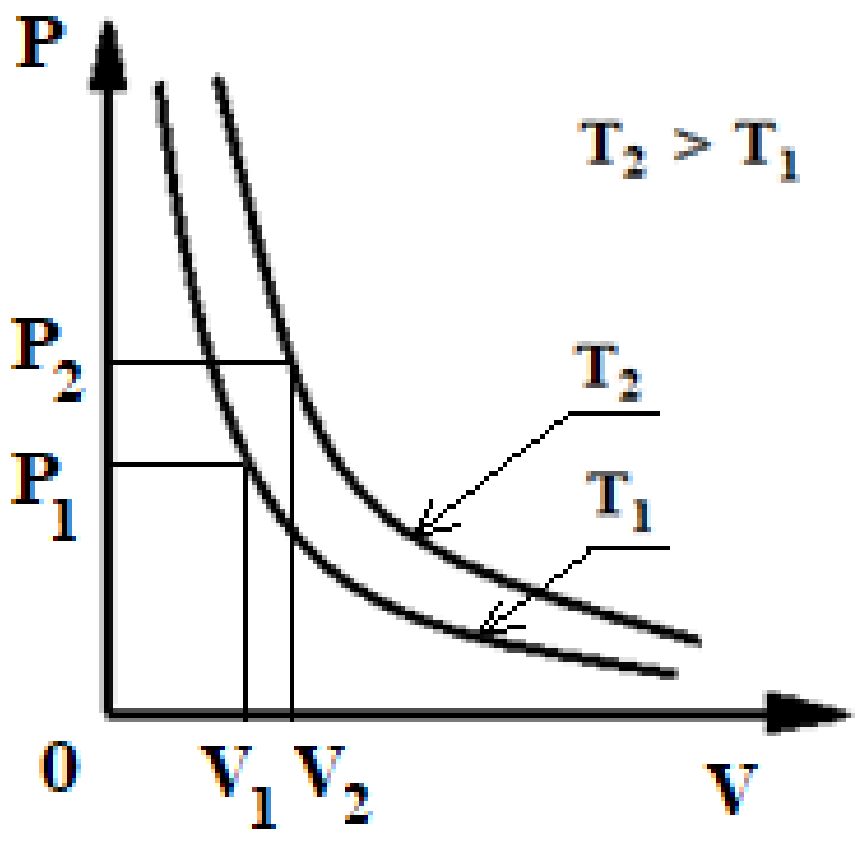

Відомо, що ізотерма, яка відповідає більш високої температурі $T_2$, лежить вище ізотерми, яка відповідає більш низькій температурі $T_1$. Аналітично зрозуміло, що найбільшим значенням тиску, а це точки *4* і *1*, буде відповідати більша температура, але цім точкам з близьким значенням тиску відповідають з ще більшою різницею об’єми. Аналізуючи $p$-$V$ діаграму зміни стану ідеального газу можна припустити, що всі точки *1,2,3* і *4* лежать на різних ізотермах, тобто характеризують стани з різною температурою. Приблизний характер залежності тиску і об’єму при постійній температурі відомий і ми можемо накласти цю криву так, щоб вона проходила по черзі через всі точки.

Точки *4* та *2* лежать на ізотермах з близькими по значенню температурах і відповідають середнім значенням, тобто треба визнааботись між ізотермой

$T_1$ – точка 3 і ізотермой $T_3$ – точка 1. Маючі на увазі, що при однаковому

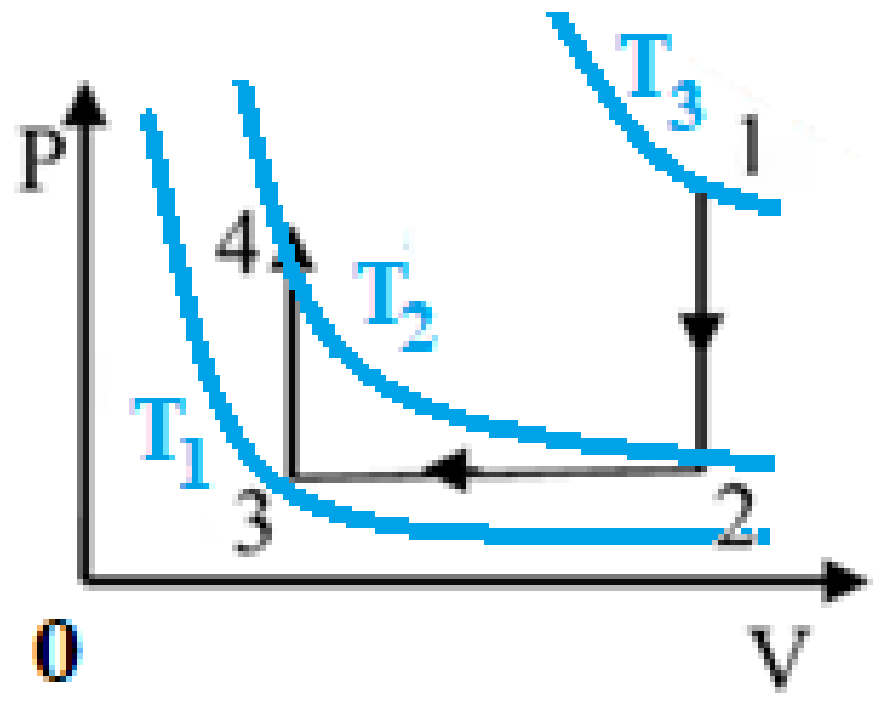

об'ємі більшому тиску відповідає більша температура, розуміємо, що абом вище температура, тім далі від координптних осей розташовані відповідні ізотерми. Таким абоном, мінімальній температурі відповідає точка *3*.

**Відповідь:** точка *3*.

**Задача 2.17.** Ідеальний газ здійснює замкнений цикл, показаний на рисунку. Визнааботь на яких ділянках зменшується тиск із запропонованих варіантів відповідей:

1) *1-2*; 2) *1-2* і *2-3*; 3) *2-3*; 4) *2-3* і *3-1*; 5) *3-1*

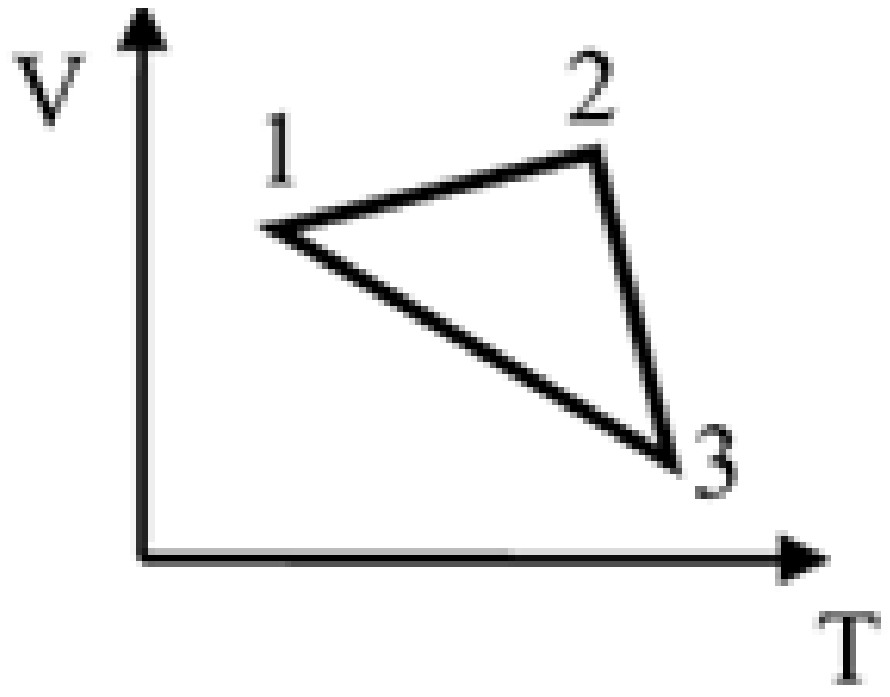

**Розв'язання.**

Це тестове завдання і так саме, як у попередніх задачах аналізуемо кожний варіант відповіді із запропонованих по черзі, маюабо на увазі, що напрямок на інтервалі (збільшується або зменшується) вказує позначення інтервалу – від першої цифри до другої.

Інтервал *1–2*, маємо збільшення температури (призводить до підвищення тиску) і об’єму (призводить до зниження тиску), а зміна цих двох показників одночасно є взаімозворотня і може призводити до того, що тиск не зміниться, для однозначної відповіді тут необхідні конкретні абосельні розрахунки.

Наступний варіант відповіді *1–2* і *2–3*. Розглянемо інтервал *2–3* на якому незначно збільшується температура і значно зменшується об’єм, що призведе до збільшення тиску, тобто цей і два наступних, в які також входить інтервал *2–3*, варіанти відповіді не підходять.

Інтервал *3–1* аналізуємо в напрямку від стану, який характерізує точка *3* до стану в точці *1*. На цьому інтервалі суттєво зменшується температура і суттєво збільшується об’єм. Така поведінка показників стану ідеального газу призводить к суттєвому зменшенню тиска.

**Відповідь:** інтервал *3–1*.

**Задача 2.18.** На рисунку зображено замкнений процес *1-2-3-4*, точки *1* і *3* якого відповідають ізотермам з температурами $t_1$= 27°C і $t_2$=177°C. У стані *1* об’єм газу дорівнює $V_1$= 3 дм$^3$. При якому значенні об’єму $V_2$ точки *2* і *4* будуть знаходитися на одній ізотермі? Чому дорівнює температура *T* цієї ізотерми?

**Дано:**

$V_1 = 3\ дм^3 = 3 \cdot 10^{-3}\ м^3$

$T_1 = 27°C + 273 = 330\ K$

$T_2 = 177°C + 273 = 450\ K$

**Знайти:**

$V_2$ - ? $T$ - ?

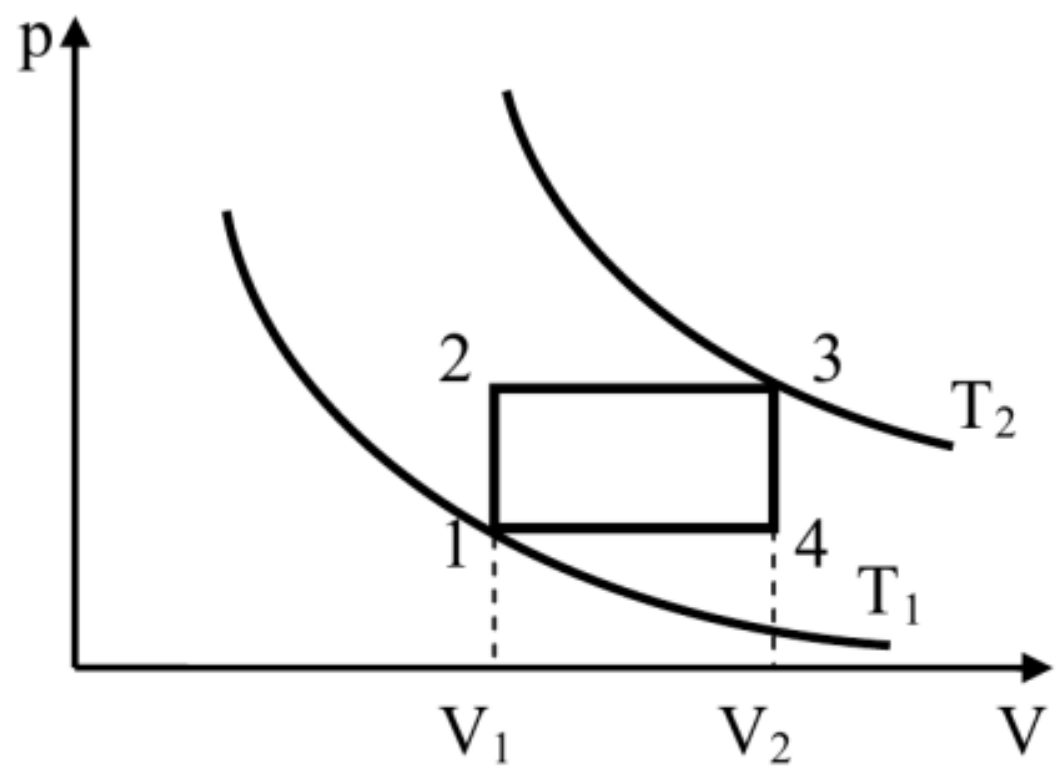

**Розв’язання.**

На даному рисунку замкненого процесу інтервали *2–3* та *1–4* відповідають зміні станів при ізобаричному процесі, тобто при $p = \text{const}$.

Точки *2* і *3* знаходяться на різних ізотермах і, за умовою задачі, точка *2* повинна бути на тієї ж, що і точка *4*, якій відповідає температура *T*.

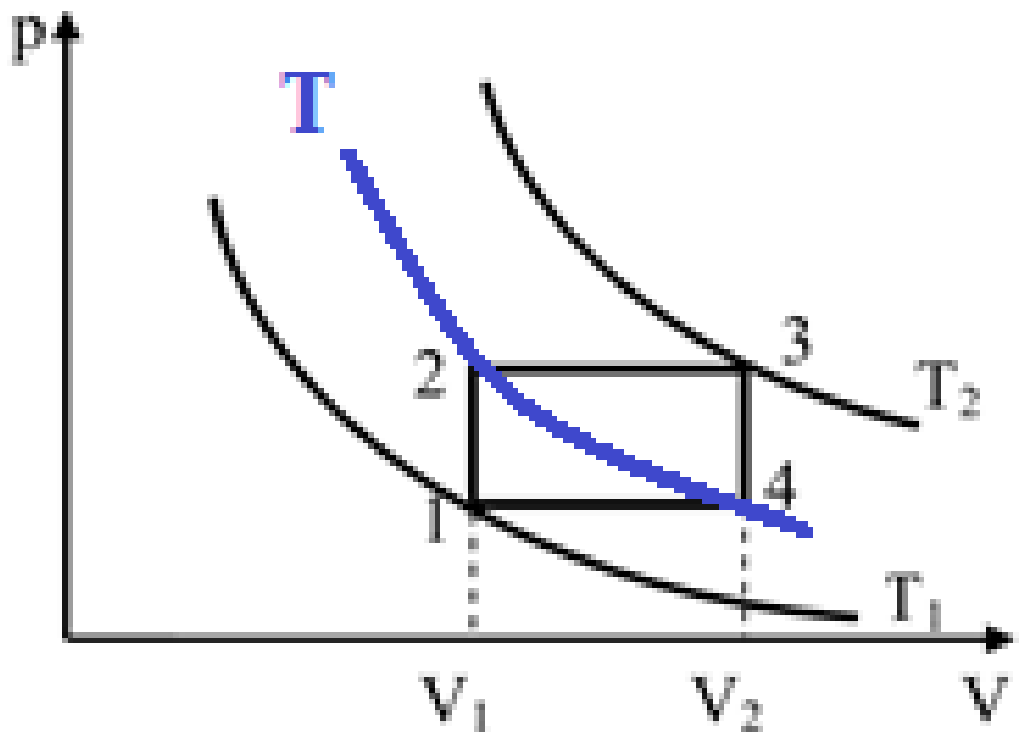

На основі закону Гей-Люссака для станів *2* і *3* маємо:

$$\frac{V_1}{T} = const \quad \text{і} \quad \frac{V_2}{T_2} = const \quad \text{або} \quad \frac{V_1}{T} = \frac{V_2}{T_2}$$

Перепишемо так, щоб температура була з однієї сторони рівняння

$$\frac{V_1}{V_2} = \frac{T}{T_2}$$

Аналогічно зробимо для станів *1* та *4*

$$\frac{V_1}{\mathrm{T}_1} = const \quad \text{и} \quad \frac{V_2}{\mathrm{T}} = const \quad \text{або} \quad \frac{V_1}{\mathrm{T}_1} = \frac{V_2}{\mathrm{T}} \quad \text{и} \quad \frac{V_1}{V_2} = \frac{\mathrm{T}_1}{\mathrm{T}}$$

Маємо об’єднати отримані співвідношення

$$\frac{V_1}{V_2} = \frac{T}{T_2} \qquad \text{і} \qquad \frac{V_1}{V_2} = \frac{\mathrm{T}_1}{\mathrm{T}} \qquad \text{тобто} \quad \frac{T}{T_2} = \frac{\mathrm{T}_1}{\mathrm{T}}$$

Із останнього рівняння виразимо $T$ і отримаємо:

$$\mathrm{TT} = \mathrm{T}_1\mathrm{T}_2 \qquad \mathrm{T} = \sqrt{\mathrm{T}_1\mathrm{T}_2}$$

Знайдемо $V_2$ із співвідношення:

$$\frac{V_1}{V_2} = \frac{T}{T_2}\ ;$$

$$V_1T_2 = V_2T; \qquad V_2 = \frac{V_1T_2}{\mathrm{T}} \quad \text{і підставимо} \quad \mathrm{T} = \sqrt{\mathrm{T}_1\mathrm{T}_2}$$

$$V_2 = \frac{V_1 T_2}{\sqrt{\mathrm{T_1 T_2}}} = V_1\sqrt{\frac{\mathrm{T_2^2}}{\mathrm{T_1 T_2}}} = V_1\sqrt{\frac{\mathrm{T_2}}{\mathrm{T_1}}}$$

Підставимо числові значення в отримані рівняння.

$$\mathrm{T} = \sqrt{\mathrm{T_1 T_2}} = \sqrt{300\cdot 450} = 367{,}4\ (\mathrm{К})$$

$$V_2 = V_1\sqrt{\frac{\mathrm{T_2}}{\mathrm{T_1}}} = 3\cdot 10^{-3}\cdot\sqrt{\frac{450}{150}} = 5{,}2\cdot 10^{-3}\ (\mathrm{м}^3)$$

**Відповідь**: $\mathrm{T} = 367{,}4$ К; $V_2 = 5{,}2\cdot 10^{-3}$ м$^3$

**Задача 2.19.** Теплова машина виконує за один цикл роботу $A$=4,82 кДж і віддає холодильнику кількість теплоти $Q_2 = 23{,}48$ кДж. Визнааботи коефіцієнт корисної дії (ККД) циклу.

| **Дано:** |
|---|
| $A = 4{,}82$ кДж |
| $Q_2 = 23{,}48$ кДж |
| **Знайти:** |
| $\eta - ?$ |

**Розв'язання.**

Коефіцієнт корисної дії (ККД):

$$\eta = \frac{A}{Q_н} = \frac{Q_1 - Q_2}{Q_1}$$

Теплота нагрівача $Q_\mathrm{н}$ або $Q_1 -$ це теплота, яку отримало робоче тіло машини за один цикл; $Q_2$ – теплота, яку робоче тіло віддало за цикл, а $A = Q_1 - Q_2$ корисна робота, яку виконало робоче тіло. Знайдемо $Q_2$ :

$$Q_1 = \mathrm{A} + Q_2 = 4{,}82 + 23{,}48 = 28{,}3\ \text{кДж}\quad \text{тоді}$$

Підставимо $Q_1$ в формулу:

$$\eta = \frac{A}{Q_н}\cdot 100\% = \frac{4{,}82}{28{,}3}\cdot 100\% = 0{,}17\cdot 100\% = 17\%$$

**Відповідь:** $\eta = 17\%$

**Задача 2.20.** У циклі Карно ідеальний газ дістав від нагрівача кількість теплоти $Q_1 = 4{,}2$ кДж і виконав роботу $A = 590$ Дж. Визнааботи ККД цього циклу. У скільки разів температура $T_1$ нагрівача більша за температури $T_2$ холодильника?

**Дано:**

$Q_1 = 4{,}2$ кДж $= 4{,}2 \cdot 10^3$ Дж.

$A = 590$ Дж

**Знайти:**

$\Delta\eta$–? і $T_1/T_2$ –?

**Розв'язання.**

Спочатку визначимо коефіцієнт корисної дії:

$$\eta = \frac{A}{Q_н} = \frac{590}{4{,}2 \cdot 10^3} = 140{,}5 \cdot 10^{-3} = 0{,}14$$

Також коефіцієнт корисної дії можна визначити через температури нагрівача та холодильника:

$$\eta = \frac{T_1 - T_2}{T_1}$$

За умовою задачі друге питання «У скільки разів температура $T_1$ нагрівача більша за температури $T_2$ холодильника?», тобто треба знайти відношення $\frac{T_1}{T_2}$, тому перетворимо цю формулу наступним чином:

$$\eta = 1 - \frac{T_2}{T_1}; \quad \frac{T_2}{T_1} = 1 - \eta = 1 - 0{,}14 = \frac{86}{100}$$

але треба знайти відношення $\frac{T_1}{T_2}$, тобто обернене знайденому.

$$\frac{T_1}{T_2} = \frac{100}{86} = 1{,}16$$

**Відповідь:** $\eta = 0{,}14$; $\frac{T_1}{T_2} = 1{,}16$.

**Задача 2.21.** Теплова машина працює по оборотному процесу Карно. Температура нагрівача $T_1 = 500$ К. Визначити термічний коефіцієнт корисної дії (ККД) циклу і температуру $T_2$ охолоджувача теплової машини, якщо за рахунок кожного кілоджоуля теплоти, отриманої від нагрівача, машина здійснює роботу $A = 350$ Дж.

| Дано: | Розв’язання. |
|---|---|
| $T_1 = 500$ К | Термічний коефіцієнт корисної дії (ККД), що також має назву коефіцієнту використання теплоті, показує яка доля теплоти, отриманої від нагрівача, перетворюється в механічну роботу і виражається формулою: |
| $A = 350$ Дж | |
| **Знайти:** | |
| $\eta$ – ?; $T_2$ – ? | |

$$\eta = \frac{A}{Q_н}$$

де $Q_н$ – теплота, що отримана від нагрівача; $A$ – робота, яку здійснює робоче тіло теплової машини. По умовам задачі дано значення роботи, отримане від нагрівача за один кілоджоуль теплоти, тобто маємо записати: $Q_н$=1000 Дж. Підставимо числові значення в формулу:

$$\eta = \frac{A}{Q_н} = \frac{350\,Дж}{1000\,Дж} = 0{,}35$$

Коефіцієнт корисної дії ККД для циклу також визначається формулою:

$$\eta = \frac{T_1 - T_2}{T_1}$$

Знаючи $\eta$ підставимо в цю формулу, але спочатку виразимо з цієї формули $T_2$

$$\eta T_1 = T_1 - T_2; \qquad T_2 = T_1 - \eta T_1 = T_1(1 - \eta)$$

$$T_2 = 500(1 - 0{,}35) = 325\,К$$

**Відповідь:** $\eta = 0{,}35$; $T_2 = 325$ К.

## Задачі для самостійного розв’язання

1. В балоні ємністю $V$=5 л міститься кисень масою $m$=20 г. Визначити концентрацію молекул в балоні. [$n \approx 7.53 \times 10^{25}$ молекул/м$^3$]

2. Середня квадратична швидкість деякого газу за нормальних умов дорівнює 480 м/с. Скільки молекул містить 1 г цього газу? [$N \approx 2.038 \times 10^{23}$]
3. У балоні ємністю $V$=20 л, знаходиться аргон під тиском $p_1$=0,6 МПа і температурі $t_1$=50 °С. Коли з балона узяли деяку кількість аргону, тиск в балоні знизився до $p_2$=0,3 МПа, а температура знизилася до $t_2$=30°С. Визначити масу $m$ аргону, узятого з балона. [ $\Delta m$=0.083 кг]
4. Визначити густину $\rho$ азоту за температури $T$=400 К і тиску $p$=2 МПа. [$\rho \approx 16.85$ кг/м$^3$].
5. Визначити повну енергію молекул кисню, що знаходиться при температурі 47 °С, якщо його маса 64 г. [$E$=13.32 кДж].
6. Визначити середню кінетичну енергію «$\varepsilon_0$» поступального руху молекул газу, що знаходиться під тиском 0,1 Па. Концентрація молекул газу рівна $10^{13}$ см$^{-3}$. [$\varepsilon_0 = 1.5 \times 10^{-20}$ Дж].
7. Визначити повну кінетичну енергію, а також кінетичну енергію обертального руху однієї молекули аміаку $NH_3$ при температурі $t$=27 °С. [$E_{повна} \approx 1.243 \times 10^{-20}$ Дж; $E_{оберт} \approx 6.213 \times 10^{-21}$ Дж].
8. Азот масою $m$=280 г розширюється в результаті ізобарного процесу під тиском $p$=1 МПа. Визначити роботу розширення, кінцевий об'єм газу, якщо на розширення витрачена теплота $Q$=5 кДж, а початкова температура азоту $T_1$ = 290 К. [А≈1,43 кДж; $V_2 \approx 0.026$ м$^3$]
9. Деякий газ масою 1 кг знаходиться при температурі $T$=300 К і під тиском $p_1$=0,5 МПа. В результаті ізотермічного стискання тиск газу збільшився в два рази. Робота, витрачена на стискання, $A$=432 кДж. Визначите який це газ, первинний питомий об'єм газу. [Молярна маса 4 г/моль відповідає **Гелію (Не);** $v_1 \approx 1.25$ м$^3$/кг].
10. У циліндрі під поршнем знаходиться азот масою $m$=20 г, який нагрівали від температури $t_1$=20 °С до температури $t_2$=180 °С при постійному тиску. Визначити кількість теплоти $Q$, передану газу, виконану газом роботу $A$ і приріст $\Delta U$ внутрішньої енергії. [Q≈3.33 кДж; A≈0.95кДж; ΔU≈2.38 кДж].

11. Газ здійснює цикл Карно. Температура охолоджувача $t$=0°C. Яка температура нагрівача, якщо за рахунок $4{,}2\cdot 10^3$ Дж теплоти, отриманої від нагрівача, газ здійснює роботу $A$=1200 Дж? [$T_2$=382.4 K]

## Розділ 3. ЕЛЕКТРОСТАТИКА.

### 3.1. Характеристики електричного поля.

Закон Кулона: сила взаємодії двох нерухомих точкових зарядів $q_1$ і $q_2$ прямо пропорційна велиабоні цих зарядів і обернено пропорційна квадрату відстані між ними:

$$F = \frac{q_1 q_2}{4\pi\varepsilon_0 \varepsilon r^2}$$

де $r$ – відстань між зарядами; $\varepsilon_0$= 8,85 $10^{-12}$ Ф/м – електрична стала; ε – відносна діелектрична проникність середовища.

**Кулонівські сили** – центральні, тобто вони спрямовані вздовж прямої, яка сполучає точкові заряди. Однойменні заряди відштовхуються, а різнойменні притягуються. Із спостережень встановлено, що сила взаємодії двох даних зарядів не змінюється, якщо поблизу них помістить ще якийсь заряди. Наприклад, маємо заряд $q_a$ та $N$ зарядів $q_1$, $q_2$, … , $q_N$, тоді результуюча сила $F$, з якою діють на $q_a$ усі $q_i$ заряди кількістю $N$ має вигляд:

$$F = \sum_{i=1}^{N} F_{ai},$$

де $F_{ai}$ - сила з якою діє на $q_a$ заряд $q_i$ при відсутності інших $N$-1 зарядів.

Кожний електричний заряд завжди змінює властивості простору, який його оточує, створюють в ньому електричне поле. Це поле проявляється таким абоном, що при вміщенні в ньому в будь-якій точці електричного заряду на нього буде діяти сила. Силовою характеристикою електричного поля є вектор напруженості $\vec{E}$, який дорівнює силі, що діє на одиничний позитивний заряд, поміщений у дану точку поля.

$$\vec{E} = \frac{\vec{F}}{q}$$

де $\vec{F}$– сила, що діє на точковий заряд $q$, вміщений у дану точку поля.

Напруженість електричного поля точкового заряду за допомогою закону Кулона можна представити у вигляді :

$$\vec{E} = \frac{q}{4\pi\varepsilon_0\varepsilon r^2}\cdot\frac{\vec{r}}{r} \text{ і за модулем: } E = \frac{q}{4\pi\varepsilon_0\varepsilon r^2}.$$

Тобто $\vec{E}$ - вектор напруженості поля точкового заряду спрямований уздовж радіальної прямої від заряду, якщо він позитивний, та до заряду, якщо поле утворюється негативним зарядом.

Фізична велиабона, яка абосельно дорівнює потенціальній енергії, яку має одиничний додатний заряд, вміщений в певну точку електростатичного поля, називається потенціалом поля в цій точці. Потенціал є енергетичною характеристикою поля:

$$\varphi = \frac{W_{\text{п}}}{q_{+\text{пр}}}$$

В полі точкового заряду $q$ потенціальна енергія $W_{\text{п}}$ пробного заряду $q_{+np}$ визначається наступним співвідношенням:

$$W_{\text{п}} = \frac{q\cdot q_{+\text{пр}}}{4\pi\varepsilon_0\varepsilon r},$$

Для поля точкового заряду маємо:

$$\varphi = \frac{q}{4\pi\varepsilon_0\varepsilon r},$$

отже, такий потенціал створює точковий заряд $q$ на відстані $r$. За принципом суперпозиції потенціал електростатичного поля створеного системою зарядів, в довільній точці поля дорівнює алгебраїчній сумі потенціалів, створених кожним із зарядів в цій точці:

$$\varphi = \sum_{i=1}^{n} \varphi_i, \qquad \text{або} \quad \varphi = \frac{1}{4\pi\varepsilon_0\varepsilon} \sum_{i=1}^{n} \frac{q_i}{r_i}$$

Геометричне місце точок однакового потенціалу називають еквіпотенціальною поверхнею, тобто еквіпотенціальна поверхня - це поверхня, в усіх точках якої потенціал електростатичного поля має однакове значення. Лінії напруженості електричного поля завжди перпендикулярні до еквіпотенціальної поверхні. Абом менше змінюється потенціал електростатичного поля на даній відстані, тим меншою є напруженість цього поля, якщо потенціал не змінюється, то напруженість поля дорівнює нулю. При переміщенні пробного заряду з однієї точки поля в іншу матимемо роботу сил електричного поля, яка виконується при переміщенні цього заряду:

$$A_{12} = W_{\Pi 1} - W_{\Pi 2} = q_{+\text{пр}}(\varphi_1 - \varphi_2)$$

Різниця потенціалів між двома точками:

$$U = \varphi_1 - \varphi_2 = \frac{A}{q},$$

де $A$ – робота сил поля при переміщенні заряду з однієї точки поля в іншу. Тоді робота сил поля по переміщенню заряду:

$$A = q(\varphi_1 - \varphi_2) = qU$$

У випадку однорідного поля:

$$E = \frac{\varphi_1 - \varphi_2}{d},$$

де $d$ – відстань між еквіпотенціальними поверхнями, до яких належать точки *1* і *2*.

Еквіпотенціальні поверхні тісно пов’язані із силовими лініями електричного поля. Якщо електричний заряд переміщується по еквіпотенціальній поверхні, то робота поля дорівнює нулю, оскільки:

$$A = -q\Delta\varphi$$

а на еквіпотенціальній поверхні

$$\Delta\varphi = 0.$$

Цю роботу можна також подати через силу F, що діє на заряд з боку електростатичного поля:

$$A = Fs\cos\alpha,$$

де $s$ – модуль переміщення заряду; $\alpha$ – кут між векторами $\vec{F}$ і $\vec{s}$.

Характеристика безперервно розподіленого заряду - густина заряду - це кількість електричного заряду, що припадає на одиницю довжини, площі або об'єму, тому визначаються лінійна, поверхнева і об'ємна щільність заряду.

**Задача 3.1.** Чотирі заряди однакової велиабони $q$ розташовані у вершинах квадрату зі стороною $a$. Визнааботи потенціал поля у центрі квадрата.

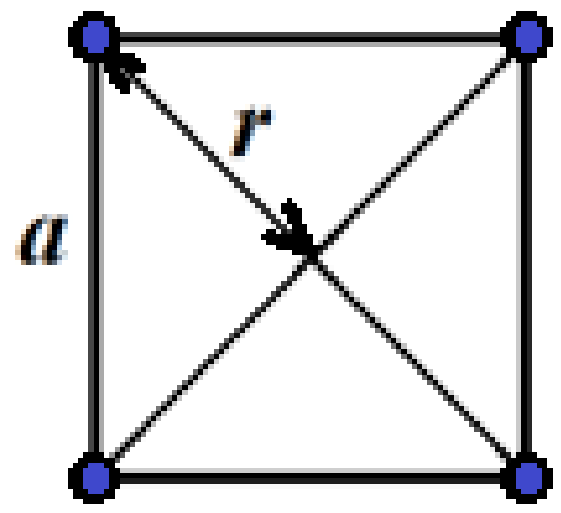

**Розв’язання.**

Відповідно з принципом суперпозиції електричного поля, потенціал $\varphi$ результуючого поля, що створюється двома зарядами $q_1$ та $q_2$ , дорівнюється алгебраїчній сумі потенціалів, тобто

$$\varphi = \varphi_1 + \varphi_2.$$

Запишемо формулу для потенціалу, що створюється зарядом на відстані r від заряду:

$$\varphi = \frac{q}{4\pi\varepsilon_0\varepsilon r}$$

Тоді для цієї задачі маємо, що потенціал у центрі квадрата, створений всіма чотирьма зарядами:

$$\varphi = \varphi_1 + \varphi_2 + \varphi_3 + \varphi_4$$

тобто:

$$\varphi = \frac{1}{4\pi\varepsilon_0\varepsilon}\left(\frac{q_1}{r_1} + \frac{q_2}{r_2} + \frac{q_3}{r_3} + \frac{q_4}{r_4}\right).$$

По умовам задачі заряди однакової велиабони, тому $\varphi = \frac{4q}{4\pi\varepsilon_0\varepsilon r}$ але замість $r$, нам надана сторона квадрата $a$. Тому виразимо $r$ через $a$. В формулі $r$ це відстань між зарядом та $\varphi$, тобто маємо розглянути прямокутний трикутник з діагональю $a$ і катетами $r$.

Тоді

$$a^2 = r^2 + r^2\,; \quad 2r^2 = a^2\,; \quad r = \frac{a}{\sqrt{2}}.$$

Підставимо $r$ виражений через $a$.

$$\varphi = \frac{4q}{4\pi\varepsilon_0\varepsilon r} = \frac{q}{\pi\varepsilon_0\varepsilon\frac{a}{\sqrt{2}}} = \frac{q\sqrt{2}}{\pi\varepsilon_0\varepsilon a}.$$

**Відповідь:** $\varphi = \frac{q\sqrt{2}}{\pi\varepsilon_0\varepsilon a}$.

**Задача 3.2.** Два точкові електричні заряди $q_1$= 1 нКл і $q_2$= –2 нКл знаходяться в повітрі на відстані $d$=10 см один від одного. Визначити потенціал поля $\varphi$, який створюється цими зарядами в точці $A$, яка знаходиться від заряду $q_1$ на відстані $r_1$ =9 см, і від заряду $q_2$ на відстані $r_2$ = 7 см.

| **Дано:** | **Розв'язання.** |
|---|---|
| $q_1 = 1$ нКл $= 10^{-9}$Кл<br>$q_2 = -2$ нКл $= -2 \cdot 10^{-9}$Кл<br>$d = 10$ см $= 0{,}1$ м<br>$r_1 = 9$ см $= 0{,}09$ м<br>$r_2 = 7$ см $= 0{,}07$ м | $$\varphi = \frac{q}{4\pi\varepsilon_0 r}$$ при $\varepsilon = 1$ - відносна діелектрична проникність повітря приймають рівною 1 ($\varepsilon = 1$), як і для вакууму. $\varepsilon_0 = 8{,}85 \cdot 10^{-12}$Ф/м |
| **Знайти:**<br>$\varphi -?$ | Відровідно до принципу суперпозиції електричних полів потенціал $\varphi$ результуючого поля, яке створюється двома зарядами $q_1$ і $q_2$ |

дорівнюється алгебраїчній сумі потенціалів:

$$\varphi = \varphi_1 + \varphi_2$$

Потенціал електричного поля, який створюється в вакуумі точковим зарядом $q$ на відстані $r$ від нього, виражається формулою:

$$\varphi = \frac{q}{4\pi\varepsilon_0 r}$$

Згідно з цією формулою запишемо формулу для визначення потенціалу $\varphi_1$, який створюється зарядом $q_1$ і потенціалу $\varphi_2$, створеного зарядом $q_2$:

$$\varphi_1 = \frac{q_1}{4\pi\varepsilon_0 r_1} \qquad \varphi_2 = \frac{q_2}{4\pi\varepsilon_0 r_2}$$

$$\varphi = \frac{q_1}{4\pi\varepsilon_0 r_1} + \frac{q_2}{4\pi\varepsilon_0 r_2} \quad \text{або} \quad \varphi = \frac{1}{4\pi\varepsilon_0}\left(\frac{q_1}{r_1} + \frac{q_2}{r_2}\right)$$

Підставляємо абослові значення і отримаємо:

$$\varphi = \frac{1}{4 \cdot 3{,}14 \cdot 8{,}85 \cdot 10^{-12}}\left(\frac{10^{-9}}{0{,}09} + \frac{-2 \cdot 10^{-9}}{0{,}07}\right) = -157 \text{ В}.$$

**Відповідь:** $\varphi = -157$ В.

**Задача 3.3.** Два точкових однойменних заряди $Q_1$=2 нКл і $Q_2$=5 нКл знаходяться в вакуумі на відстані $r_1$= 20 см. Визначить роботу $A$, яку треба здійсніти, щоб наблизити їх до відстані $r_2$= 5 см.

| **Дано:** | **Розв'язання.** |
| --- | --- |
| $Q_1$=2 нКл=$2\cdot10^{-9}$Кл<br>$Q_2$=5 нКл=$5\cdot10^{-9}$Кл<br>$r_1$= 20 см=0,2 м<br>$r_2$= 5 см=0,05 м | Робота, що здійснюється силами електростатичного полю при переміщенні заряду $Q$ з точки поля , що має потенціал $\varphi_1$, в точку з потенціалом $\varphi_2$ знаходиться за формулою: |
| **Знайти:**<br>$A-?$ | $$A_{12}=Q(\varphi_1-\varphi_2)$$<br>тобто, для першого заряду потенціал |

$$\varphi_1=\frac{q_1}{4\pi\varepsilon_0 r_1};\ \varphi_2=\frac{q_1}{4\pi\varepsilon_0 r_2}$$

Також запишемо для другого заряду:

$$\varphi_1=\frac{q_2}{4\pi\varepsilon_0 r_1};\ \varphi_2=\frac{q_2}{4\pi\varepsilon_0 r_2}$$

Але в нашій задачі не важливо як рухаються заряди, тому будемо вважати, що один заряд рухається, а другий, наприклад, $Q_2$ залишився на місті, тоді

$$\varphi_1=\frac{Q_1}{4\pi\varepsilon_0 r_1}\qquad \varphi_2=\frac{Q_1}{4\pi\varepsilon_0 r_2}$$

Підставимо у формулу

$$A_{12}=Q(\varphi_1-\varphi_2)$$

При наближенні однойменних зарядів роботу здійснюють зовнішні сили, тому роботи цих сил рівна по модулю, але протилежна по знаку роботі кулонівських сил, тому :

$$A=-Q(\varphi_1-\varphi_2)=Q(\varphi_2-\varphi_1)$$

Тобто, в нашому випадку маємо записати:

$$A=Q_2(\varphi_2-\varphi_1)=Q_2\left(\frac{Q_1}{4\pi\varepsilon_0 r_2}-\frac{Q_1}{4\pi\varepsilon_0 r_1}\right)=\frac{Q_2Q_1}{4\pi\varepsilon_0}\left(\frac{1}{r_2}-\frac{1}{r_1}\right)$$

Підставимо абослови значення:

$$A=\frac{5\cdot10^{-9}\cdot2\cdot10^{-9}}{4\cdot3{,}14\cdot8.85\cdot10^{-12}}\left(\frac{1}{0{,}05}-\frac{1}{0{,}2}\right)=9\cdot10^{-8}\cdot15=135\cdot10^{-8}\text{Дж}$$

**Відповідь:** $A=1{,}35\cdot10^{-6}$Дж.

**Задача 3.4.** Два точкові електричні заряди різних знаків ($q_1$>0; $q_2$<0) закріплені на прямій на відстані $L$ один від одного $|q_2|$< $q_1$ (див. рис.). Третій заряд $q_0$ може переміщуватися тільки вздовж прямої, що з'єднує заряди. Визнааботи положення заряду $q_0$, при якому він буде знаходитися у рівновазі.

**Дано:**

$q_1 > 0$

$q_2 < 0$

$|q_2| < q_1$

$L_{12}$

**Знайти:**

$x$ - ?

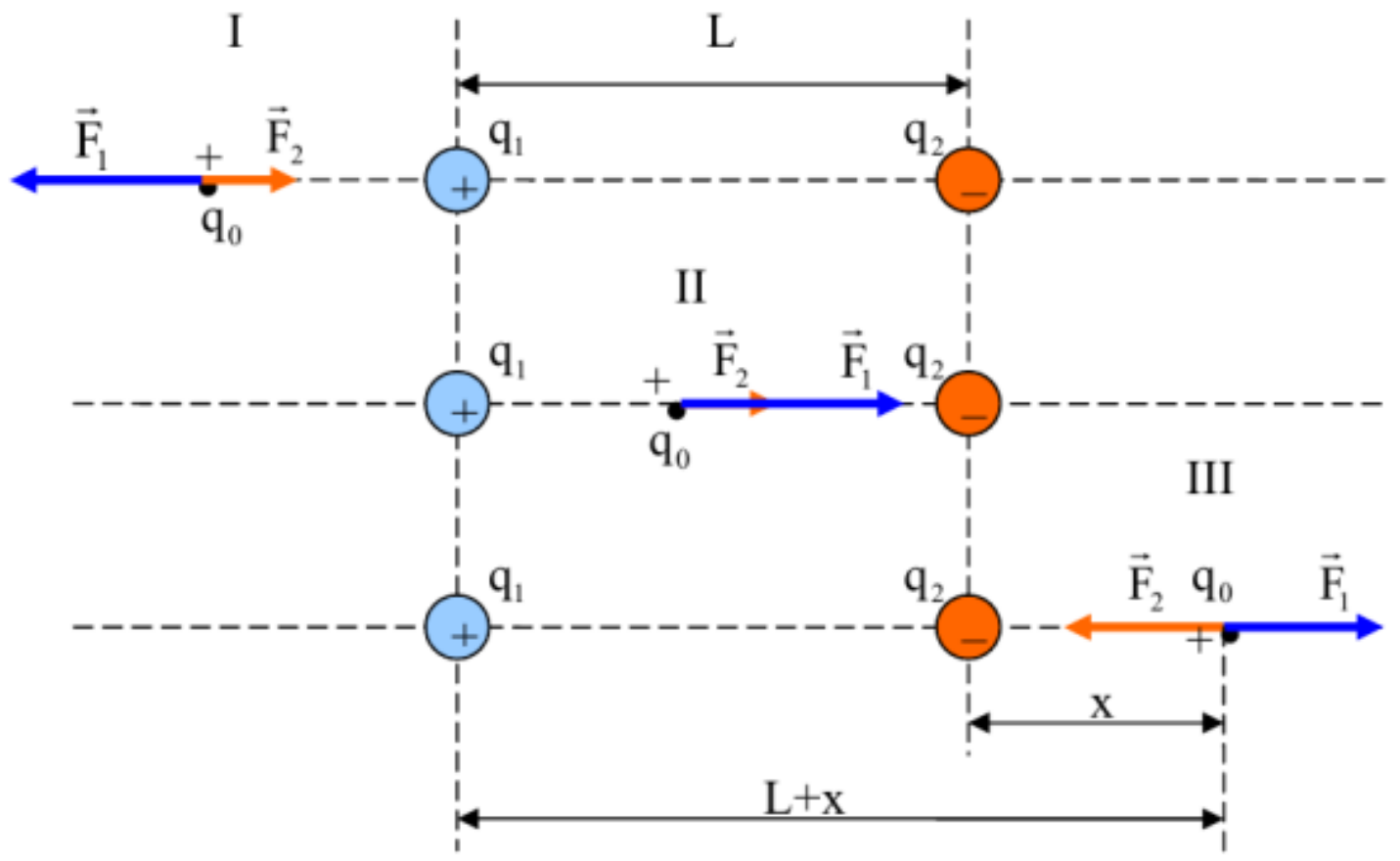

**Розв'язання.**

Заряд $q_0$ буде знаходитися у рівновазі, якщо геометрична сума сил $\vec{F}_1$ і $\vec{F}_2$, що діють на нього з боку зарядів $q_1$ і $q_2$ дорівнюватиме нулю:

$$\vec{F}_1 + \vec{F}_2 = 0$$

Визнаабомо, на якій з ділянок, які показані на рис. може виконуватися ця умова. Припустимо для визначеності, що заряд $q_0$ є позитивним.

1) Заряд $q_0$ розташований на *1* ділянці. Сили $\vec{F}_1$ і $\vec{F}_2$ спрямовані протилежно, причому сила $\vec{F}_1$ буде більшою за силу $\vec{F}_2$ оскільки більший за велиабоною заряд $q_1$ буде знаходитись ближче до заряду $q_0$.

2) Заряд $q_0$ розташований на *2* ділянці. Сили $\vec{F}_1$ і $\vec{F}_2$ спрямовані в один бік до негативного заряду $q_2$, тобто рівновага неможлива.

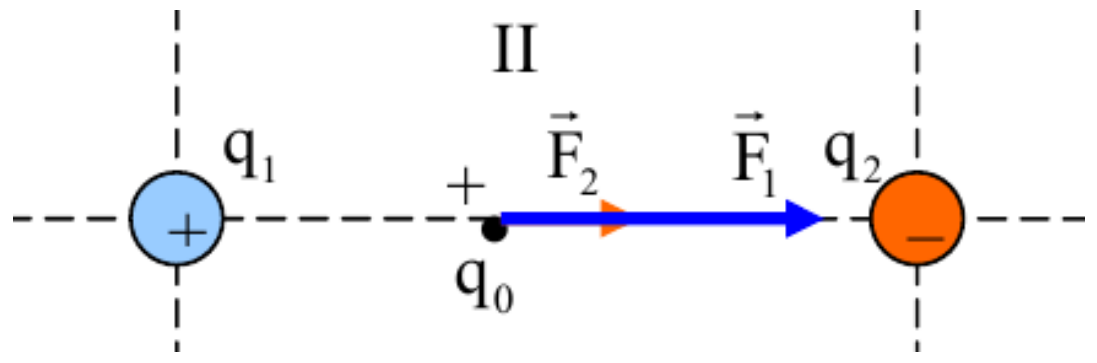

3) Заряд $q_0$ розташований на *3* ділянці. Сили $\vec{F}_1$ і $\vec{F}_2$ спрямовані протилежно і менший за велиабоною заряд $q_2$ буде знаходитись ближче до заряду $q_0$, що означає можливість існування точки в якій сили $\vec{F}_1$ і $\vec{F}_2$ будуть однаковими за модулем $\vec{F}_1 = \vec{F}_2$.

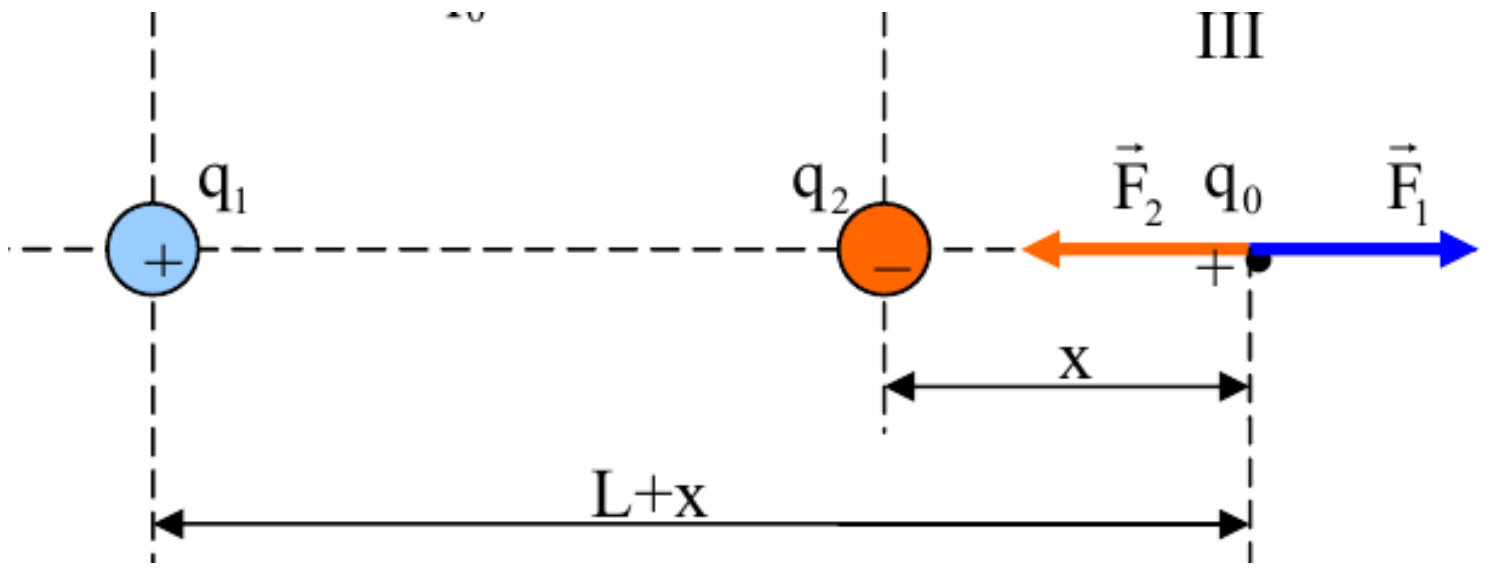

З умовою задачі електричні заряди закріплені на прямій на відстані $L$ один від одного, через $x$ познаабомо відстань від заряду $q_2$ до заряду $q_0$, де $\vec{F}_1 = \vec{F}_2$ за модулем, тоді відстань від заряду $q_1$ до заряду $q_0$ буде $L + x$.

Застосуємо закон Кулона

$$F = \frac{q_1 q_2}{4\pi\varepsilon_0 \varepsilon r^2}$$

до сил $\vec{F}_1$ і $\vec{F}_2$:

$$F_1 = \frac{1}{4\pi\varepsilon_0} \cdot \frac{q_1 q_0}{(L+x)^2} \; ; \; F_2 = \frac{1}{4\pi\varepsilon_0} \cdot \frac{q_0 q_2}{(x)^2}$$

за умовою задачі $F_1 = F_2$, тому:

$$\frac{1}{4\pi\varepsilon_0} \cdot \frac{q_1 q_0}{(L+x)^2} = \frac{1}{4\pi\varepsilon_0} \cdot \frac{q_2 q_0}{(x)^2}$$

Після скорочення коефіцієнтів та $q_0$ отримаємо:

$$\frac{q_1}{(L+x)^2} = \frac{q_2}{(x)^2},$$

дістанемо корінь з обох частин рівності:

$$\sqrt{\frac{q_1}{(L+x)^2}} = \sqrt{\frac{q_2}{(x)^2}}, \quad \text{або} \quad \frac{\sqrt{q_1}}{L+x} = \pm\frac{\sqrt{q_2}}{x}$$

$$\sqrt{q_2}(L+x) = \pm\sqrt{q_1}x;$$

$$\sqrt{q_2}L = \sqrt{q_1}x - \sqrt{q_2}x \quad \text{і} \quad \sqrt{q_2}L = -\sqrt{q_1}x - \sqrt{q_2}x$$

$$x_1 = \frac{\sqrt{q_2}L}{\sqrt{q_1} - \sqrt{q_2}} \quad \text{та} \quad x_2 = -\frac{\sqrt{q_2}L}{\sqrt{q_1} + \sqrt{q_2}}$$

Корінь рівняння $x_2$ визначає точку, що належить ділянці II. В цій точці сили $F_1$ і $F_2$ рівні, але направлені в один бік. Таким абоном, значення $x_2$ не відповідає умові задачі.

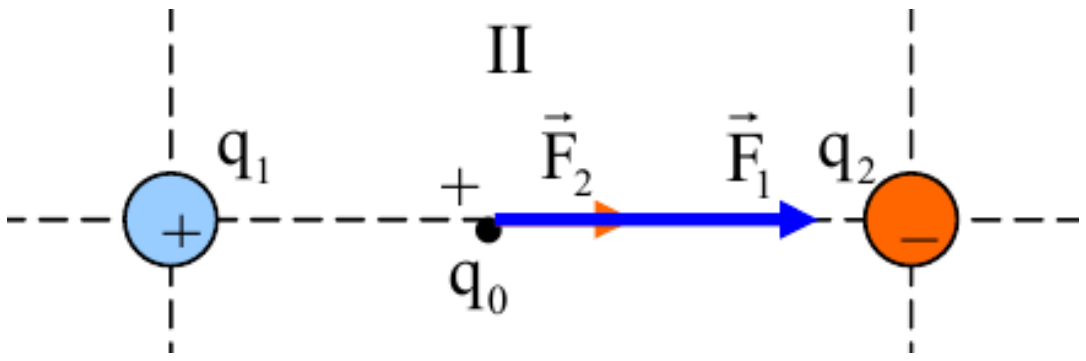

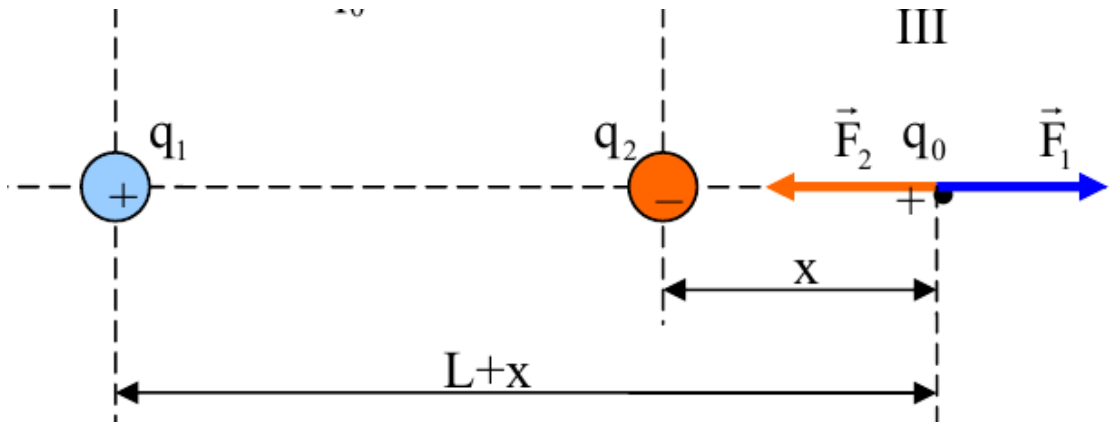

**Відповідь:** $x_1 = \frac{\sqrt{q_2}L}{\sqrt{q_1} - \sqrt{q_2}}$

**Задача 3.5.** Три точкових заряди $q_1 = q_2 = q_3 = 1$ нКл розташовані в вершинах рівностороннього трикутника. Який заряд $q_4$ потрібно помістити в центрі трикутника, щоб зазначена система зарядів знаходилася в рівновазі?

**Дано:**

$q_1 = q_2 = q_3 = 1$ нКл

**Знайти:**

$q_4$–?

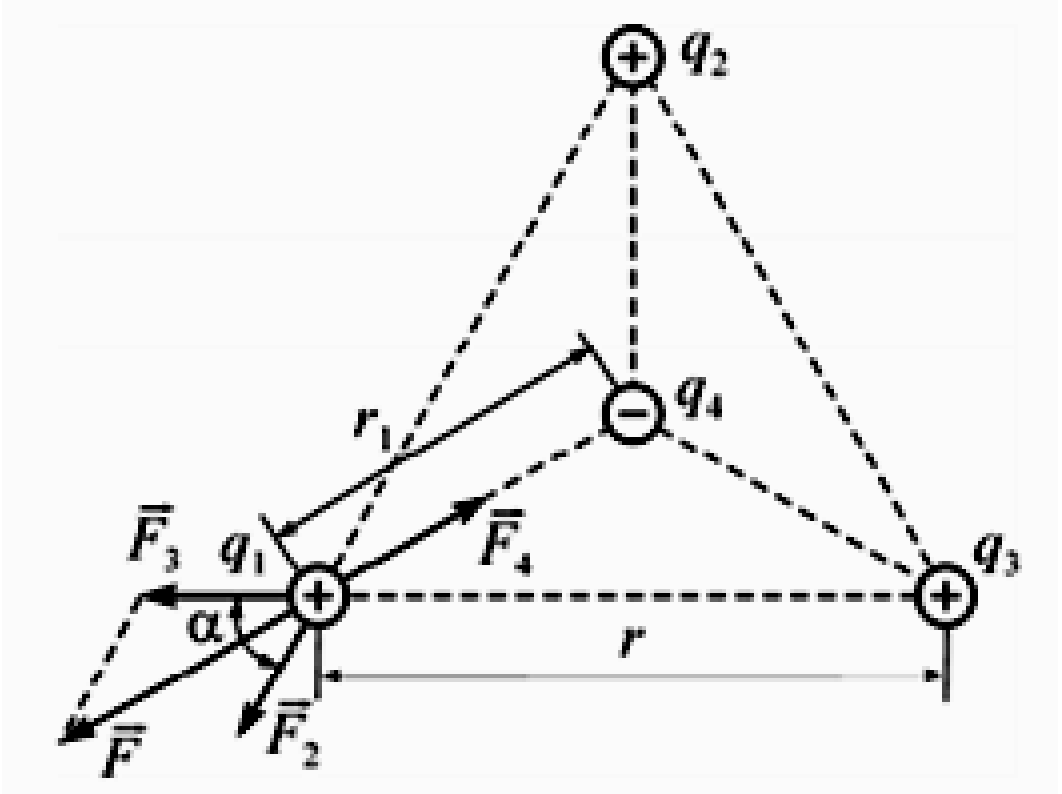

**Розв'язання.**

Всі три заряда розташовані по вершинам рівнобічного трикутника, тому знаходяться в однакових умовах і щоб з'ясувати, який заряд треба розмістити у центрі, достатньо з'ясувати які сили діють між одним із трьох зарядів і тим, що у центрі. Візьмемо заряд, наприклад, $q_1$ і розгланемо рішення відносно цього заряда.

Заряд $q_1$ буде знаходитися в рівновазі, якщо векторна сума діюабох на нього сил дорівнюється нулю.

$$\vec{F}_2 + \vec{F}_3 + \vec{F}_4 = \vec{F} + \vec{F}_4 = 0$$

де $\vec{F}_2; \vec{F}_3; \vec{F}_4 -$ сили з якими відповідно діють на заряд $q_1$ заряди $q_2, q_3, q_4$; сила $\vec{F} -$ рівнодіюча сил $\vec{F}_2 + \vec{F}_3$.

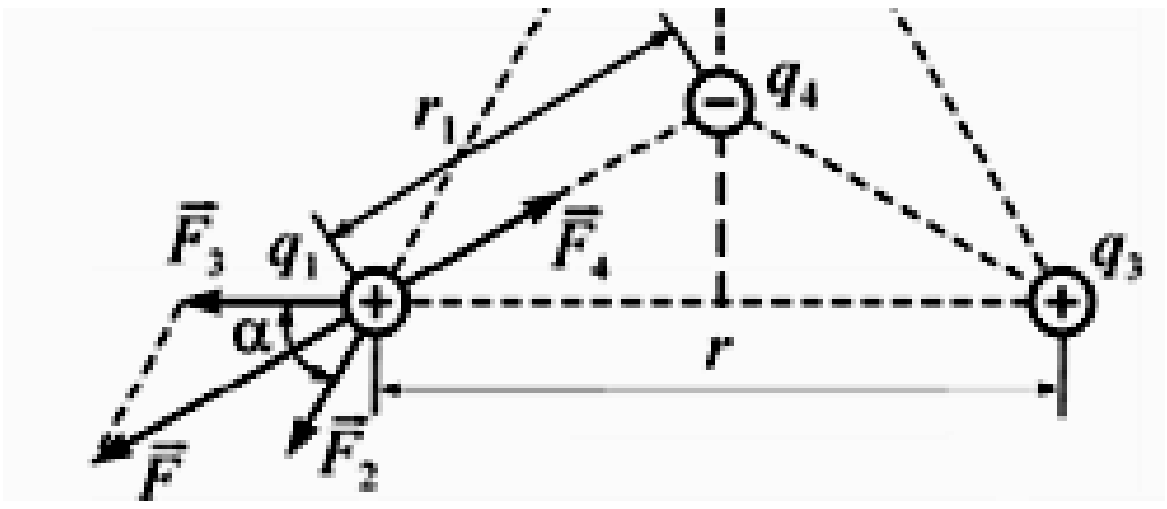

Відстань між позитивними зарядами познаабомо через *r*, відстань між позитивним та від'ємним зарядом через $r_1$, а кут між векторами сил через $\alpha$.

Цей кут $\alpha$ і кут рівнобічного трикутника є рівними і дорівнюється 180:3=60$^o$

Із рисунка ми бачимо, що сили $\vec{F}$ і $\vec{F}_4$ розташовані на одній прямій і спрямовані у протилежні сторони, тому векторне рівняння

$$\vec{F}_2 + \vec{F}_3 + \vec{F}_4 = \vec{F} + \vec{F}_4 = 0$$

можна замінити на скалярне рівняння

$$F - F_4 \quad \Rightarrow \quad F = F_4.$$

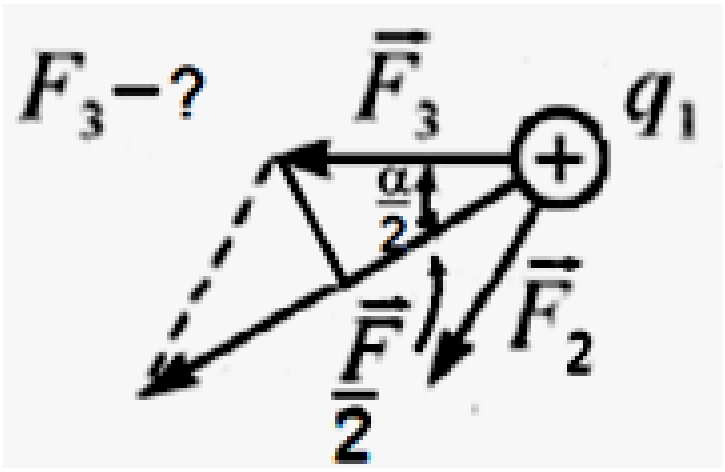

Виразимо $F$ через $F_2$ або $F_3$, не має значення тому що $F_2 = F_3$ і отримаємо:

$$\cos\frac{\alpha}{2} = \frac{\frac{F}{2}}{F_3} \ ; \qquad \frac{F}{2} = F_3 \cdot \cos\frac{\alpha}{2}$$

За законом Кулона:

$$F = \frac{q_1 q_2}{4\pi\varepsilon_0 \varepsilon r^2}$$

В нашому випадку $F$ повинна бути врівноважена силою $F_4$, тому маємо записати, що $F = F_4$.

$$F = F_4 = \frac{q_1 q_4}{4\pi\varepsilon_0 \varepsilon {r_1}^2}$$

В отриманому рівнянні невідомо $r_1$, яке можемо визнаaботи із виразу:

$$\cos\frac{\alpha}{2} = \frac{\frac{r}{2}}{r_1}; \qquad r_1 = \frac{r}{2\cdot\cos\frac{\alpha}{2}}$$

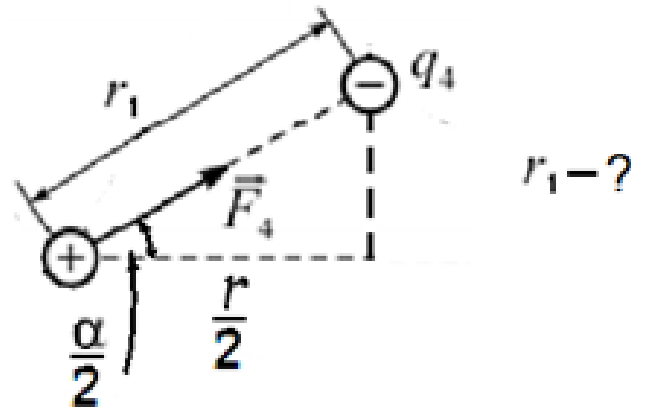

Маюабо на увазі, що $\cos 30^{\circ} = \frac{\sqrt{3}}{2}$ отримаємо, що $r_1 = \frac{r}{\sqrt{3}}$

Дивимось $F = 2\cdot F_3 \cdot \cos\frac{\alpha}{2}$, в свою чергу за законом Кулона

$$F_3 = \frac{q_1 q_3}{4\pi\varepsilon_0 \varepsilon r^2}$$

Маюабо на увазі, що $q_1 = q_2 = q_3$, можемо записати:

$$F_3 = \frac{q_1^2}{4\pi\varepsilon_0 \varepsilon r^2} \quad \text{тоді} \quad F = 2\frac{q_1^2}{4\pi\varepsilon_0 \varepsilon r^2}\cos\frac{\alpha}{2} \qquad \text{також}$$

$$F = F_4 = \frac{q_1 q_4}{4\pi\varepsilon_0 \varepsilon {r_1}^2} \quad \text{і можна прирівняти два рівняння, тобто}$$

$$2\frac{q_1^2}{4\pi\varepsilon_0 \varepsilon r^2}\cos\frac{\alpha}{2} = \frac{q_1 q_4}{4\pi\varepsilon_0 \varepsilon {r_1}^2}$$

$$2\frac{q_1}{r^2}\cos\frac{\alpha}{2} = \frac{q_4}{{r_1}^2}$$

Виразимо $q_4$ і підставимо $r_1 = \frac{r}{\sqrt{3}}$ та $\cos\frac{\alpha}{2} = \frac{\sqrt{3}}{2}$

$$q_4 = 2\frac{q_1 {r_1}^2}{r^2}\cos\frac{\alpha}{2} = 2\frac{q_1\left(\frac{r}{\sqrt{3}}\right)^2}{r^2}\cos\frac{\alpha}{2} = 2\frac{q_1}{3}\frac{\sqrt{3}}{2} = \frac{q_1}{\sqrt{3}}$$

Підставимо $q_1 = q_2 = q_3 = 1$ нКл $= 10^{-9}$Кл і отримаємо:

$$q_4 = \frac{q_1}{\sqrt{3}} = \frac{10^{-9}}{\sqrt{3}} = 5{,}77 \cdot 10^{-10} = 577\ \text{пКл}$$

**Відповідь:** $q_4 = 577$ пКл.

**Задача 3.6.** Два точкові електричні заряди $q_1$=–1 нКл і $q_2$=2 нКл знаходяться в повітрі на відстані $r$=10 см один від одного. Визнааботи напруженість $\vec{E}$ у точці, яка знаходиться на відстані $r_1$=9 см від першого заряду та $r_2$=7 см від другого.

**Дано:**

$q_1 = -1\ \text{нКл} = 1 \cdot 10^{-9}\text{Кл}$

$q_2 = 2\ \text{нКл} = 2 \cdot 10^{-9}\text{Кл}$

$r = 10\ \text{см} = 0{,}1\ \text{м}$

$r_1 = 9\ \text{см} = 0{,}09\ \text{м}$

$r_2 = 7\ \text{см} = 0{,}07\ \text{м}$

**Знайти:**

E–?

**Розв'язання.**

Згідно принципу суперпозиції полів напруженість поля може бути визначена як геометрична сума напруженості полів $\vec{E}_1$ і $\vec{E}_2$ кожного із зарядів:

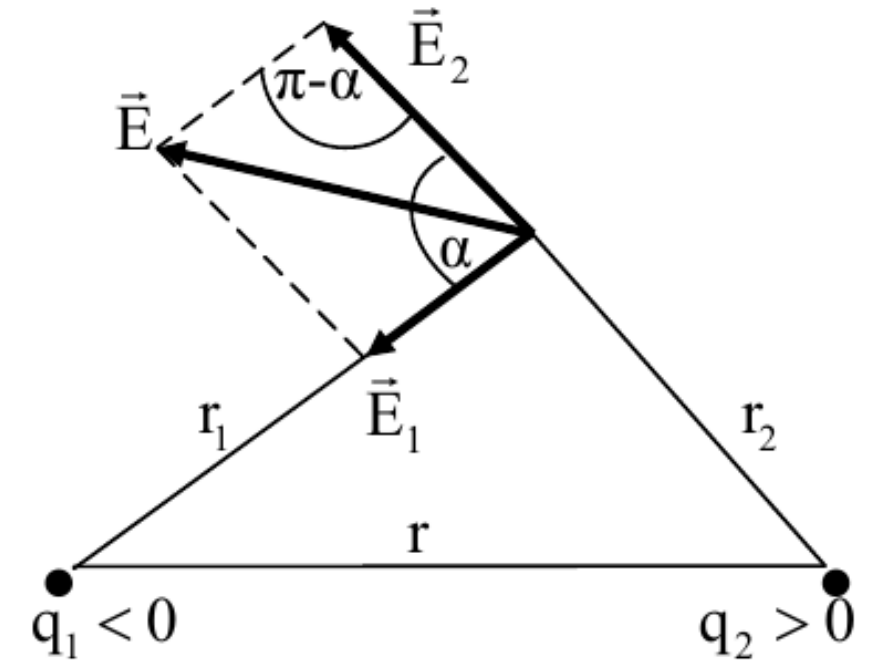

$$\vec{E} = \vec{E}_1 + \vec{E}_2$$

Напруженість електричного поля, створеного у повітрі (ε =1) зарядом $q_1$ і $q_2$ дорівнює:

$$E_1 = \frac{q_1}{4\pi\varepsilon_0 r_1^2} \quad \text{та} \quad E_2 = \frac{q_2}{4\pi\varepsilon_0 r_2^2}$$

Вектор $\vec{E}_1$ спрямований вздовж прямої до заряду $q_1$, оскільки він негативний; відповідно, вектор $\vec{E}_2$ спрямований від позитивного заряду $q_2$.

Абсолютне значення вектора $\vec{E}$ знаходимо за теоремою косинусів, яка є узагальненням теореми Піфагора: *«Квадрат будь-якої сторони трикутника дорівнює сумі квадратів двох інших його сторін без подвоєного добутку цих сторін на* ***косинус*** *кута між ними».*

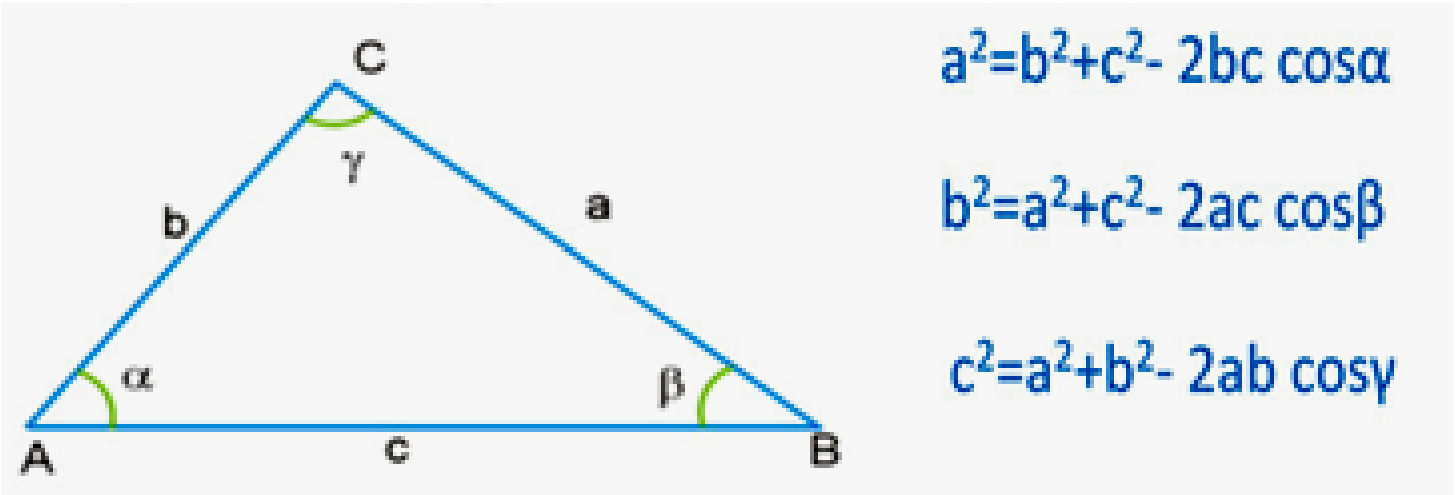

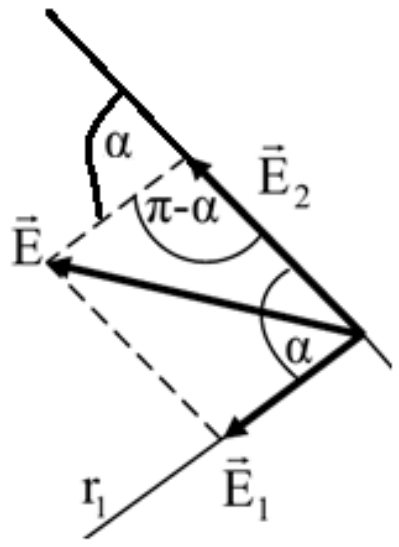

$$E^2 = E_1^2 + E_2^2 - 2E_1E_2\cos(\pi - \alpha)$$

$$E = \sqrt{E_1^2 + E_2^2 - 2E_1E_2\cos(\pi - \alpha)}$$

де α – кут між векторами $E_1$ і $E_2$, який можна визнаaботи з трикутника із сторонами $r$, $r_1$ і $r_2$ також за теоремою косинусів:

$$r^2 = r_1^2 + r_2^2 - 2r_1r_2\cos(\pi - \alpha)$$

Маюaбо на увазі формули зведення, а саме: $\cos(\pi - \alpha) = -\cos\alpha$ перепишемо рівняння:

$E = \sqrt{E_1^2 + E_2^2 - 2E_1E_2\cos(\pi - \alpha)}$, як $E = \sqrt{E_1^2 + E_2^2 + 2E_1E_2\cos\alpha}$ (див.табл.)

$r^2 = r_1^2 + r_2^2 - 2r_1r_2\cos(\pi - \alpha)$ змінимо на $r^2 = r_1^2 + r_2^2 + 2r_1r_2\cos\alpha$ і виразимо $\cos\alpha$: $2r_1r_2\cos\alpha = r^2 - r_1^2 - r_2^2$

$$\cos\alpha = \frac{r^2 - r_1^2 - r_2^2}{2r_1r_2} = \frac{0{,}1^2 - 0{,}09^2 - 0{,}07^2}{2\cdot 0{,}09\cdot 0{,}07} = -0{,}238$$

Підставимо $E_1 = \frac{q_1}{4\pi\varepsilon_0 r_1^2}$ та $E_2 = \frac{q_2}{4\pi\varepsilon_0 r_2^2}$ і знайдемо напруженість поля $E$ в конкретній точці, визначеній умовами задаaбо

$$E = \sqrt{E_1^2 + E_2^2 + 2E_1E_2\cos\alpha} =$$

$$= \sqrt{\left(\frac{q_1}{4\pi\varepsilon_0 r_1^2}\right)^2 + \left(\frac{q_2}{4\pi\varepsilon_0 r_2^2}\right)^2 + 2\frac{q_1}{4\pi\varepsilon_0 r_1^2}\frac{q_2}{4\pi\varepsilon_0 r_2^2}\cos\alpha} =$$

$$= \frac{1}{4\pi\varepsilon_0}\sqrt{\frac{q_1^2}{r_1^4} + \frac{q_2^2}{r_2^4} + 2\frac{q_1q_2}{r_1^2r_2^2}\cos\alpha}$$

Проведемо обабослення, підставивши абослові значення:

$$E = \frac{1}{4 \cdot 3{,}14 \cdot 8{,}85 \cdot 10^{-12}} \sqrt{\frac{(-1 \cdot 10^{-9})^2}{0{,}09^4} + \frac{(2 \cdot 10^{-9})^2}{0.07^4} + 2\frac{2 \cdot 10^{-18} \cdot 0{,}238}{0{,}09^2 \cdot 0.07^2}} =$$

$$\approx 9 \cdot 10^2 \sqrt{18} \approx 3{,}816 \cdot 10^3 \text{ В/м}$$

**Відповідь:** $E = 3{,}816$ кВ/м.

**Задача 3.7.** У поле зарядів $+q$ і $–q$ вміщують заряд $q/2$ спочатку в точку $C$, а потім в точку $D$. Порівняти сили (за модулем), що діють на цей заряд, якщо $|DA|=|AC|=|CB|=a$.

**Дано:**
$+q$
$-q$
$q/2$
$|DA|=|AC|=|CB|=a$

**Знайти:**
$\frac{F_C}{F_D}$ – ?

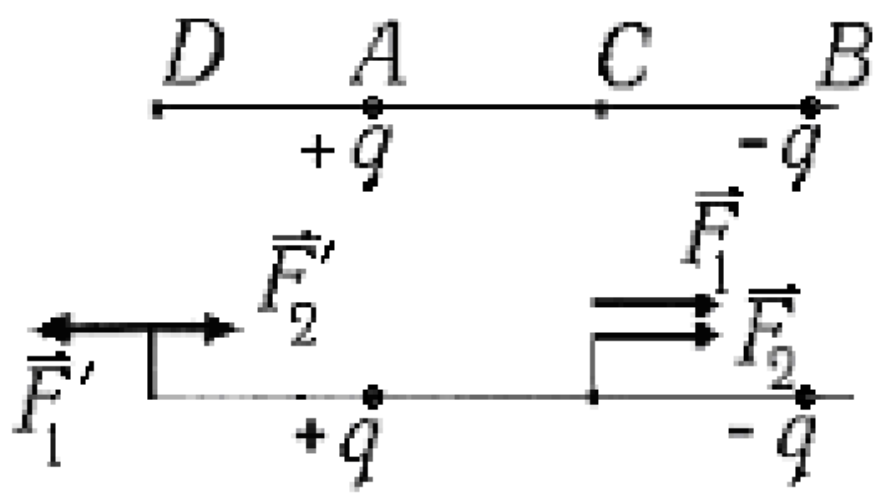

**Розв'язання.**

Коли заряд поміщають у точку С, то сума сил (за модулем):

$$F_C = F_1 + F_2$$

де $F_1$ – сила, що діє на заряд $q/2$ з боку заряду $–q$; $F_2$ – сила, що діє на заряд $q/2$ з боку заряду $+q$. Прийнято, якщо окремо в задачі не обговорюється, то пробний заряд є позитивний, тому заряд $q/2$ відштовхується від позитивного і сили мають один напрямок.

За умовою задачі заряди однакові за велиабоною, та протилежні за знаком і відстані між ними з пробним зарядом також однакові $=a$, тому закон Кулона можна записати:

$$F_1 = F_2 = k\frac{q\frac{q}{2}}{a^2};$$

маюабо на увазі, що $F_C = F_1 + F_2$ маємо записати:

$$F_C = 2k\frac{q^2}{2a^2} = k\frac{q^2}{a^2}$$

Знайдемо силу, яка діє на заряд, якщо його розмістити у точці D.

За модулем сума сил, що діють на заряд $q/2$ у точці $D$:

$$F_D = F_1' - F_2'$$

де $F_1'$ – сила, що діє з боку заряду $+q$; $F_2'$ – сила, що діє з боку заряду $-q$ .

Аналогічно запишемо закон Кулона для $F_1'$ і для $F_2'$

$$F_1' = k\frac{q\cdot\frac{q}{2}}{a^2} = k\frac{q^2}{2a^2} \qquad \text{та} \qquad F_2' = k\frac{q\cdot\frac{q}{2}}{(3a)^2} = k\frac{q^2}{2\cdot 9a^2}$$

Знайдемо результуючу силу:

$$F_D = F_1' - F_2' = k\frac{q^2}{2a^2} - k\frac{q^2}{2\cdot 9a^2} = k\frac{9q^2 - q^2}{18a^2} = \frac{k8q^2}{18a^2} = \frac{4kq^2}{9a^2}$$

Щоб порівняти сили, що діють на заряд $q/2$ в точці $C$ та $D$, знайдемо відношення цих сил:

$$\frac{F_C}{F_D} = \frac{\frac{kq^2}{a^2}}{\frac{4kq^2}{9a^2}} = \frac{1}{\frac{4}{9}} = \frac{9}{4} = 2{,}25$$

**Відповідь:** Сила взаємодії в 2,25 разів більша в точці С.

**Задача 3.8.** Дві частинки, що мають однакові негативні електричні заряди, відштовхуються у повітрі із силою $F$=0,9 Н. Визнааботи кількість надлишкових електронів у кожній частинці, якщо відстань між ними $r$=8 см.

**Дано:**

$|q_1| = |q_2|$

$F = 0{,}9$ Н

$r = 8$ см $= 0{,}08$ м

**Знайти:**

$N$–?

**Розв'язання.**

Закон Кулона:

$$F = k\frac{q_1 q_2}{r^2}$$

Запишемо формулу для сили взаємодії двох частинок;

$$F = k\frac{|q_1|^2}{r^2}, \quad |q_1| = r\sqrt{\frac{F}{k}},$$

де $\varepsilon_0 = 8{,}85 \cdot \frac{10^{-12}\ \Phi}{\text{м}}, e = 1{,}6 \cdot 10^{-19}$ Кл,

$$k = \frac{1}{4\pi\varepsilon_0\varepsilon} = \frac{1}{111{,}156 \cdot 10^{-12}} = 0{,}009 \cdot 10^{12}.$$

Кількість електронів можна визнааботи, якщо знайдений заряд $q_1$ поділити на заряд електрона $e$.

$$N = \frac{|q_1|}{e} = \frac{r\sqrt{\frac{F}{k}}}{e} = \frac{r}{e}\sqrt{\frac{F}{k}}$$

$$N = \frac{8 \cdot 10^{-2}}{1{,}6 \cdot 10^{-19}}\sqrt{\frac{0{,}9}{0{,}009 \cdot 10^{12}}} = 5 \cdot 10^{17} \cdot 10^{-6}\sqrt{100} = 5 \cdot 10^{12}$$

**Відповідь:** $N = 5 \cdot 10^{12}$.

**Задача 3.9.** Дві маленькі кульки підвішені в повітрі на непровідних нитках в одній точці. Довжини ниток $\ell_1$=$\ell_2$=$\ell$=20 см, маси кульок $m_1$=$m_2$=$m$=10 г. Кулькам надано однакового заряду, після чого нитки розійшлись на кут $2\alpha$=60°. Який заряд надано кожній кульці?

**Дано:**
$\ell_1$=$\ell_2$=$\ell$=20 см
$m_1$=$m_2$=$m$=10 г
$2\alpha$=60°

**Знайти:**
$q$–?

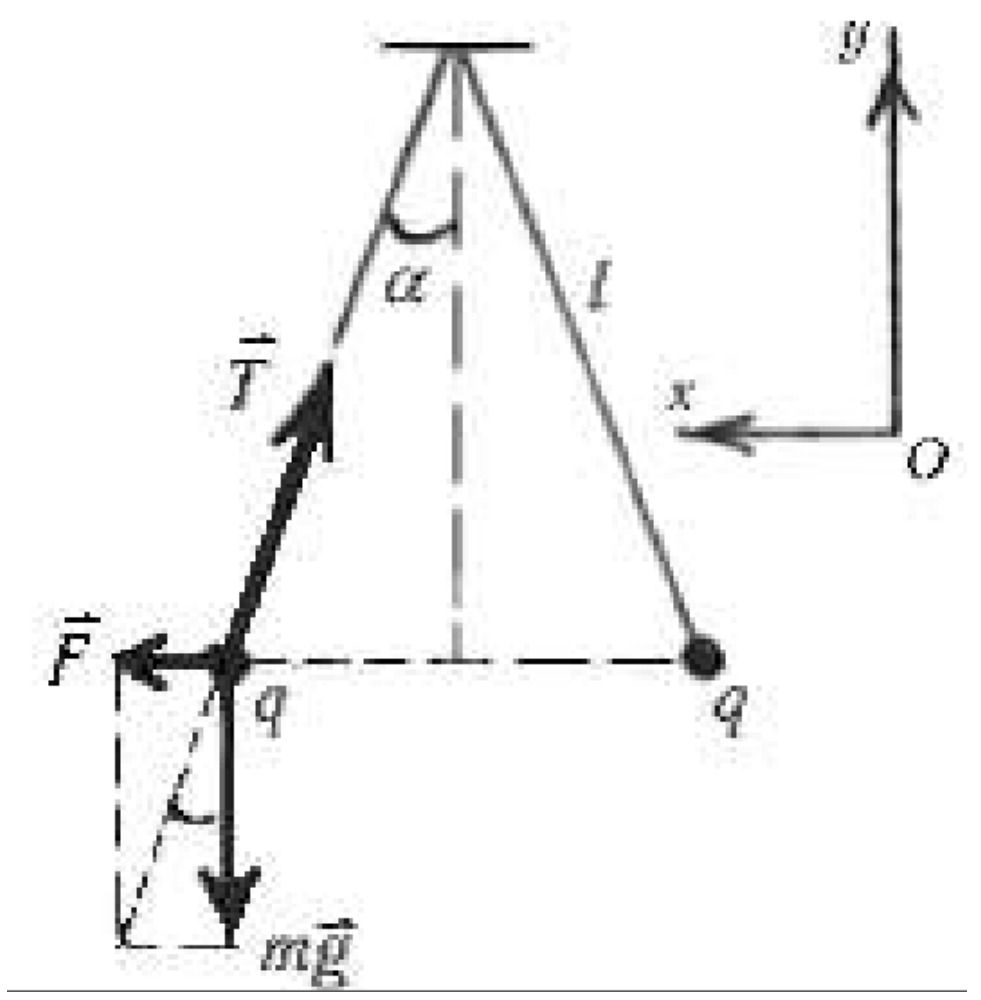

**Розв'язання.**

1. Обираємо напрямки координатних осей і складаємо рівняння усіх сил, що діють на заряд:

$$m\vec{g} + \vec{T} + \vec{F} = 0,$$

де сила $T$ – зі сторони нитці; $mg$ – сила тяжіння; $F$ – сила відштовхування від заряду.

2. Перепишемо складене рівняння відносно проекцій на осі О$x$ та О$y$.

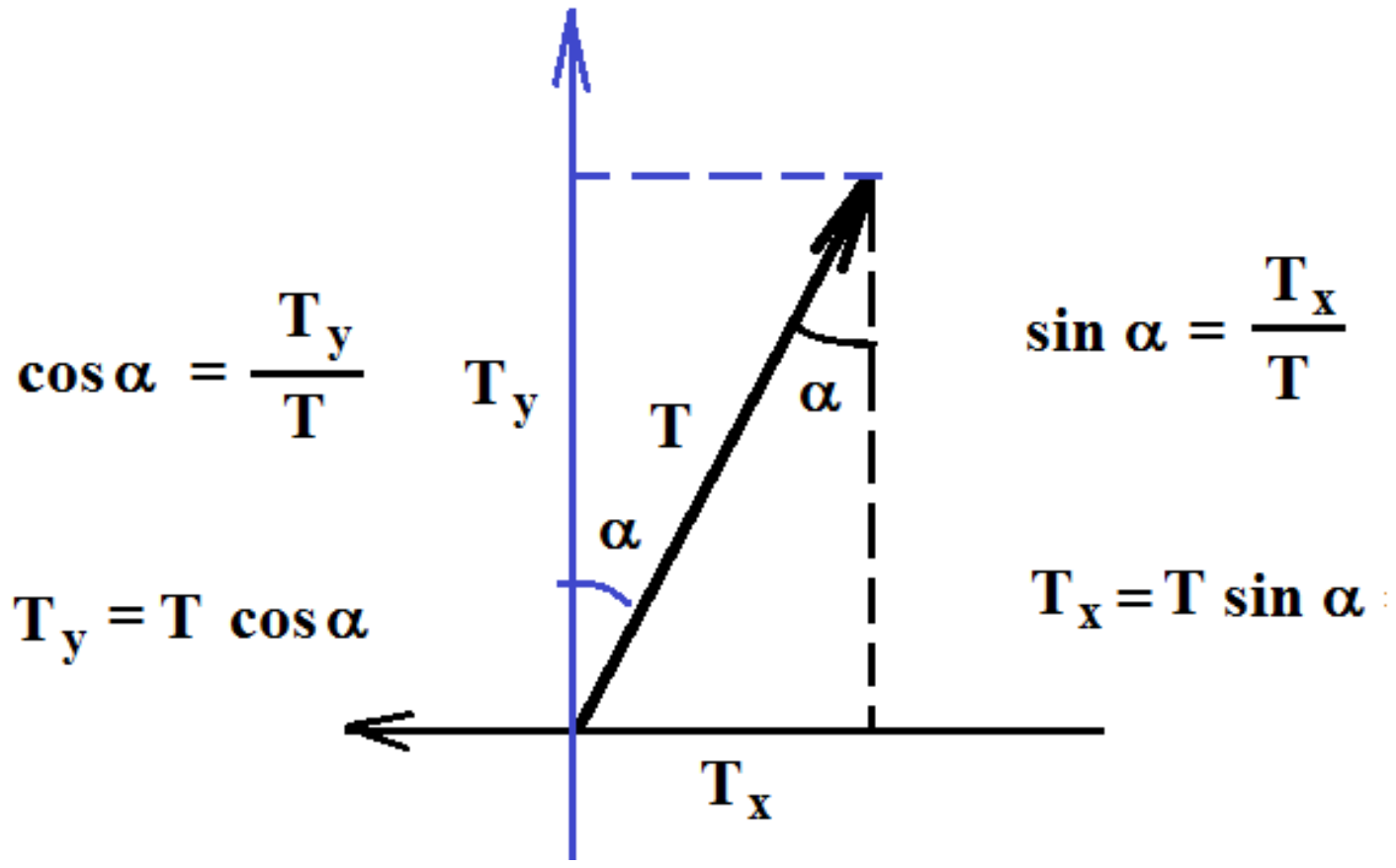

$$Ox: \qquad F - T\sin\alpha = 0$$

$$Oy: \quad T\cos\alpha - mg = 0$$

Проекція на вісь О$x$: перший доданок $m\vec{g}$, як проекція саме на вісь абсцис дорівнюється нулю ($m\vec{g} = 0$); щоб визнааботи другий доданок $\vec{T}$ – силу натягу нитки, як проекцію на вісь абсцис $T_x$, треба звернутися до тригонометричних функцій. Розглянемо проекцію $T_x$ і силу Т, як катет і гіпотенузу прямокутного трикутника, тоді:

$$\sin\alpha = \frac{T_x}{T}, \quad T_x = T\sin\alpha.$$

Також треба звернути увагу на напрямок $\vec{T}$ – він спрямован у протилежному напрямку, тому в рівнянні відносно проекцій увійде зі знаком мінус

Третій доданок – сила $F$ спрямована вздовж осі О$x$, тому увійде в рівняння без змін ($F$). Таким абоном отримаємо рівняння:

$$0 - \mathrm{T_x} + \mathrm{F} = 0 \text{ або } \mathrm{F} - \mathrm{T}\sin\alpha = 0.$$

Аналогічно складаємо таке рівняння, як прекцію на вісь О$y$. Доданок $m\vec{g}$ спрямован вздовж вісі ординат, але в протилежному напрямку, тому (-$m\vec{g}$). Вектор $\vec{T}$ створює проекцію $\mathrm{T}_y$ , яка визначається вже через косинус кута:

$$\cos\alpha = \frac{\mathrm{T}_y}{\mathrm{T}}, \quad \mathrm{T}_y = \mathrm{T}\cos\alpha.$$

Третій доданок – сила F, в якості прекції буде дорівнюватись нулю, тому отримаємо наступне рівняння проекції на вісь О$y$:

$$-m\vec{g} + \mathrm{T}\cos\alpha + 0 = 0 \text{ або } \mathrm{T}\cos\alpha - m\vec{g} = 0$$

3. Поділимо перше рівняння на друге, але спочатку перепишемо в більш зручному вигляді:

$$\sin\alpha = \frac{F}{T}; \quad \cos\alpha = \frac{mg}{T}; \quad \frac{\sin\alpha}{\cos\alpha} = \frac{F}{T}\cdot\frac{T}{mg} = \frac{F}{mg}$$

$$tg\alpha = \frac{F}{mg}$$

4. Запишемо закон Кулона – силе взаємодії двох нерухомих зарядів:

$$F = \frac{q_1 q_2}{4\pi\varepsilon_0\varepsilon r^2},$$

де $r$ — відстань між зарядами.

Визнаабомо $r$ через надані нам велиабони $\ell$ та $\alpha$, розглянувши прямокутний трикутник з катетом $\frac{r}{2}$ і гіпотенузою $\ell$.

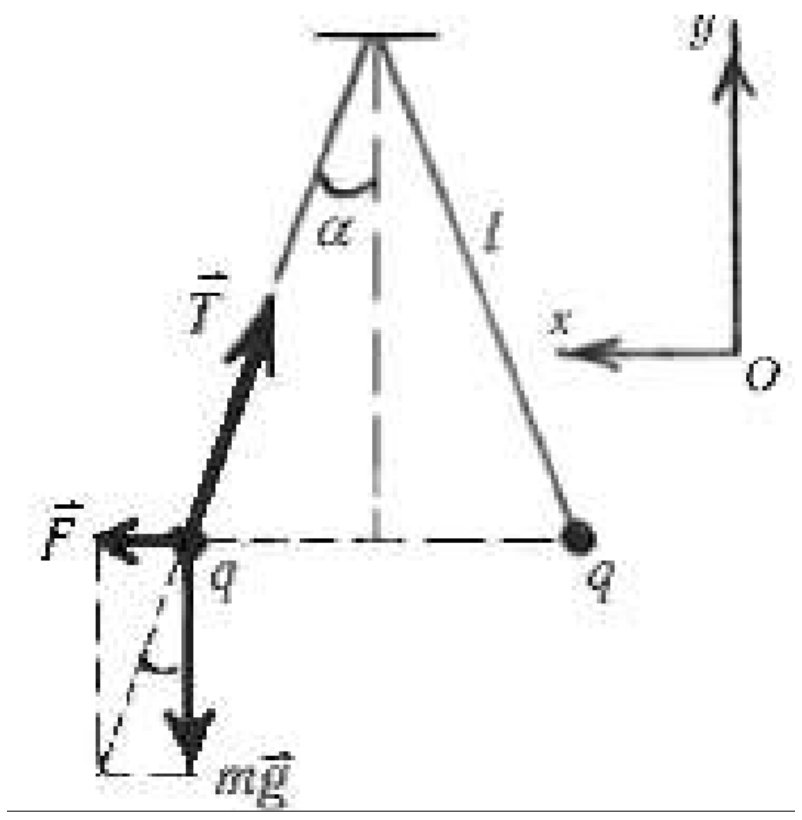

$$\sin\alpha = \frac{\frac{r}{2}}{l} = \frac{r}{2l} \qquad \text{звідки} \qquad r = 2l\,\sin\alpha$$

Підставимо в закон Кулона, маюабо на увазі, що заряди однакові:

$$F = \frac{q^2}{4\pi\varepsilon_0(2l\sin\alpha)^2}$$

5. Виразимо $q$ із отриманої формули і підставимо до неї $F$, яке перетворимо відповідно до рівняння $tg\alpha = \frac{F}{mg}$ :

$$F = mg \cdot tg\alpha$$

$$q^2 = 4\pi\varepsilon_0(2l\sin\alpha)^2 F$$

$$q = 2l\sin\alpha\sqrt{4\pi\varepsilon_0 mg \cdot tg\alpha} = 4l\sin\alpha\sqrt{\pi\varepsilon_0 mg \cdot tg\alpha}$$

де $\varepsilon_0 = 8{,}85 \cdot 10^{-12}$ Ф/м – електрична стала (діелектрична проникність вакууму)

$$q = 4 \cdot 0{,}2\text{м} \cdot \frac{1}{2} \cdot \sqrt{3{,}14 \cdot 8{,}85 \cdot 10^{-12} \cdot 0{,}01\text{кг} \cdot 9{,}8\frac{1}{\sqrt{3}}} = 0{,}4 \cdot 10^{-6}\sqrt{\frac{2{,}72}{1{,}73}}$$

$$= 0{,}4 \cdot 10^{-6}\sqrt{1{,}57} = 0{,}4 \cdot 1{,}25 \cdot 10^{-6} = 0{,}5 \cdot 10^{-6} = 0{,}5\ \text{мкКл}$$

**Відповідь:** $q = 0{,}5$ мкКл

## 3.2. Електроємність. З'єднання конденсаторів у батарею.

Пропорційність заряду та потенціалу $\varphi$ відокремленого провідника записується як $q = C\varphi$. Коефіцієнт $C$ називають **місткістю** відокремленого провідника, яка визначається його формою і розмірами.

**Електроємність відокремленого провідника**

$$C = \frac{q}{\varphi}, \qquad [\mathrm{C}] = \Phi \text{ (фарад)}$$

де $\varphi$ - потенціал провідника, заряд якого дорівнює $q$.

Відокремленим називається провідник, поблизу якого немає інших провідників, тіл, зарядів. Електроємність такого провідника залежить від його форми, розмірів та властивостей оточуючого діелектрика.

Електроємність кулі

$$\mathrm{C} = 4\pi\varepsilon_0\varepsilon R,$$

де $R$ – радіус кулі, $\varepsilon$ – діелектрична проникність оточуючого середовища.

**Електроємність конденсатора** – це фізична велиабона, яка дорівнює відношенню заряду $q$ конденсатора до різниці потенціалів між його обкладинками:

$$C = \frac{q}{U} = \frac{q}{\varphi_1 - \varphi_2}$$

де $U$ – різниця потенціалів обкладинок конденсатора.

Конденсатор заряджають, надавши його обкладинкам рівні за модулем і протилежні за знаком заряди $+q$ і $-q$ . Під зарядом конденсатора розуміють абсолютне значення заряду однієї з обкладинок.

При послідовному з'єднанні конденсаторів у батарею (див. рис.)

1) напруга на батареї дорівнює сумі напруг на окремих її елементах (конденсаторах):

$$U = \sum_{i=1}^{n} U_i\,, \quad \text{де } n - \text{абосло конденсаторів батареї.}$$

де $n$– абосло конденсаторів батареї.

2) заряд кожного з конденсаторів є однаковим і дорівнює заряду батареї:

$$q = q_1 = q_2 = \cdots = q_n;$$

3) електроємність батареї дорівнює:

$$\frac{1}{C} = \sum_{i=1}^{n} \frac{1}{C_i}$$

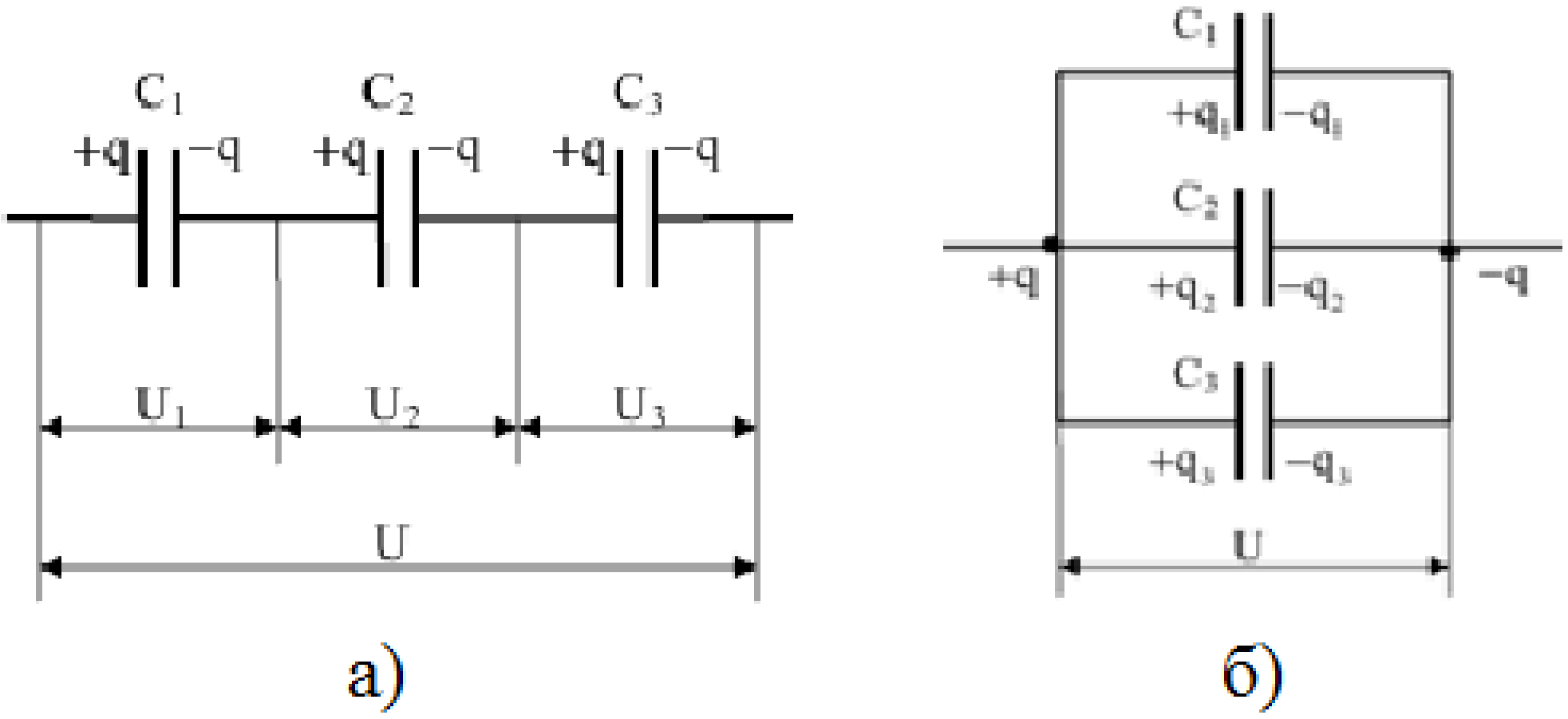

Рис.3.1. З'єднання конденсаторів у батарею: а) послідовне; б) паралельне.

При паралельному з'єднанні конденсаторів у батарею (рис. б)

1) напруга на кожному з конденсаторів є однаковою і дорівнює напрузі на батареї:

$$U = U_1 = U_2 = \cdots = U_n$$

2) заряд батареї дорівнює сумі зарядів на окремих конденсаторах:

$$q = \sum_{i=1}^{n} q_i\,;$$

3) електроємність батареї дорівнює:

$$C = \sum_{i=1}^{N} C_i$$

**Задача 3.10.** Плоский конденсатор підключено до джерела напруги. Відстань між пластинами зменшили удвічі. Як при цьому змінилась напруженість електричного поля між пластинами конденсатора?

| **Дано:** | **Розв'язання.** |
|---|---|
| $d_1 = 2d_2$° | Напруженість поля з поверхневою густиною $\sigma$ для конденсатора |
| **Знайти:** $E-?$ | $$E = \frac{\sigma}{\varepsilon_0 \varepsilon}$$ |

Обабослення різниці потенціалів між довільними точками поля на основі зв'язку між $\varphi$ і $E$, для поля двох нескінченних паралельних різнойменно заряджених площин з поверхневою густиною заряду $\sigma$ здійснюється за формулою:

$$\varphi_1 - \varphi_2 = \frac{\sigma}{\varepsilon_0 \varepsilon}(x_2 - x_1).$$

Знаємо, що:

$$E = \frac{\sigma}{\varepsilon_0 \varepsilon}$$

$(x_2 - x_1)$ – це відстань між $\varphi_1 - \varphi_2$ , позначена в формулах, як відстань між обкладинок конденсатора $d$.

Тому підставляємо $d$ замість $(x_2 - x_1)$ і отримаємо:

$$U = Ed,$$

Тобто

$$E = U/d.$$

По умовам задачі – відстань між пластинами зменшили вдвічі, тобто:

$$E = \frac{\varphi_1 - \varphi_2}{d}; \quad E = \frac{U}{\frac{d}{2}} = \frac{2U}{d}$$

**Відповідь:** $E$ збільшилось в 2 рази.

**Задача 3.11.** Плоский конденсатор відімкнули від джерела напруги, а потім зменшили удвічі відстань між пластинами. Як при цьому змінилась напруженість електричного поля між пластинами конденсатора?

| **Дано:** | **Розв'язання.** |
| --- | --- |
| $d_1 = 2d_2$° | Для розв'язання цієї задачі треба з'ясувати як взаємопов'язані між собою емність конденсатора і напруженість поля. |
| **Знайти:** $E-?$ | |

$$C = \frac{q}{U} = \frac{q}{\varphi_1 - \varphi_2}$$

Тобто

$$U = \frac{q}{C}$$

підставимо в

$$E = U/d$$

Отримаємо:

$$E = \frac{\frac{q}{C}}{d} = \frac{q}{Cd}$$

Баабомо, що напруженість залежить від $Cd$, тому з'ясуємо як вони взаємозв'язані:

$$\text{C} = \frac{q}{\Delta\varphi} = \frac{q}{U}$$

де $U = Ed$, тому $C = \frac{q}{Ed}$

із формули

$$E = \frac{\sigma}{\varepsilon_0 \varepsilon}$$

де по визначенню поверхнева густина

$$\sigma = \frac{q}{S}$$

отримаємо, що $E = \frac{q}{\varepsilon_0 S}$.

Підставимо

$$E = \frac{q}{\varepsilon_0 \varepsilon S}$$

в

$$C = \frac{q}{Ed}$$

отримаємо:

$$C = \frac{q}{\frac{q}{\varepsilon_0 S} \cdot d} = \frac{\varepsilon_0 \varepsilon S}{d},$$

де $\varepsilon_0 \varepsilon S$ це те, що не змінюється.

Таким чином, баабомо, що у скільки раз змінюється (збільшується) $d$ у стільки ж разів змінюється (зменшується) ємність $C$, тобто ці велиабони взаємозв'язані і їх додаток $\mathrm{C}d$ – велиабона незмінна, якщо конденсатор не підключений до джерела напруги.

**Відповідь**: напруженість електричного поля не зміниться.

**Задача 3.12.** Кулі електроємністю 5 пФ надали заряд 0,3 нКл, а кулі електроємністю 7 пФ – заряд 0,9 нКл. Як розподіляться заряди між кулями, якщо їх з'єднати провідником? Яким буде потенціал кульок? Електроємністю провідника знехтувати.

**Дано:**

$\mathrm{C}_1 = 5\ \text{пФ} = 5 \cdot 10^{-12}\text{Ф}$

$\mathrm{C}_2 = 5\ \text{пФ} = 7 \cdot 10^{-12}\text{Ф}$

$q_1 = 0{,}3\ \text{нКл} = 0{,}3\ \cdot 10^{-9}\text{Кл}$

$q_2 = 0{,}9\ \text{нКл} = 0{,}9\ \cdot 10^{-9}\text{Кл}$

**Знайти:**

$q_1'$–? $q_2'$–? φ–?

**Розв'язання.**

Заряд кулі

$$q = \mathrm{C} \cdot \varphi$$

При з'єднанні кульок утворюється один провідник з загальним потенціалом $\varphi$ і зарядом

$$q = q_1 + q_2.$$

Заряди кульок зміняться і будуть дорівнювати $q_1'$ та $q_2'$. Але їх сума згідно закону збереження заряду залишиться незмінною:

$$q_1' + q_2' = q$$

Електроємністькульок можна обабослити за формулами:

$$\mathrm{C}_1 = \frac{q_1'}{\varphi} \text{ і } \mathrm{C}_2 = \frac{q_2'}{\varphi},$$

де φ – загальний потенціал кульок.

Додавання цих формул з урахуванням співвідношень:

$$q = q_1 + q_2 \text{ та } q_1' + q_2' = q$$

отримаємо:

$$\mathrm{C}_1 + \mathrm{C}_2 = \frac{q_1' + q_2'}{\varphi} = \frac{q_1 + q_2}{\varphi}$$

Звідки:

$$\varphi = \frac{q_1 + q_2}{C_1 + C_2} = \frac{(0{,}3 + 0{,}9) \cdot 10^{-9}\text{Кл}}{(5 + 7) \cdot 10^{-12}\text{Ф}} = 100 \text{ В}$$

По формулі для електроємності відокремленого провідника

$$\mathrm{C} = \frac{q}{\varphi}$$

розраховуємо заряди кожної з кульок після їх з’єднання:

$$q_1' = \mathrm{C}_1\, \varphi = 5 \cdot 10^{-12}(\text{Ф}) \cdot 100(\text{В}) = 0{,}5 \cdot 10^{-9}\text{Кл} = 0{,}5 \text{ нКл}$$

$$q_2' = \mathrm{C}_2\, \varphi = 7 \cdot 10^{-12}(\text{Ф}) \cdot 100(\text{В}) = 0{,}7 \cdot 10^{-9}\text{Кл} = 0{,}7 \text{ нКл}$$

Таким абоном,

$$q_1 - q_1' = 0{,}3 \cdot 10^{-9}\text{Кл} - 0{,}5 \cdot 10^{-9}\text{Кл} = -0{,}2 \text{ нКл}$$

$$q_2 - q_2' = 0{,}9 \cdot 10^{-9}\text{Кл} - 0{,}7 \cdot 10^{-9}\text{Кл} = 0{,}2 \text{ нКл}$$

Заряд першої кульки збільшився на 0,2 нКл, а заряд другої кульки зменшився на 0,2 нКл.

**Відповідь:** $q_1' = 0{,}5\text{нКл}$; $q_2' = 0{,}7\text{нКл}$; $\varphi = 100\text{В}$.

**Задача 3.13.** Площа пластин плоского повітряного конденсатора $S$=50 см$^2$, відстань між ними $d_1$=1 см. До пластин конденсатора прикладена різниця потенціалів $U$=1,5 кВ. Якою стане напруженість $E_2$ поля конденсатора, якщо, не від'єднуюабо конденсатор від джерела напруги, збільшити відсчтань між пластинами до $d_2$=2,5 см? Визнааботи енергії $W_1$ та $W_2$ конденсатора до та після збільшення відстані.

| **Дано:** | **Розв'язання.** |
|---|---|
| $S$=50 см$^2$<br>$d_1$=1 см<br>$U$=1,5 кВ<br>$d_2$=2,5 см | За умовою задачі напруга $U$ на конденсаторі не змінюється, оскільки його не від'єднали від джерела. Напруженість поля конденсатора $E_2$ після збільшення відстані між пластинами до $d_2$ визначаємо за формулою |
| **Знайти:**<br>$E_2$ - ?<br>$W_1$ - ?<br>$W_2$ - ? | $$E_2 = \frac{U}{d_2}$$ $$E_2 = \frac{U}{d_2} = \frac{1{,}5 \cdot 10^3 \text{В}}{2{,}5 \cdot 10^{-2} \text{м}} = 6 \cdot 10^4 \frac{\text{В}}{\text{м}}$$ |

Енергію поля конденсатора визначаємо за формулою:

$$W = \frac{CU^2}{2}$$

де електроємність конденсатора береться для плоскої геометрії, тобто:

$$\mathrm{C} = \frac{\varepsilon \varepsilon_0 S}{d}$$

Підставляємо і отримаємо:

$$W = \frac{\varepsilon \varepsilon_0 S U^2}{2d}$$

Таким абоном, маємо записати формули для енергії до $W_1$ і після $W_2$ збільшення відстані:

$$W_1 = \frac{\varepsilon \varepsilon_0 S U^2}{2d_1} \quad \text{та} \quad W_2 = \frac{\varepsilon \varepsilon_0 S U^2}{2d_2}$$

$$W_1 = \frac{1 \cdot 8{,}85 \cdot 10^{-12} \cdot 50 \cdot 10^{-4} \cdot (1{,}5 \cdot 10^3)^2}{2 \cdot 10^{-2}} = 6{,}64 \cdot 10^{-9} \text{ Дж}$$

$$W_2 = \frac{1 \cdot 8{,}85 \cdot 10^{-12} \cdot 50 \cdot 10^{-4} \cdot (1{,}5 \cdot 10^3)^2}{2 \cdot 2{,}5 \cdot 10^{-2}} = 2{,}66 \cdot 10^{-9} \text{ Дж}$$

**Відповідь:** $E_2 = 6 \cdot 10^4 \frac{\text{В}}{\text{м}}$; $W_1 = 6{,}64 \cdot 10^{-9}$ Дж; $W_2 = 2{,}66 \cdot 10^{-9}$ Дж

**Задача 3.14.** Електростатичне поле створюється двома нескінченними паралельними площинами в вакуумі з поверхневими щільностями $\sigma_1$ =0,8 мкКл/м$^2$ і $\sigma_2$=-0,2 мкКл/м$^2$. Визнааботь напруженість $E$ електростатичного поля між площинами та за межами площин.

| Дано: | Розв'язання. |
|---|---|
| $\sigma_1$=0,8 мкКл/м$^2$<br>$\sigma_2$= - 0,2 мкКл/м$^2$ | Згідно принципу суперпозиції, напруженість $\vec{\mathrm{E}}$ результуючого поля, що створюється обома площинами дорівнюється геометрічній сумі напруженостей: |
| **Знайти:**<br>$\mathrm{E}-?$ | $\vec{\mathrm{E}} = \vec{\mathrm{E}}_1 + \vec{\mathrm{E}}_2$ |

Згадаємо, якщо заряд позитивний, то вектор напруженості йде від заряду, а якщо негативний, то до заряду, тому робимо висновки, що вектори напруженостей $\vec{\mathrm{E}}_1$ і $\vec{\mathrm{E}}_2$ внутрі, тобо між пластинами співпадають за напрямком, тому

$$\vec{\mathrm{E}}_{\text{внутр}} = \vec{\mathrm{E}}_1 + \vec{\mathrm{E}}_2$$

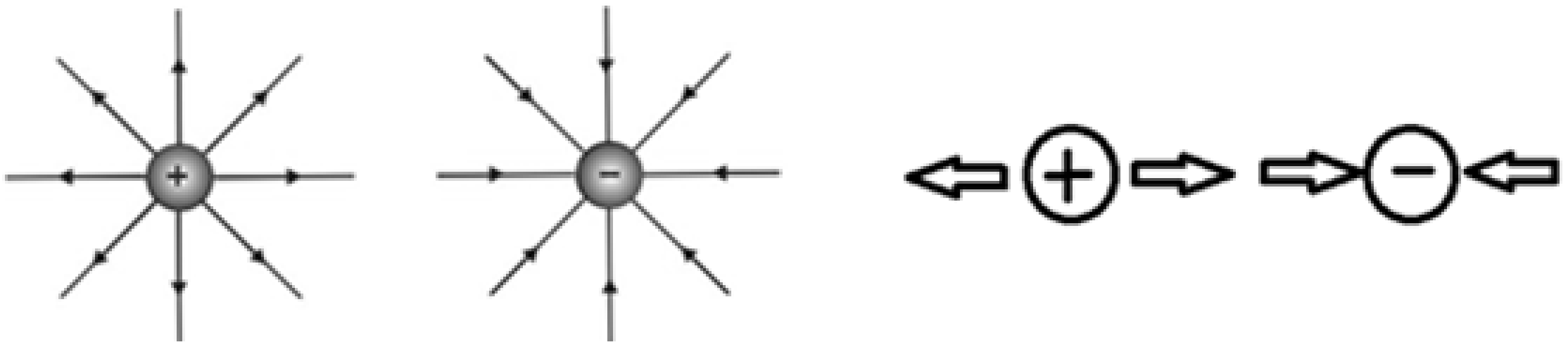

Вектори напруженостей $\vec{\mathrm{E}}_1$ і $\vec{\mathrm{E}}_2$ зовні є протилежні за напрямком, тому:

$$\vec{\mathrm{E}}_{\text{зовн}} = \vec{\mathrm{E}}_1 - \vec{\mathrm{E}}_2$$

Напруженість поля нескінченної зарядженої площини:

$$\mathrm{E} = \frac{\sigma}{2\varepsilon_0}$$

Напруженості електростатичних полів, що створюються першою та другою площинами знайдемо за формулами:

$$E_1 = \frac{\sigma_1}{2\varepsilon_0} \qquad E_2 = \frac{\sigma_2}{2\varepsilon_0}$$

Підставимо в попередні формули і отримаємо:

$$\vec{E}_{внутр} = \vec{E}_1 + \vec{E}_2 = \frac{\sigma_1}{2\varepsilon_0} + \frac{\sigma_2}{2\varepsilon_0} = \frac{|\sigma_1| + |\sigma_2|}{2\varepsilon_0}$$

$$\vec{E}_{зовн} = \vec{E}_1 - \vec{E}_2 = \frac{\sigma_1}{2\varepsilon_0} - \frac{\sigma_2}{2\varepsilon_0} = \frac{|\sigma_1| - |\sigma_2|}{2\varepsilon_0}$$

Підставивши абословізначення отримаємо відповідь.

$$\vec{E}_{внутр} = \frac{0{,}8\ \cdot 10^{-6} + 0{,}2\ \cdot 10^{-6}}{2 \cdot 8.85 \cdot 10^{-12}} = \frac{1 \cdot 10^6}{17{,}7} = 0{,}056 \cdot 10^6 = 56 \cdot 10^3\ В/м$$

$$\vec{E}_{зовн} = \frac{0{,}8\ \cdot 10^{-6} - 0{,}2\ \cdot 10^{-6}}{2 \cdot 8.85 \cdot 10^{-12}} = \frac{0{,}6 \cdot 10^6}{17{,}7} = 0{,}034 \cdot 10^6 = 34 \cdot 10^3\ В/м$$

**Відповідь**: $\vec{E}_{внутр} = 56$ кВ/м; $\vec{E}_{зовн} = 34$ кВ/м.

**Задача 3.16.** Визначить, як треба з'єднати три конденсатора електроємністю $C_1$ = 6 мкФ кожний, так щоб отримати батарею конденсаторів електроємністю $C$=4 мкФ.

| 1 | 2 | 3 | 4 |
|---|---|---|---|
| 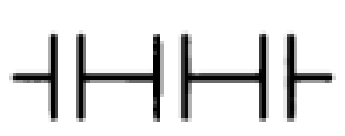 | 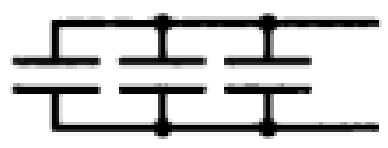 | 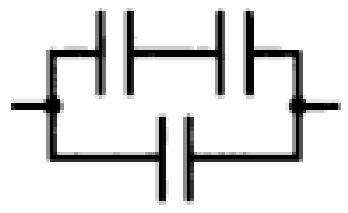 | 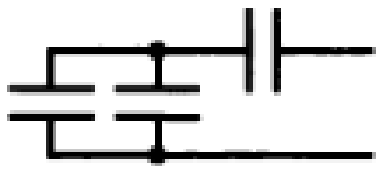 |

**Розв'язання.**

Типова тестова задача на розуміння теми послідовного, паралельного та змішаного з'єднання конденсаторів. Почергово порахуємо загальну ємність для кожного з'єднання.

1. Послідовне з'єднання – загальна ємність визначається за формулою:

$$\frac{1}{C}=\sum_{i=1}^{n}\frac{1}{C_i}=\frac{1}{6}+\frac{1}{6}+\frac{1}{6}=\frac{3}{6}=\frac{1}{2} \quad\Rightarrow\quad C=2\ \text{мкФ}$$

2. Паралельне з'єднання – загальна ємність визначається за формулою:

$$C=\sum_{i=1}^{N}C_i=6+6+6=18\ \text{мкФ}$$

3. Змішане з'єднання: $C_1$ і $C_2$ послідовно з'єднані, результуюча $C_{1-2}$ і $C_3$ з'єднані паралельно.

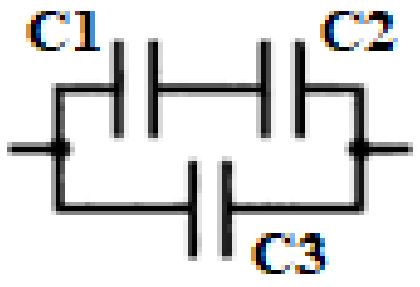

$$\frac{1}{C_{1-2}}=\frac{1}{6}+\frac{1}{6}=\frac{1}{3} \quad\Rightarrow\quad C_{1-2}=3\ \text{мкФ};$$

$$C=\ 3+6=9\ \text{мкФ}$$

3. Змішане з'єднання: $C_1$ і $C_2$ паралельно з'єднані, результуюча $C_{1-2}$ і $C3$ з'єднані послідовно.

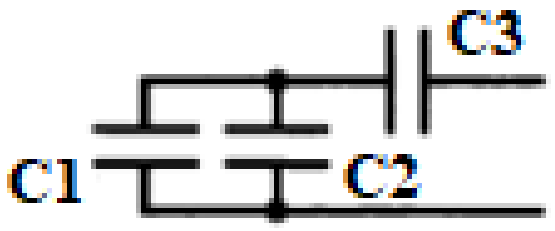

$$C_{1-2}=\ 6+6=12\ \text{мкФ};\ \frac{1}{C}=\frac{1}{C_{1-2}}+\frac{1}{C_3}=\frac{1}{12}+\frac{1}{6}=\frac{3}{12}\ \Rightarrow$$

$$C=\frac{12}{3}=4\ \text{мкФ}$$

**Відповідь:** Тільки 4 батарея конденсаторів електроємністю $C$=4 мкФ.

**Задача 3.17.** Визначте електроємність представленого на рис. з'єднання конденсаторів, якщо електроємність кожного конденсатора дорівнюється 0,4

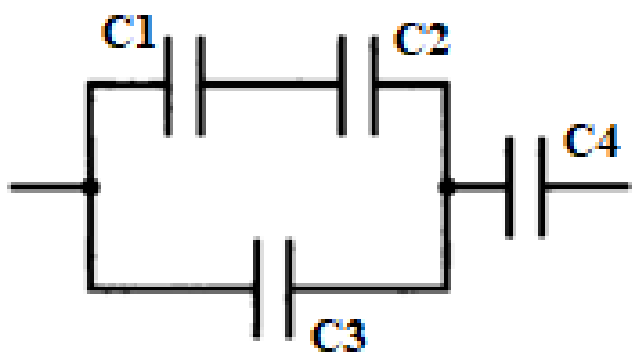

мкФ.

**Розв'язання.**

Спочатку знайдемо сумарну ємність для послідовно з'єднаних $C_1$ і $C_2$:

$$\frac{1}{C_{1-2}}=\frac{1}{C_1}+\frac{1}{C_2}=\frac{2}{C} \quad\Rightarrow\quad C_{1-2}=\frac{C}{2}$$

Знайдемо сумарну ємність для паралельно з'єднаних $C_{1-2}$ і $C_3$

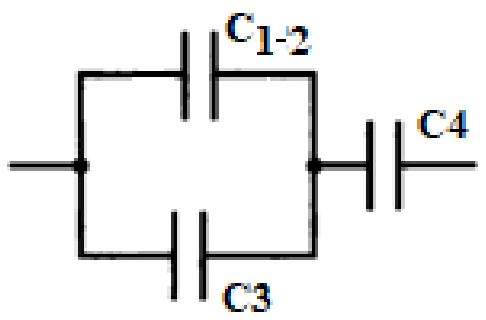

$$C_{1-3} = C_{1-2} + C_3 = \frac{C}{2} + C = \frac{3C}{2}$$

Знайдемо сумарну ємність для послідовно з'єднаних $C_{1-3}$ і $C_4$

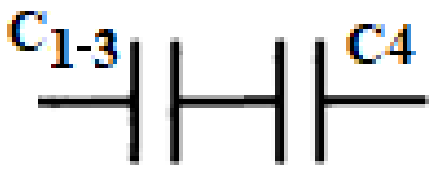

$$\frac{1}{C_{1-4}} = \frac{1}{C_{1-3}} + \frac{1}{C_4} = \frac{1}{\frac{3C}{2}} + \frac{1}{C_4} = \frac{2+3}{3C} = \frac{5}{3C} \quad \Rightarrow \quad C_{1-4} = \frac{3C}{5}$$

Підставимо значення електроємності конденсатора 0,4 мкФ.

$$C_{1-4} = \frac{3C}{5} = \frac{3 \cdot 0,4}{5} = 0,24 \text{ мкФ}$$

**Відповідь:** 0,24 мкФ.

**Задача 3.18.** Визначте електроємність представленої на рис. системи конденсаторів, якщо електроємкість кожного конденсатора дорівнюється 0,4 мкФ.

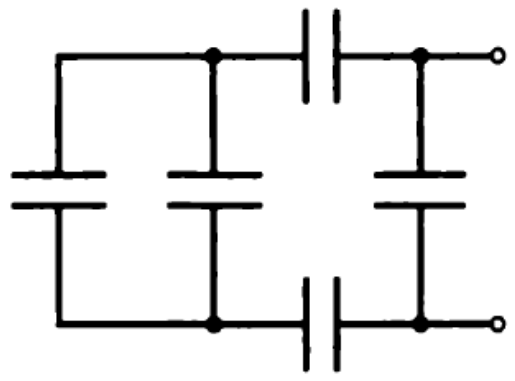

**Розв'язання.**

Для розв'язання задачі має сенс, для більшої наочності, спочатку перетворити рисунок даного змішаного з'єднання конденсаторів у батарею, наприклад, перевернути на 90$^{o}$ і розташувати елементи батареї так, щоб краще розуміти яким способом з'єднані всі конденсатори.

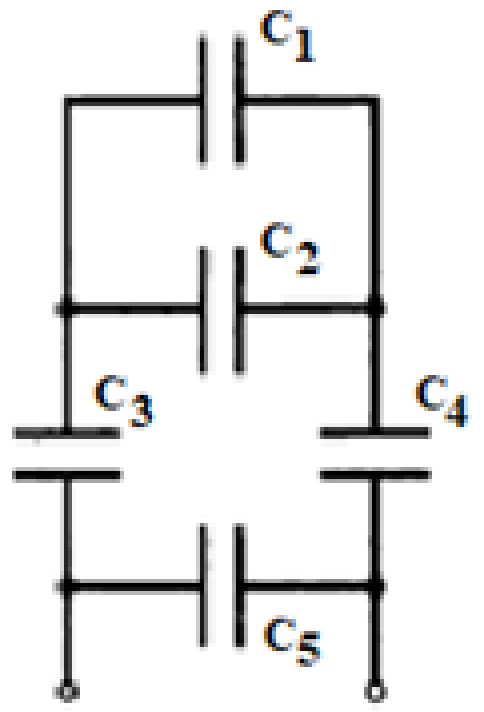

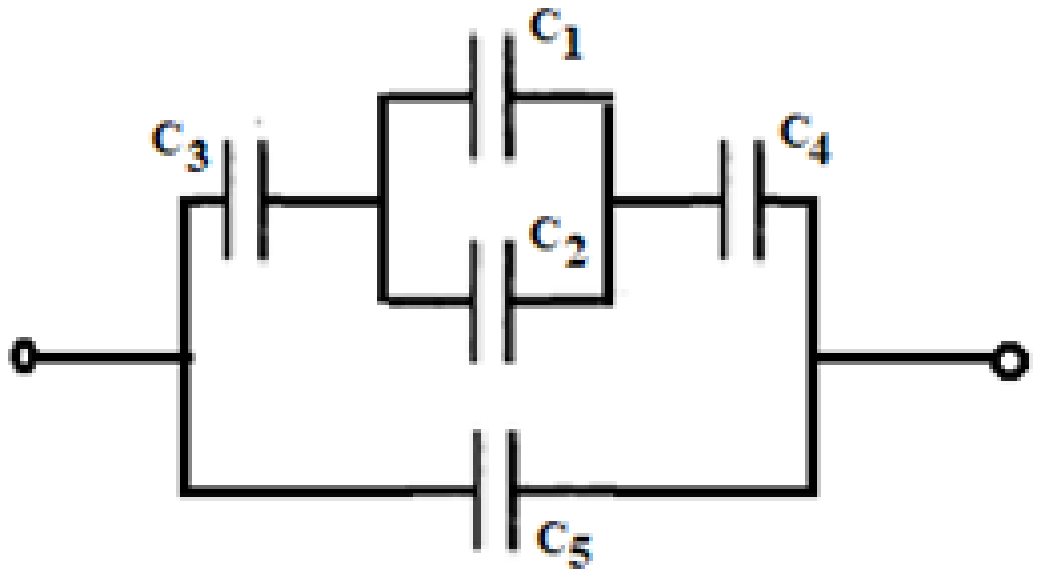

Почнемо спрощувати схему з паралельно з’єднаних конденсаторів $C_1$ і $C_2$, знайдемо їх суму:

$$C_{1-2} = C_1 + C_2 = 2C$$

Після першого шагу спрощення отримали наступну схему:

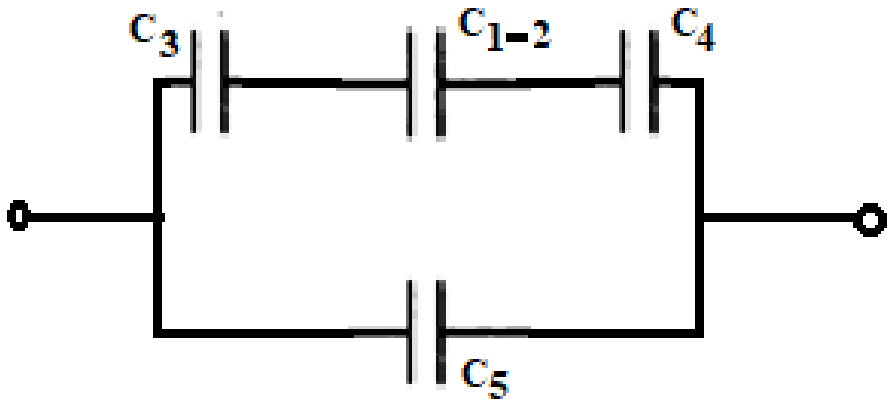

Другим шагом є знаходження ємності трьох послідовно з’єднаних конденсаторів:

$$\frac{1}{C_{1-4}} = \frac{1}{C_3} + \frac{1}{C_{1-2}} + \frac{1}{C_4} = \frac{2}{C} + \frac{1}{2C} = \frac{4+1}{2C} = \frac{5}{2C} \Rightarrow C_{1-4} = \frac{2C}{5}$$

На останньому етапі маємо два паралельно з’єднаних елемента:

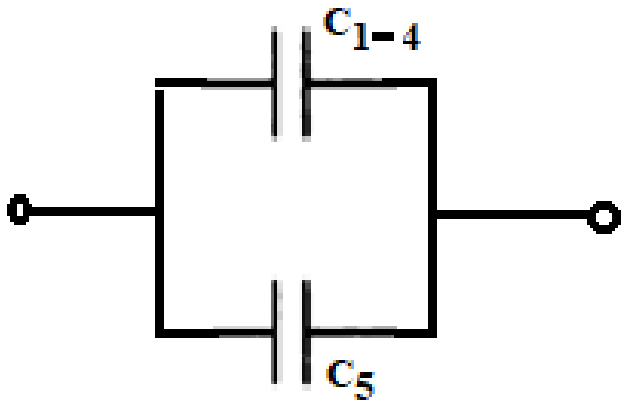

$$C_{1-5} = C_{1-4} + C_5 = \frac{2C}{5} + C = \frac{2C + 5C}{5} = \frac{7C}{5}$$

Щоб отримати кінцеву відповідь підставимо значення ємності конденсатора в отриманий вираз:

$$C_{1-5} = \frac{7C}{5} = \frac{7 \cdot 0{,}4}{5} = 0{,}56 \text{ мкФ}.$$

**Відповідь:** $C_{1-5} = 0{,}56$ мкФ.

**Задача 3.19.** Батарея з трьох послідовно з'єднаних конденсаторів $C_1$=2 мкФ; $C_2$=3 мкФ і $C_3$= 6 мкФ приєднані до джерела ЕРС. Заряд батареї конденсаторів $Q$=40 мкКл. Визначте:1)Напруження $U_1$, $U_2$, і $U_3$ на кожному конденсаторі; 2)Електрорухівну силу джерела; 3)Електроємність батареї конденсаторів.

**Дано:**

$C_1 = 2$ мкФ

$C_2 = 3$ мкФ

$C_3 = 6$ мкФ

$Q$=40 мкКл

**Знайти:**

$U_1$–? $U_2$–?

$U_3$–? $\varepsilon$–?

$C$-?

**Розв'язання.**

Напруження на конденсаторі знайдемо за формулою:

$$U = \frac{Q}{C}$$

При послідовному з'єднанні конденсаторів заряди всіх обкладок рівні по модулю, тому:

$$Q_1 = Q_2 = Q_3 = Q$$

і ми можемо записати напруження для кожного конденсатора:

$$U_1 = \frac{Q}{C_1} = 20 \text{ В}; \quad U_2 = \frac{Q}{C_2} = 13{,}3 \text{ В}; \quad U_3 = \frac{Q}{C_3} = 6{,}67 \text{ В}$$

Електрорухівну силу джерела знайдемо за формулою:

$$\varepsilon = U_1 + U_2 + U_3 = 20 + 13{,}3 + 6{,}67 = 40 \text{ В}$$

При послідовному з'єднанні сумуються велиабони, обернені емкостям кожного із конденсаторів:

$$\frac{1}{C} = \frac{1}{C_1} + \frac{1}{C_2} + \frac{1}{C_3} = \frac{C_2C_3 + C_1C_3 + C_1C_2}{C_1C_2C_3}$$

Звідси виразимо шукану ємність батареї конденсаторів:

$$C = \frac{C_1C_2C_3}{C_2C_3 + C_1C_3 + C_1C_2}$$

Після підстановки у отримане рівняння всіх значень ємностей отримаємо

$$C = 1\ мкФ$$

**Відповідь:** $U_1 = 20$ В; $U_2 = 13{,}3$ В; $U_3 = 6{,}67$ В; $\varepsilon = 40$ В; $C = 1$ мкФ.

**Задача 3.20.** На рис. зображена батарея конденсаторів, електроємності яких, віражені у мікрофорадах. Визнааботи повну електроємність батареї $C_{AB.}$

**Дано:**

$C_1 = 3мкФ$

$C_2 = 4$ мкФ

$C_3 = 5$ мкФ

$C_4 = 4$ мкФ

$C_5 = 4$ мкФ

$C_6 = 2$ мкФ

$C_7 = 3$ мкФ

**Знайти:**

$C_{AB}$–?

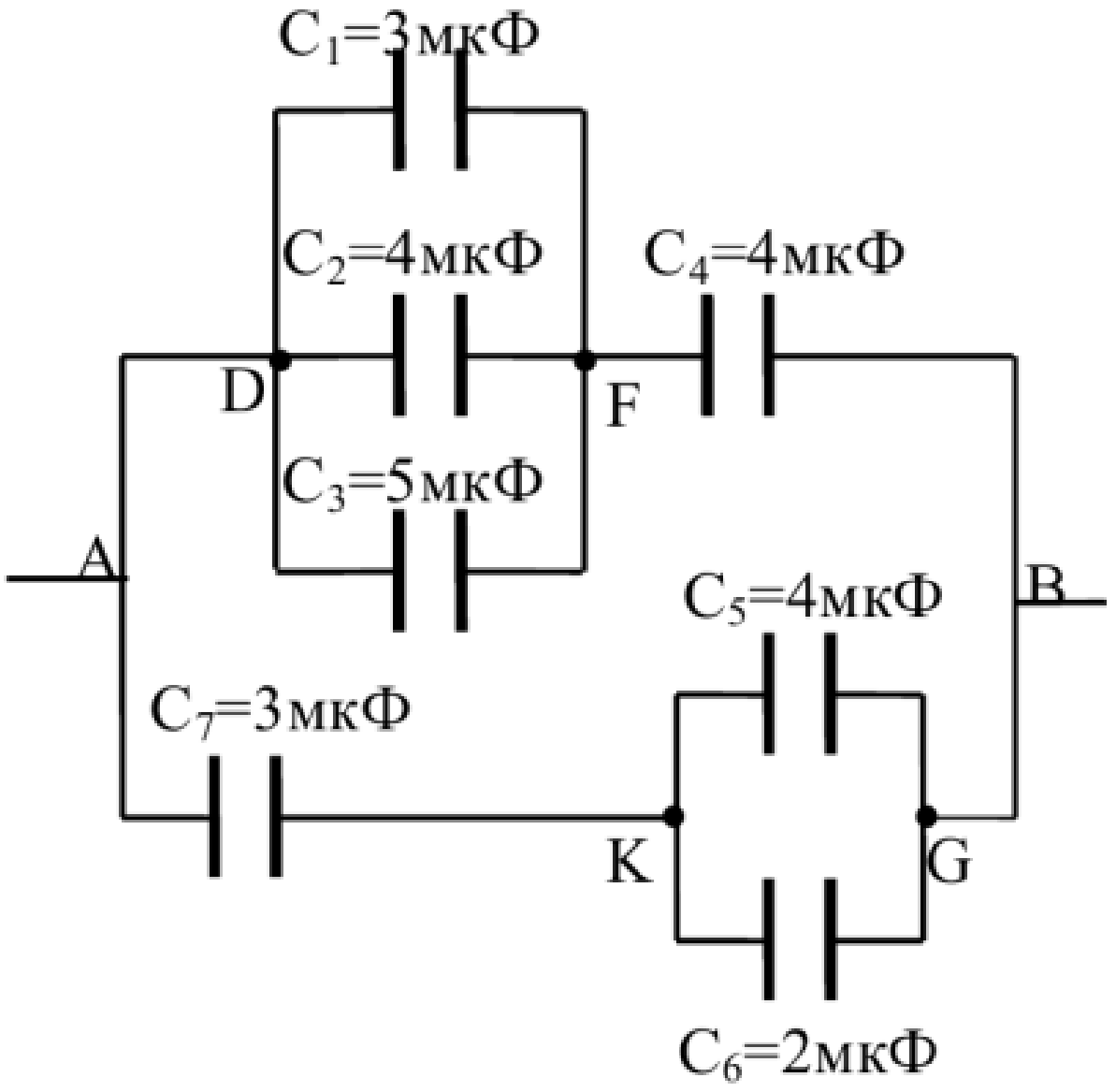

**Розв'язання.**

При послідовному з'єднані конденсаторів у батарею електроємність батареї дорівнює:

$$\frac{1}{C} = \sum_{i=1}^{n} \frac{1}{C_i}$$

При паралельному з'єднані конденсаторів у батарею електроємність батареї дорівнює:

$$C = \sum_{i=1}^{n} C_i$$

Спочатку розраховуємо електроємність конденсаторів, які з'єднані паралельно – це $C_1, C_2, C_3$ . При паралельному з'єднанні електроємності додаються:

$$C_{DF} = C_1 + C_2 + C_3 = 3 + 4 + 5 = 12 \text{ мкФ}$$

$$C_{KG} = C_5 + C_6 = 4 + 2 = 6 \text{ мкФ}$$

Після спрощення отримали еквівалентну схему:

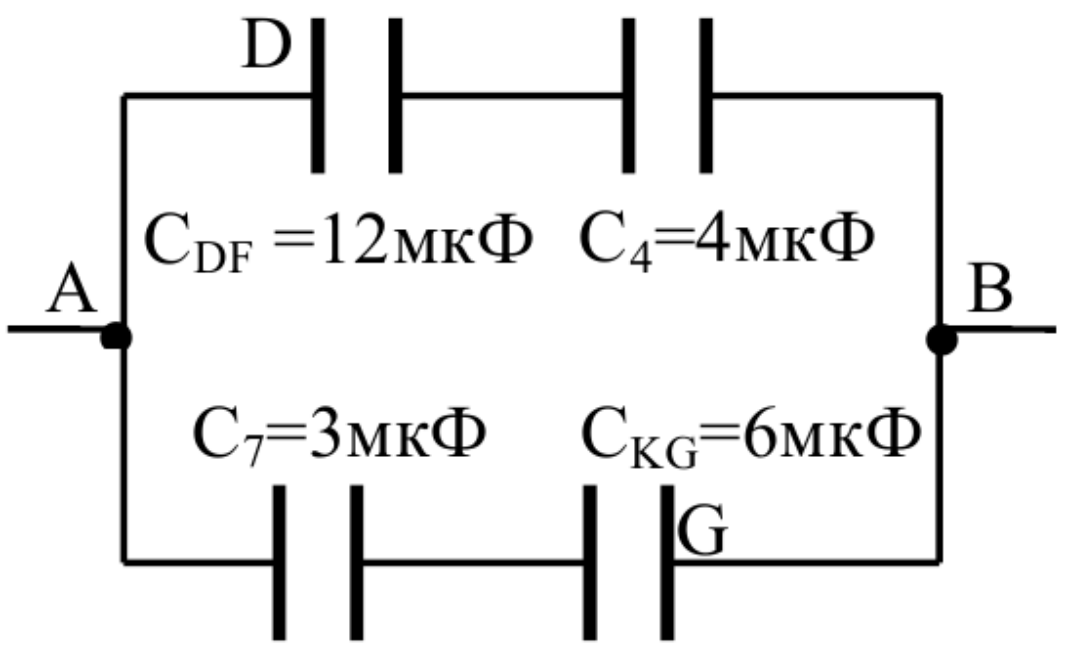

на якій замінюємо відповідні ділянки конденсаторами з розрахованою ємністю.

На другому етапі розраховуємо електроємність ділянок DB та AG з послідовно з'єднаними конденсаторами. При послідовному з'єднанні додаються величини, обернені ємностям, тобто:

$$\frac{1}{C_{DB}} = \frac{1}{C_{DF}} + \frac{1}{C_4} = \frac{1}{12} + \frac{1}{4} = \frac{1}{3}; \qquad C_{DB} = 3 \text{ мкФ}$$

$$\frac{1}{C_{AG}} = \frac{1}{C_7} + \frac{1}{C_{KG}} = \frac{1}{3} + \frac{1}{6} = \frac{1}{2}; \qquad C_{AG} = 2 \text{ мкФ}$$

Оскільки ділянки DB та AG між собою з'єднані паралельно, остаточно знаходимо:

$$C_{AB} = C_{DB} + C_{AG} = 3 + 2 = 5 \text{ мкФ}$$

**Відповідь:** $C_{AB} = 5$ мкФ.

## Задачі для самостійного розв'язання

1. В центр квадрату, в кожній вершині якого знаходиться заряд $q$=2,33 нКл, поміщений негативний заряд $q_0$. Знайти цей заряд, якщо на кожний заряд $q$ діє результуюча сила $F$=0. [$q_0$=2,23 нКл]

2. Дві кулі однакового радіусу і маси підвішені на нитках однакової довжини, так що їх поверхні торкаються. Після передачі кулям заряду $q_0$=0,4 мкКл вони відштовхнулися одна від одної і розійшлися на кут 2α=60°. Знайти масу *m* кожної кулі, якщо відстань від центру кулі до точки підвісу ℓ=20 см. [*m*=1,6 г]

3. Свинцева куля ($\rho_1$=11300 кг/м$^3$) діаметром *d*=0,5 см знаходиться в гліцерині ($\rho_2$=1260 кг/м$^3$). Визначити заряд кулі, якщо в однорідному електростатичному полі куля опинилася в рівновазі. Електростатичне поле спрямоване вертикально вгору, і його напруженість *E*=4 кВ/см. [*q*=1,6 нКл]

4. Знайти потенціал $\varphi$ точки поля, яка знаходиться на відстані *r*=10 см від центру зарядженої кулі радіусом *R*=1 см. Задачу розв'язати, якщо: а) відома поверхнева густина заряду на кулі σ=0,1 мкКл/м$^2$; б) відомий потенціал кулі $\varphi_0$=300 В. [$\varphi_1$=11,3 В, $\varphi_2$=30 В]

5. Куля масою *m*=1 г і зарядом *q*=10 нКл переміщується з точки 1, потенціал якої $\varphi_1$=600 В, в точку 2, потенціал якої $\varphi_2$=0. Знайти її швидкість в точці 1, якщо в точці 2 вона стала дорівнювати $\upsilon_2$=20 см/с. [$\upsilon_1$=16,7 см/с]

6. Електрон летить від однієї пластини плоского конденсатора до іншої. Різниця потенціалів між пластинами *U*=3кВ; відстань між пластинами *d*=5 мм. Знайти силу *F*, яка діє на електрон, прискорення а електрона, швидкість $\upsilon$, з якою електрон приходить до другої пластини, і поверхневу густину заряду $\sigma$ на пластинах. [$F=9{,}6\cdot10^{-14}$ Н, $a=1{,}05\cdot10^{17}$ м/с$^2$, $\upsilon=32{,}5\cdot10^6$ м/с, $\sigma$=5,3 мкКл/м$^2$]

7. Куля, яка знаходиться в керосині, має потенціал $\varphi$=4,5 кВ і поверхневу густину заряду $\sigma$=11,3 мкКл/м$^2$. Знайти радіус *r*, заряд *q*, ємність *C* і енергію *W* кулі.[*r*=7 мм, *q*=7 нКл, *C*=1,56 пФ, *W*=15,7 мкДж]

**8.** Конденсатори ємностями $C_1$=2 мкФ і $C_2$=4 мкФ, що мають заряди $q_1$=8 мкКл і $q_2$=6 мкКл, з'єднали паралельно. Визначити зміну ΔW енергії конденсаторів. [Δ*W*=4,16 мкДж].

# Розділ 4. ЗАКОНИ ПОСТІЙНОГО СТРУМУ.

## 4.1. Основні характеристики постійного струму.

Якщо через деяку уявну поверхню переноситься сумарний заряд, відмінний від нуля, то говорять, що через цю поверхню тече електричний струм. Інтенсивність і напрям струму характеризується: велиабоною струму $I$ і щільністю струму $\vec{j}$. Струм $I$ – велиабона інтегральна, він визначається як заряд, що проходить за секунду через весь перетин або через задану площадку, якщо за час dt через поверхню $S$ переноситься заряд dq, то сила струму дорівнюється:

$$I = \frac{dq}{dt}.$$

Постійний струм – струм, сила і напрямок якого незмінні:

$$I = \frac{q}{t},$$

де $q$- електричний заряд, що проходить за час $t$ крізь розглянуту поверхню.

Струм $I$ скалярна велиабона, тоді як густина – вектор.

Густина струму визначається за формулою:

$$j = \frac{dI}{dS_{\perp}}$$

Вектор $\vec{\jmath}$ середньої густини потоку зарядів співпадає за напрямом із вектором швидкості $\vec{v}$ упорядкованого руху позитивно заряджених частинок.

$$\vec{j} = qn\vec{v} = \rho\vec{v}$$

Сила струму дорівнює потоку вектора густини струму через переріз провідника:

$$I = \int \vec{j} \cdot d\vec{S}$$

Якщо густина постійного струму однакова по всьому поперечному перерізі S провідника, то у цьому випадку:

$$I = j \cdot S$$

**Електрорушійна сила (ЕРС) джерела струму** – характеристика джерела струму, яка абосельно дорівнює роботі сторонніх сил по переносу (переміщенні) одиничного позитивного заряду вздовж електричного кола:

$$\varepsilon = \frac{A_{cm}}{q}$$

**Закон Ома** полягає в тому, що сила струму пропорціональна різниці потенціалів на кінцях ділянки:

$$I = \frac{(\varphi_1 - \varphi_2)}{R} = \frac{U}{R}$$

де велиабону $R$ називають опором, а $U = \varphi_1 - \varphi_2$ напругою.

Напрямом струму вважається напрям руху позитивних зарядів, тому, якщо потенціал початкової точки ділянки $\varphi_1$ більше потенціалу кінцевої точки, то струм тече від точки *1* до точки *2*. Насправді в металевих провідниках струм переносять електрони, які рухаються від точки *2* до *1*, якщо $\varphi_1 > \varphi_2$.

При виникненні електричного поля на носій заряду діє сила

$$F = qE,$$

де $E$ – напруженість електричного поля. Під дією сили носії заряду (електрони провідності, дірки, іони) розганяються і збільшують свою кінетичну енергію. Але такий процес є обмежений за рахунок зіткнень носіїв заряду з атомами матеріалу, що мають коливальний рух навколо положень рівноваги в кристалічній гратці внаслідок теплового руху. При зіткненнях носіїв заряду їх надлишкова кінетична енергія буде передаватися атомам у вузлах кристалічної гратки та виділиться у вигляді тепла. Тобто носії заряду мають швидкість, яка визначається частотою зіткнень. При опису цього процесу використовують такі характеристики як час розсіяння (час між зіткненнями, що змінюють напрямок руху), довжина вільного пробігу, а середня швидкість носіїв заряду є пропорційною прикладеному електричному полю або напрузі.

**Електричне коло** — це сукупність електронних компонент – джерел струму й напруги, резистори, перемиквчі тощо, які зв’язані між собою провідниками по яким може проходитиелектричний струм. Тобто до складу електричного кола обов’язково входить джерело електромагнітної енергії, а також споживачі електромагнітної енергії.

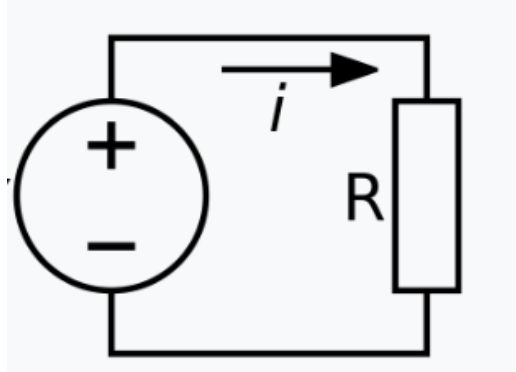

Рис. 4.1. Схема найпростішого електричного кола.

Напруга $U_{12}$ на ділянці кола – велиабона роботи сил електричного поля та сторонніх сил по переносу одиничного позитивного заряду вздовж ділянки кола:

$$U_{12} = \frac{A_{12} + A_{12}^{*}}{q} = \varphi_1 - \varphi_2 + \varepsilon_{12}$$

Закон Ома для неоднорідної ділянки кола, що містить джерело струму:

$$I = \frac{\varphi_1 - \varphi_2 + \varepsilon_{12}}{R_{12}}$$

де $\varepsilon_{12}$ – електрорушійна сила джерела струму; $R_{12}$- повний опір ділянки.

Закон Ома для замкненого (повного) нерозгалуженого кола: сила струму в замкненому колі прямо пропорційна алгебраїчній сумі всіх ЕРС (електрорушійних сил), що діють у колі, і обернено пропорційна його повному опору, який дорівнює сумі опорів зовнішньої ($R$) та внутрішньої ($r$) ділянок:

$$I = \frac{\mathcal{E}}{R + r},$$

де $R$ зовнішній опір – це загальний опір усіх ділянок електричного кола, що розташовані ззовні від джерела струму, а опір внутрішньої ділянки кола $r$ – це є опір ділянки кола по якій електричні заряди рухаються під впливом сторонніх сил, тобто $r -$ сума внутрішніх опорів джерел струму.

Ще однією із найважливіших фізичних характеристик є провідність провідника, яка залежить від наявності електронів – носіїв заряду.

Опір провідника $R$ та його провідність $G$ являються взаємно оберненими велиабонами:

$$R = \frac{1}{G}$$

Опір залежить від форми і розміру провідника, а також від матеріалу з якого він зроблений, та температури. У простому випадку однорідного провідника з постійним перерізом опір цього провідника визначається за формулою:

$$R = \rho \frac{\ell}{S},$$

де $\ell$ – довжина провідника; $S$ – площа перерізу провідника; $\rho$ – питомий опір матеріалу провідника.

Питомий опір матеріалу і його питома провідність $\gamma$ є взаємно оберненими велиабонами:

$$\rho = \frac{1}{\gamma}.$$

**Задача 4.1.** По залізному провіднику ($\rho$ = 7,87 г/см$^3$, $M$ = 56·10$^{-3}$ кг/моль) з перетином $S$ = 0,5 мм$^2$ тече струм $I$ = 0,1 А. Визначте середню швидкість упорядкованого (спрямованого) руху електронів, вважаючи, що число $n$ вільних електронів в одиниці об'єму провідника дорівнює кількості атомів $n'$ в одиниці об'єму провідника.

| **Дано:** | **Розв'язання.** |
|---|---|
| $\rho$ = 7,87 г/см$^3$=7,87· $10^3 \frac{кг3}{м}$<br>$M$ = 56·10$^{-3}$ кг/моль<br>$S$ = 0,5 мм$^2$= 0,5· $10^{-6}$м$^2$<br>$I$ = 0,1 А | За умовою задачі треба визначити середню швидкість спрямованого руху електронів, для цього можна скористатися формулою густини струму:<br>$$j = en < \upsilon >,$$ |
| **Знайти:**<br>$\langle \upsilon \rangle$–? | де $< \upsilon >$ – середня швидкість упорядкованого |

руху носіїв заряду, $n$ – концентрація носіїв заряду, $e$ – елементарний заряд ($e = 1{,}6 \cdot 10^{-19}$ Кл).

Згідно умовам задачі концентрація вільних електронів в одиниці об'єму провідника, тобто концентрація $n = n'$ концентрації атомів в одиниці об'єму:

$$n = n' = \frac{N(\text{кількість})}{V(\text{об'єм})}$$

Кількість атомів $N$ можна розрахувати по формулі:

$$N = \frac{m}{M} \cdot N_A,$$

де $m$ – маса провіднику; $M$ – його молярна маса; $N_A = 6{,}02 \cdot 10^{23} моль^{-1}$ – стала Авогадро.

Підставимо рівняння кількості носіїв в формулу концентрації і отримаємо:

$$n = n' = \frac{N}{V} = \frac{\frac{m}{M} \cdot N_A}{V}$$

В умовах задачі відсутні дані маси і об'єму, але є інформація, що провідник залізний і відомі дані о щільність заліза. Густина визначається:

$$\rho = \frac{m(\text{маса})}{V(\text{об'єм})}; \quad m = \rho V$$

Підставимо отриманий вираз у рівняння концентрації:

$$n = \frac{m \cdot N_A}{M \cdot V} = \frac{\rho V \cdot N_A}{M \cdot V} = \frac{\rho \cdot N_A}{M}$$

Для знаходження середньої швидкості із рівняння $j = en < v >$ ще треба визначити густину електричного струму. Визначити густину електричного струму можна через силу електричного струму $I$ площу поперечного перерізу провідника $S$:

$$j = \frac{I}{S}$$

Тоді маємо прирівняти обидва рівняння:

$$en\langle v \rangle = \frac{I}{S}$$

В це рівняння підставимо отриманий вираз для кількості носіїв заряду в одиниці об’єму:

$$n = \frac{\rho N_A}{M}$$

де $\rho$ – густина заліза) і отримаємо:

$$\frac{I}{S} = \frac{\rho N_A}{M} e\langle v\rangle$$

Звідси виразимо $\langle v\rangle$

$$\langle v\rangle = \frac{IM}{\rho N_A eS}$$

Всі значення надані в умовах задачі і підставивши в формулу отримаємо

$$\langle v\rangle = 14{,}8 \cdot 10^{-6} \text{ м/с}$$

**Відповідь**: $\langle v\rangle = 14{,}8 \cdot 10^{-6}$ м/с.

**Задача 4.2.** Визначити густину струму в мідному дроті довжиною 10 м, якщо різниця потенціалів $\varphi_1 - \varphi_2$ на його кінцях дорівнює 8,5 В.

**Дано:**
$\rho$ = 1,7 $10^{-8}$ Ом · м
$\ell$=10 м
$\varphi_1$-$\varphi_2$=8,5 В

**Знайти:**
$j$–?

**Розв’язання.**

Цю задачу можна розв’язати двома способами.

*1 спосіб.* За законом Ома у диференціальній формі

$$j = \sigma E, \qquad \sigma = \frac{1}{\rho} \qquad \Longrightarrow \qquad \vec{j} = \frac{1}{\rho}\vec{E}$$

де ε – електрична провідність речовини провідника; ρ – питомий електричний опір (характеризує здатність матеріалу перешкоджати проходженню струму).

Значення питомого опору мідного провідника $\rho = 1{,}7 \cdot 10^{-8}$ Ом·м дістаємо з таблиці.

Напруженість електричного поля всередині провідника виражаємо через різницю потенціалів на його кінцях та довжину $\ell$:

$$E = \frac{\varphi_1 - \varphi_2}{\ell} \quad \text{тоді} \quad J = \frac{\varphi_1 - \varphi_2}{\rho\ell}$$

*II спосіб.* Густину струму $j$ виражаємо через силу струму $I$ за означенням:

$$j = \frac{I}{S}$$

де $S$ – площа поперечного перерізу провідника.

Силу струму $I$ знаходимо із закону Ома для однорідної ділянки кола:

$$I = \frac{U}{R} = \frac{\varphi_1 - \varphi_2}{R}$$

Підстановка цього виразу у формулу густини електричного струму

$$j = \frac{\text{I} \quad \text{(сила струму)}}{S \ \text{(площа перерізу)}}$$

з урахуванням виразу для опору однорідного провідника сталого перерізу

$$R = \rho \frac{\ell \text{(довжина провідника)}}{S \quad \text{(площа перерізу)}}$$

де ρ – питомий електричний опір, приводить до того самого результату як і у першому способі:

$$j = \frac{I}{S} = \frac{\frac{\varphi_1 - \varphi_2}{R}}{S} = \frac{\varphi_1 - \varphi_2}{RS} = \frac{\varphi_1 - \varphi_2}{\frac{\rho\ell}{S} S} = \frac{\varphi_1 - \varphi_2}{\rho\ell}$$

$$j = \frac{\varphi_1 - \varphi_2}{\rho\ell} = \frac{8{,}5}{1{,}7 \cdot 10^{-8} \cdot 10} = 5 \cdot 10^7 \ \text{А/м}^2$$

**Відповідь:** $j = 5 \cdot 10^7$ А/м$^2$.

**Задача 4.3.** Визначте густину $j$ електричного струму в мідному дроті (питомий електричний опір $\rho = 17$ нОм $\cdot$ м), якщо питома теплова потужність струму $\omega = 1{,}7$ Дж/(м$^3 \cdot$ с).

| **Дано:** | **Розв’язання.** |
|---|---|
| $\rho = 1{,}7\ 10^{-8}\,\text{Ом}\cdot\text{м}$ | |
| $\omega = 1{,}7\dfrac{\text{Дж}}{\text{м}^3\cdot\text{с}}$ | |
| **Знайти:** | |
| $j - ?$ | |

Напишемо формулу визначення питомої теплової потужності струму і густини електричного струму. Згідно з законами Джоуля-Ленца і Ома в диференціальній формі маємо:

$$\omega = \gamma E^2 = \frac{E^2}{\rho};$$

$$j = \gamma \mathrm{E} = \frac{E}{\rho},$$

де $\gamma$ і $\rho$ - відповідно питомі провідність і опір провідника.

З останнього виразу отримаємо, що

$$E = \rho j$$

Підставимо його у рівняння:

$$\omega = \frac{E^2}{\rho} = \frac{(\rho j)^2}{\rho} = \rho j^2$$

Звідси знайдемо шукану густину струму:

$$j = \sqrt{\frac{\omega}{\rho}} = \sqrt{\frac{1{,}7}{17\cdot 10^{-9}}} = 10\ \text{кА/м}^3.$$

**Відповідь:** $j = 10\ \text{кА/м}^3$.

**Задача 4.4.** Обмотка реостату зроблена з мідного дроту діаметром $d$=0,5 мм, який щільно намотано на циліндр радіусом $r$=2 см. Скільки витків $N$ слід зробити, щоб опір реостата дорівнював $R$=50 Ом?

| **Дано:** | **Розв'язання.** |
| --- | --- |
| $R$=50 Ом<br>$d = 0{,}5$ мм $= 0{,}5 \cdot 10^{-3}$м<br>$r = 2$ см $= 2 \cdot 10^{-2}$м<br>$\rho = 1{,}7 \cdot 10^{-8}$ Ом $\cdot$ м | Кількість витків обмотки реостату $N$ можна визнааботи поділивши довжину дроту $L$ на довжину одного витка<br>$\ell_1 = 2\pi r$ |
| **Знайти:**<br>$N$–? | де $r$ – радіус витка.<br>$N = \frac{L}{\ell_1} = \frac{L}{2\pi r}$ |

Довжину дроту визначимо із формули опору:

$$R = \rho \frac{L}{S}$$

де $\rho$ — питомий опір провідника; $S$ — площа перерізу провідника.

Визначити $S$ можна за формулою:

$$S = \frac{\pi d^2}{4}$$

Відомо, що $S_{кр} = \pi r^2$ , якщо замість радіусу підставити діаметр то отримаємо саме цю формулу.

$$Sкр = \pi r^2 = \pi \left(\frac{d}{2}\right)^2 = \frac{\pi d^2}{4}$$

Виразимо довжину дроту із формули опору:

$$L = \frac{RS}{\rho} = \frac{R\frac{\pi d^2}{4}}{\rho} = \frac{\pi d^2 R}{4\rho}$$

Після підстановки виразу в формулу

$$N = \frac{L}{2\pi r}$$

визначаємо кількість витків:

$$N = \frac{\frac{\pi d^2 R}{4\rho}}{2\pi r} = \frac{\pi d^2 R}{4\rho \cdot 2\pi r} = \frac{d^2 R}{8\rho r} = \frac{0{,}25 \cdot 10^{-6} \cdot 50}{8 \cdot 1{,}7 \cdot 10^{-8} \cdot 2 \cdot 10^{-2}} = 4595 \text{ витків.}$$

**Відповідь:** $N = 4595$ витків.

**Задача 4.5.** Сила струму у провіднику змінюється за законом $I = 2 + 2t$, де $I$ вимірюється в амперах, $t$ – в секундах. Знайти заряд $q$, який проходить крізь переріз провідника за проміжок часу від $t_1 = 1$ с до $t_2 = 3$ с.

**Дано:**

$I = 2 + 2t$

$t_1$=1 с

$t_2$=3 с

**Знайти:**

$q$–?

**Розв'язання.**

Кількість заряду, що пройшла за певний час можна визначити знайшовши визначений інтеграл:

$$q = \int_{t_1}^{t_2} I(t)\, dt$$

За формулою Ньютона-Лейбніца:

$$\int_a^b f(x)dx = F(X)\Big|_a^b = F(b) - F(a)$$

Застосуємо до нашого завдання:

$$q = \int_1^3 (2 + 2\mathrm{t})\, dt = \int_1^3 2\, dt + \int_1^3 2\mathrm{t}\, dt = 2\mathrm{t}\Big|_1^3 - t^2\Big|_1^3 =$$

Замість $t$ підставляємо його значення: $t_1 = 1$ с і $t_2 = 3$ с

$$= 2(t_2 - t_1) + (t_2^2 - t_1^2) = 2(3 - 1) + (9 - 1) = 4 + 8 = 12 \text{ Кл.}$$

**Відповідь:**$q = 12$ Кл.

**Задача 4.6.** Джерелу струму має електрорушійну силу $\varepsilon$ =1,5 В і внутрішній опір $r$, який у $\eta$=15 разів менше зовнішнього опору $R$. Визначити напругу $U$ на затискачах джерела.

| **Дано:** | **Розв'язання**. |
|---|---|
| $\varepsilon$=1,5 В | Закон Ома для неоднорідної ділянки кола |
| $\eta = \frac{R}{r} = 15$ | $$I = \frac{(\varphi_1 - \varphi_2) \pm \varepsilon}{R + r}$$ |
| **Знайти:** | але при $\varphi_1 = \varphi_2$(замкнене електричне коло) приймає вид: |
| $U$–? | $$I = \frac{\varepsilon}{R + r}$$ |

За законом Ома для однорідної ділянки кола $I = \frac{U}{R}$ тобто

$$U = IR$$

$$U = \frac{\varepsilon R}{R + r}$$

За умовою задачі нам дано відношення $\frac{R}{r}$ тому поділимо та помножимо на $r$

$$U = \frac{\varepsilon \frac{Rr}{r}}{\frac{Rr}{r} + \frac{rr}{r}} = \frac{\varepsilon r \frac{R}{r}}{r\left(\frac{R}{r} + \frac{r}{r}\right)} = \frac{\varepsilon \frac{R}{r}}{1 + \frac{R}{r}}$$

Підставимо, надане в умовах задачі, відношення:

$$\frac{R}{r} = \eta = 15$$

і отримаємо:

$$U = \frac{\varepsilon\eta}{1 + \eta} = \frac{1{,}5 \cdot 15}{1 + 15} = \frac{22{,}5}{16} = 1{,}4 \text{ В}.$$

**Відповідь:** $U = 1{,}4$ В.

## 4.2. Послідовне і паралельне з'єднання.

**При послідовному з'єднанні** ділянок кола на провіднику, що їх з'єднує, вузли (точки розгалуження) – відсутні, струм на цих ділянках (опорах) є однаковим, напруга дорівнює сумі напруг на кожній з них, а загальний опір $n$ провідників при послідовному з'єднанні дорівнюється сумі опорів:

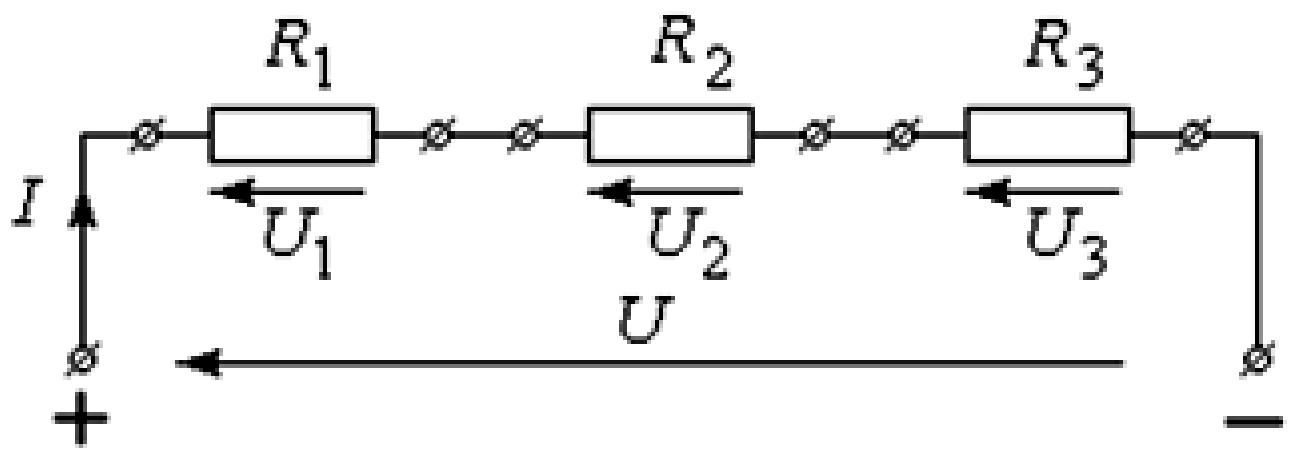

$$R = \sum_{i=1}^{n} R_i$$

$$U = \sum_{i=1}^{n} U_i$$

$$I = I_1 = I_2 = \cdots = I_n$$

Рис. 4.2. Послідовне з’єднання опорів.

На рис.4.2. маємо послідовно з’єднані опори $R_1$, $R_2, R_3$, які підключені до джерела енергії з напругою $U$. Згідно наведеним вище формулам по всім ділянкам послідовного кола протікає однаковий струм *I*, тому, за законом Ома можемо записати, що напруга на окремих елементах кола:

$$U_1 = IR_1; \quad U_2 = IR_2; \quad U_3 = IR_3$$

Тобто спади напруги на послідовно з’єднаних опорах пропорційні велиабонам опорів, а сума напруг на окремих елементах дорівнює напрузі на вхідних клемах кола:

$$U_1 + U_2 + U_3 = U$$

Якщо всі елементи цього рівняння помножити на струм *I,* то отримаємо потужність всього кола $P$, яке дорівнює сумі потужностей окремих ділянок:

$$IU_1 + IU_2 + IU_3 = IU; \qquad P_1 + P_2 + P_3 = P$$

**При паралельному з’єднанні** ділянок кола на провіднику, що їх з’єднує, існують вузли (точки розгалуження A і B), тобто кінці усіх ділянок підключені к точкам кола, наприклад, A і B (див. рис.4.3.), напруга на цих ділянках (опорах) є однаковою, сума струмів в усіх ділянках дорівнює струму в колі до розгалуження:

$$I = \sum_{i=1}^{n} I_i$$

Загальний опір n провідників при паралельному з'єднанні знаходиться за формулою:

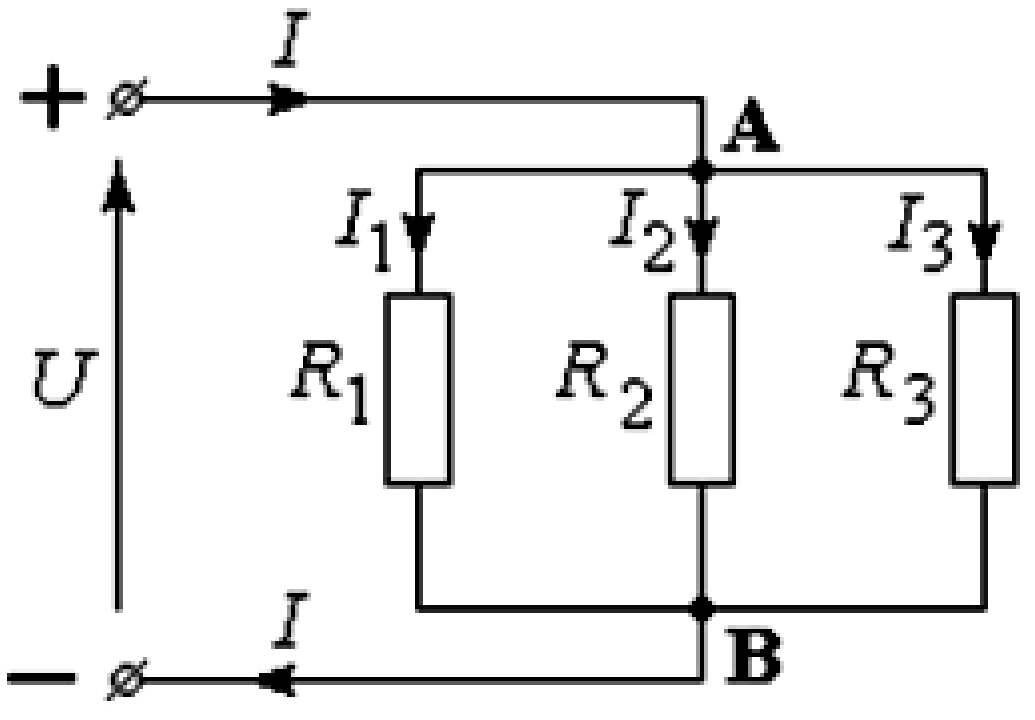

$$\frac{1}{R} = \sum_{i=1}^{n} \frac{1}{R_i}$$

$$U = U_1 = U_2 = \cdots = U_n$$

Рис. 4.3. Паралельне з'єднання опорів.

Опір провідника $R$ та його провідність $G$ являються взаємно оберненими велиабонами:

$$R = \frac{1}{G}$$

При паралельному з'єднанні загальна провідність $G$ складається з додавання провідності $G_i$ окремих провідників:

$$G = \sum_{i=1}^{n} G_i$$

Таким абоном, для паралельного з'єднання ділянок кола маємо записати:

$$I = I_1 + I_2 + I_3 = \frac{U}{R_1} + \frac{U}{R_2} + \frac{U}{R_3} = U\left(\frac{1}{R_1} + \frac{1}{R_2} + \frac{1}{R_3}\right) = \frac{U}{R_E}$$

$$I = UG_1 + UG_2 + UG_3 = U((G_1 + G_2 + G_3) = UG_E$$

Якщо вираз:

$$\frac{U}{R_1}+\frac{U}{R_2}+\frac{U}{R_3}=\frac{U}{R_E}$$

помножити на $U$, то отримаємо потужність, тобто потужність, споживана розгалуженим колом, дорівнює сумі потужностей, споживаних окремими елементами, або одним еквівалентним елементом:

$$\frac{U^2}{R_1}+\frac{U^2}{R_2}+\frac{U^2}{R_3}=\frac{U^2}{R_E} \quad \text{або} \quad P_1+P_2+P_3=P$$

Опір еквівалентного елемента $R_E$ , у випадку двох паралельно включених опорів $R_1$ та $R_2$ знаходиться за формулою:

$$\frac{1}{R_E}=\frac{1}{R_1}+\frac{1}{R_2}=\frac{R_2+R_1}{R_1R_2}, \text{ тобто } R_E=\frac{R_1R_2}{R_1+R_2}$$

При утворенні електричного кола поєднанням послідовно і паралельно включених елементів маємо змішану схему з'єднань.

**Задача 4.7.** Опір однорідного дроту $R$=36 Ом. Визначте, на скільки різних відрізків $r$ розрізали дріт, якщо після їх паралельного з'єднання опір став рівним $R_1$=1Ом.

| **Дано:** | **Розв'язання.** |
|---|---|
| $R$=36 Ом | Послідовне з'єднання: |
| $R_1$=1Ом | $R=R_1+R_2+\cdots+R_n$ |
| **Знайти:** | Паралельне з'єднання: |
| $N$–? | $\frac{1}{R}=\frac{1}{R_1}+\frac{1}{R_2}+\cdots+\frac{1}{R_n}$ |

Нерозрізану проволоку можна представити як $N$ послідовно з'єднаних опорів $r$. Тоді:

$$R=Nr,$$

де $r$ – опір кожного відрізку.

В випадку паралельного з'єднання $N$ відрізків проволоки (опорів $r$)

$$\frac{1}{R_1} = \frac{N}{r} \quad \text{або} \quad R_1 = \frac{r}{N}$$

Виразимо $r$ з обох отриманих рівнянь, для послідовного та паралельного з'єднання:

$$r = \frac{R}{N} \quad \text{та} \quad r = R_1 \cdot N, \quad \text{тобто} \quad \frac{R}{N} = R_1 \cdot N \quad \text{або} \quad N^2 = \frac{R}{R_1}$$

Звідси знайдемо $N$:

$$N = \sqrt{\frac{R}{R_1}} = \sqrt{\frac{36}{1}} = \sqrt{36} = 6$$

**Відповідь:** $N$=6 відрізків.

**Задача 4.8.** Опір другого провідника в п'ять разів більше, ніж опір першого. Їх спочатку включають в ланцюг послідовно, а потім - паралельно. Визначте відношення кількостей теплоти, що виділилися в цих провідниках, для обох випадків.

| **Дано:** | **Розв'язання.** |
|---|---|
| $R_1$ | В випадку послідовного з'єднання провідників $I = const.$ |
| $R_2=5R_1$ | Кількість теплоти, яка виділяється в провіднику при |
| **Знайти:** | проходженні струму, обабослюється за законом Джоуля-Ленца: |
| $\frac{Q_1}{Q_2}$–? | $Q = I^2Rt = IUt = \frac{1}{R}U^2t,$ |

Кількість теплоти, що виділилася у першому та другому провідниках:

$$Q_1 = \frac{U_1^2}{R_1}t \quad \text{і} \quad Q_2 = \frac{U_2^2}{R_2}t,$$

де $t$ – час проходження струму через провідник; $U_1$ і $U_2$ – відповідно різниця потенціалів між кінцями першого та другого провідників.

Маючи на увазі, що $I_1$= $I_2$ отримаємо, що

$$\frac{U_1}{R_1} = \frac{U_2}{R_2} \quad \Longrightarrow \quad \frac{U_1}{U_2} = \frac{R_1}{R_2}$$

Знаходимо відношення $\frac{Q_1}{Q_2}$ підставивши замість відношення напруги

$$\frac{U_1}{U_2} = \frac{R_1}{R_2}$$

відношення опорів.

$$\frac{Q_1}{Q_2} = \frac{U_1^2 R_2}{U_2^2 R_1} = \frac{R_1^2 R_2}{R_2^2 R_1} = \frac{R_1}{R_2}$$

В випадку паралельного з'єднання провідників $U = const.$ Аналогічно пишемо закон Джоуля-Ленца для кількості теплоти, що виділилася у

$$Q_1 = \frac{U^2}{R_1} t, \qquad Q_2 = \frac{U^2}{R_2} t,$$

Знаходимо відношення

$$\frac{Q_1}{Q_2} = \frac{R_2}{R_1}$$

Підставивши значення отримаємо:

1) для послідовного з'єднання провідників $\frac{Q_1}{Q_2} = 0{,}2$

2) для паралельного з'єднання провідників $\frac{Q_1}{Q_2} = 5$

**Відповідь:** $\frac{Q_1}{Q_2} = 0{,}2$ для послідовного і $\frac{Q_1}{Q_2} = 5$ для паралельного з'єднань

**Задача 4.9.** Коли ключ *К* замкнутий, опір між точками *А* і *В* схеми дорівнюється $R_1 = 60$ Ом. Визнаaботи опір $R_2$ між цими точками, коли ключ розімкнутий.

**Дано:**

$R_1$=60 Ом

**Знайти:**

$R_2$–?

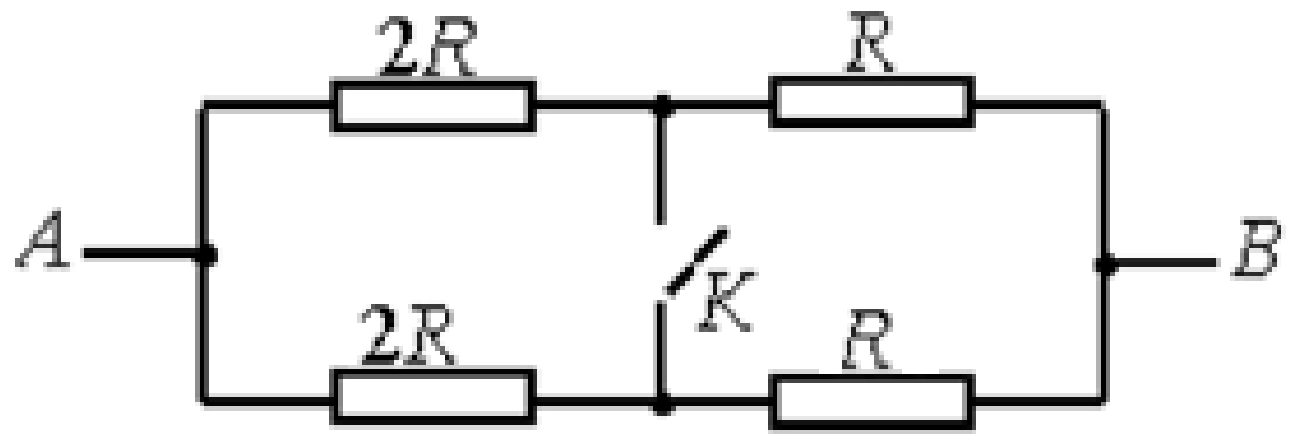

**Розв'язання.**

Розглянемо схему з'єднання опорів при замкненому ключі:

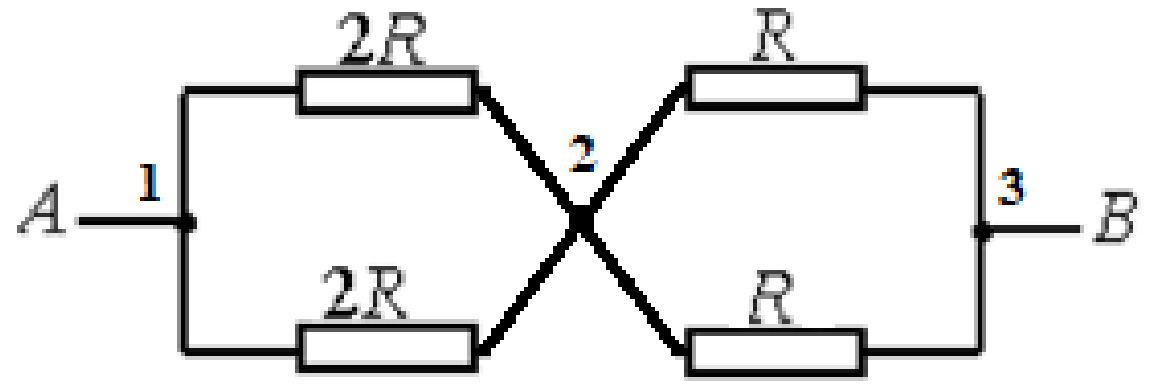

Саме для такого з'єднання в умовах задачі дан опір між точками $A$ і $B$, якій дорівнюється $R_1 = 60$ Ом. Якщо скласти рівняння для визначення загального опору при такому з'єднані, то зможемо знайти значення $R$.

На схемі маємо три вузли 1,2 і 3. Наявність вузлів – це паралельне з'єднання, для якого загальний опір визначається за формулою:

$$\frac{1}{R} = \sum_{i=1}^{n} \frac{1}{R_i}$$

Спочатку розглянемо гілки між вузлами 1 і 2, знайдемо опір на цієї ділянці:

$$\frac{1}{R_{1-2}} = \frac{1}{2R} + \frac{1}{2R} = \frac{1}{R} \;\Rightarrow\; R_{1-2} = R$$

Розглянемо гілки між вузлами 2 і 3, знайдемо опір на цієї ділянці:

$$\frac{1}{R_{2-3}} = \frac{1}{R} + \frac{1}{R} = \frac{2}{R} \;\Rightarrow\; R_{2-3} = \frac{R}{2}$$

Ця схема спростилась до послідовного з'єднаня двох еквівалентних опорів $R_{1-2} = R$ і $R_{2-3} = \frac{R}{2}$, загальне для яких визначається за формулою:

$$R = \sum_{i=1}^{n} R_i$$

Сума опорів $R_{1-2}$ і $R_{2-3}$ буде дорівнюватись наданому $R_1 = 60\,\text{Ом}$, тобто

$$R_1 = R + \frac{R}{2} = \frac{2R + R}{2} = \frac{3R}{2} \text{ звідси виразимо } R$$

$$R = \frac{2R_1}{3} = \frac{2 \cdot 60}{3} = \frac{120}{3} = 40\,\text{Ом}$$

Тепер розглянемо схему з’єднання опорів при розімкненому ключі:

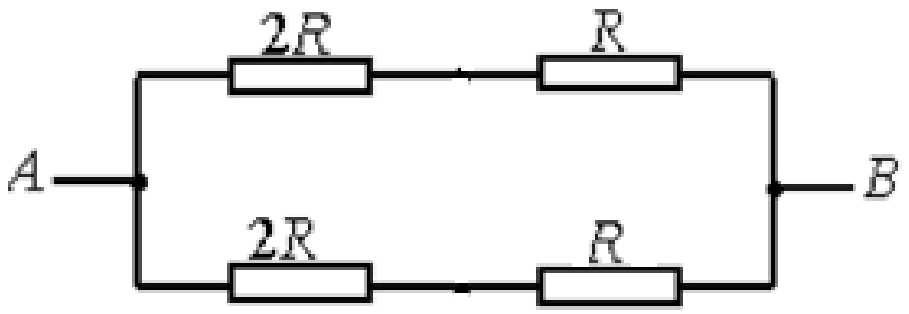

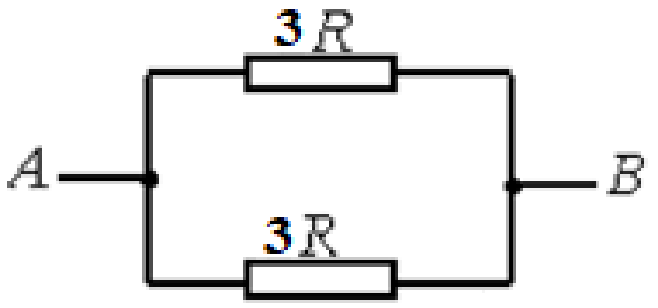

Маємо дві однакові гілки з двома, послідовно з’єднаними, елементами. Спростимо схему склавши їх: $2R + R = 3R$. Отримали спрощену схему з двома паралеотно з’єднаними опорами і можемо написати рівняння для визначення загального опору при з’єднані з розімкненим ключем, який в умовах задачі позначений як $R_2$.

$$\frac{1}{R_2} = \frac{1}{3R} + \frac{1}{3R} = \frac{2}{3R} \quad \Rightarrow \quad R_2 = \frac{3R}{2} = \frac{3 \cdot 40}{2} = \frac{120}{2} = 60\,\text{Ом}$$

**Відповідь:** $R_2 = 60\,\text{Ом}$.

**Задача 4.10.** Визнааботи опір $R_{\text{AB}}$ ділянки кола між точками A і B, якщо $R = 30\,\text{Ом}$.

**Дано:**

$R$=30 Ом

**Знайти:**

$R_{\text{AB}}$–?

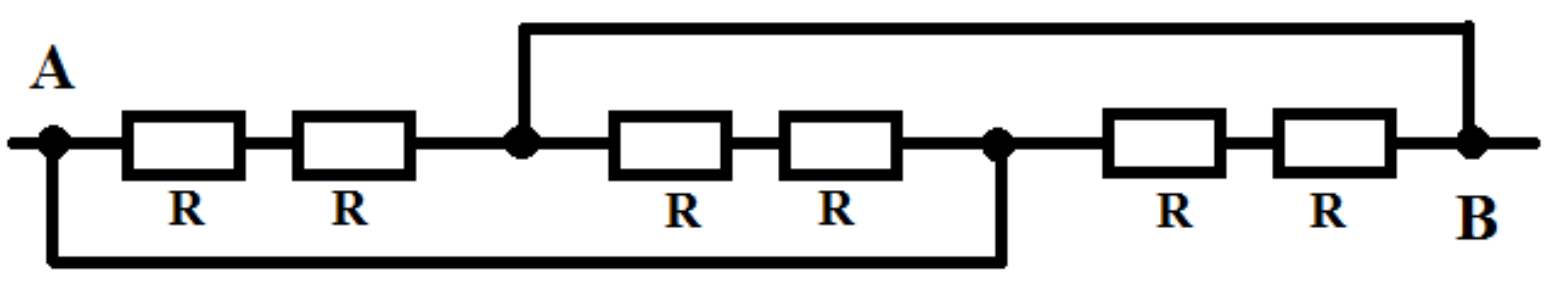

**Розв’язання.**

Першим кроком для розв'язання задаабо зробимо спрощення цієї ділянки. Маємо на схемі точки, що з'єднані провідниками, опором яких нехтують, отже потенціали таких точок будуть однаковими і можна перерисувати схему так, щоб поєднати ці точки.

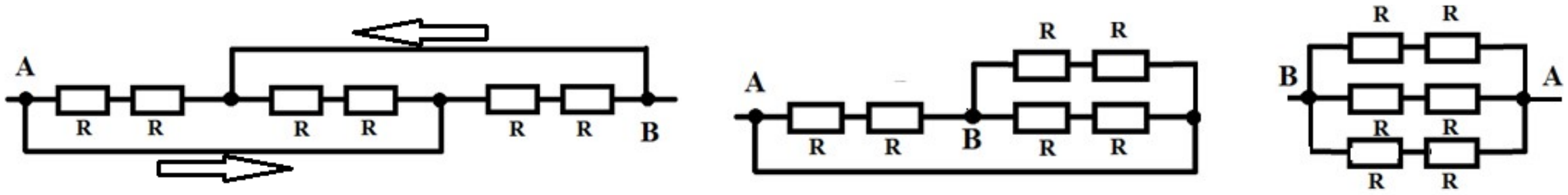

Після поєднання точок з однаковим потенціалом отримали наочну схему, яку також можна спростити, просумувавши на кожній гілці послідовно з'єднані опори. Отримали еквівалентну схему з трьома паралельно з'єднаними опорами:

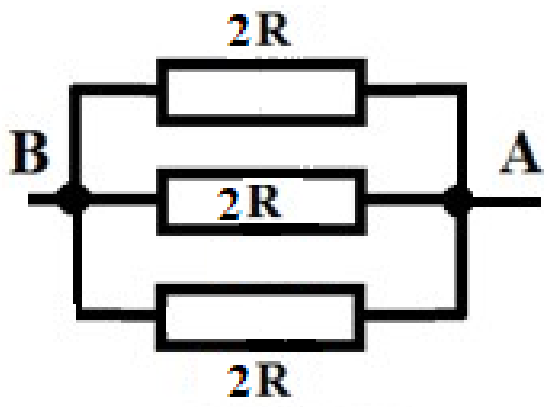

$$\frac{1}{R} = \sum_{i=1}^{n} \frac{1}{R_i}$$

Порахуємо загальний опір $R_{\text{AB}}$ за формулою для паралельно з'єднаних опорів при $R = 30$ Ом:

$$\frac{1}{R_{\text{AB}}} = \frac{1}{2R} + \frac{1}{2R} + \frac{1}{2R} = \frac{3}{2R}$$

$$R_{\text{AB}} = \frac{2R}{3} = \frac{2 \cdot 30}{3} = \frac{60}{3} = 20 \text{ Ом}$$

**Відповідь:** $R_{\text{AB}} = 20$ Ом.

**Задача 4.11.** Визнааботи загальний опір $R$ участка цепі, якщо опір кожного резистора $r = 2$ Ом.

| **Дано:** | **Розв'язання.** |
|---|---|
| $r = 2$ Ом | Маємо змішане з'єднання і щоб схему зробити більш наочною |
| **Знайти:** | трохи її перетворимо, розгорнувши на $90^{\circ}$, змінивши довжину |
| $R-?$ | |

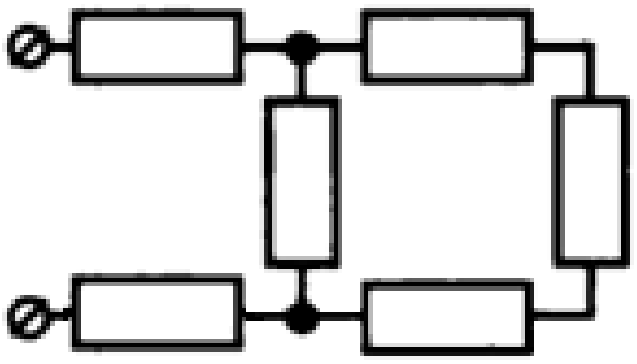

провідничків, що з’єднують опори (їх опором нехтують) і пронумерувавши всі опори.

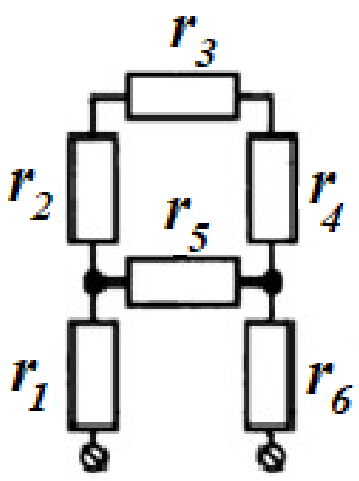

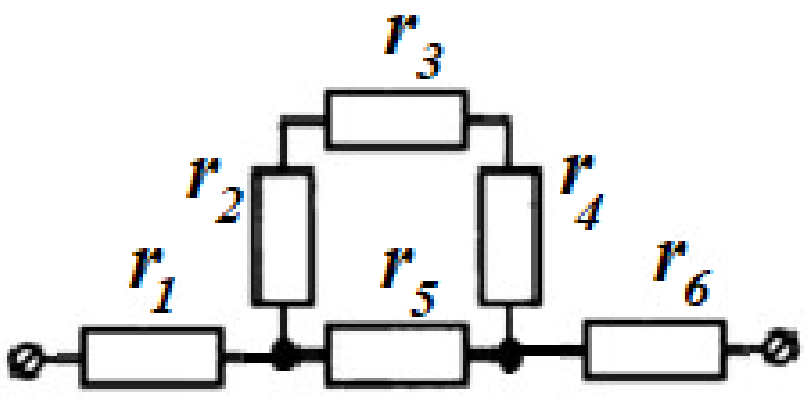

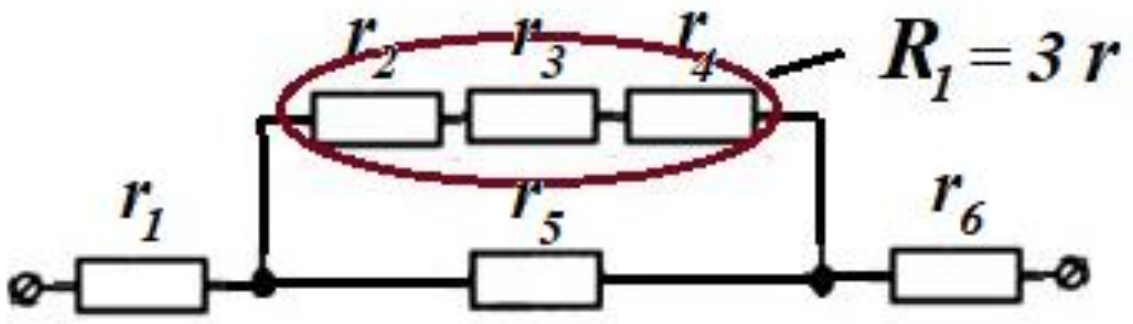

Останній рисунок вже достатньо наочний і можемо почати визначати загальний опір цього участка цепі. Почнемо з послідовно з’єднаних трьох опорів $r_2, r_3, r_4$ для якіх еквівалентним опором буде: $R_1 = 3r$.

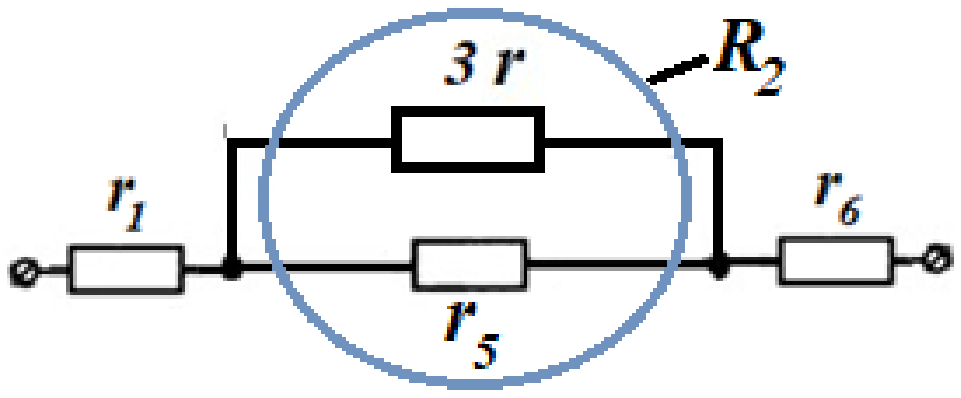

Наступним кроком знайдемо еквівалентний опір $R_2$ для двох паралельно з’єднаних опорів:

$$\frac{1}{R_2} = \frac{1}{r} + \frac{1}{3r} = \frac{3+1}{3r} = \frac{4}{3r} \qquad \Longrightarrow \qquad R_2 = \frac{3r}{4}$$

Тепер маємо три послідовно з’єднинних елемента і можемо визнаaботи загальний опір для заданого участка цепі $r = 2$ Ом .

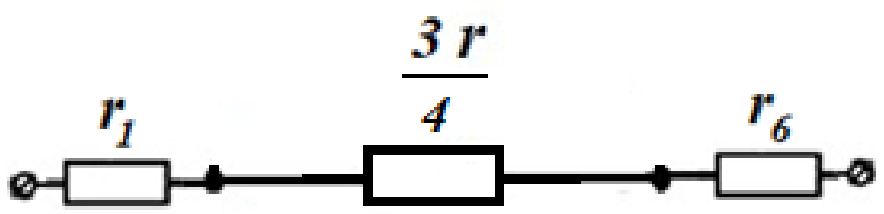

$$R = r + \frac{3r}{4} + r = \frac{4r + 3r + 4r}{4} = \frac{11r}{4} = \frac{11 \cdot 2}{4} = \frac{22}{4} = 5{,}5\text{ Ом}$$

**Відповідь:** $R = 5{,}5$ Ом.

**Задача 4.12.** Надайте еквівалентну схему і визнааботь загальний опір між точками A і B цепі провідників у вигляді прямокутника. Опір кожної проволоки $r = 2{,}5$ Ом.

**Дано:**

$r = 2{,}5$ Ом.

**Знайти:**

$R_{AB}$–?

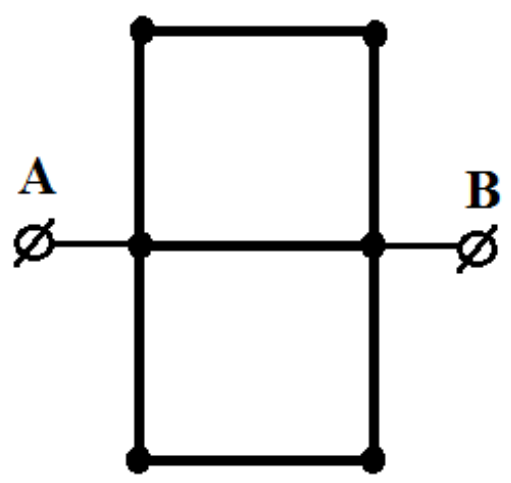

або

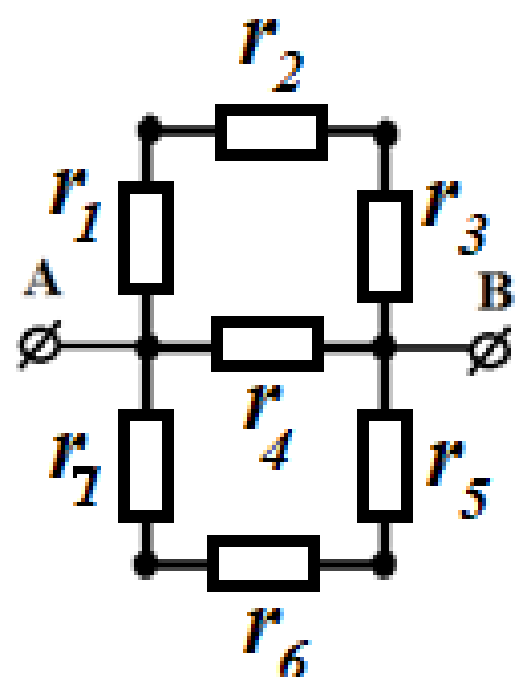

Для більшої наочності приведемо надану схему до більш звичного вигляду – замінемо проволочки на резистори з $r = 2{,}5$ Ом і пронумеруємо їх. Маємо змішане з'єднання і щоб краще розуміти де є послідовне а де паралельне з'єднання зробимо перетворення схеми, змінивши довжину провідничків, що з'єднують елементи. Після спрощення отримали еквівалентні опори $r_{1-3} = 3r$ і $r_{5-7} = 3r$.

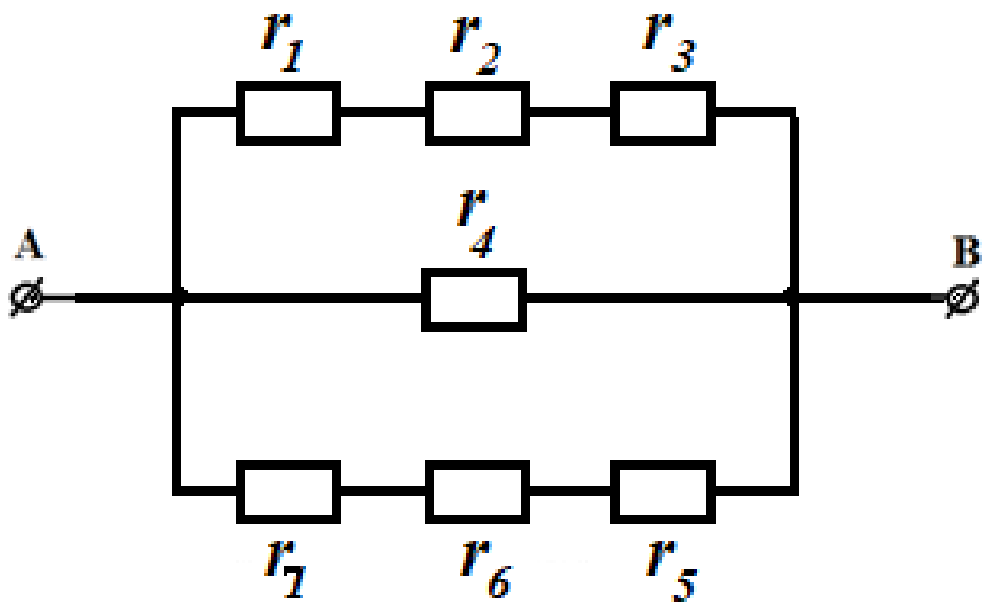

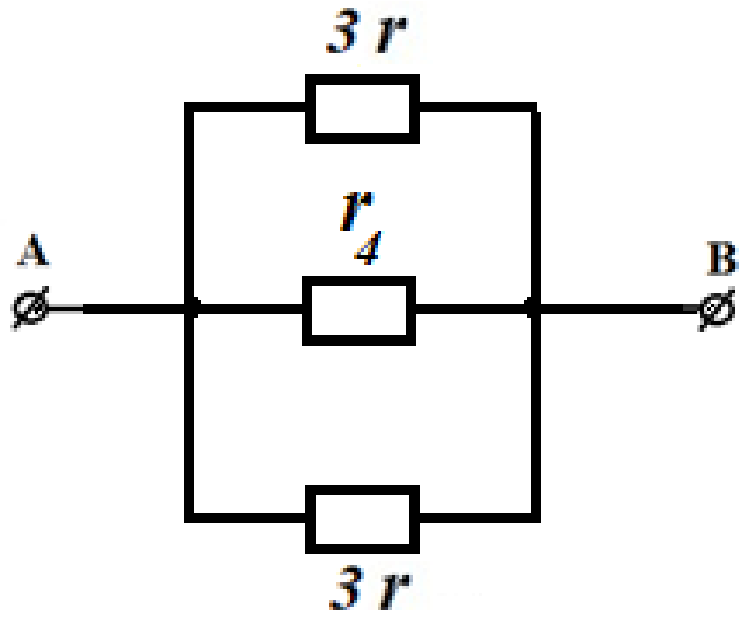

Тепер маємо три паралельно з'єдниних елемента і можемо визнааботи загальний опір для заданого участка цепі при $r = 2{,}5$ Ом:

$$\frac{1}{R_{\text{AB}}} = \frac{1}{3r} + \frac{1}{r} + \frac{1}{3r} = \frac{1 + 3 + 1}{3r} = \frac{5}{3r}$$

$$R_{AB} = \frac{3r}{5} = \frac{3 \cdot 2{,}5}{5} = 1{,}5\ \text{Ом}$$

Відповідь: $R_{AB} = 1{,}5$ Ом.

**Задача 4.13.** Надайте еквівалентну схему і визнааботь загальний опір між точками А і В цепі провідників у вигляді прямокутника. Опір кожної проволоки $r = 3$ Ом

**Дано:**

$r = 3$ Ом

**Знайти:**

$R_{AB}$–?

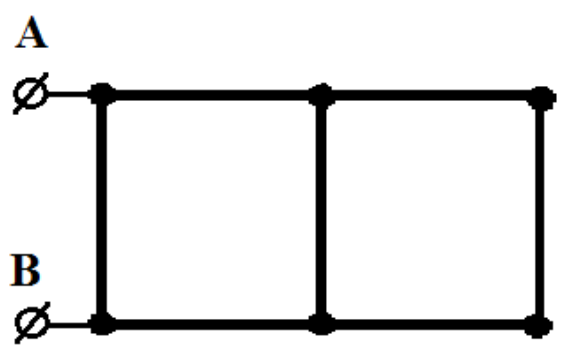

**Розв'язання.**

Зробимо схему більш наочною, розгорнувши на $90^{\circ}$, замінемо проволочки на резистори з $r = 3$ Ом і пронумеруємо їх. Наступним кроком щоб краще розуміти де є послідовне а де паралельне з'єднання зробимо перетворення схеми, змінивши довжину провідничків, що з'єднують елементи.

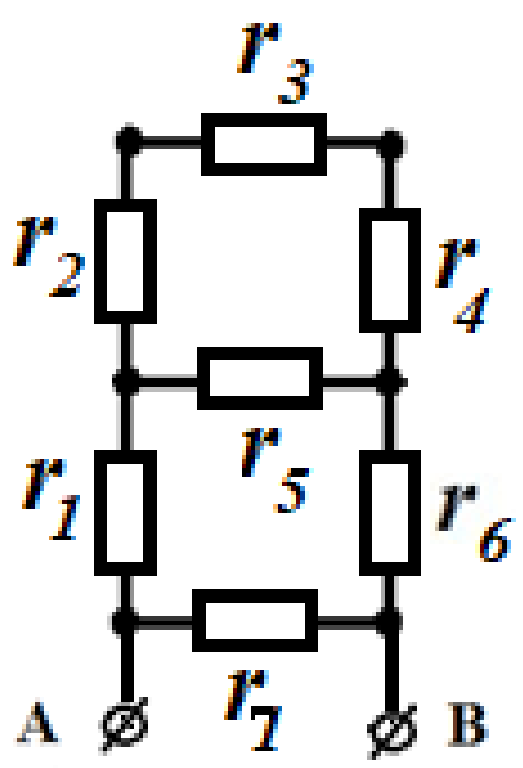

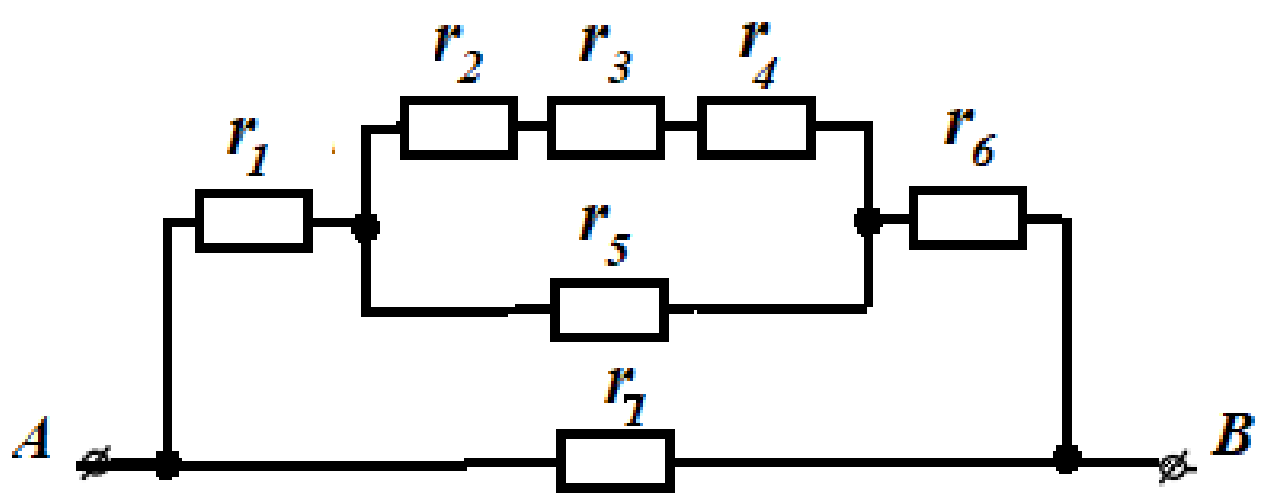

Знаходження загального опору почнемо з послідовно з'єднаних трьох опорів $r_2, r_3, r_4$ для якіх еквівалентним опором буде: $R_1 = 3r$. Далі знайдемо еквівалентне значення опору $R_2$ для двох паралельно з'єднаних елементів $r_5$ і $R_1 = 3r$:

$$\frac{1}{R_2} = \frac{1}{3r} + \frac{1}{r} = \frac{1+3}{3r} = \frac{4}{3r} \quad \Rightarrow \quad R_2 = \frac{3r}{4}$$

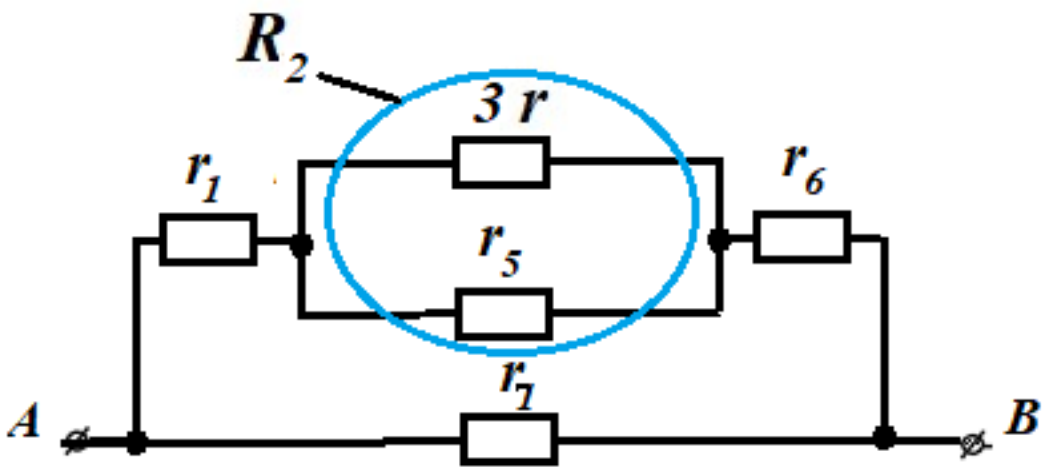

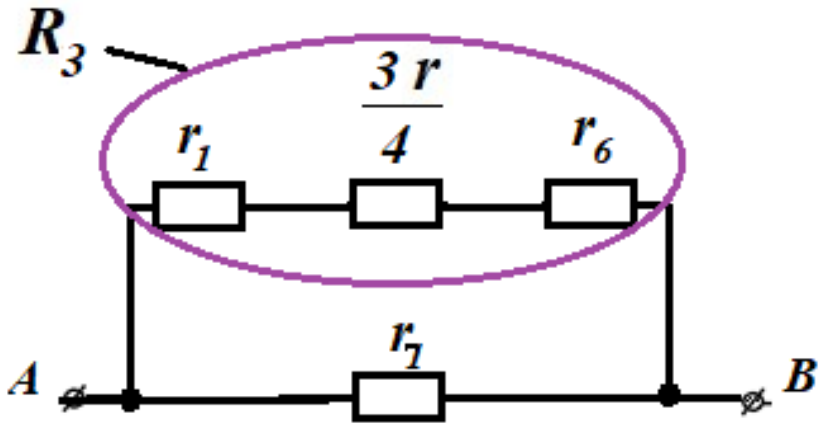

Еквівалентний опір для трьох послідовно з'єднаних елементів $r_1$, $R_2$, $r_6$ :

$$R_3 = r + \frac{3r}{4} + r = \frac{4r + 3r + 4r}{4} = \frac{11r}{4}$$

Тепер маємо два паралельно з'єднинних елемента і можемо визнааботи загальний опір $\mathrm{R_{AB}}$ для заданого участка цепі при $r = 3$ Ом:

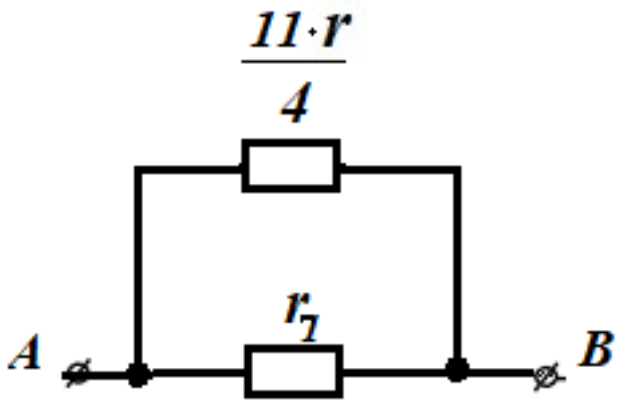

$$\frac{1}{R_{\text{AB}}} = \frac{1}{\frac{11r}{4}} + \frac{1}{r} = \frac{4}{11r} + \frac{1}{r} = \frac{4 + 11}{11r} = \frac{15}{11r}$$

$$R_{\text{AB}} = \frac{11r}{15} = \frac{11 \cdot 3}{15} = \frac{33}{15} = 2{,}2 \text{ Ом}.$$

**Відповідь:** $R_{\text{AB}} = 2{,}2$ Ом.

**Задача 4.14.** Визнааботи загальний опір $R$ між точками A і B цепі провідників у вигляді шестикутника. Опір кожної проволоки $r = 1$ Ом.

**Дано:**

$r = 1$ Ом

**Знайти:**

$R - ?$

**Розв'язання.**

Опір провідників при послідовному і паралельному з'єднанні:

$$R = \sum_{i=1}^{n} R_i, \qquad \frac{1}{R} = \sum_{i=1}^{n} \frac{1}{R_i}$$

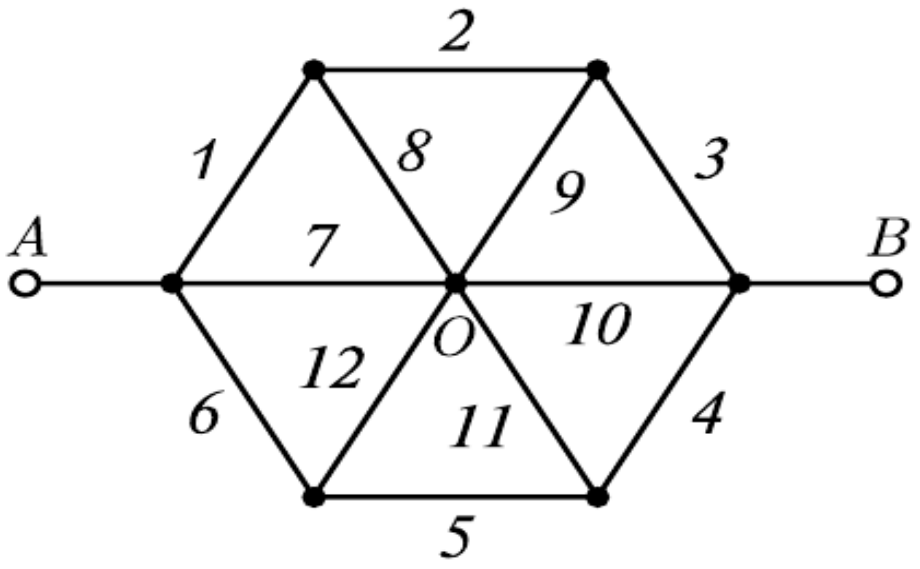

В силу симетрії струми, що течуть по опорам 8, 9, 11 і 12, однакові. Тому струм через вузол О дорівнюється нулю. Тоді схемі, що представлена у вигляді шестикутника на рис. буде відповідати еквівалентна схема, яка задана наступним абоном:

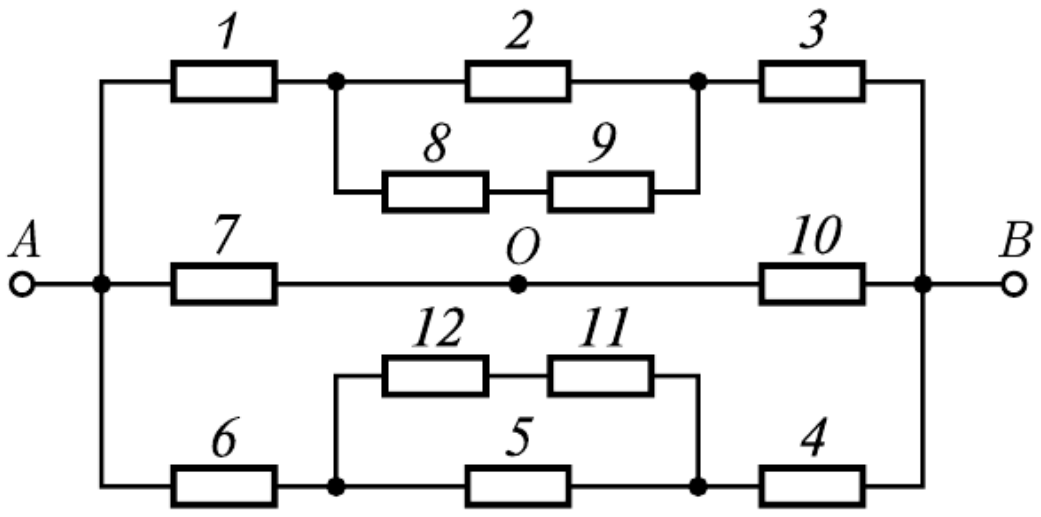

Опори 8 і 9 з'єднані між собою послідовно а також паралельно з опором 2.

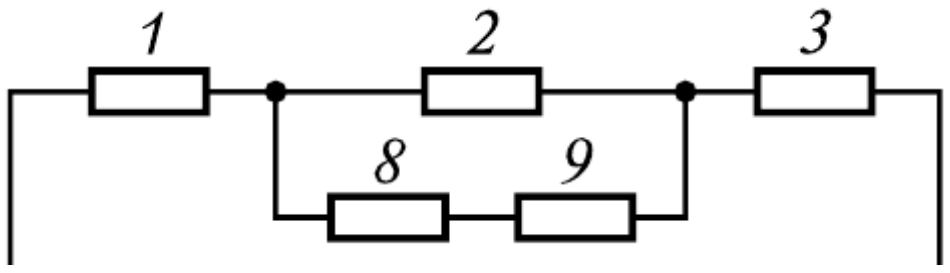

Тоді $R_{8,9} = r + r = 2r$ ;

$$\frac{1}{R_{8,9,2}} = \frac{1}{2\text{r}} + \frac{1}{r} = \frac{1+2}{2r} = \frac{3}{2r}$$

тобто $R_{8,9,2} = \frac{2}{3}r$

Для всієї цепі від 1 по 3

$$R_{1-3} = \frac{2}{3}r + r + r = \frac{8}{3}r$$

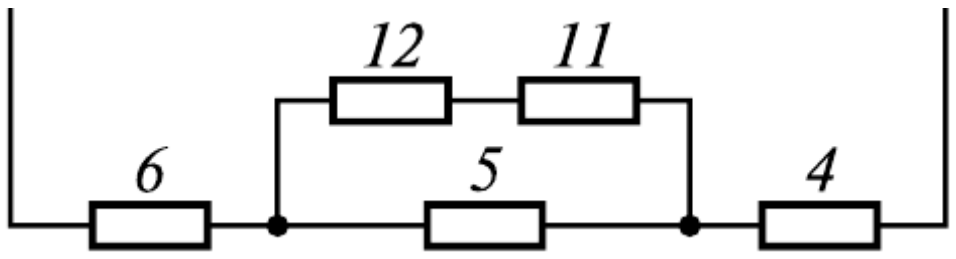

Із схеми ми баабомо, що опір $R_{4-6} = \frac{8}{3}r$ тому що опір $R_{4\to6}$ еквівалентний опору $R_{1\to3}$.

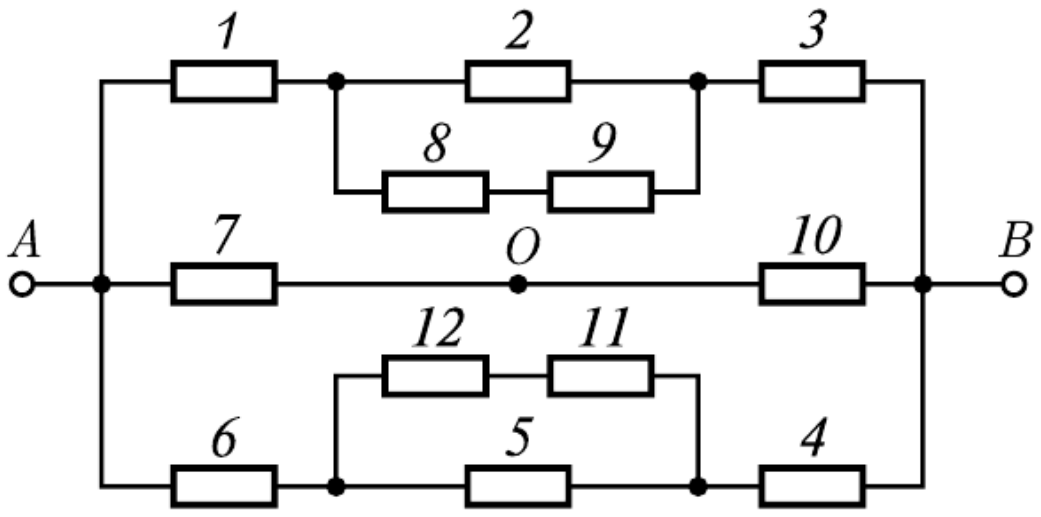

Середня гілка – опір $R_{7\to10} = R_7 + R_{10} = 2r$

Опір $R_{4\to6}$ , опір $R_{1\to3}$ та $R_{7\to10}$ з'єднані паралельно, тому маємо записати:

$$\frac{1}{R} = \frac{1}{R_{1\to3}} + \frac{1}{R_{4\to6}} + \frac{1}{R_{7\to10}}$$

Підставимо значення $R_{4\to6}$ , та $R_{1\to3}$ :

$$\frac{1}{R} = \frac{3}{8r} + \frac{3}{8r} + \frac{1}{2r} = \frac{5}{4r}$$

Звідки загальний опір $R = \frac{4}{5}r = \frac{4}{5} \cdot 1\ \text{Ом} = 0{,}8\ \text{Ом}$

**Відповідь:** $R = 0{,}8$ Ом.

## 4.3. Розрахунок резистивних кіл на основі використання законів Ома і Кірхгофа

**Визначення:**

«Вузол – точка на схемі, де з'єднуються не менше абом три і більше вітки».

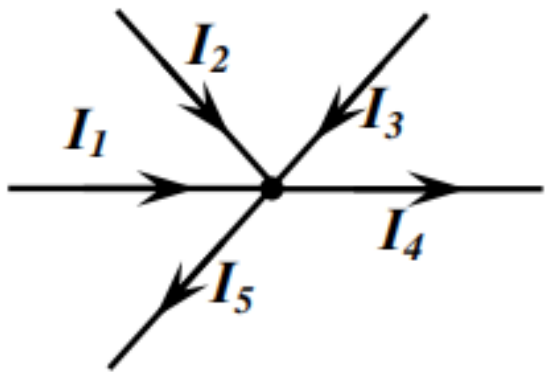

вузол

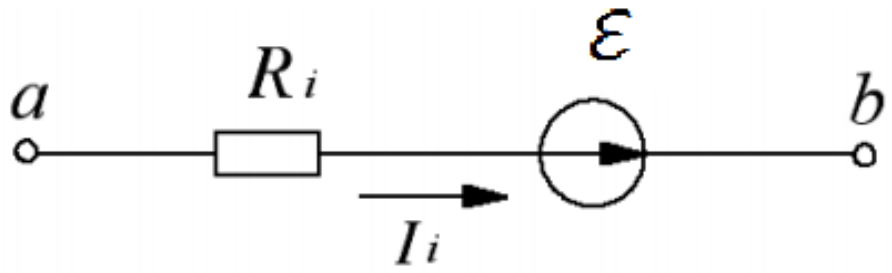

вітка

«Вітка – ділянка кола, утворена кількома послідовно з'єднаними елементами».

«Вітки, приєднані до одних і тих самих вузлів, називаються паралельними».

«Контур – будь-який замкнутий шлях, що проходить по декількох вітках».

«Електричні кола, в яких діє тільки одне джерело електричної енергії, називають простими колами».

«Електричні кола, що складаються тільки з однієї вітки утворюють тільки один контур називаються нерозгалуженими колами.

**Перше правило Кірхгофа:**

Алгебраїчна сума струмів, що сходяться в будь-якому вузлі, дорівнює нулю:

$$\sum_{k=1}^{m} \mathrm{I}_k = 0\,, \qquad m - \text{кількість віток, що сходятьс у даному вузлі}$$

Якщо обрати напрямок до вузла додатним, то згідно до рис. маємо записати:

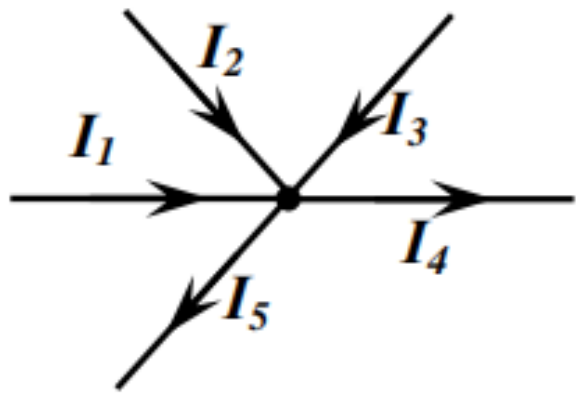

$$I_1 + I_2 + I_3 - I_4 - I_5 = 0$$

Тобто сума струмів, які втікають до вузла, дорівнює сумі струмів, які з нього витікають.

**Друге правило Кірхгофа:**

Для будь якого замкненого контуру, виділеного в розгалуженому колі, алгебраїчна сума спадів напруг, тобто добутків сили струму на опір відповідної ділянки кола дорівнює алгебраїчній сумі всіх ЕРС, які діють в цьому контурі:

$$\sum_{k=1}^{n} R_k \mathrm{I}_k = \sum_{k=1}^{q} E_k,$$

де $n$, $q$ – кількості пасивних елементів і джерел ЕРС у контурі.

Або друге правило записується у вигляді:

$$\sum_{i} I_i R_i = \sum_{k} \varepsilon_k$$

Друге правилю Кірхгофо це є узагальнення закона Ома для неоднорідної ділянки кола.

**Задача 4.15.** У схемі на рисунку $R_1$=30 Ом, $R_2$=20 Ом, $R_3$= 40 Ом. Напруга між точками $A$ і $B$ $U$ =65 В. Визнааботи силу струму, що проходить крізь кожний опір.

**Дано:**
$R_1$=30 Ом
$R_2$=20 Ом
$R_3$= 40 Ом
$U$ =65 В

**Знайти:**
$I_1$-? $I_2$-? $I_3$-?

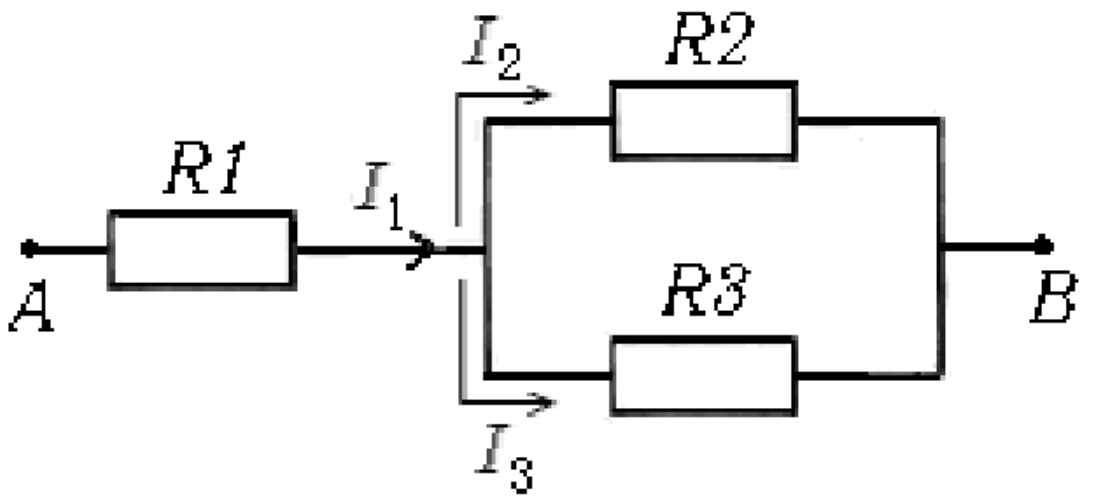

**Розв'язання.**

Зазвичай такі задачі поабонають розв'язувати з пошуку загального опору для всього участку між точками $A$ і $B$. Для визначення опору для паралельно з'єднаних двох елементів $R_2$ і $R_3$ скористаємось формулой:

$$R_E = \frac{R_1 R_2}{R_1 + R_2}$$

Тоді загальний опір для нашої задачі знайдемо за рівнянням:

$$R_{\text{Заг}} = R_1 + \frac{R_2 R_3}{R_2 + R_3} = 30 + \frac{20 \cdot 40}{20 + 40} = 30 + 13{,}33 = 43{,}33 \text{ Ом}$$

Визнаабовши загальний опір можемо знайти $I_1$ за законом Ома:

$$I = \frac{U}{R}$$

$$I_1 = \frac{65}{43,33} = 1,5 \text{ A}$$

За першим правилом Кірхгофа - сума струмів, які втікають до вузла, дорівнює сумі струмів, які з нього витікають (струми, що втікають записуються зі знаком «+», а ті, що витікають зі знаком «-»). Складемо таке рівняння для даної задачі:

$$I_1 - I_2 - I_3 = 0, \quad \text{або} \quad I_1 = I_2 + I_3$$

Щоб визнааботи $I_2$ і $I_3$ цього рівняння замало, треба скласти ще одно. Розглянемо напругу на кожному з елементів: при послідовному напруга складається $U_{\text{Заг}} = U_1 + U_2$ , а при паралельному з'єднані маємо

$$U_{\text{Заг}} - U_1 = U_2 = U_3$$

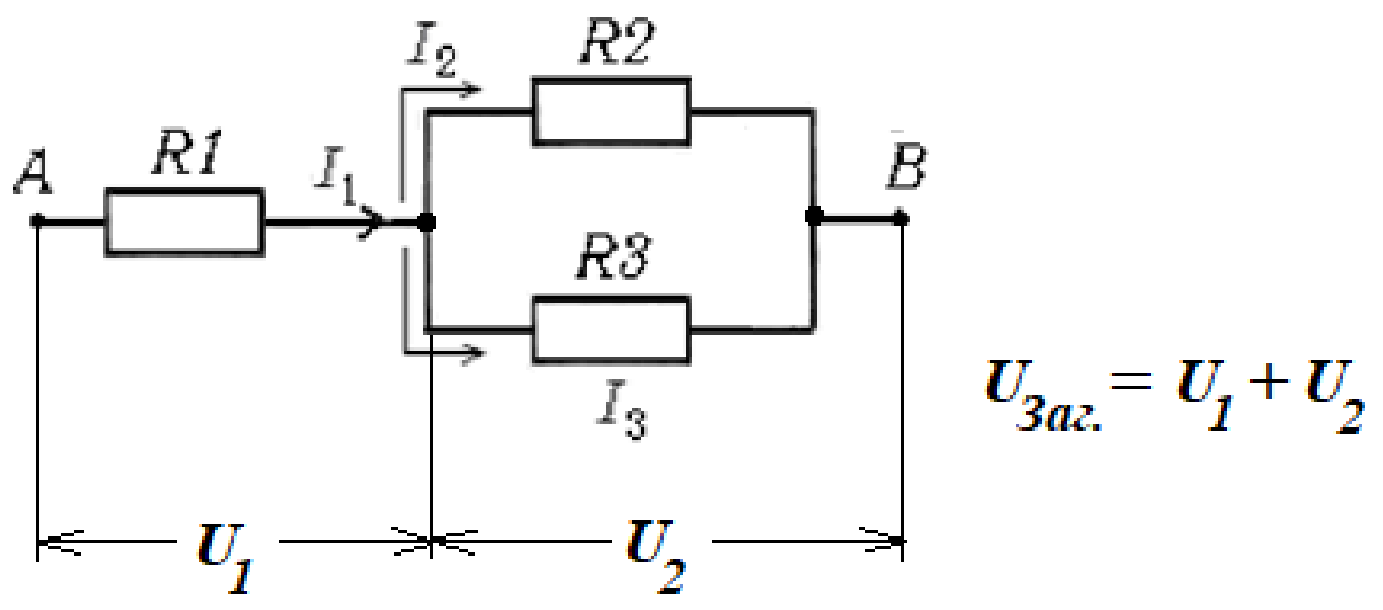

Скористаємося інформацією, що при паралельному з'єднані напруга на кожній з віток є однаковою, тобто $U_2 = U_3$. Тоді за законом Ома маємо:

$$I_2 R_2 = I_3 R_3$$

Використовуємо $I_1 = I_2 + I_3$, виразивши $I_3$ і підставимо $I_3 = I_1 - I_2$ .

$$I_2 R_2 = (I_1 - I_2) R_3$$

Розкриємо скобки: $I_2 R_2 = I_1 R_3 - I_2 R_3$ і виразимо $I_2$:

$$I_2 R_2 + I_2 R_3 = I_1 R_3; \quad I_2 (R_2 + R_3) = I_1 R_3$$

$$I_2 = \frac{I_1 R_3}{R_2 + R_3} = \frac{1,5 \cdot 40}{20 + 40} = \frac{60}{60} = 1 \text{ A}$$

Знайшовши $I_1$ і $I_2$, розрахуємо $I_3$:

$$I_3 = I_1 - I_2 = 1{,}5 - 1 = 0{,}5\ \text{А}$$

**Відповідь:** $I_1 = 1{,}5$ А; $I_2 = 1$ А; $I_3 = 0{,}5$ А.

**Задача 4.16.** У схемі електрорушійна сила $\varepsilon_1$=130 В, $\varepsilon_2$=117 В, $R_1$=1 Ом, $R_2$ =0,6 Ом, $R_3$ =24 Ом. Визнаaботи сили струму в усіх ділянках кола. Внутрішнім опором елементів нехтувати.

**Дано:**

$R_1$=1 Ом

$R_2$=0,6 Ом

$R_3$= 24 Ом

$\varepsilon_1$=130 В

$\varepsilon_2$=117 В

**Знайти:**

$I_1$-? $I_2$-?

$I_3$-?

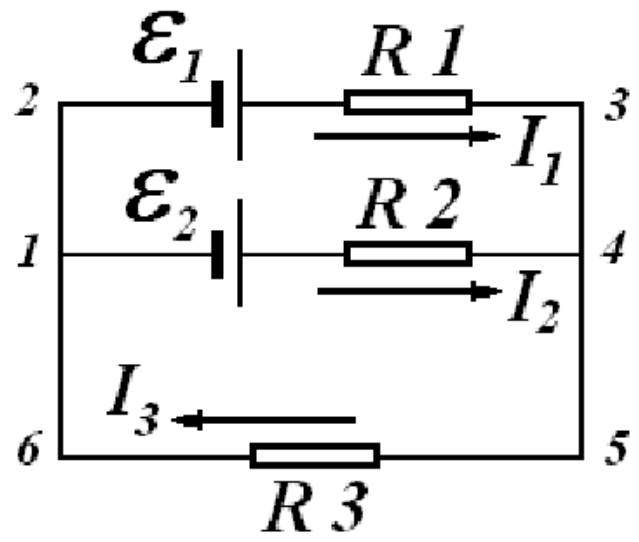

**Розв'язання.**

Задачу розв'язують у декілька етапів.

**1-й етап**: обираємо довільні, але, за можливістю, правдоподібні напрями струмів на всіх ділянках кола і позначаємо їх на схемі. Якщо вибір деяких з них буде невірним, у розв'язку цей струм вийде зі знаком мінус, що дасть можливість наприкінці виправити напрям струму.

**2-й етап.** Щоб застосувати правила Кірхгофа, обираємо вузли і контури. «Вузол – точка на схемі, де з'єднуються не менше абом три і більше вітки».

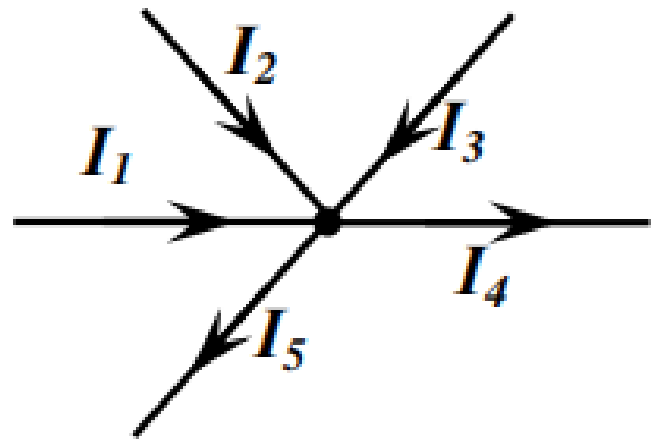

Маємо два вузли – точки 1 та 4.

Контуром називається замкнута частина розгалуженого електричного кола. Контури обирають за принципом – кожний новий контур повинен містити нові ЕРС і опори. У нашому прикладі незалежними будуть рівняння для верхнього позначеного цифрами 12341 і нижнього контурів - 14561.

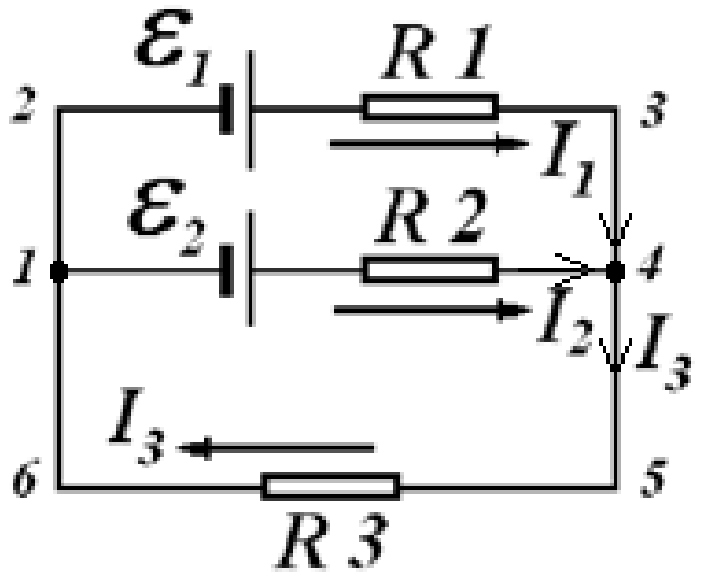

Напрям обходу контурів – за або проти годинникової стрілки – значення не має. Але треба обходити контури якимось одним абоном.

**3-й етап.** Складаємо рівняння за правилами Кірхгофа.

1. Алгебраїчна сума всіх сил струмів, що сходяться у вузлі розгалуженого кола, дорівнює нулю:

$$\sum_{k=1}^{m} I_k = 0$$

Струми, що входять у вузол, вважають додатними, ті, що виходять – від'ємними. Згідно до цього правила маємо записати:

$$I_1 + I_2 - I_3 = 0$$

2. У будь якому замкненому контурі розгалуженого електричного кола алгебраїчна сума добутків сил струмів на опори відповідних ділянок контуру дорівнює алгебраїчній сумі електрорушійних сил, що діють у ньому:

$$\sum_{k=1}^{n} R_k I_k = \sum_{i=1}^{m} \varepsilon_i,$$

Добуток $I_k R_k$ вважається додатним, якщо напрям струму збігається з напрямом обходу по замкненому контуру, і від'ємним у протилежному випадку. Електрорушійні сили вважаються додатними, якщо їхній власний струм збігається з напрямом обходу, тобто ті ЕРС, для яких напрям обходу збігається з переходом від низького потенціалу (негативний полюс) до високого (позитивний полюс). У противному випадку ЕРС входять у суму від'ємними.

Складемо рівняння за другим правилом Кірхгофа для контуру 12341: добуток $I_1 R_1$ є додатним томущо напрям струму співпадає з обраним напрямом струму, тоді добуток $I_2 R_2$ є від'ємний – струм тече «назустріч» обраному напряму. Так само визначаємо знаки для електрорушійних сил: $\varepsilon_1$ –додатна , $\varepsilon_2$ – від'ємна.

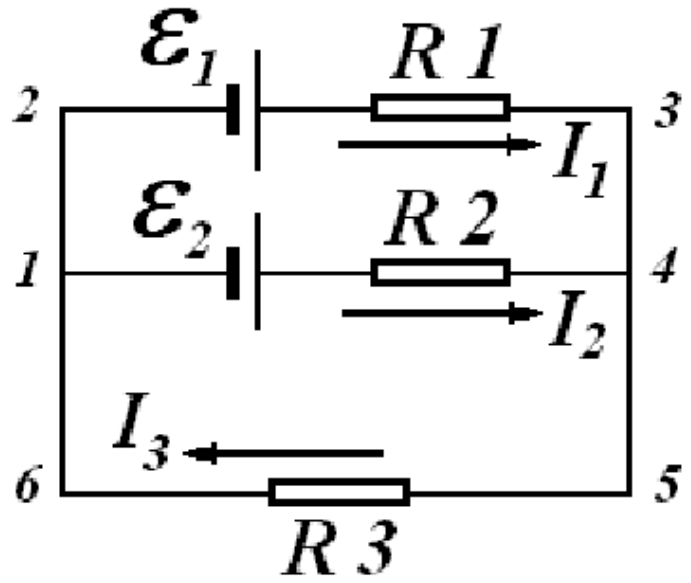

Рівняння для контуру 12341:

$$I_1 R_1 - I_2 R_2 = \varepsilon_1 - \varepsilon_2$$

Для другого контуру 14561 добутки $I_2 R_2$ і $I_3 R_3$ та електрорушійна сила $\varepsilon_1$ є додатними - напрями струму співпадають з обраним напрямом струму. Рівняння для другого контуру 14561:

$$I_2 R_2 + I_3 R_3 = \varepsilon_2$$

**4-й етап.** Складаємо систему рівнянь із трьох отриманих:

$$\begin{cases} I_1 + I_2 - I_3 = 0 \\ I_1 R_1 - I_2 R_2 = \varepsilon_1 - \varepsilon_2 \\ I_2 R_2 + I_3 R_3 = \varepsilon_2 \end{cases}$$

Підставимо у рівняння (2) та (3) значення опорів і ЕРС, отримаємо наступну систему рівнянь:

$$\begin{cases} I_1 + I_2 - I_3 = 0 \\ 1 \cdot I_1 - 0{,}6 \cdot I_2 = 13 \\ 0{,}6 \cdot I_2 + 24 \cdot I_3 = 117 \end{cases}$$

З першого рівняння виразимо $I_3$:

$$I_3 = I_1 + I_2$$

З другого рівняння виразимо $I_1$:

$$I_1 = 13 + 0{,}6 \cdot I_2$$

Отриманий вираз для $I_1$ підставимо в $I_3$

$$I_3 = 13 + 0{,}6 \cdot I_2 + I_2 = 13 + 1{,}6 I_2$$

і результат підставимо в третє рівняння:

$$0{,}6 \cdot I_2 + 24(13 + 1{,}6 I_2) = 117$$

$$39 I_2 = 117 - 312 = -195$$

$$I_2 = -5$$

Підставивши $I_2 = -5$ А в рівняння $I_1 = 13 + 0{,}6 \cdot I_2$ отримаємо $I_1 = 10$ А, після чого значення $I_1$ та $I_2$ підставимо в $I_3 = I_1 + I_2 = 5$ А.

Таким абоном, розв’язок системи дає $I_1$ = 10 А, $I_2$ = –5 А, $I$= 5 А.

Мінус у значенні другого струму означає, що під час довільного вибору напряму цього струму був вказаний протилежний напрям. Насправді струм проходить від вузла 4 до вузла 1.

**Відповідь:** $I_1$ = 10 А, $I_2$ = –5 А, $I$= 5 А.

**Задача 4.17.** Визнааботи струми усіх віток кола на рис., якщо $E$=11 В; $R_1$=1 Ом ; $R_2$ = 2 Ом; $R_3$ = 3 Ом.

**Дано:**
$R_1$=1 Ом
$R_2$=2 Ом
$R_3$= 3 Ом
$E$ =11 В

**Знайти:**
$I_1$-? $I_2$-?
$I_3$-?

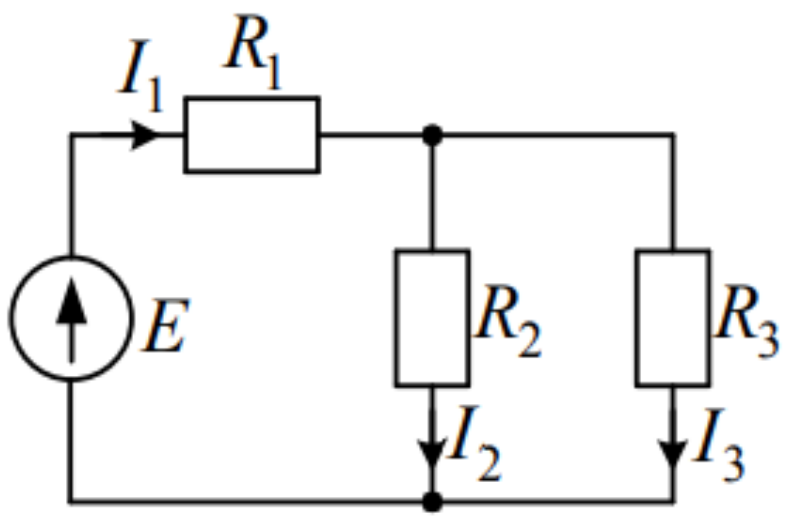

**Розв’язання.**

Знайдемо еквівалентний опір схеми, але спочатку визнаабомо загальний опір для паралельно з’єднаних ; $R_2$ = 2 Ом та $R_3$ = 1 Ом.

$$R_{23} = \frac{R_2 R_3}{R_2 + R_3} = \frac{2 \cdot 3}{2 + 3} = 1{,}2 \text{ Ом}$$

тоді схему можна спростити до

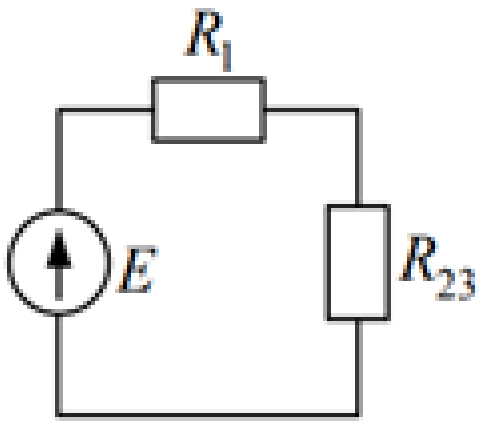

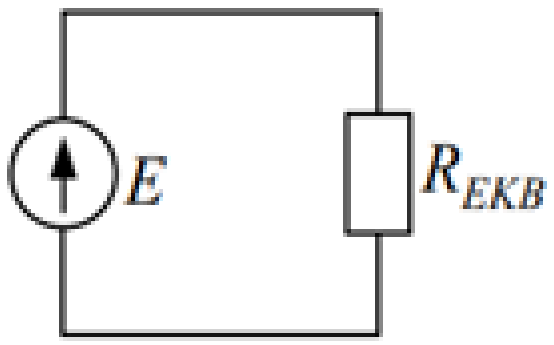

$$R_{\text{ЕКВ}} = R_1 + R_{23} = 1 + 1{,}2 = 2{,}2 \text{ Ом}$$

і знайти еквівалентний опір $R_{\text{екв}}$ для всієї схеми.

Вхідний струм джерела напруги розраховуємо за законом Ома:

$$I_{\text{Е}} = \frac{E}{R_E} = \frac{11}{2{,}2} = 5\ A$$

За першим законом Кірхгофа визнаабомо струми у 1 вузлі

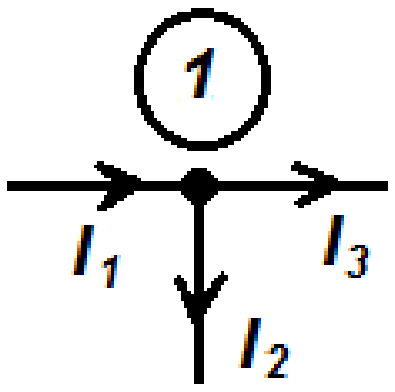

$$I_1 - I_2 - I_3 = 0 \quad \text{або} \quad I_1 = I_2 + I_3$$

Для послідовного з’єднання маємо:

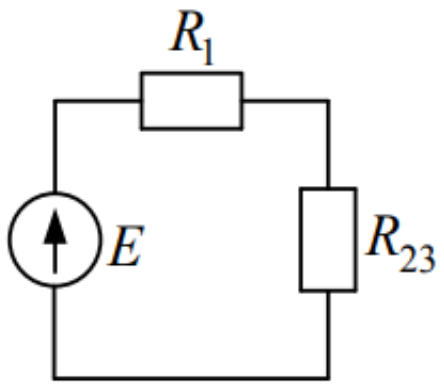

$$I_1 = I_{\text{Е}} = I_{23} = 5\text{А};$$

При паралельному з’єднанні $U = U_1 = U_2 = \cdots = U_n$ , тобто маємо записати:

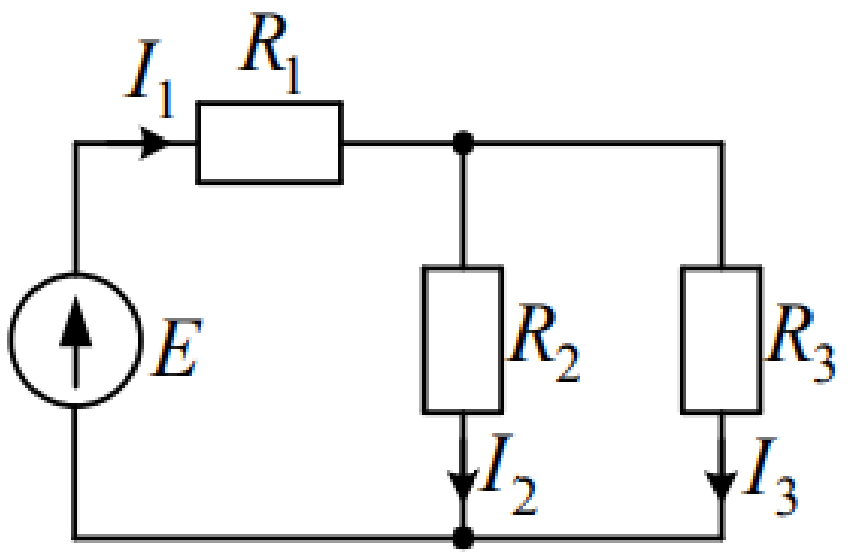

$$U_{23} = U_2 = U_3$$

тобто

$$I_{23}R_{23} = I_2R_2,$$

Знайдемо $I_2$

$$I_2 = \frac{I_{23}R_{23}}{R_2} = I_{23}\frac{R_3}{R_2 + R_3} = 5\text{ А}\cdot\frac{3\text{ Ом}}{(2+3)\text{ Ом}} = 3\text{ А}$$

як з'ясували з 1 закону Кірхгофа

$$I_1 = I_2 + I_3$$

тоді

$$I_3 = I_1 - I_2\text{ , але}$$

$$I_1 = I_\text{Е} = I_{23} = 5\text{А}$$

тому

$$I_3 = I_{23} - I_2 = 5 - 3 = 2\text{ А}$$

**Відповідь:** $I_1 = 5$ А; $I_2 = 3$ А; $I_3 = 2$ А.

**Задача 4.18.** Задано електричне резистивне коло. Визнааботи: $R_{\text{екв}}$, струми в гілках, падіння напруги на елементах, якщо $E$=11 В; $R_1 = 1$ Ом; $R_2 = 2$ Ом; $R_3 = 3$ Ом; $R_4 = 4$ Ом.

**Дано:**

$R_1$=1 Ом

$R_2$=2 Ом

$R_3$=3 Ом

$R_4$=4 Ом

$E$ =11 В

**Знайти:**

$I_1$-? $I_2$-?

$I_3$-?$R_{екв}$ -?

$U_1$-? $U_2$ -?

$U_3$-? $U_4$-?

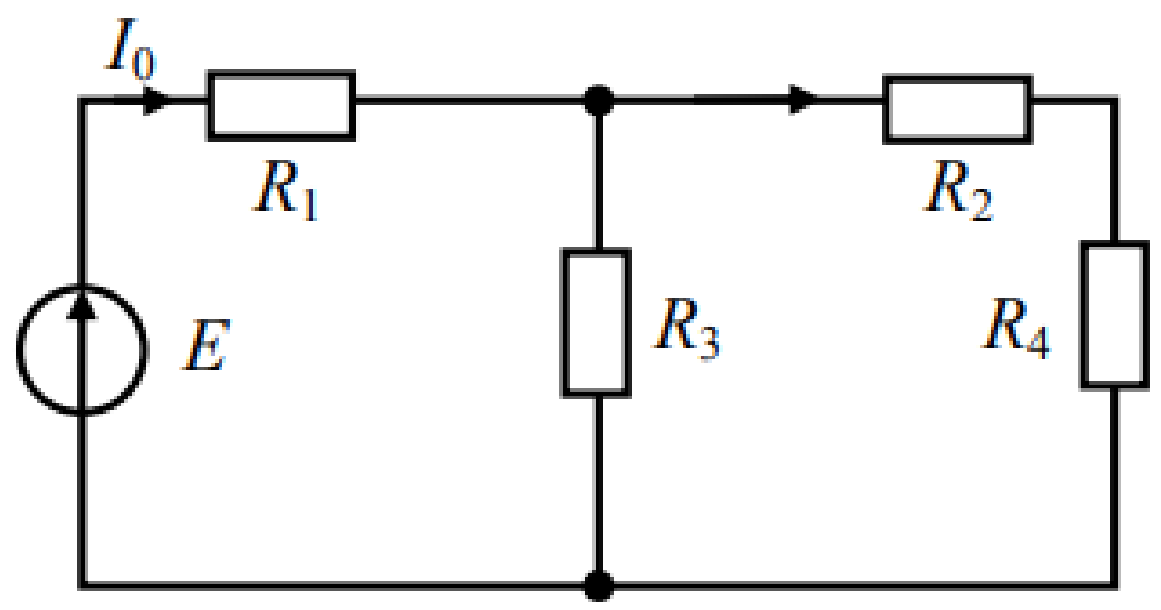

**Розв’язання.**

Для визначення параметрів електричного резистивного кола застосуємо методику розрахунку резистивного кола драбинної структури. Спочатку визнаабомо $R_{екв}$ шляхом ряду послідовно-паралельних перетворень із кінця електричного кола; наступна дія – визнааботи струм у гілках і падіння напруги на кожному елементі, використовуюабо закони Кірхгофа.

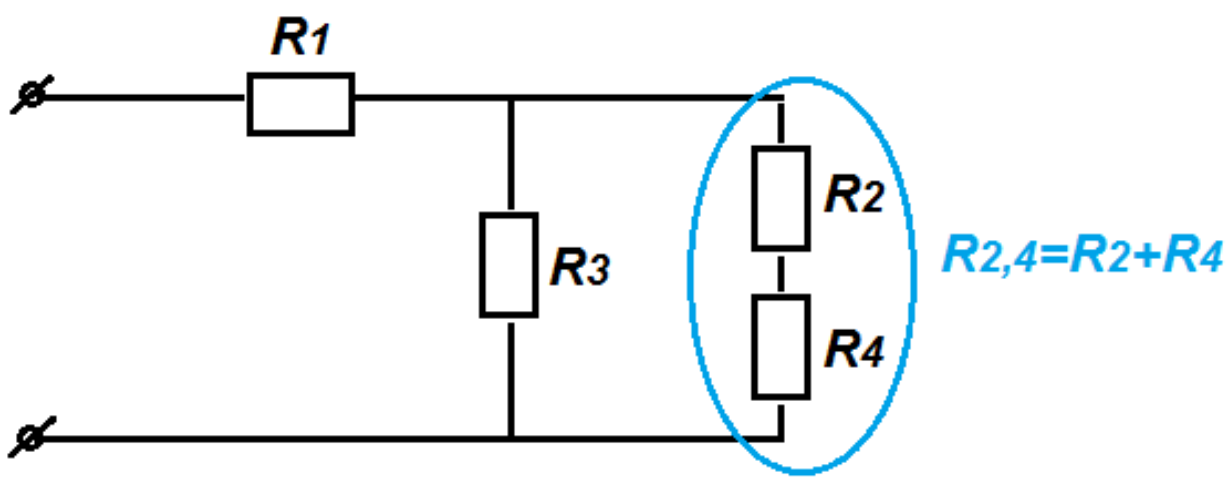

1. Визнаабомо $R_{екв}$

$$\frac{1}{R_{2-4}} = \frac{1}{R_3} + \frac{1}{R_{2,4}} = \frac{R_{2,4} + R_3}{R_{2,4} \cdot R_3}$$

$$R_{2-4} = \frac{R_{2,4} \cdot R_3}{R_{2,4} + R_3} = \frac{6 \cdot 3}{6+3} = 2\ \text{Ом}$$

Кінцеве значення загального опору схеми $R_е$

$$R_е = R_1 + \frac{R_{2,4} \cdot R_3}{R_{2,4} + R_3} = 1 + \frac{6 \cdot 3}{6+3} = 1 + 2 = 3\ \text{Ом}$$

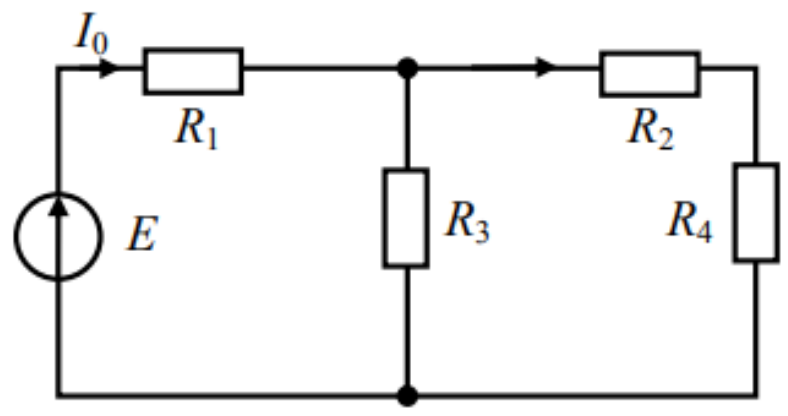

Визнаабовши еквівалентний опір усього кола, можемо визнааботи струми в гілках і падіння напруги на опорах.

Загальний струм в колі:

$$I_0 = I_1, \qquad I_1 = \frac{\mathcal{E}}{R_e} = \frac{11}{3} = 3{,}66\ \text{А}$$

Падіння напруги на $R_1$:

$$U_1 = I_1 \cdot R_1 = 3{,}66 \cdot 1 = 3.66\ \text{В}$$

Розглянемо як поділяється струм $I_0 = I_1$ в першому вузлі

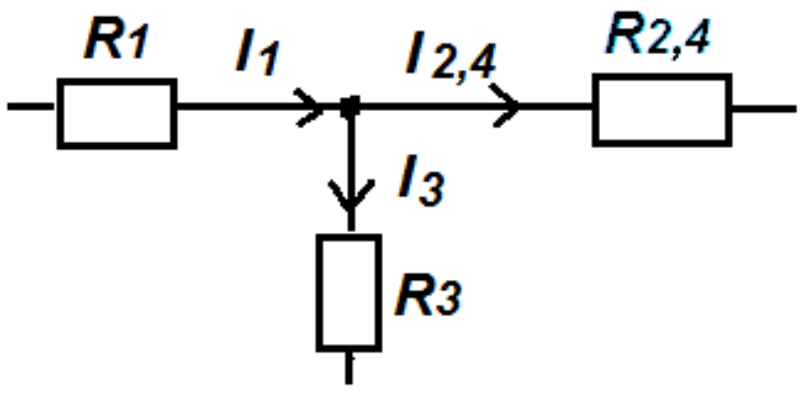

$$I_1 = I_{2,4} + I_3$$

Тоді падіння напруги на загальному опорі

$$R_{2-4} = \frac{R_{2,4} \cdot R_3}{R_{2,4} + R_3}$$

буде дорівнюватись:

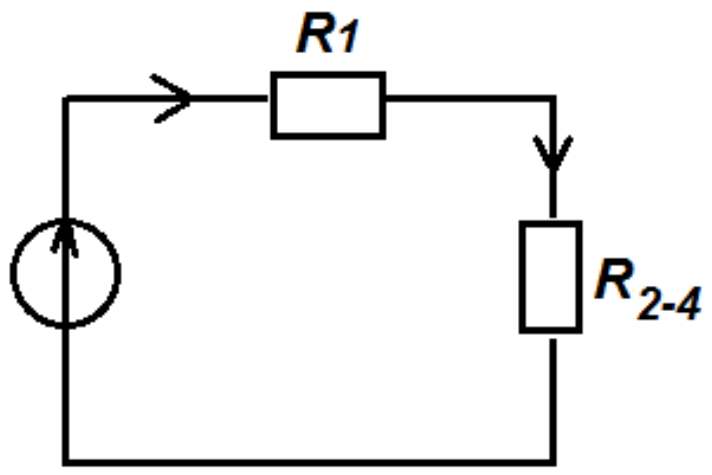

$$U_{2-4} = I_0 \cdot R_{2-4,} \ (I_0 = I_1)$$

Маємо послідовно з'єднані елементи, тобто при послідовному з'єднані напруга на окремих елементах кола:

$$U_1 = I_1 \cdot R_1 \ldots\ldots U_n = I_1 \cdot R_n$$

Тобто спади напруги на послідовно з'єднаних опорах пропорційні велиабонам опорів, а сума напруг на окремих елементах дорівнює напрузі на вхідних клемах кола:

$$U_1 + U_2 + U_3 = U$$

Тому для нашої задачі маємо записати, що:

$$\mathcal{E} = U_1 + U_{2-4}$$

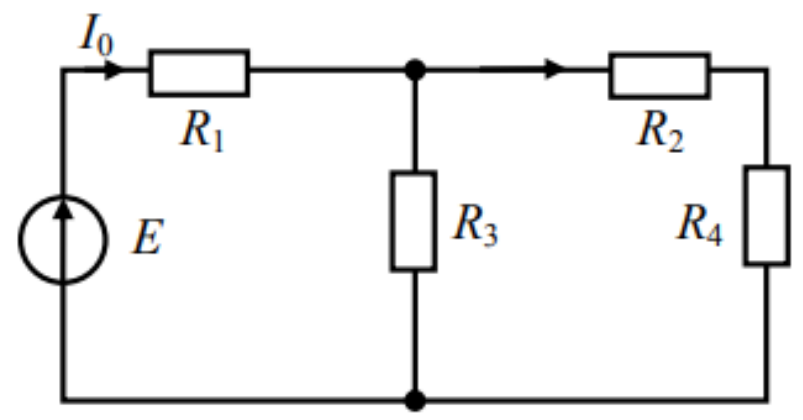

тобто падіння напруги на опорі $R_{2-4}$ буде:

$$U_{2-4} = \mathcal{E} - U_1 = 11 - 3.66 = 7{,}34 \text{ В}$$

Для наочності перетворимо надану схему на наступну:

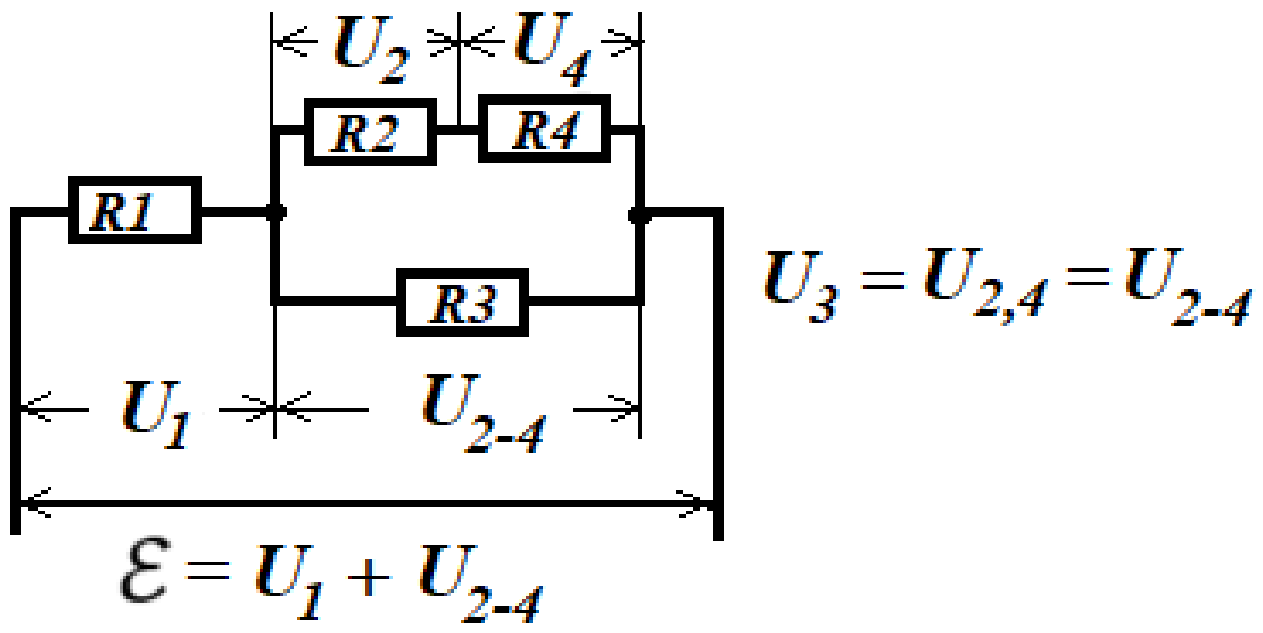

Струм через опір $R_3$ і через $R_{2,4}$ можна записати, враховуюабо, що ці опори є паралельно з'єднані, а для паралельного з'єднання маємо, що

$$U = U_1 = U_2 = \cdots = U_n$$

тобто

$$U_{2-4} = U_3 = U_{2,4} = 7{,}34 \text{ В}$$

тоді:

$$I_3 = \frac{\mathcal{E} - U_1}{R_3} = \frac{7{,}34}{3} = 2{,}44 \text{ А}$$

$$I_{2,4} = \frac{\mathcal{E} - U_1}{R_{2,4}} = \frac{\mathcal{E} - U_1}{R_2 + R_4} = \frac{7{,}34}{6} = 1{,}22\ \text{А}$$

При визначенні падіння напруги на $R_2$ та $R_4$ враховуємо, що ці резистори з'єднанні послідовно, а при послідовному з'єднані маємо, що

$$U = \sum_{i=1}^{n} U_i\,; \qquad I = I_1 = I_2 = \cdots = I_n$$

тому струм, що йде через $R_2$ або $R_4$ є той самий, що через обидва $R_{2,4}$, який ми вже знайшли ($I_{2,4} = I_2 = I_4$), тому маємо записати:

$$U_2 = I_{2,4} \cdot R_2 = 1{,}22 \cdot 2 = 2{,}44\ \text{В}; \quad U_4 = I_{2,4} \cdot R_4 = 1{,}22 \cdot 4 = 4{,}88\ \text{В}$$

**Відповідь:** $R_е = 3$ Ом; $I_1 = 3{,}66$ А; $I_2 = 1{,}22$ А; $I_3 = 2{,}44$ А; $I_4 = 1{,}22$ А;

$U_1 = 3.66$ В; $U_2 = 2{,}44$ В; $U_3 = 7{,}34$ В; $U_4 = 4{,}88$ В.

**Задача 4.19.** Складіть рівняння для розрахунку резистивного кола за правилами Кірхгофа при відомих значеннях опорів $R_1 - R_{10}$.

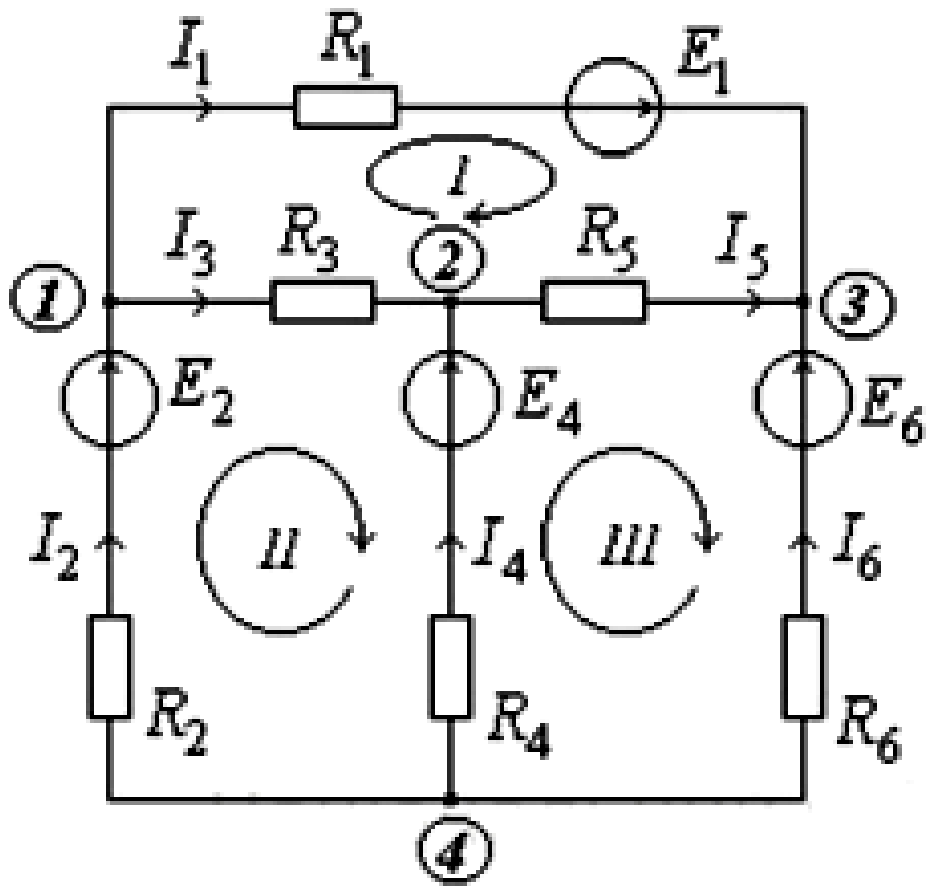

**Розв'язання:**

За алгоритмом розрахунку електричного кола методом рівнянь Кірхгофа спочатку визнаабом вузли $n_{вз}$, вітки $n_в$ та напрями струмів у вітках.

В цьому електричному колі маємо 4 вузла, 6 віток і 3 незалежних кола.

Напрями струму у вітках вибираємо довільно і складаємо $(n_{\text{вз}}-1)$ рівнянь за 1 законом Кірхгофа - «Алгебраїчна сума струмів віток, що сходяться у вузлі електричного кола, дорівнює нулю, тобто сума струмів, які втікають до вузла, дорівнює сумі струмів, які з нього витікають».

За 1 законом Кірхгофа маємо записати наступні рівняння:

1. $I_2 - I_1 - I_3 = 0$
2. $I_3 + I_4 - I_5 = 0$
3. $I_1 + I_5 + I_6 = 0$

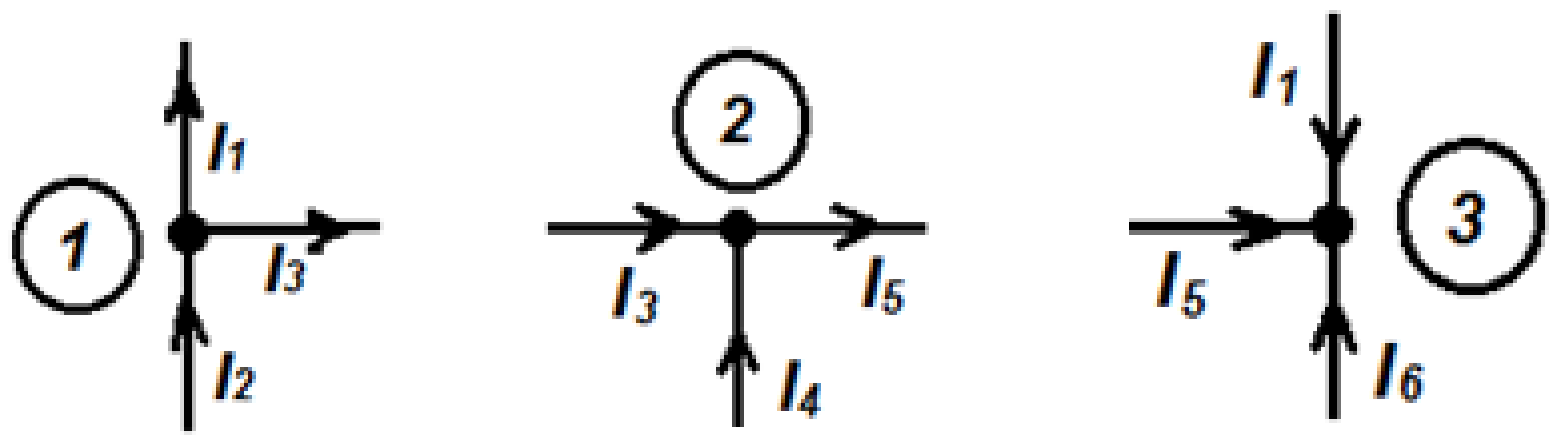

Вибираємо незалежні контури і довільно задаємо напрям їх обходу, наприклад, за годинковою стрілкою. Складаємо, для кожного з незалежних контурів, рівняння за другим правилом Кірхгофа – «Алгебраїчна сума спадів напруг, тобто добутків сили струму на опір відповідної ділянки кола дорівнює алгебраїчній сумі всіх ЕРС, які діють в цьому контурі»:

$$\sum_{k=1}^{n} R_k \mathrm{I}_k = \sum_{k=1}^{q} E_k,$$

де $n, q$ – кількості пасивних елементів і джерел ЕРС у контурі.

Складається $n = n_{\text{в}} - n_{\text{вз}} + 1 = 3$ рівнянь = кількості незалежних контурів, в ході розв’язання яких визначаємо невідомі струми.

1 рівняння складаємо за 1 простим контуром, де напрям обходу співпадає з напрямом ЕРС.

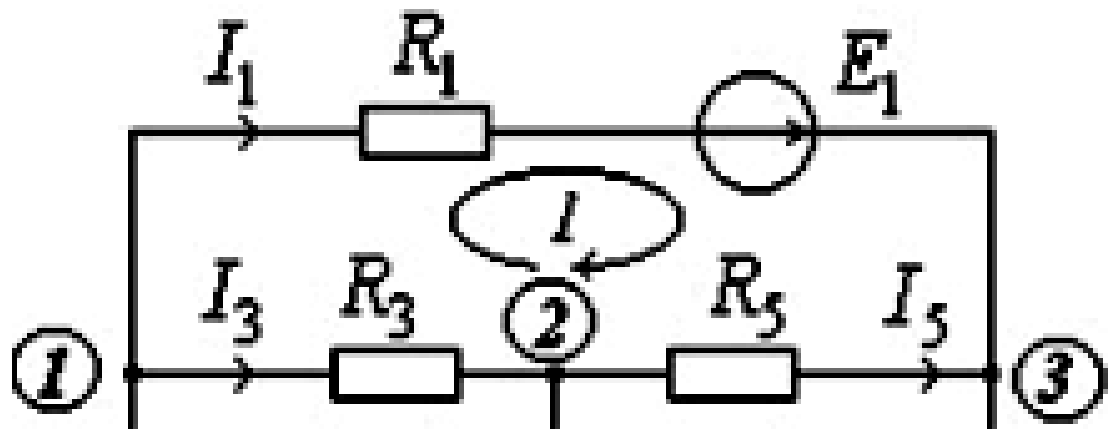

$$I_1R_1 - I_5R_5 - I_3R_3 = E_1$$

2 рівняння складаємо за наступним простим контуром, в якому знаходиться вже два ЕРС джерела.

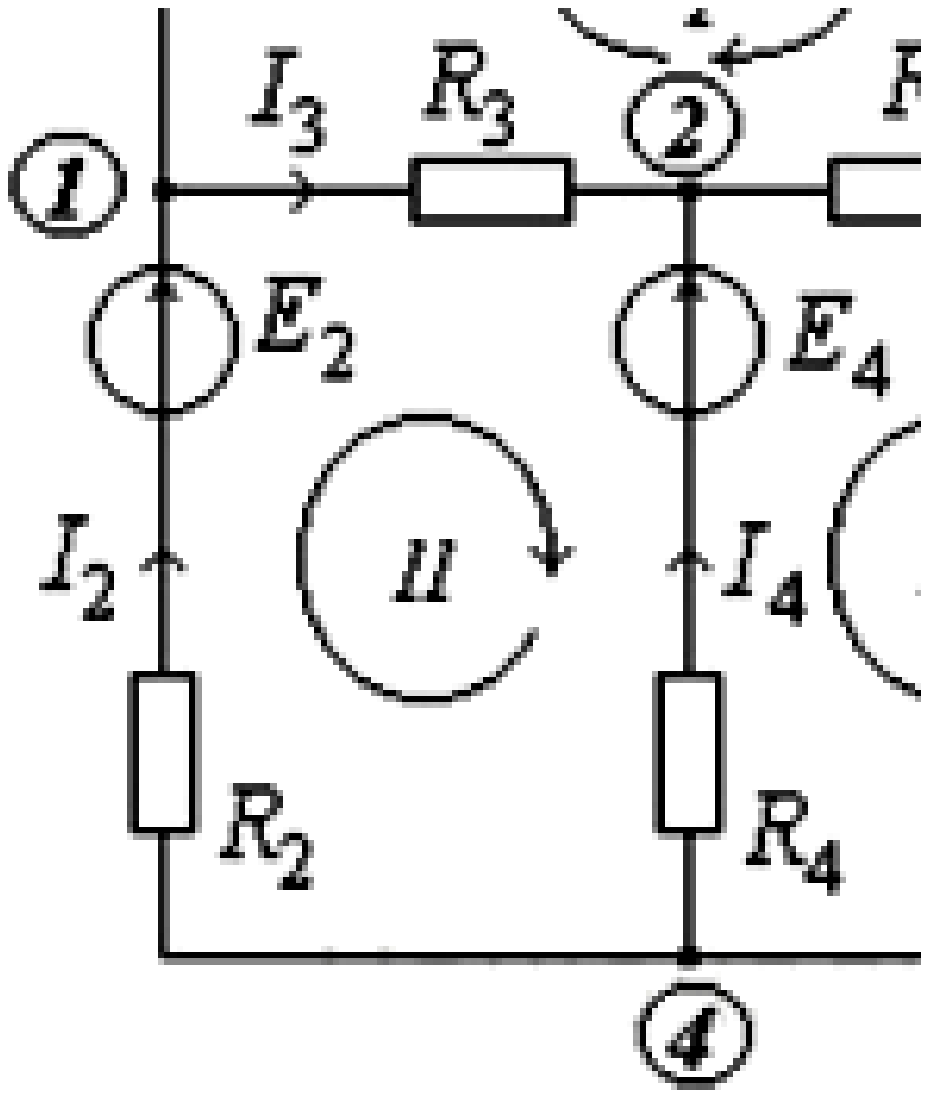

$$I_2R_2 + I_3R_3 - I_4R_4 = E_2 - E_4$$

3 рівняння складаємо за останнім простим контуром також з двома ЕРС джерелами.

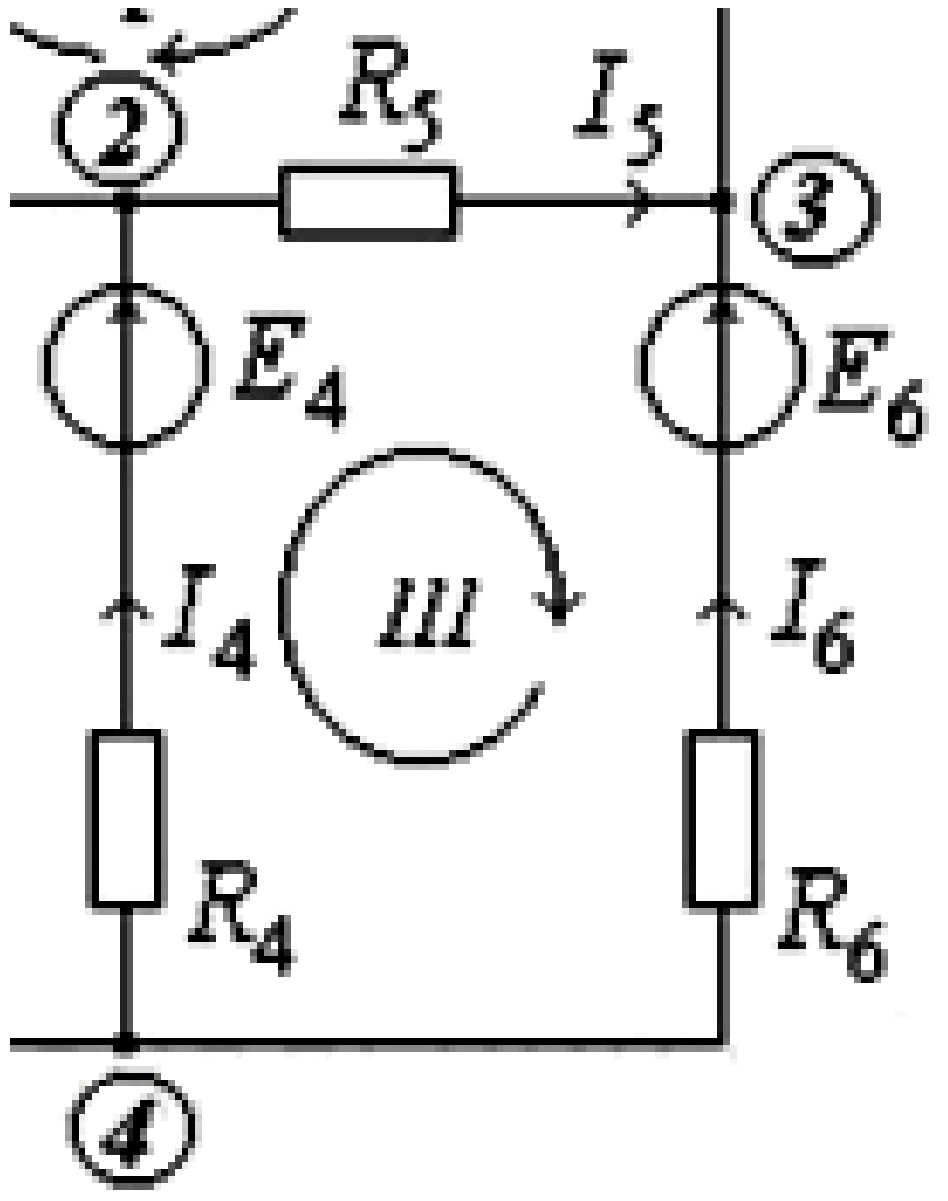

$$I_4R_4 + I_5R_5 - I_6R_6 = E_4 - E_6$$

Таким абоном отримали систему з 6 рівнянь, щоб знайти 6 невідомих струмів.

**Відповідь:** .$I_2 - I_1 - I_3 = 0; I_3 + I_4 - I_5 = 0; I_1 + I_5 + I_6 = 0;$
$I_1R_1 - I_5R_5 - I_3R_3 = E_1;\ I_2R_2 + I_3R_3 - I_4R_4 = E_2 - E_4;$
$I_4R_4 + I_5R_5 - I_6R_6 = E_4 - E_6$

**Задача 4.20.** Складіть рівняння для розрахунку резистивного кола методом рівнянь Кірхгофа при відомих значеннях опорів $R_1 - R_{10}$.

**Дано:**
$R_1$-$R_{10}$

**Знайти:**
$I_1 \dots I_{10}$
$U_1 \dots U_2$

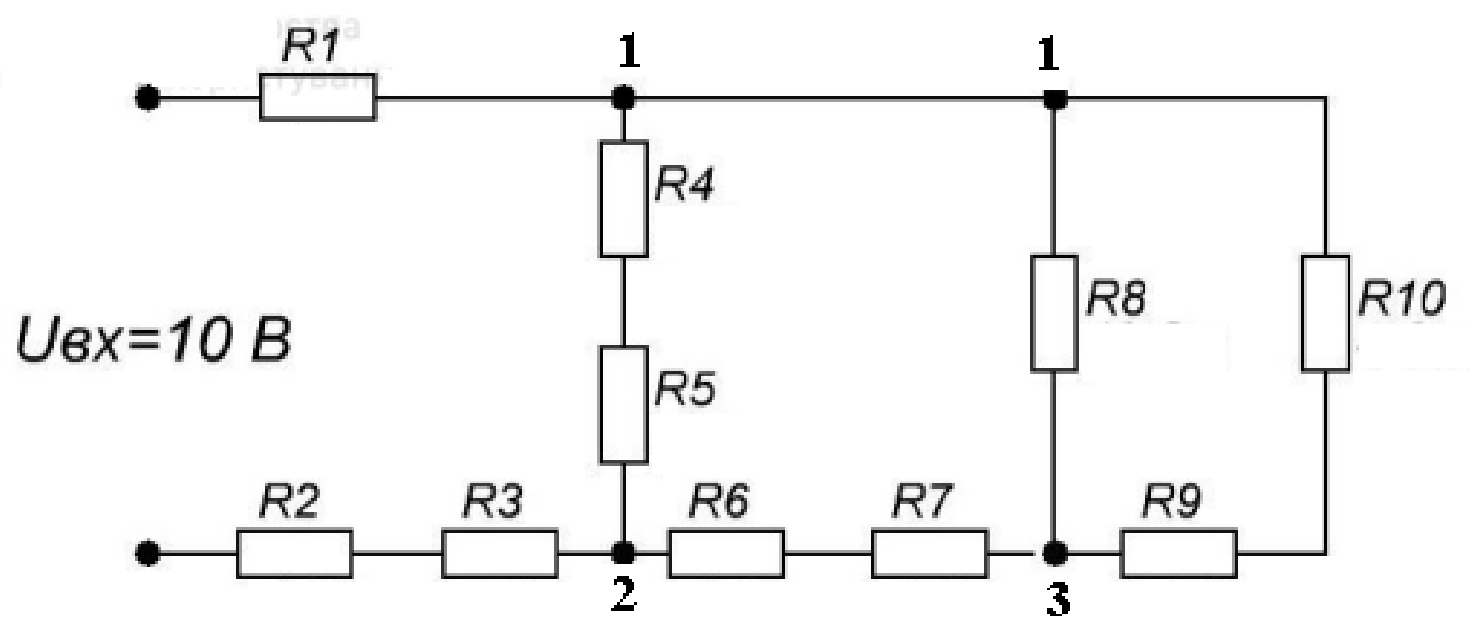

**Розв'язання.**

Маємо однорідну ділянку кола (ділянку кола називають однорідною, якщо на заряди діють тільки сили електричного походження і тоді напруга буде співпадати з різницею потенціалів на її кінцях:

$$U_{12} = \varphi_1 - \varphi_2$$

Для інформації - ділянку кола називають неоднорідною, якщо на заряди крім сил електричного походження також діють сторонні сили (джерела електричної енергії) і напруга на неоднорідній ділянці буде визначатися за формулою

$$\mathrm{U}_{12} = \varphi_1 - \varphi_2 + \mathcal{E}_{12}.$$

Почнемо з того, що поставимо на схемі напрямки струму та визнаабомо вузли. Другим шагом запишемо рівняння для струмів за першим правилом Кірхгофа.

На відміну від попередніх задач, коли вхідний струм:

$I_0 = I_1$, знайшовши $R_е$, підставляли у формулу $I_1 = \frac{U}{R_е}$, зараз треба тільки скласти рівняння, тому шукати $R_е$ не будемо.

Маємо три контури та три вузла.

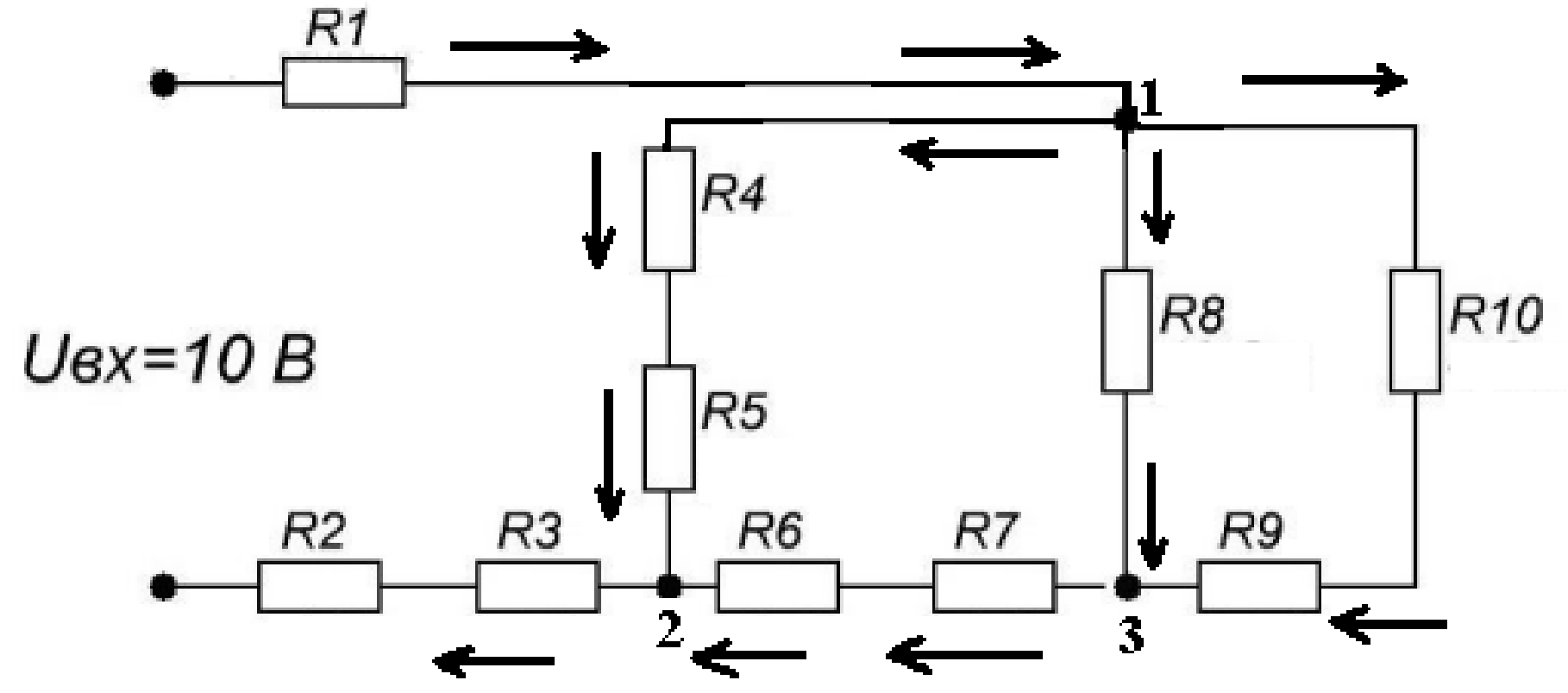

Струм 2 резистора є такими, що втікає у джерело вхідного сигналу, тому $I_1 = I_2$, але $R_2$ та $R_3$ з'єднані послідовно, тому $I_2 = I_3$ , тобто маємо:

$$I_1 = I_2 = I_3$$

Складемо рівняння за першим правилом Кірхгофа:

Для вузла 1 маємо:

$$I_1 - I_4 - I_8 - I_{10} = 0 \quad \text{тобто} \quad I_1 = I_4 + I_8 + I_{10}$$

Для вузла 2 маємо:

$$I_5 + I_6 - I_3 = 0 \quad \text{тобто} \quad I_3 = I_5 + I_6$$

Для вузла 3 маємо:

$$I_8 + I_9 - I_7 = 0 \quad \text{тобто} \quad I_7 = I_8 + I_9$$

Далі за другим законом Кірхгофа складаємо рівняння для трьох контурів:

Для 1 контуру: з'єднання послідовне, тобто

$$U = \sum_{i=1}^{n} U_i\,; \quad I = I_1 = I_2 = \cdots = I_n$$

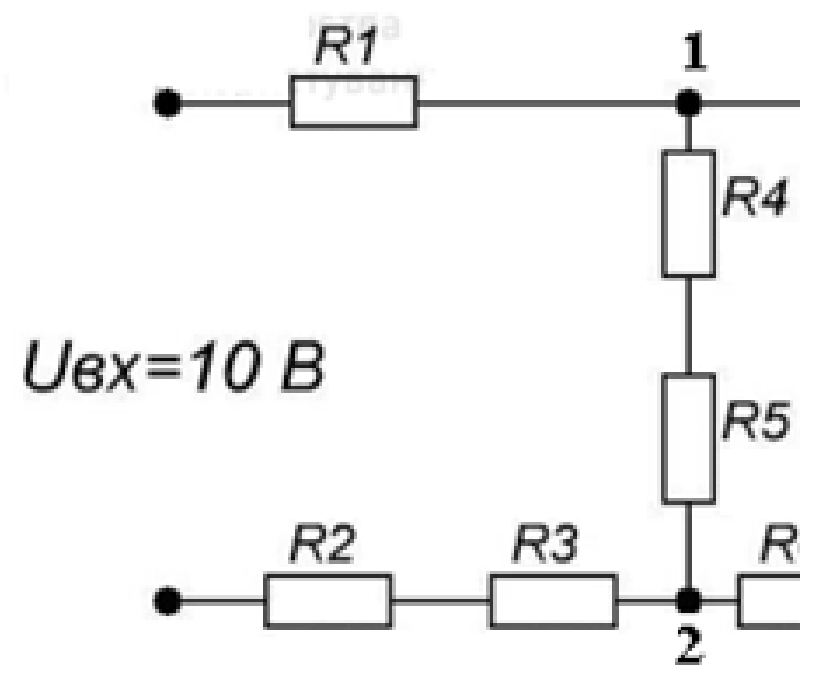

маємо:

$$U_{\text{вх}} = U_1 + U_4 + U_5 + U_3 + U_2$$

Для 2 контуру - напруга між вузлами 1 та 2.

Маємо дві паралельні вітки між вузлами 1 і 2 – до складу першої вітки входять резистори $R_4$ і $R_5$, до складу другої – резистори $R_8$ , $R_7$,$R_6$.

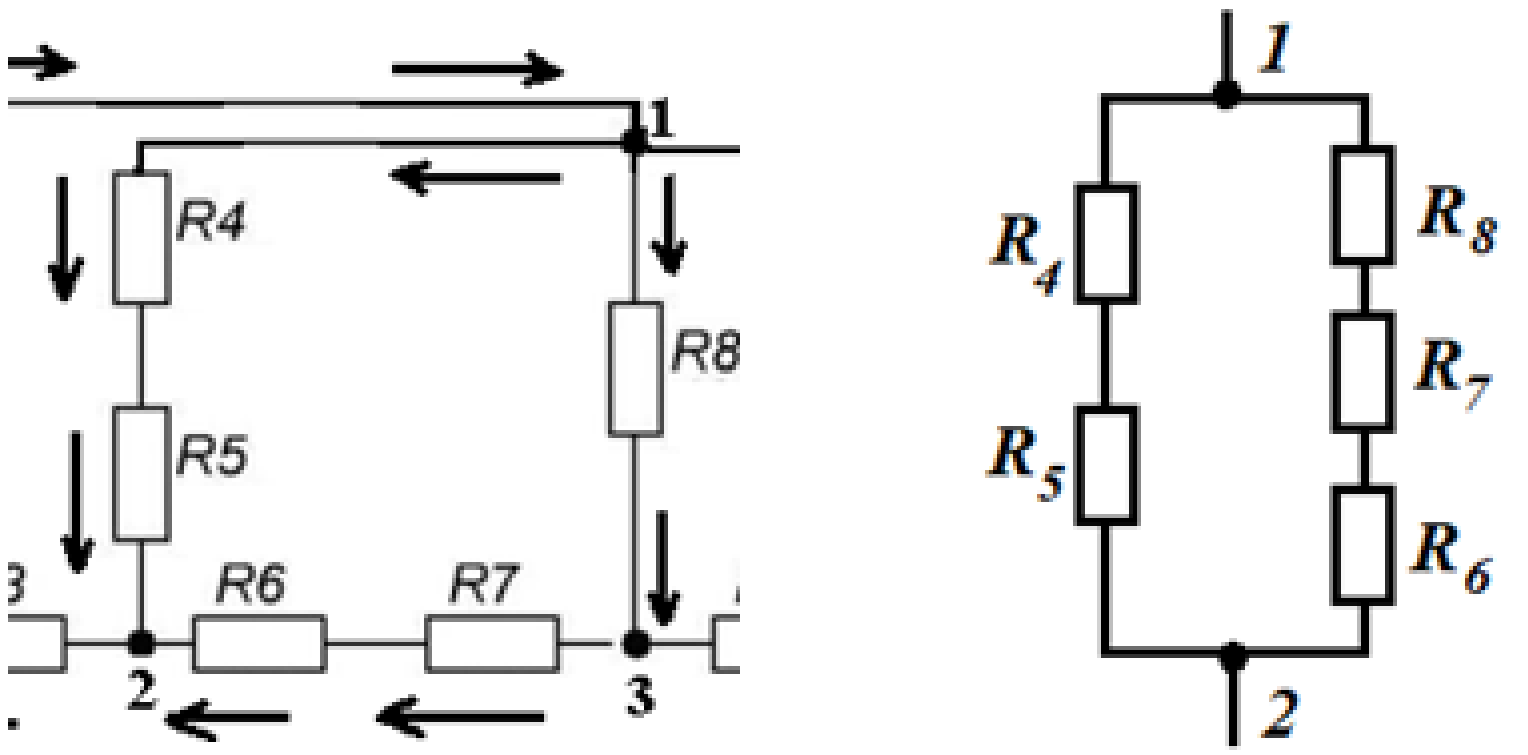

Тобто маємо послідовно з'єднані опори $R_4$ та $R_5$ , які є паралельно з'єднані з сумою послідовно з'єднаних опорів $R_6$, $R_7$ та $R_8$, тому запишемо, пам'ятаюабо, що при паралельному з'єднанні $U = U_1 = U_2 = \cdots = U_n$ тобто:

$$U_4 + U_5 = U_6 + U_7 + U_8$$

Для 3 контуру - напруга між вузлами 1 та 3, тому:

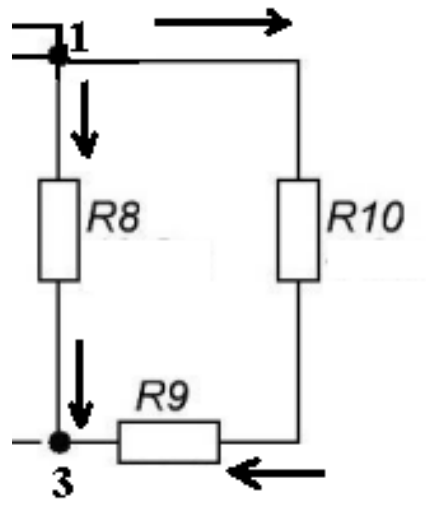

$$U_8 = U_9 + U_{10}$$

За законом Ома напруга є добуток струму та опору, тому ці рівняння можна переписати:

$$I_1R_1 + I_2R_2 + I_3R_3 + I_4R_4 + I_5R_5 = U_{вх}$$

$$I_4R_4 + I_5R_5 = I_6R_6 + I_7R_7 + I_8R_8$$

$$I_8R_8 = I_9R_9 + I_{10}R_{10}$$

Додамо ще рівняння для струмів, що отримали раніше і отримаємо наступну систему рівнянь:

$$\begin{cases} I_1 = I_4 + I_8 + I_{10} \\ I_7 = I_8 + I_9 \\ I_1R_1 + I_2R_2 + I_3R_3 + I_4R_4 + I_5R_5 = U_{вх} \\ I_4R_4 + I_5R_5 = I_6R_6 + I_7R_7 + I_8R_8 \\ I_8R_8 = I_9R_9 + I_{10}R_{10} \end{cases}$$

Маюаби на увазі, що через послідовно з'єднані елементи електричного кола протікають однакові струми, тобто $I_1 = I_2 = I_3$; $I_4 = I_5$; $I_6 = I_7$; $I_9 = I_{10}$ можна переписати систему рівнянь:

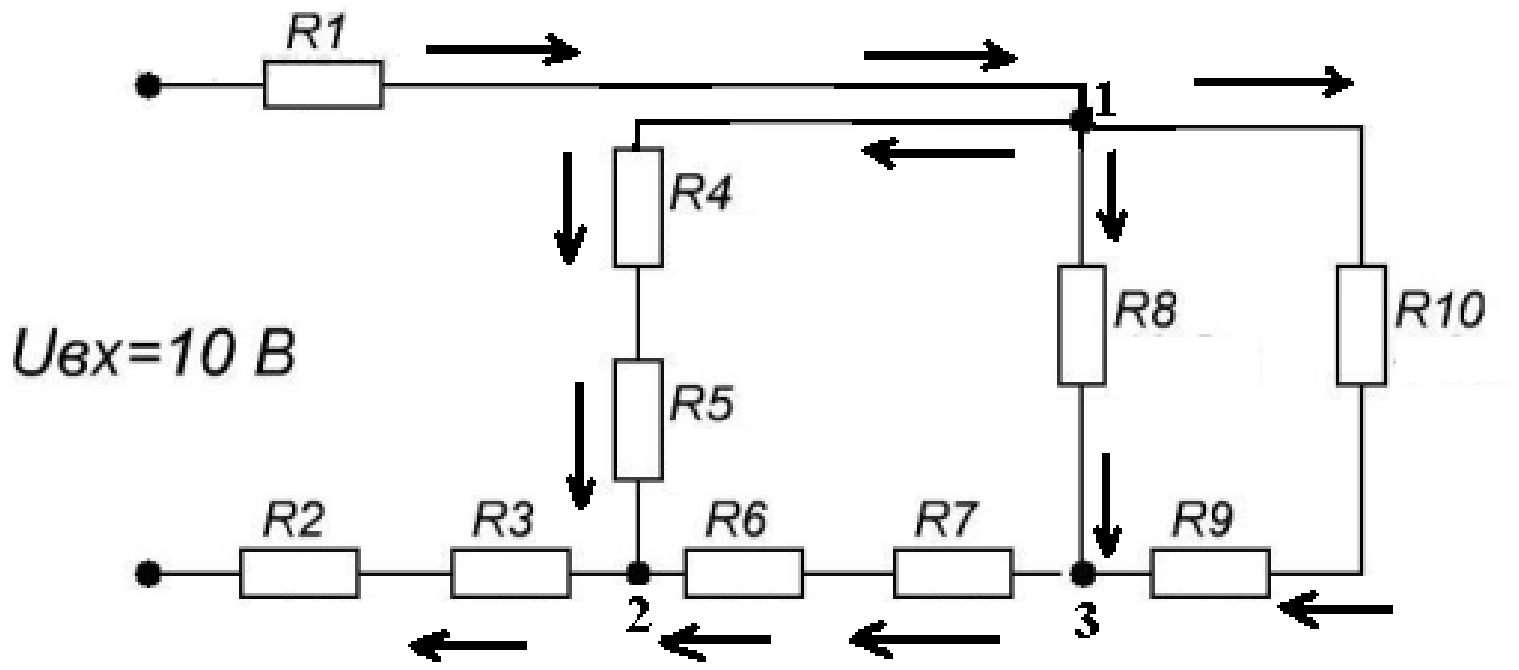

$$I_1 = I_4 + I_8 + I_{10}$$

$$I_7 = I_8 + I_9$$

$$I_1(R_1 + R_2 + R_3) + I_4(R_4 + R_5) = U_{вх}$$

$$I_4(R_4 + R_5) = I_6(R_6 + R_7) + I_8R_8$$

$$I_8R_8 = I_9(R_9 + R_{10})$$

За умовою , що в задачі надані значення опорів та вхідна напруга, то підставивши абослові значення, система розв'язується – знаходяться всі струми. Після цього знаходяться напруги на всіх елементах схеми за формулами:

$$U_1 = I_1 R_1$$

$$U_2 = I_2 R_2$$

………………..

$$U_{10} = I_{10} R_{10}$$

## 4.4. Способи з'єднання елементів електричного кола зіркою та трикутником.

Сполучення трьох резисторів, які утворюють сторони трикутника, називають «трикутником» - рис. 4.5.б). «Зірка» – це з'єднання трьох резисторів, яке має вигляд трипроменевої зірки - рис. 4.5.а). Приведені схеми будуть еквівалентними якщо при однакових відповідних напругах $U_{12}, U_{23}, U_{31}$ вхідні струми $I_1, I_2, I_3$ в обох схемах також будуть відповідно однаковими.

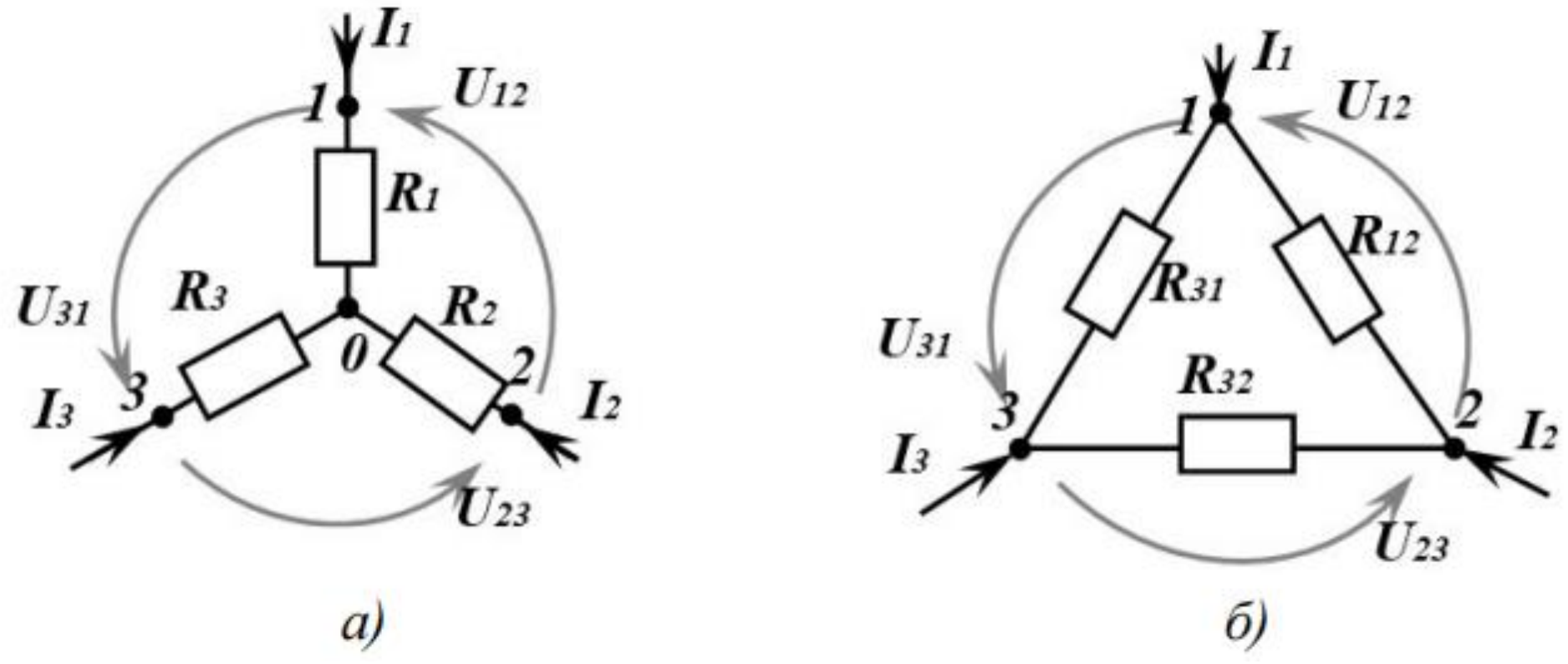

Рис. 4.5. Способи з'єднання елементів електричного кола:
а) зіркою; б) трикутником.

Правило еквівалентного перетворення для опорів трикутника:

«Опір сторони трикутника дорівнює сумі двох відповідних примикаюабох опорів зірки та їх добутку, поділеному на третій опір зірки.

$$R_{12} = R_1 + R_2 + \frac{R_1 \cdot R_2}{R_3}$$

$$R_{23} = R_2 + R_3 + \frac{R_2 \cdot R_3}{R_1}$$

$$R_{31} = R_3 + R_1 + \frac{R_3 \cdot R_1}{R_2}$$

Правило еквівалентного перетворення для опорів зірки:

«Опір променю зірки дорівнює добутку двох відповідних примикаюабох опорів трикутника, поділеному на суму опорів усіх сторін трикутника».

$$R_1 = \frac{R_{12} \cdot R_{31}}{\Delta R}; \qquad R_2 = \frac{R_{23} \cdot R_{12}}{\Delta R};$$

$$R_3 = \frac{R_{13} \cdot R_{32}}{\Delta R};$$

$$\Delta R = R_{12} + R_{23} + R_{31}$$

**Задача 4.21.** Напишить рівняння для знаходження загального опору між точками А і В ланцюга з провідників, вказаних на рис.

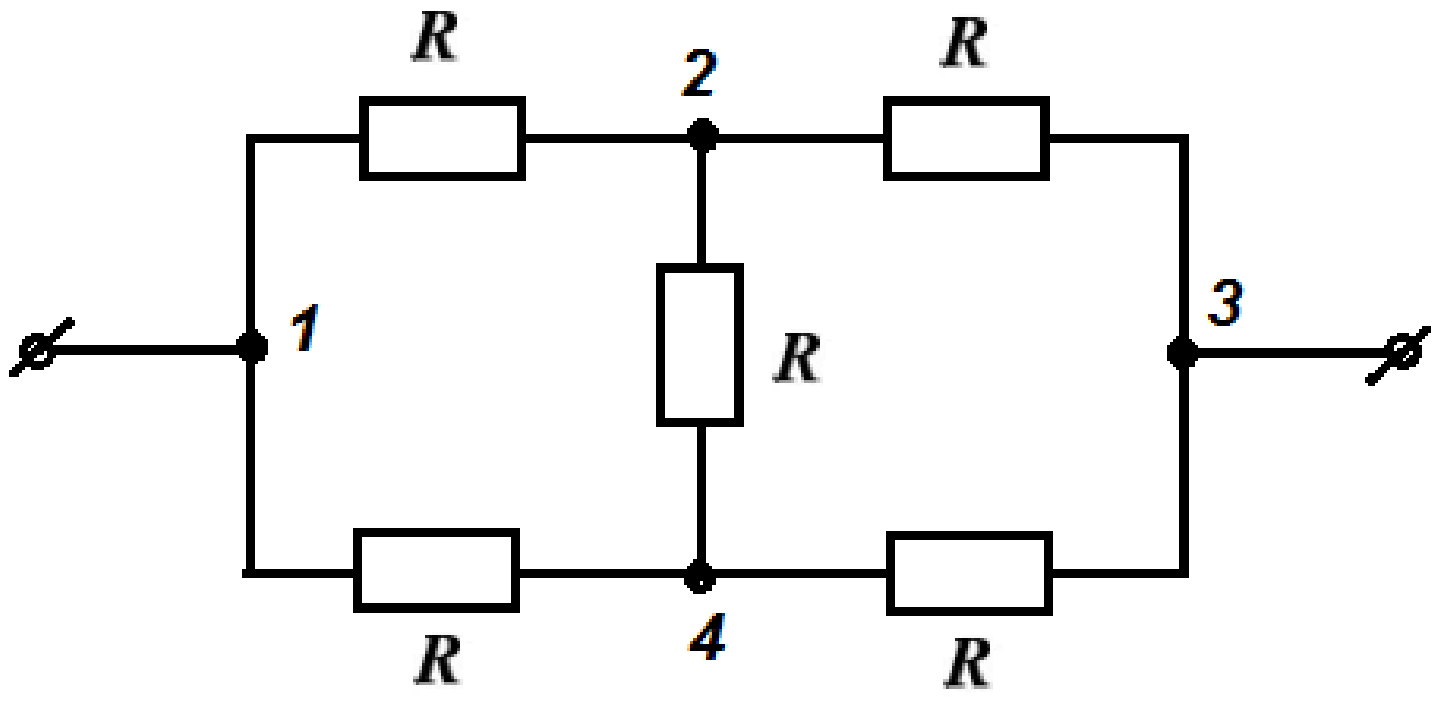

**Розв'язання.**

Еквівалентний опір кола зі змішаним з'єднанням резисторів відносно вхідних затискачів розраховується поетапно – схема нібито згортається до одного еквівалентного елемента. Розпоабонати згортання слід з еквівалентних перетворень елементів, увімкнених якнайдалі від вхідних затискачів.

В цієї задачі вхідні затискачі розташовані так, що схему можна спростити використовуюабо правилю зірки. Це правило можна застосувати до вузлів 2 або 4, не має значення. Розглянемо відносно 2 вузла

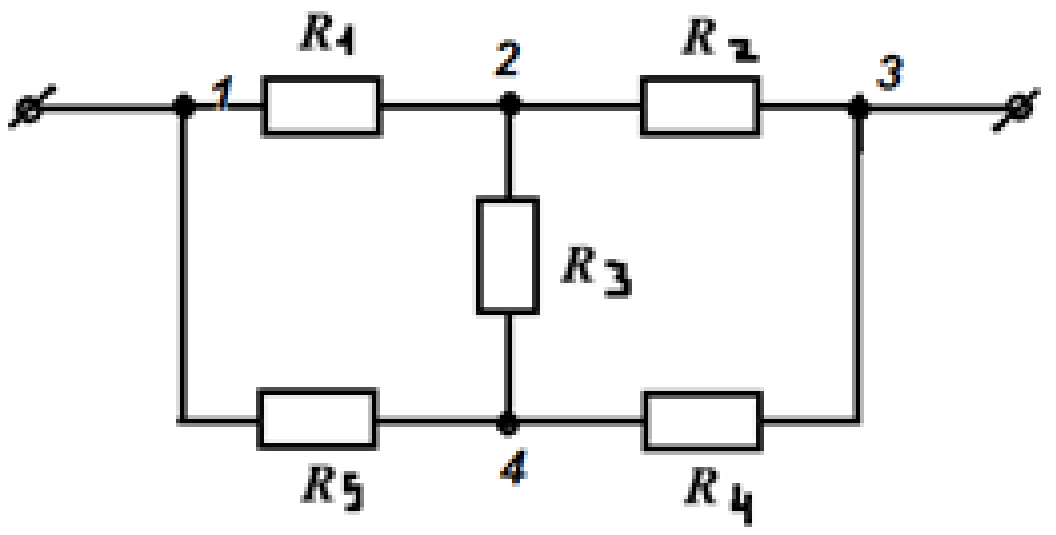

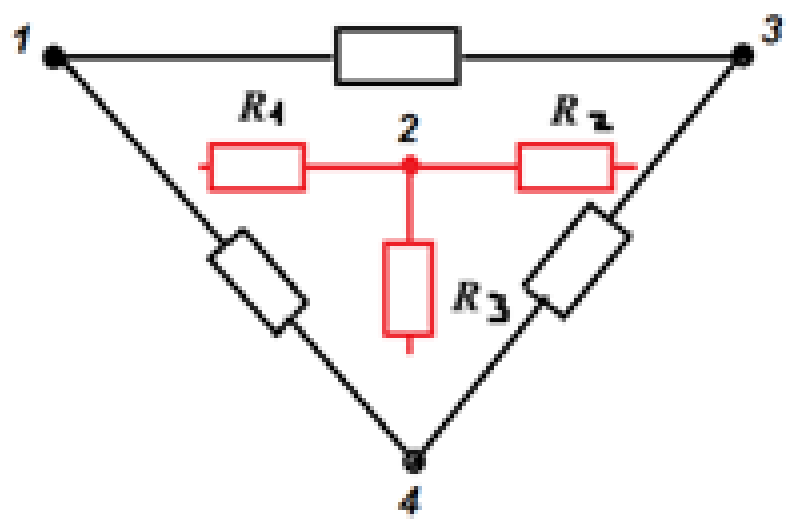

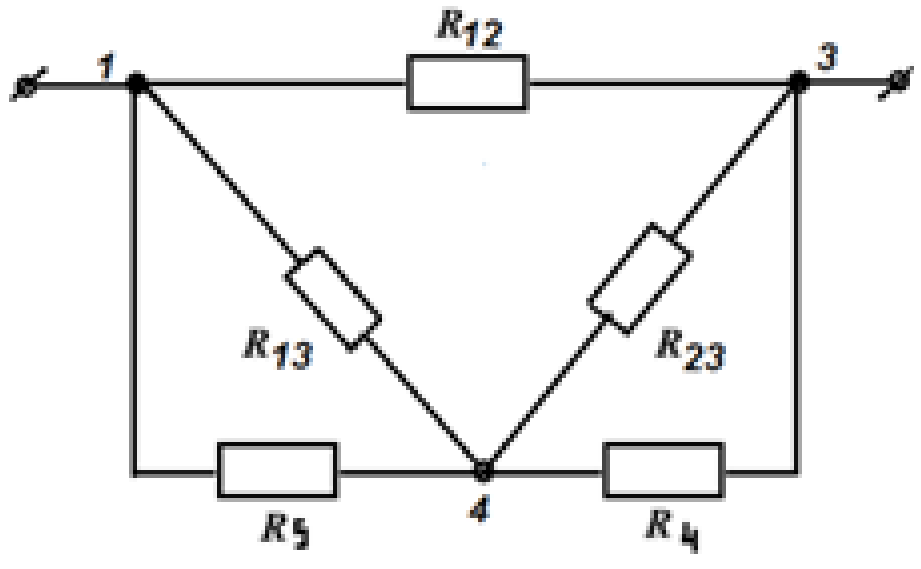

За формулами знайдемо:

$$R_{12} = R_1 + R_2 + \frac{R_1 \cdot R_2}{R_3}; \; R_{23} = R_2 + R_3 + \frac{R_2 \cdot R_3}{R_1}; \quad R_{31} = R_3 + R_1 + \frac{R_3 \cdot R_1}{R_2}$$

Отриману схему після перетворення зірки у трикутник, перерисуємо для більшої наочності наступним абоном:

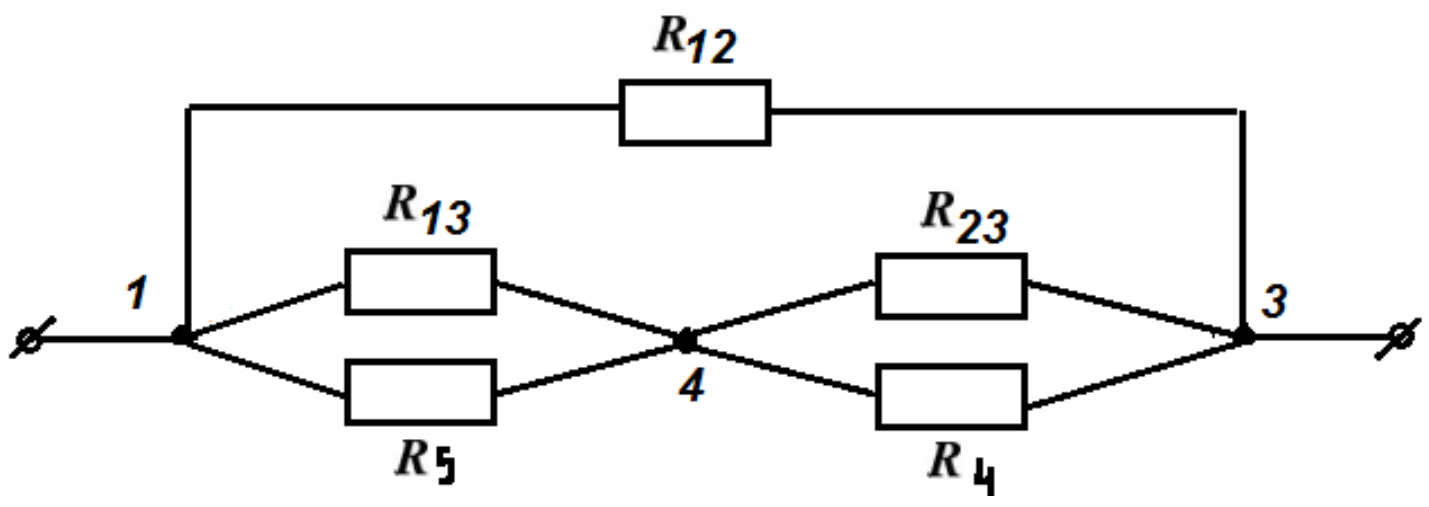

Щоб знайти загальний опір спочатку спростимо схему, знайшовши опір паралельно з'єднаних пар $R_{13}$ та $R_5$ і $R_{23}$ та $R_4$

$$\frac{1}{R_{13-5}} = \frac{1}{R_{13}} + \frac{1}{R_5} \quad \text{та} \quad \frac{1}{R_{23-4}} = \frac{1}{R_{23}} + \frac{1}{R_4}$$

$$R_{13-5} = \frac{R_{13} \cdot R_5}{R_{13} + R_5} \quad \text{та} \quad R_{23-4} = \frac{R_{23} \cdot R_4}{R_{23} + R_4}$$

Після спрощення схем має вигляд:

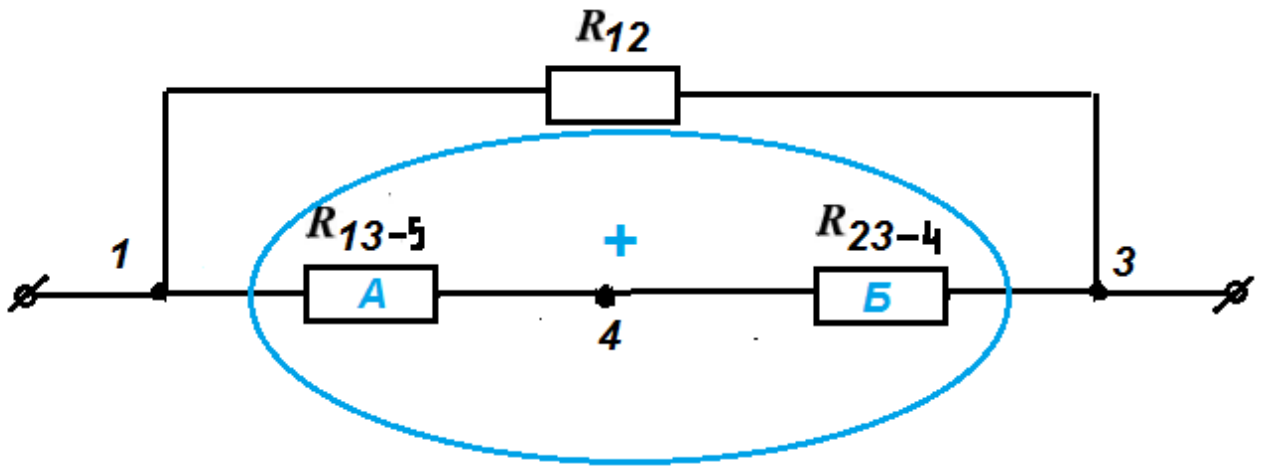

Тепер загальний опір буде складатися із паралельно з’єднаних опорів $R_{12}$ та суми опорів, які ми познааболи через А та Б:

$$\frac{1}{R_{заг}} = \frac{1}{R_{12}} + \frac{1}{R_А + R_Б}; \qquad R_{заг} = \frac{R_{12}(R_А + R_Б)}{R_{12} + R_А + R_Б}$$

**Відповідь:** $R_{заг} = \frac{R_{12}(R_А+R_Б)}{R_{12}+R_А+R_Б}$.

**Задача 4.22.** Визначити еквівалентний опір схеми відносно точок А і В, якщо:

**Дано:**

$R_1$=20 Ом

$R_2$=20 Ом

$R_3$=10 Ом

$R_4$=20 Ом

$R_6$=4 Ом

$R_7$=4 Ом

**Знайти:**

$R_{екв}$ -?

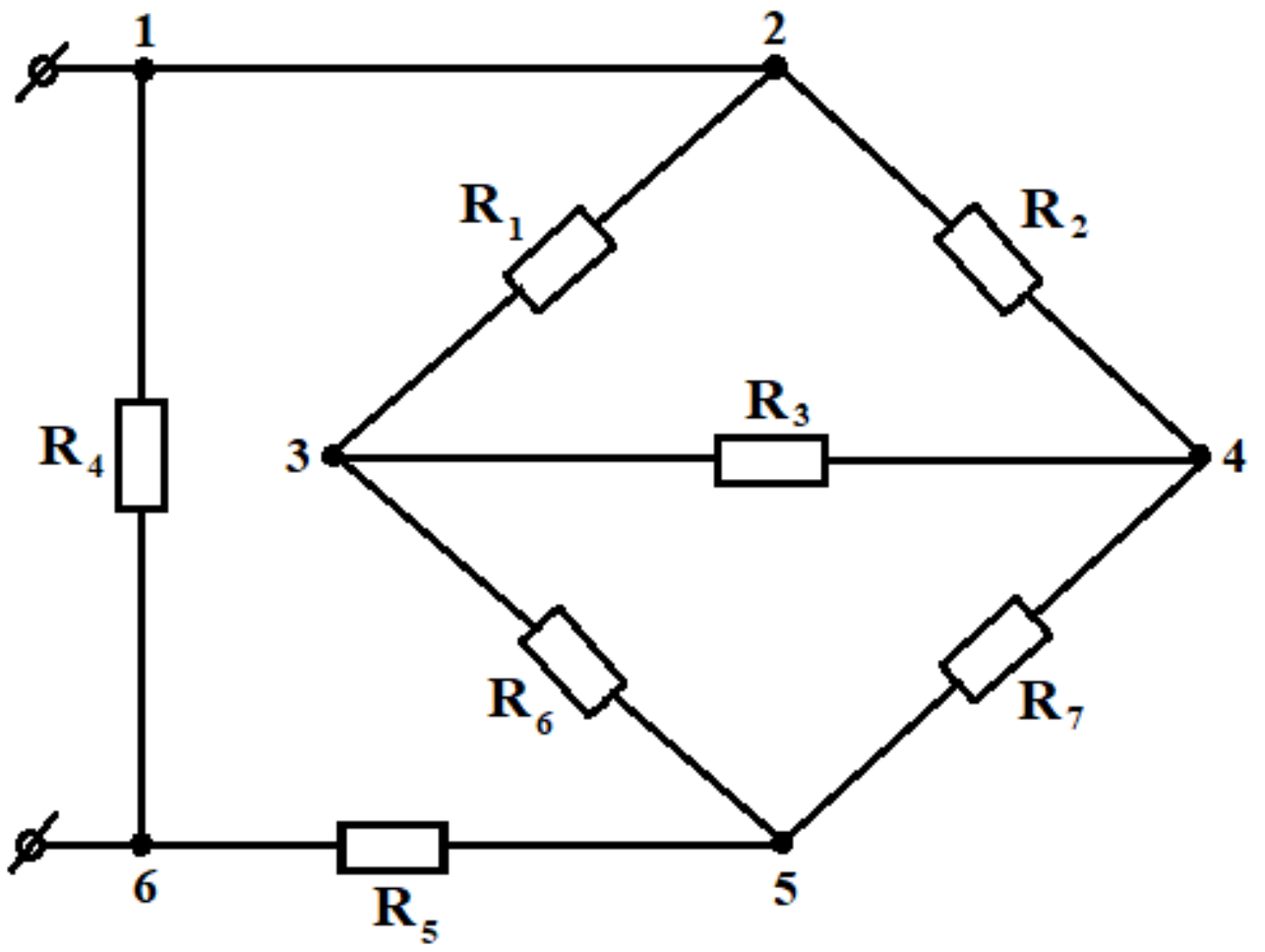

**Розв’язання.**

Еквівалентний опір кола з мішаним з’єднанням резисторів відносно вхідних затискачів в такому вигляді без попереднього перетворення неможливо розрахувати. Тому, до поетапного спрощення, спочатку застосуємо правило перетворення трикутника на зірку. Це правило можна застосувати до вузлів 2 або 5, не має значення. Розглянемо відносно 2 вузла.

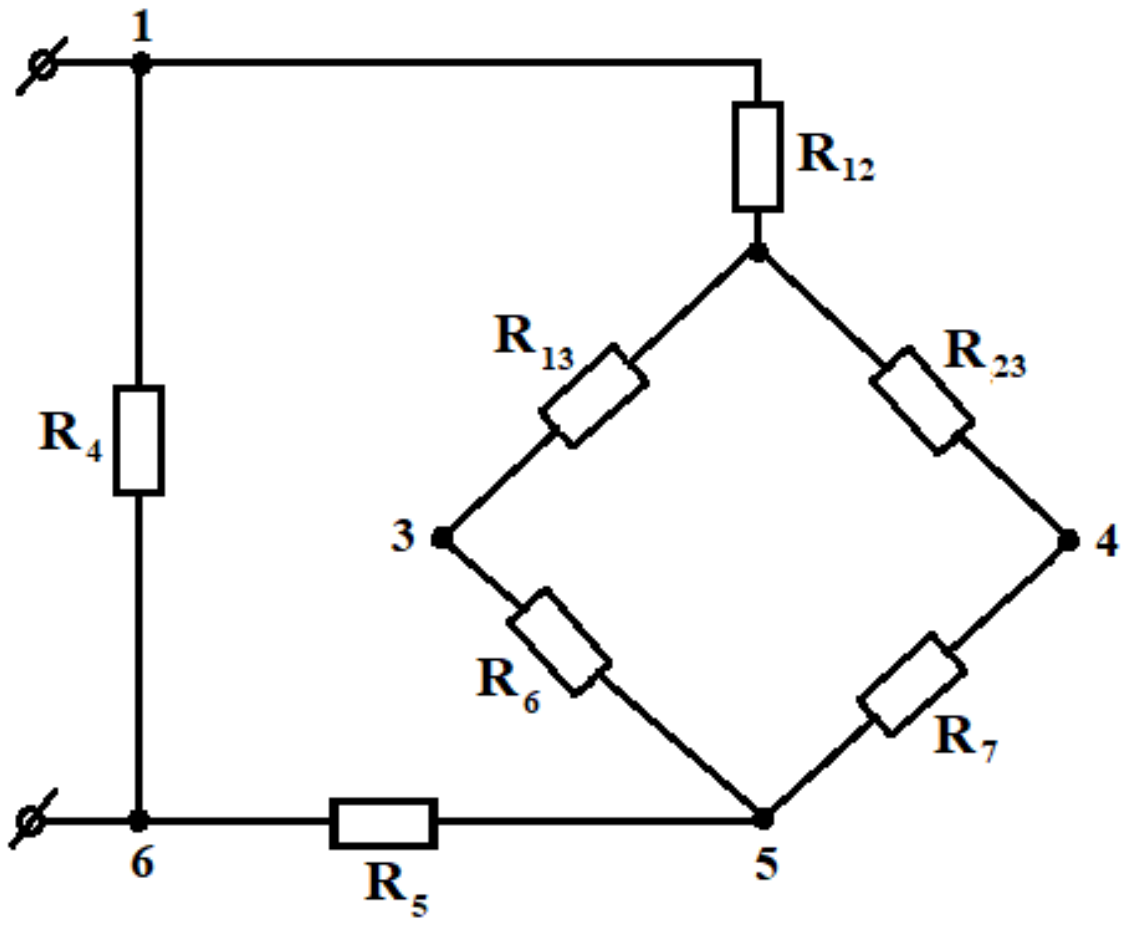

Після перетворення отримали резистори: $R_{12}$; $R_{13}$: $R_{23}$. Розрахуемо їх за формулами:

$$R_{12} = \frac{R_1 \cdot R_2}{R_1 + R_2 + R_3} = \frac{20 \cdot 20}{50} = \frac{40}{5} = 8 \text{ Ом}$$

$$R_{13} = \frac{R_1 \cdot R_3}{R_1 + R_2 + R_3} = \frac{20 \cdot 10}{50} = \frac{20}{5} = 4 \text{ Ом}$$

$$R_{23} = \frac{R_2 \cdot R_3}{R_1 + R_2 + R_3} = \frac{20 \cdot 10}{50} = \frac{20}{5} = 4 \text{ Ом}$$

Після перетворення можна зробити спрощення:

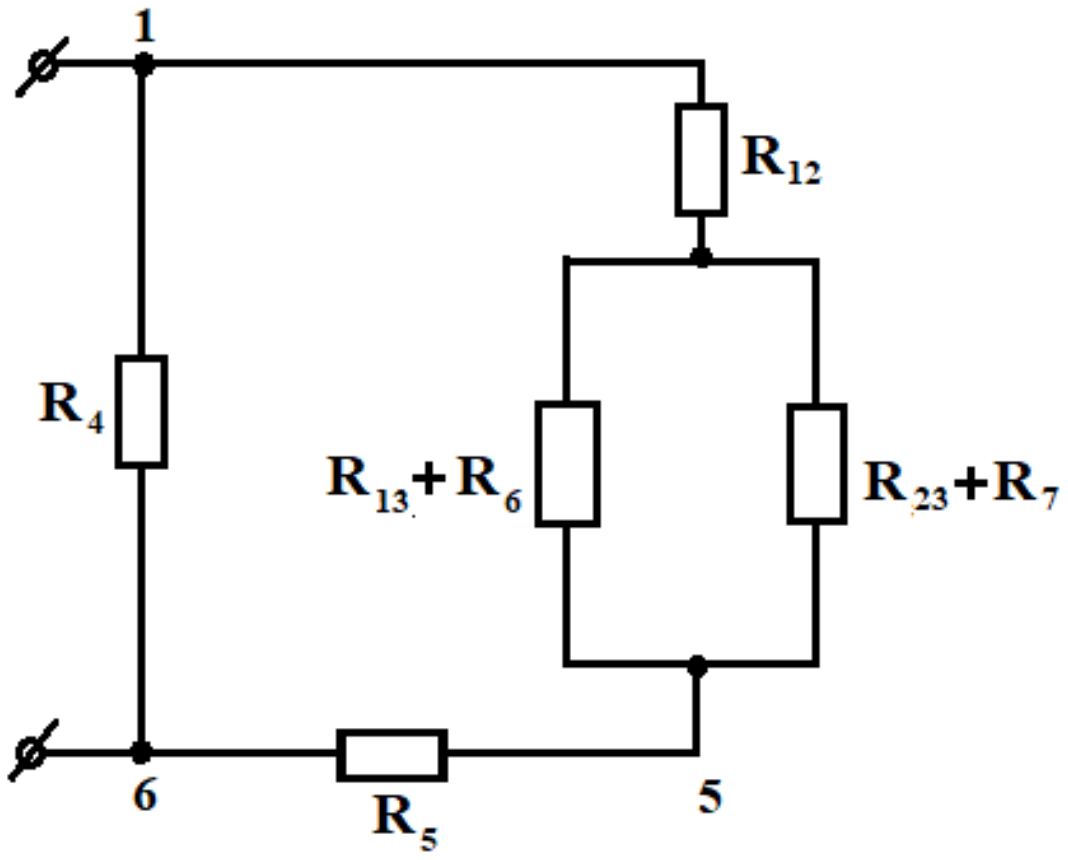

Розрахуемо: $R_{13} + R_6 = 4 + 4 = 8$ Ом; $R_{23} + R_7 = 4 + 4 = 8$ Ом

Знайдені опори з'єднані паралельно, познаабомо результат складання, як $R_\Sigma$

$$\frac{1}{R_\Sigma} = \frac{1}{8} + \frac{1}{8} = \frac{2}{8} \implies R_\Sigma = \frac{8}{2} = 4 \text{ Ом}$$

Продовжуємо спрощення схеми:

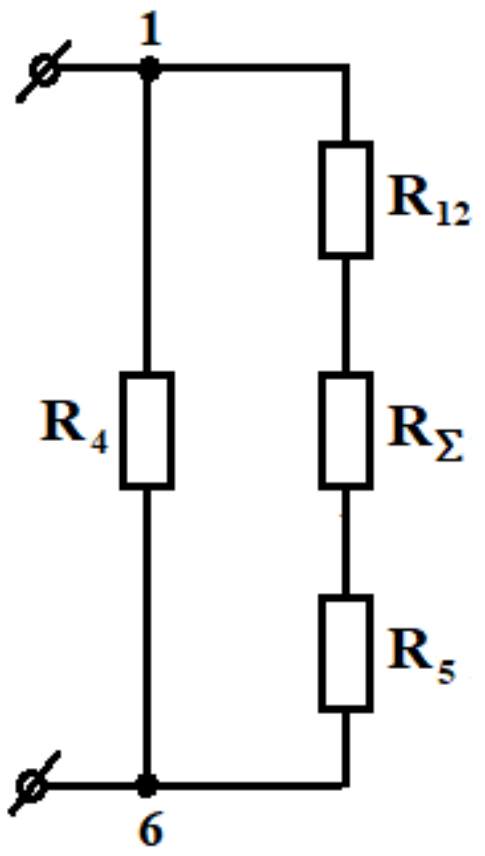

$$R_{\text{посл}} = R_{12} + R_{\Sigma} + R_5 = 8 + 4 + 8 = 20\text{ Ом}$$

Тепер можемо знайти $R_{\text{екв}}$ як результат паралельно з'єднаних резисторів $R_{\text{посл}}$ і $R_4$ :

$$\frac{1}{R_{\text{екв}}} = \frac{1}{R_{\text{посл}}} + \frac{1}{R_4} = \frac{1}{20} + \frac{1}{20} = \frac{2}{20} \quad \Rightarrow \quad R_{\text{екв}} = \frac{20}{2} = 10\text{ Ом}$$

**Відповідь:** $R_{\text{екв}} = 10$ Ом.

## Задачі для самостійного розв'язання

1. Сила струму в провіднику рівномірно зростає від $I_0$=0 до $I$=2 А протягом часу $t$=5 с. Визначити заряд, який пройшов по провіднику. [$q$=5 Кл].

2. Визначити густину стуму $j$, якщо за $t$=2 с через провідник перерізом $S$=1,6 мм$^2$ пройшло $N=2\cdot10^{19}$ електронів. [ $j=10^6$ А/м$^2$]

3. Два провідники виготовлені з одного матеріалу. У скільки разів довжина одного провідника більша від довжини іншого, коли відомо, що $R_1=8R_2$, а $m_1=2m_2$? [$l_1/l_2$=4].

4. Визначити опір ділянки кола між точками $A$ і $B$ (див.рис.), якщо опори $R_1=R_3=R_5=R_7=R_8=R_9$=1 Ом, $R_2=R_4=R_6$=20 Ом. [$R$=10,9 Ом].

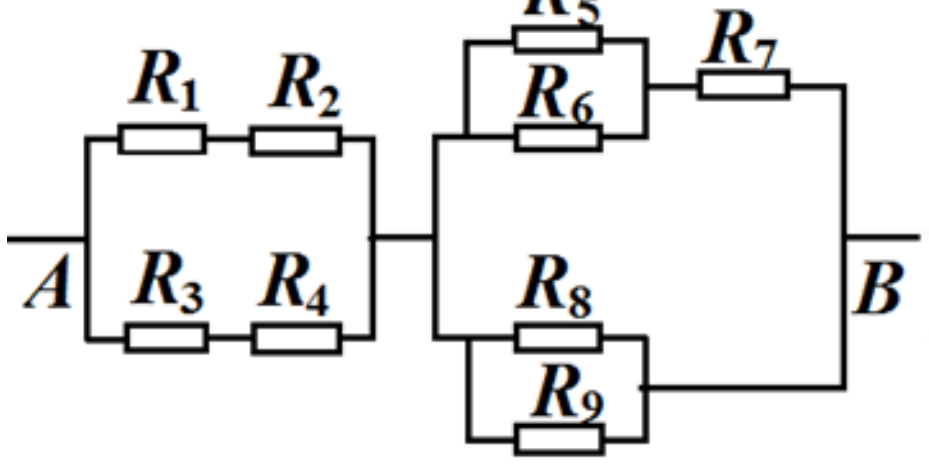

5. Сила струму в провіднику, опір якого $R$=100 Ом, рівномірно збільшується від нуля до $I_{max}$=10 А протягом часу $t_1$=30 С. Визначити кількість теплоти $Q$, яка виділяється за цей час у провіднику. [$Q$=100 кДж].

6. У схемі електрорушійна сила ε=5 В, опори $R_1$=4 Ом, $R_2$=6 Ом. Внутрішній опір джерела струму $r$=0,1 Ом. Визначити сили струмів, що проходять крізь резистори $R_1$ і $R_2$ . [$I_1$=1,2 А; $I_2$=0,8 А].

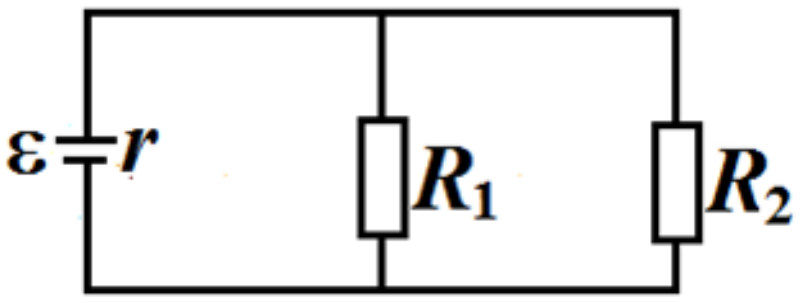

7. Напруга на шинах електростанції $U$=10 кВ, відстань від споживача $l$=500 км. Станція передає споживачеві потужність $P$=100 кВт, причому втрата напруги не повинна перевищувати $\eta$=5 %. Обчислити: силу струму $I$ в дротах; площу $S$ їх поперечного перерізу; масу $m_1$ міді, яка потрібна для проведення проводки. Як зміниться необхідна кількість $m_2$ міді, якщо напругу на вході лінії збільшити у $n$=10 разів? [$I$=10,5 А; $S$=3,58 см$^2$; $m_1$=4,19·10$^6$ кг; $m_2$=$m_1$/100].

8. Встановити розподіл струмів у колі, якщо електрорушійні сили $\varepsilon_1$=12 В, $\varepsilon_2$=8 В, опори $r_1$=4 Ом, $r_2$=3 Ом, $R_1$=20 Ом, $R_2$=40 Ом, $R_3$=29 Ом, $R_4$=8 Ом, $R_5$=16 Ом. [$I_1$=0,4 А; $I_2$=0,1 А; $I_3$=0,2 А; $I_4$=0,3 А; $I_5$=0,1 А].

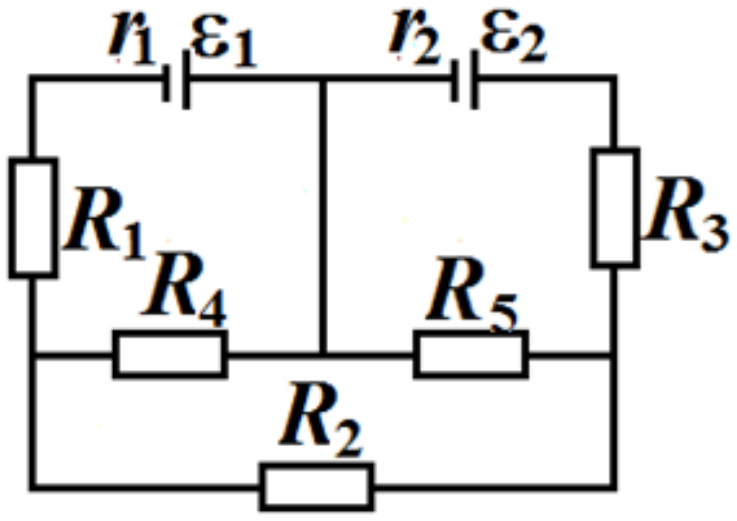

## Розділ 5. ФОТОМЕТРІЯ. ОСНОВНІ ХАРАКТЕРИСТИКИ.

**Світловий потік** (Ф) – характеристика випромінювальної здатності джерела:

$$\Phi_V = I\Omega,$$

де $I$ – сила світла джерела; Ω – тілесний кут.

Потік випромінювання (енергії)

$$\Phi_e = \frac{W - \text{кількість енергії випромінювання}}{t - \text{час за який відбулося випромінювання}}$$

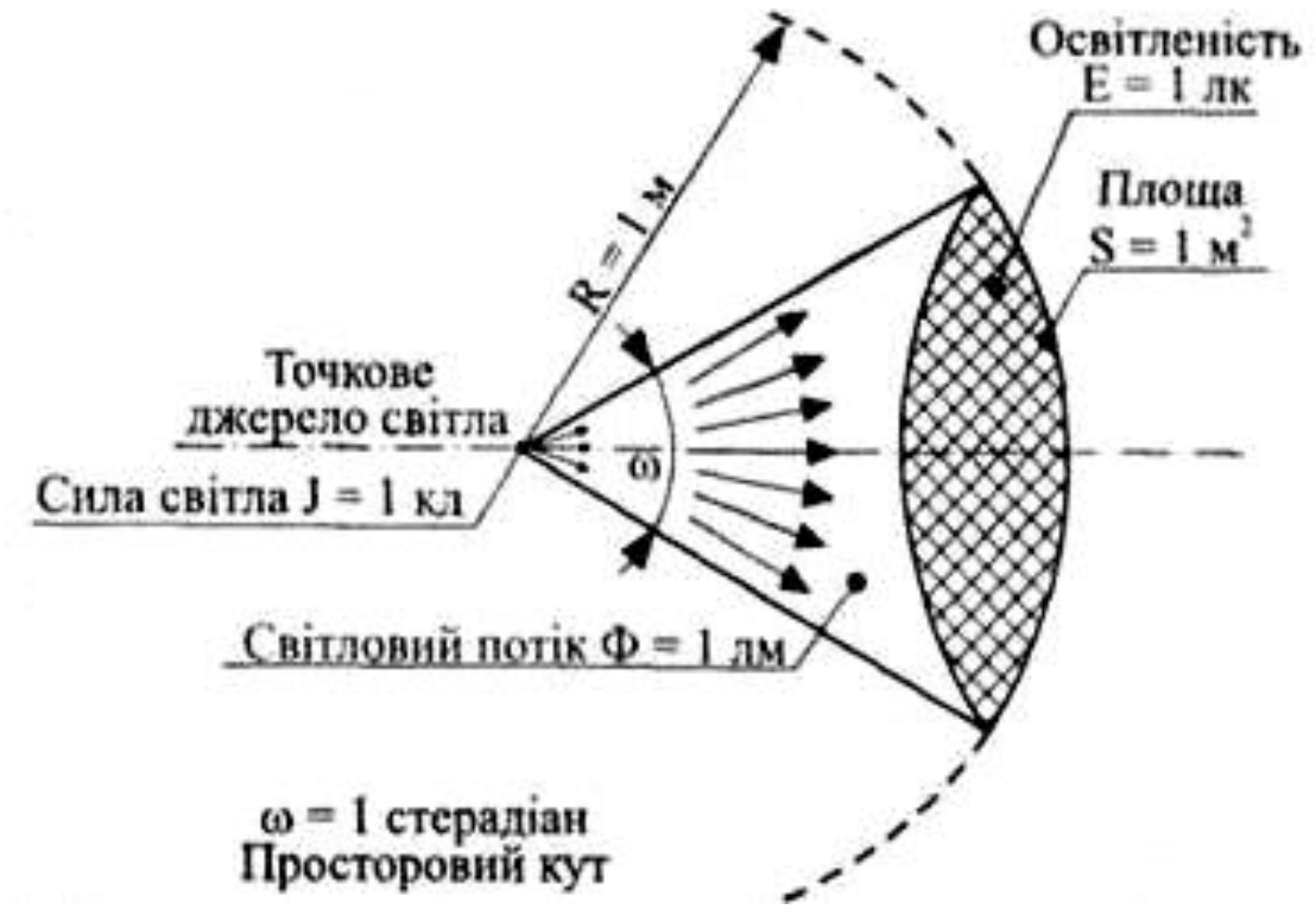

Рис. 5.1. Основні характеристики точкового джерела випромінювання.

Тілесний кут вимірюється відношенням площі той часті сфери з центром у вершині кута, яка відокремлюється цим тілесним кутом, до квадрату радіуса сфери, тобто тілесний кут це частина простору, яка є об'єднанням усіх променів, що виходять з даної точки (вершини кута) і перетинають деяку поверхню:

$$\Omega = \frac{S_0}{R^2} = \frac{4\pi R^2}{R^2} = 4\pi$$

де 4π – це є максимальний тілесний кут для повної сфери з її центру.

**Сила світла** ($I$) – це фізична велиабона, що характеризує світіння джерела в певному напрямку. Якщо джерело випромінює видиме світло рівномірно в усі боки, то сила світла обабослюється за формулою:

$$I = \frac{\Phi_V}{\Omega} = \frac{\Phi_V}{4\pi}$$

**Яскравість** ($L$, $B$) – є світловою велиабоною, яка визначається як поверхнева щільність сили світла в заданому напрямі, рівна відношенню сили світла до площі проекції поверхні, що світиться, на площину, перпендикулярну до того ж напряму. Якщо око розглядає плоску поверхню площею $S$, випромінюючу рівномірно в напрямі, перпендикулярному до неї, силу світла $I$, то яскравість поверхні у напрямі ока визначається рівнянням:

$$B = \frac{I}{S}$$

Якщо розглядати ту ж поверхню, що світить, під кутом $\theta$ до її перпендикуляра, то око побааботь частину цієї поверхні, а саме її проекцію на напрям, перпендикулярний до лінії зору площею $S\cos\theta$. В цьому випадку яскравість виражається рівнянням:

$$B_\theta = \frac{I}{S\cos\theta},$$

де $I$ – сила світла, що випромінюється елементом поверхні в даному напрямку, $S$ - площа поверхні, $\theta$ – кут між віссю виділеного світлового пучка і нормаллю до поверхні.

Відношення світлового потоку до площі поверхні, по якій розподіляється потік, називається освітленістю поверхні. При рівномірному розподілі потоку:

$$E = \frac{\Phi}{S}$$

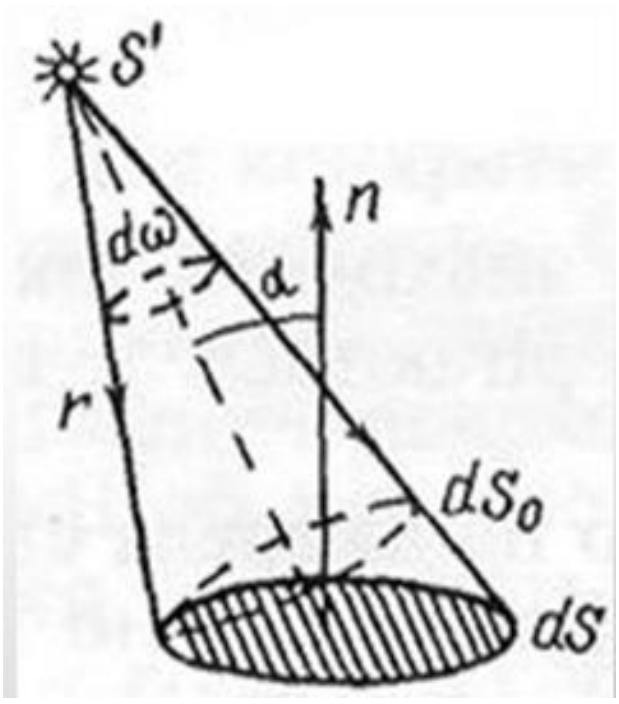

$$E = \frac{d\Phi}{dS} = \frac{I \cdot \cos\alpha}{r^2}$$

Зауважимо, що кут падіння променів відраховується від нормалі до поверхні.

**Задача 5.1.** Який світловий потік потрапляє на поверхню стола, якщо її середня освітленість дорівнює 9500 лк, а площа — 1,6 м$^2$?

| **Дано:** | **Розв'язання.** |
| --- | --- |
| $E$= 9500 лк<br>$S$=1,6 м$^2$ | В задачі питають о світловому потоці і надана освітленість, тому шукаємо формулу в якій ми виражаємо освітленість через світловий потік: |
| **Знайти:**<br>$\Phi-?$ | $$E=\frac{\Phi}{S}$$ |

$$\Phi=\mathrm{E}\cdot S=9500\text{лк}/\text{м}^2\cdot 1{,}6\text{м}^2=15200\ \text{лм}$$

**Відповідь:** $\Phi = 15200$ лм.

**Задача 5.2.** Опівдні в час весняного і осіннього рівноденнь Сонце стоїть на екваторі в зеніті. У скільки разів у цей час освітленність поверхні Землі на екваторі більше освітлености Землі в місті, який лежить на широті 45$^o$?

| **Дано:** | **Розв'язання.** |
| --- | --- |
| $\varphi=45^o$ | Світло $S$ від Сонця на екваторі падає під кутом 0$^o$ (кут між нормаллю до поверхні і промінем). Точка А розташована на екваторі, тому промінь падає під кутом 0$^o$. |
| **Знайти:**<br>$\frac{\mathrm{E_e}}{\mathrm{E_м}}-?$ | Точка $M$ знаходиться на широті $\varphi=45^o$. Побудуємо нормаль |

до точки $M$ – це буде $MP$. На основі отриманих даних зробимо рисунок:

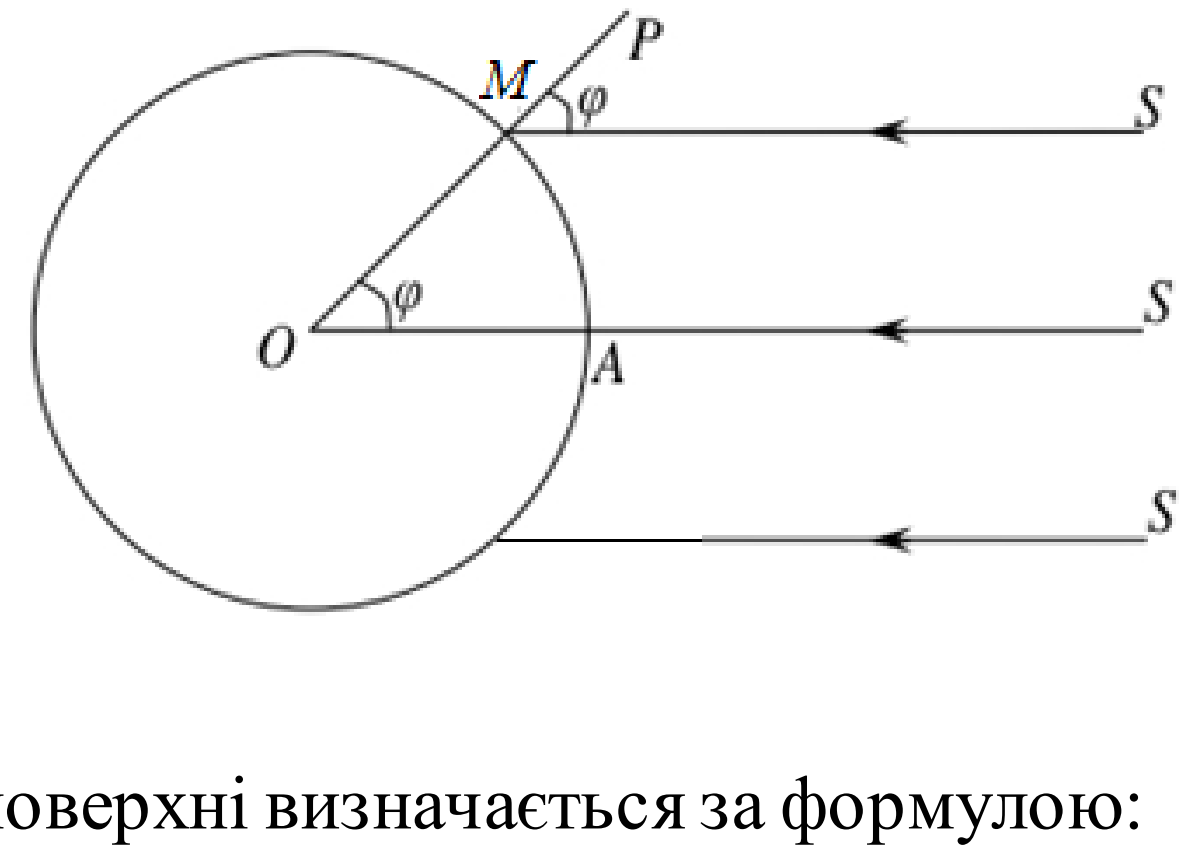

Освітленність поверхні визначається за формулою:

$$E = \frac{I\cos\varphi}{r^2}$$

Напишемо формулу освітленості поверхні Землі на екваторі, пам'ятаюабо, що $\cos 0^{\circ} = 1$:

$$E_e = \frac{I}{r^2},$$

де $I$ – сила світла Сонця; $r$ – відстань між Сонцем і Землею (їх центрами). Оскільки $r \gg R$ ($R$ – радіус Землі), то промені, що падають на Земну поверхню можна вважати паралельними між собою, отже, освітленість поверхні Землі на широті $\varphi$ визнаабомо по формулі:

$$E_{м} = \frac{I}{r^2}\cos\varphi; \qquad \cos 45^{\circ} = \frac{\sqrt{2}}{2}$$

Оскільки за умовою задачі треба знайти «У скільки разів у цей час освітленність поверхні Землі на екваторі більше освітленості Землі в місті на щироті $45^{\circ}$», то знаходимо відношення освітленностей, які ми визнааболи:

$$\frac{E_e}{E_{м}} = \frac{\frac{I}{r^2}}{\frac{I}{r^2}\cos\varphi} = \frac{I}{r^2}\cdot\frac{r^2}{I\cos\varphi} = \frac{1}{\cos\varphi} = \frac{1}{\cos 45^{\circ}} = \frac{1}{\frac{\sqrt{2}}{2}} = \frac{2}{\sqrt{2}} = 1{,}43$$

**Відповідь:** освітленність Землі на широті $45^{\circ}$ у 1,43 раза менша ніж освітленність на екваторі.

**Задача 5.3.** Визначить повний світловий потік, який надається ізотропним точковим джерелом, якщо на відстані 2 м від нього освітленість 15 лк.

| **Дано:** | **Розв'язання.** |
| --- | --- |
| $r = 2$ м<br>$E = 15$ лк | Уявимо повний світловий потік: - це світлові промені, що йдуть на всі боки, можна казати створюють сферу, яка буде обмежена відстанню 2 м, тобто маємо сферу з радіусом $r = 2$ м. |
| **Знайти:**<br>$Ф$–? | |

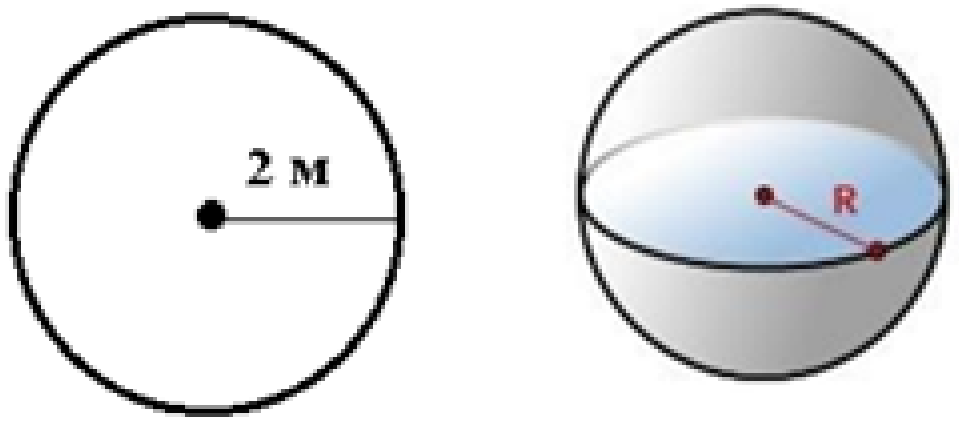

Максимальний тілесний кут для повної сфери з її центру $\omega = 4\pi$, тоді для цієї задачі світловий потік буде $\Phi = 4\pi I$. Однак, в умовах задачі надана освітленість, а не сила світла. Запишемо формулу освітленості:

$$E = \frac{I \cos\alpha}{r^2}$$

Промені йдуть від джерела по прямій і будуть перетинати умовну поверхню під прямим кутом, тобто напрямок променю співпадає з нормаллю до поверхні, отже кут $\alpha = 0^{o}$ і $\cos\alpha=1$, тому освітленість буде залежати тільки від сили світла і відстані:

$$E = \frac{I}{r^2}$$

виразимо силу світла

$$I = E \cdot r^2$$

і підставимо у

$$\Phi = 4\pi I = 4\pi \cdot E \cdot r^2 = 4 \cdot 3{,}14 \cdot 15 \cdot 4 = 753{,}6 \text{ лм}$$

**Відповідь:** $\Phi = 753{,}6$ лм.

**Задача 5.4.** Чому дорівнює повний світловий потік, що створюється джерелом світла, розміщеним на мачті висотою 12 м, якщо на відстані 16 м від основи мачти він дає освітленість 3 лк?

**Дано:**
$h$=2 м
$\ell$=16 м
$E = 3$ лк

**Знайти:**
$\Phi -?$

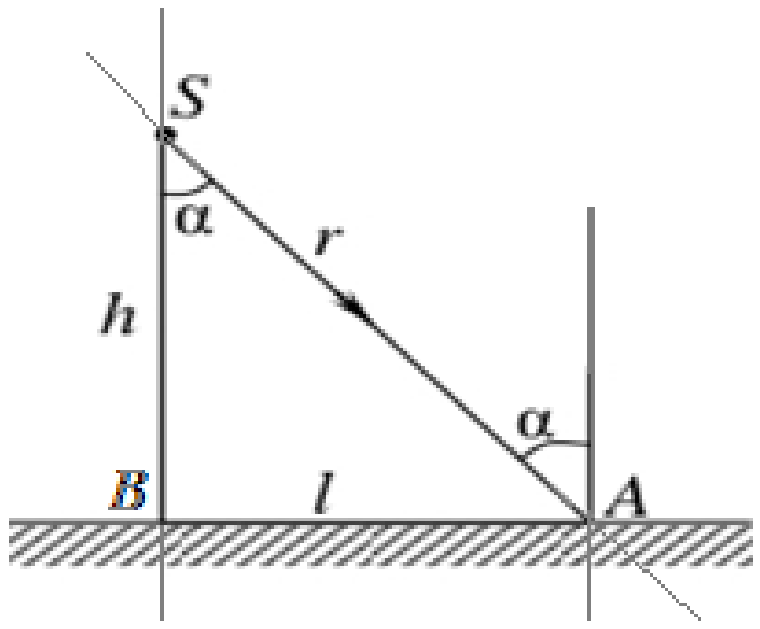

**Розв’язання.**

Сила світла джерела визначається за формулою:

$$I = \frac{\Phi}{\Omega}, \quad \text{тоді} \quad \Phi = I \cdot \Omega$$

Відомо, що повний тілесний кут $\Omega = 4\pi,$ тому маємо записати:

$$\Phi = 4\pi \cdot I$$

Силу світла $I$ знаходимо з формули освітленості для точки А:

$$E = \frac{I}{r^2} \cos \alpha,$$

де $\alpha$ - кут падіння променів.

Пам’ятаємо, що кут падіння променю, це кут між нормалю до поверхні в точці падіння променю і самим промінем. І за теоремою о кутах, утворених двома паралельними прямими та січною, яка стверджує, що при перетині двох паралельних прямих січною внутрішні різносторонні кути рівні, маємо прямокутний трикутник $\Delta$ SBA з кутом $\alpha$ у вершині S. За визначенням тригонометричної функції $\cos \alpha$ – «у прямокутному трикутнику є відношенням прилеглого катета до гіпотенузи», тобто:

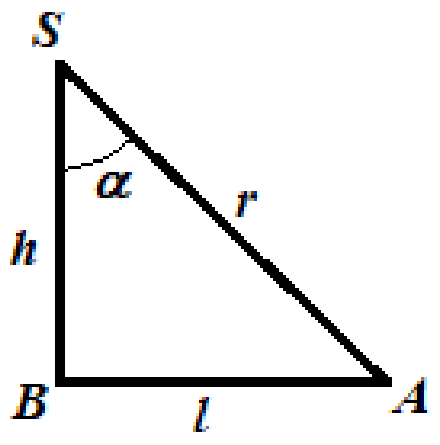

$$\cos \alpha = \frac{h}{r}$$

Тоді, підставивши це відношення в формулу освітленості отримаємо:

$$E = \frac{I}{r^2} \cdot \frac{h}{r} = \frac{I \cdot h}{r^3}$$

Але невідомим зосталося $r$, яке можна визнааботи за теоремою Піфагора «в прямокутном трикутнике квадрат гіпотенузи дорівнюється сумі квадратів катетів», тобто:

$$r^2 = h^2 + \ell^2; \qquad r = \sqrt{h^2 + \ell^2}$$

підставимо в рівняння освітленості.

$$E = \frac{I \cdot h}{r^3} = \frac{I \cdot h}{\left(\sqrt{h^2 + \ell^2}\right)^3}$$

Щоб знайти світловий потік виразимо силу світла із отриманого рівняння:

$$I = \frac{\mathrm{E}\left(\sqrt{h^2 + \ell^2}\right)^3}{h}$$

Тепер можемо знайти світловий потік:

$$\Phi = I \cdot \Omega = \frac{4\pi \mathrm{E}\left(\sqrt{h^2 + \ell^2}\right)^3}{h} = 25120 \text{ лм}$$

**Відповідь:** $\Phi = 25120$ лм.

**Задача 5.5.** Дві лампи перша 60 кд і друга 80 кд висять на висоті 2 м над поверхнею стола. Відстань між лампами 1,5 м. Знайти освітленість стола під першою і другою лампами.

**Дано:**
$I_1$=60 кд
$I_2$=80 кд
$h$ =2 м
$\ell$=1,5 м

**Знайти:**
$E_1$-? $E_2$ - ?

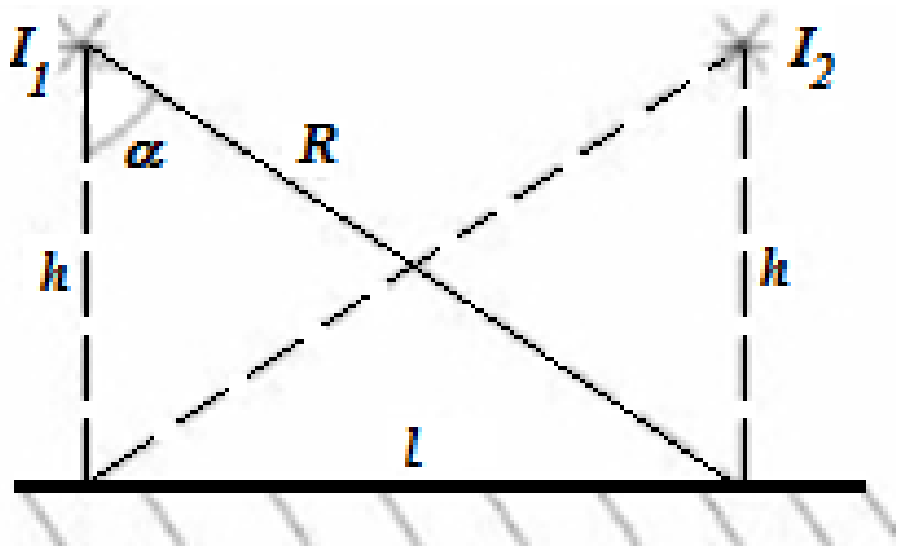

**Розв'язання.**

Освітленість під кожною лампою дорівнює сумі освітленості, яку створюють обидві лампи. Освітленість поверхні визначається за формулою:

$$E = \frac{I \cdot \cos\alpha}{r^2},$$

де $r$ – відстань від джерела до поверхні.

Розглянемо освітленість поверхні під першою лампою $E_1$, яка складається з двох джерел, від першої і від другої лампи. Познаабомо ці складові як: $\mathrm{E}_1'$ –внесок тільки від першої лампи і $\mathrm{E}_2'$ –внесок тільки від другої лампи. Отже повна освітленість під першою лампою $\mathrm{E}_1 = \mathrm{E}_1' + \mathrm{E}_2'$, тоді повна освітленість під другою лампою буде складатися з внеску тільки від першої лампи $\mathrm{E}_1''$ і внеску тільки від другої $\mathrm{E}_2''$, тобто $\mathrm{E}_2 = \mathrm{E}_1'' + \mathrm{E}_2''$. Треба звернути увагу, що $\mathrm{E}_1' \neq \mathrm{E}_1''$ і $\mathrm{E}_2' \neq \mathrm{E}_2''$. Складемо рівняння для кожного компонента:

$$\mathrm{E}_1' = \frac{I_1 \cdot \cos\alpha_1}{h^2},$$

але $\alpha_1 = 0$ (це кут між нормалю до поверхні і падаюабом променем)

$$\Rightarrow \quad \cos\alpha_1 = 1; \quad \mathrm{E}_1' = \frac{I_1 \cdot}{h^2} \quad \text{і} \quad \mathrm{E}_2' = \frac{I_2 \cdot \cos\alpha}{R^2} \quad \text{тоді:}$$

$$\mathrm{E}_1 = \frac{I_1}{h^2} + \frac{I_2}{R^2}\cos\alpha, \ \text{де} \ \cos\alpha = \frac{h}{R},$$

після підстановки отримаємо повну освітленість:

$$\mathrm{E}_1 = \frac{I_1}{h^2} + \frac{I_2 h}{R^3}$$

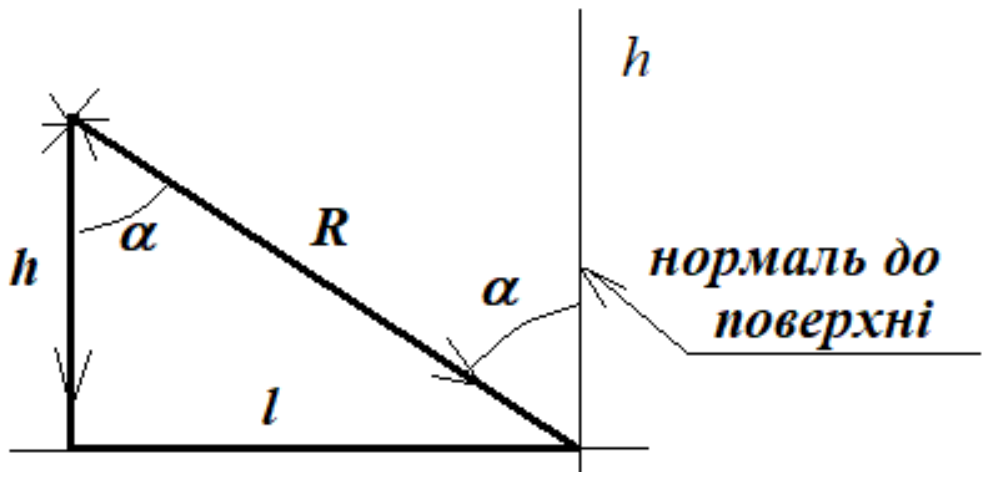

В цьому рівняння невідомим є тільки $R$ - відстань від джерела світла до точки поверхні, в якій визначаємо освітленість. Знайдемо $R$ як гіпотенузу прямокутного трикутника з катетами $h$ і $l$:

$$R^2 = h^2 + l^2 \ ; \quad R = \sqrt{h^2 + l^2} = \sqrt{2^2 + 1{,}5^2} = \sqrt{6{,}25} = 2{,}5 \text{ м}$$

Аналогічно складемо рівняння повної освітленості під другою лампою:

$$E_1'' = \frac{I_1 \cdot \cos\alpha}{R^2} \text{ – внесок від першої лампи}$$

$$E_2'' = \frac{I_2 \cdot \cos\alpha_1}{h^2}, \quad \text{але } \alpha_1 = 0 \implies \cos\alpha_1 = 1; \quad E_2'' = \frac{I_2}{h^2} \text{ – від другої}$$

Тоді повна освітленість під другою лампою:

$$E_2 = \frac{I_1 \cdot \cos\alpha}{R^2} + \frac{I_2}{h^2}, \quad \text{де } \cos\alpha = \frac{h}{R}, \quad E_2 = \frac{I_1 \cdot h}{R^3} + \frac{I_2}{h^2}$$

Підставимо абословідані в отримані рівняння:

$$E_1 = \frac{I_1}{h^2} + \frac{I_2 h}{R^3} = \frac{60}{2^2} + \frac{80 \cdot 2}{2{,}5^3} = 15 + 5{,}13 = 20{,}13 \text{ лк}$$

$$E_2 = \frac{I_1 \cdot h}{R^3} + \frac{I_2}{h^2} = \frac{60 \cdot 2}{2{,}5^3} + \frac{80}{2^2} = 7{,}7 + 20 = 27{,}7 \text{ лк}$$

**Відповідь:** $E_1 = 20{,}13$ лк; $E_2 = 27{,}7$ лк.

**Задача 5.6.** Кругла зала діаметром 30 м освічується лампою, що закріплена в центрі стелі. Знайти висоту зали, якщо відомо, що найменша освітленість стіни зали в 2 рази більше найменшої освітленості підлоги.

**Дано:**
$d$ =30 м
$n$=2

**Знайти:**
$h$-?

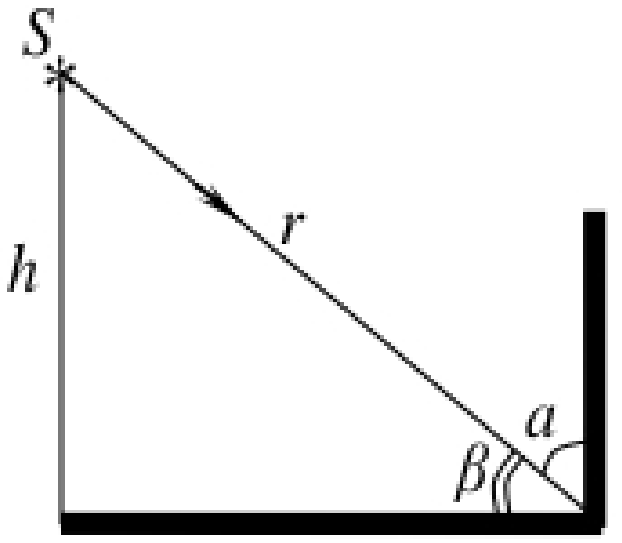

**Розв'язання.**

З рисунка можна побааботи, що з найменшою освітленостю підлоги буде точка, що знаходиться коло стінки.Такою ж найменшою освітленостю буде та частина стінки, яка стикується з підлогою. Отже освітленість підлоги визнаабомо як:

$$Е_п = \frac{I}{r^2}\cos\alpha,$$

де $\alpha$ –кут між нормалю до підлоги і промінем.

А освітленість стінки:

$$Е_{ст} = \frac{I}{r^2}\cos\beta,$$

де $\beta$ – кут між нормалю до стіни і промінем (в даному випадку нормаль співпадає з підлогою).

За умовою задачі $Е_{ст} = nЕ_п$ , тоді маємо записати:

$$\frac{I}{r^2}\cos\beta = n\frac{I}{r^2}\cos\alpha \quad \text{або} \quad \cos\beta = n\cos\alpha$$

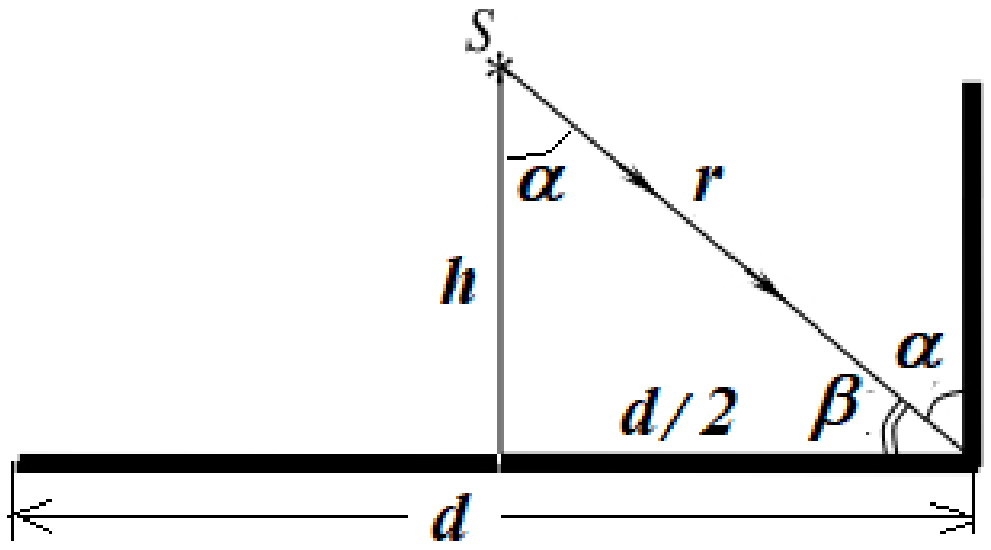

Визнаабомо $\cos\beta$ і $\cos\alpha$ із прямокутного трикутника як відношення прилеглого катета до гіпотенузи, тобто:

$$\cos\beta = \frac{d}{2r}; \quad \cos\alpha = \frac{h}{r}$$

Отримані співвідношення підставимо в рівняння $\cos\beta = n\cos\alpha$:

$$\frac{d}{2r} = n\cdot\frac{h}{r}$$

звідки отримаємо висоту зали:

$$h = \frac{d}{2n} = \frac{30}{2\cdot 2} = 7{,}5 \text{ м}$$

**Відповідь:** $h = 7{,}5$ м.

**Задача 5.7.** Дві лампи розташовані на відстані 2,4 м одна від одної. Де потрібно розмістити між ними непрозорий екран, щоб він був однаково освітлений з обох сторін? Сила світла ламп дорівнює 100 і 50 кд.

**Дано:**

$I_1$=50 кд

$I_2$=100 кд

$r$ =2,4 м

**Знайти:**

$r_1$-? $r_2$ - ?

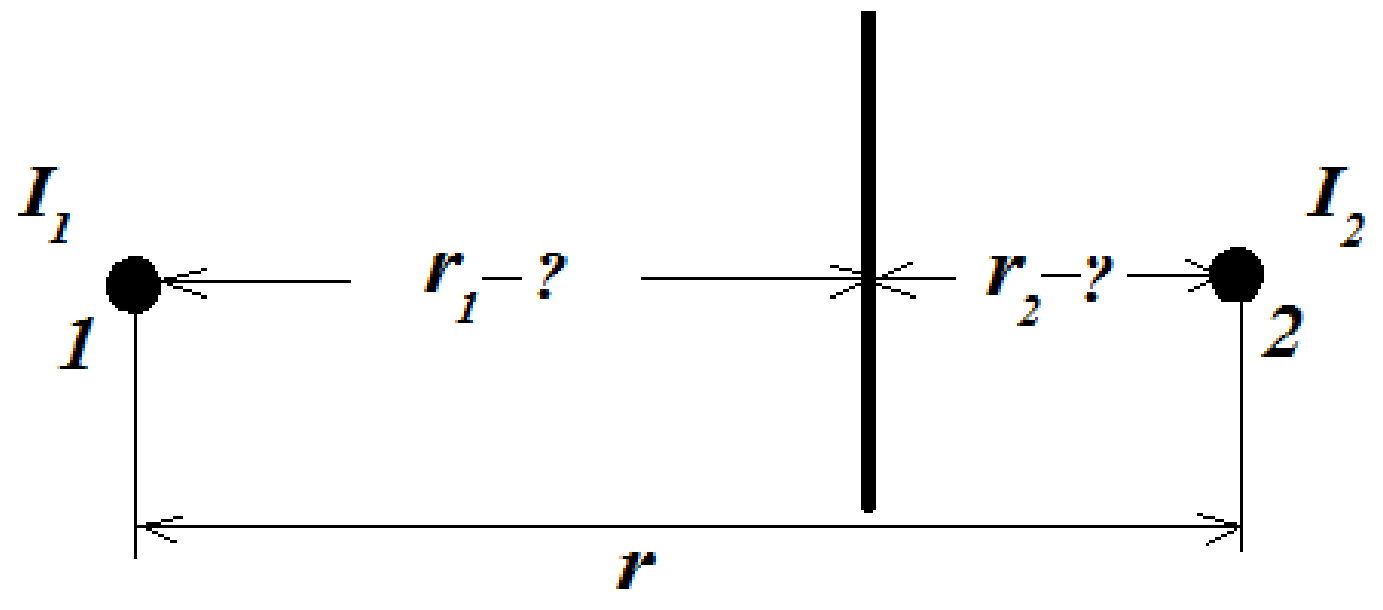

**Розв'язання.**

Освітленість визначається за формулою:

$$E = \frac{I\cos\alpha}{r^2}$$

В цієї залачі $\cos\alpha = 1$ (кут між нормаллю та променем $\alpha = 0$), тому можемо записати:

$$E_1 = \frac{I_1}{r_1^2} \quad \text{та} \quad E_2 = \frac{I_2}{r_2^2}$$

За умовою задачі $E_1 = E_2$ і маємо два невідомих показника, тому треба записати два рівняння:

$$\begin{cases} \dfrac{I_1}{r_1^2} = \dfrac{I_2}{r_2^2} \\ r = r_1 + r_2 \end{cases} \qquad \begin{cases} I_1 \cdot r_2^2 = I_2 \cdot r_1^2 \\ r_1 = r - r_2 \end{cases}$$

Друге рівняння підставляємо в перше і отримаємо квадратне рівняння.

$$I_1 \cdot r_2^2 = I_2 \cdot (r - r_2)^2$$

$$\frac{I_1}{I_2} r_2^2 = r^2 - 2rr_2 + r_2^2$$

підставимо значення надані в умові з

$$\frac{100}{50} r_2^2 = 2{,}4^2 - 2 \cdot 2{,}4 \cdot r_2 + r_2^2$$

$$2r_2^2 - r_2^2 = 5{,}76 - 4{,}8 \cdot r_2$$

$$r_2^2 + 4{,}8 \cdot r_2 - 5{,}76 = 0$$

$$r_{2_{1,2}} = \frac{-b \pm \sqrt{b^2 - 4ac}}{2a} \qquad \sqrt{b^2 - 4ac} = \sqrt{46{,}08} = 6{,}79$$

$$1)\ r_2 = \frac{-4{,}8 - 6{,}79}{2} = -5{,}8$$

$$2)\ r_2 = \frac{-4{,}8 + 6{,}79}{2} = \frac{1{,}99}{2} = 0{,}995$$

Рішенням складеного квадратного рівняння є два кореня: «$-5{,}8$» - від'ємне абосло, тому корінь не підходить та $0{,}995$, яке використаємо для знаходження

$r_1$ – відстані до екрану зі сторони лампи в 100 кд.

$$r_2 = 0{,}995 \text{ м}; \qquad r_1 = r - r_2 = 2{,}4 - 0{,}995 = 1{,}405 \text{ м}$$

**Відповідь:** $r_1 = 1{,}405$ м; $r_2 = 0{,}995$ м.

**Задача 5.8.** Над поверхнею Землі на висоті 5 м висить лампа, що має силу світла 400 кд. Визнаaботи площу дільниці, всередині якої освітленість змінюється в межах від 0,25 лк до 2 лк.

**Дано:**
$I$=400 кд
$E_1$=0,25 лк
$E_2$=2 лк
$h$ =5 м

**Знайти:**
$S$ - ?

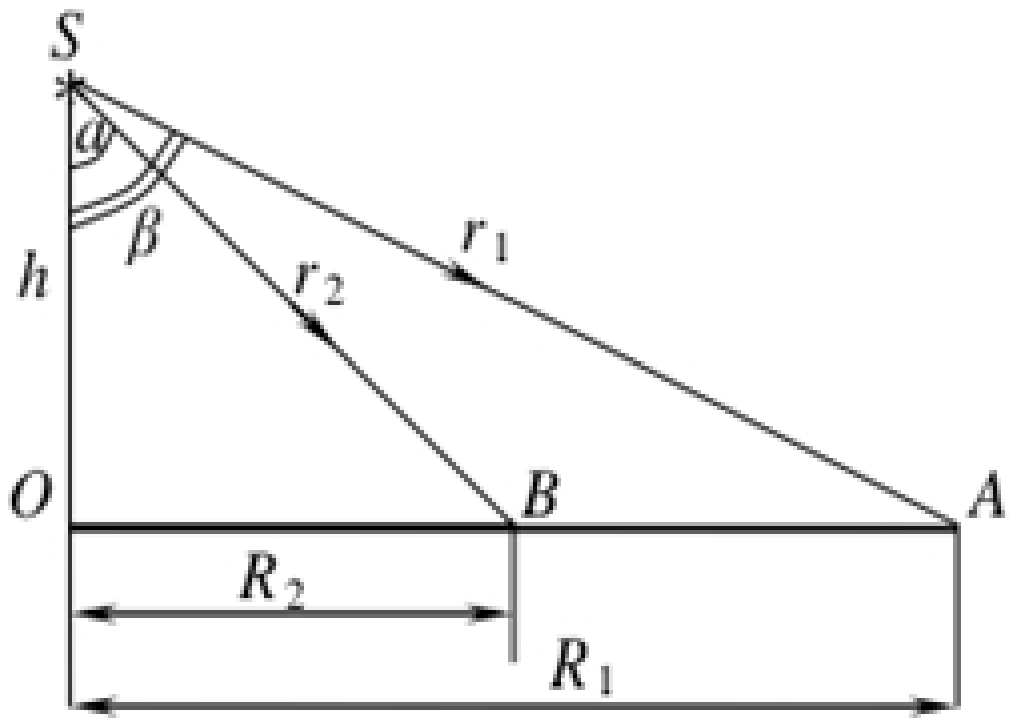

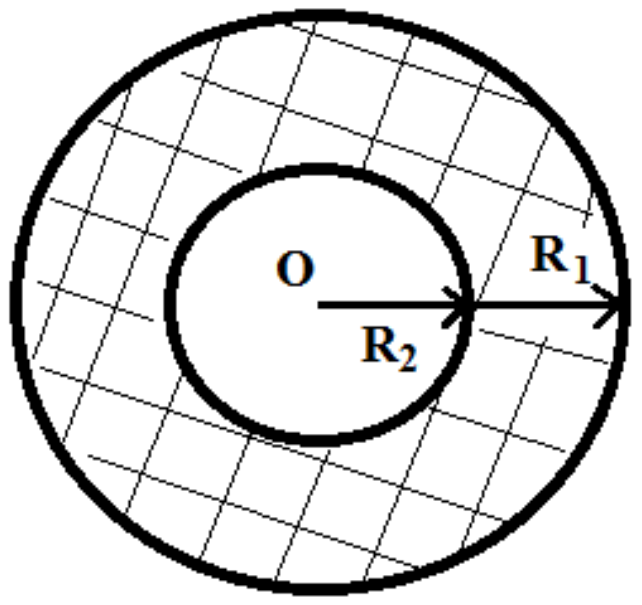

**Розв'язання.**

З рисунка видно, що меншу освітленість в 0,25 лк має поверхня, що знаходяться на відстані $OA$=$R_1$, а більшу в 2 лк – та, що знаходяться на відстані $OB$= $R_2$.

Площа дільниці, з вказаним коридором освітленості може бути знайдена за формулою:

$$S_{\text{кола}} = \pi R^2;$$

тоді для нашого сегмента

$$S = \pi(R_1^2 - R_2^2).$$

Освітленість в точці $B$ (всі точки, що на окружності з радіусом $R_2$):

$$E_2 = \frac{I}{r_2^2}\cos\alpha,$$

Визнаабомо $\cos\alpha$ із прямокутного трикутника як відношення прилеглого катета до гіпотенузи, тобто:

$$\cos\alpha = \frac{h}{r_2}$$

$$E_2 = \frac{I}{r_2^2}\cdot\frac{h}{r_2} = \frac{I\cdot h}{r_2^3}$$

Із отриманого рівняння тепер визнаабомо немідоме $r_2$.

$$r_2 = \sqrt[3]{\frac{I\cdot h}{\mathrm{E}_2}} = \sqrt[3]{\frac{400\cdot 5}{2}} = 10\ \text{м}.$$

Визначивши $r_2$ і знаюабо висоту $h$, можемо розрахувати відстань $R_2^2$ як квадрат катета в прямокутному трикутнику, для формули знаходження площі

$$S = \pi(R_1^2 - R_2^2).$$

$$R_2^2 = r_2^2 - h^2 = 100 - 25 = 75\ \text{м}^2$$

Аналогічно знайдемо $R_1^2$, але спочатку напишемо рівняння освітленості в точці А (освітленість в точці А це всі точки, що на окружності з радіусом $R_1$:

$$E_1 = \frac{I}{r_1^2}\cos\beta, \quad \text{де} \quad \cos\beta = \frac{h}{r_1}$$

$$E_1 = \frac{I}{r_1^2}\cdot\frac{h}{r_1} = \frac{I\cdot h}{r_1^3}$$

Із отриманого рівняння визнаабомо $r_1$:

$$r_1 = \sqrt[3]{\frac{I \cdot h}{E_1}} = \sqrt[3]{\frac{400 \cdot 5}{0,25}} = 20 \text{ м.}$$

Знайдемо $R_1^2$:

$$R_1^2 = r_1^2 - h^2 = 400 - 25 = 375 \text{ м}^2$$

Тепер можемо порахувати площу.

$$S = \pi(R_1^2 - R_2^2) = 3,14(375 - 75) = 942 \text{ м}^2.$$

**Відповідь:** $S = 942 \text{ м}^2$.

**Задача 5.9.** Лампа, що підвішена під стелею, дає в горизонтальному напрямку силу світла $I = 60$ кд. Який світловий потік Ф падає на картину площею $S$=0,6 м$^2$, яка висить вертикально на стіні на відстані $r$=3 м від лампи, якщо на протилежної стіні знаходиться велике дзеркало на відстані $a$=2 м від лампи?

**Дано:**
$I$=60 кд
$S$=0,6 м$^2$
$r$ =3 м
$a$=2 м

**Знайти:**
Ф - ?

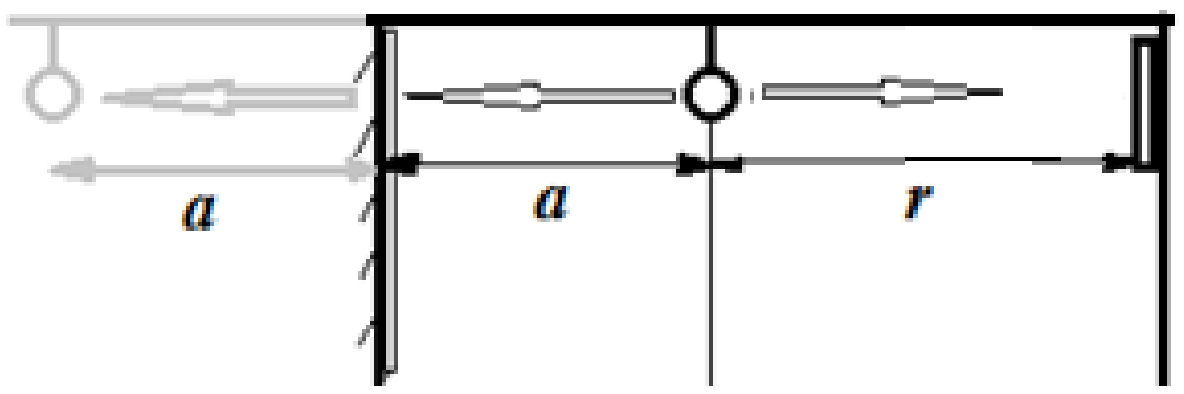

**Розв'язання.**

Освітленість $E$ характеризується світловим потоком, що приходиться на одиницю площі

$$E = \frac{\Phi}{S}$$

з цієї формули виражаємо світловий потік, який треба знайти:

$$\Phi = E \cdot S$$

Точкове джерело силою світла $I$ створює на площі, яка відстоїть від нього на відстані r, освітленість:

$$E = \frac{I}{r^2}\cos\alpha, \text{ де } \alpha - \text{ кут падіння променів.}$$

де $\alpha$ – кут падіння променів.

На картину падає світловий потік безпосередньо від лампи $\Phi_1$, а також, світловий потік, який відбитий від дзеркала $\Phi_2$:

$$\Phi = \Phi_1 + \Phi_2 = (E_1 + E_2)S$$

За умовою задачі – лампа висить навпроти картини, це означає, що кут падіння променів $\alpha = 0^{\circ} \Rightarrow \cos\alpha = 1$.

Тоді:

$$E_1 = \frac{I}{r^2};$$

$$E_2 = \frac{I}{(r+2a)^2}$$

$$\Phi = \left(\frac{I}{r^2} + \frac{I}{(r+2a)^2}\right)S = I \cdot S\left(\frac{1}{r^2} + \frac{1}{(r+2a)^2}\right)$$

$$\Phi = 60 \cdot 0{,}6\left(\frac{1}{3^2} + \frac{1}{(3+2\cdot 2)^2}\right) = 36(0{,}11 + 0{,}02) = 4{,}68 \text{ лм.}$$

**Відповідь:** $\Phi = 4{,}68$ лм.

**Задача 5.10.** Тунель циліндричної форми радіусом $R$ освітлюється світильником, встановленим у верхній точці склепіння. Порівняти освітленості найнижчої точки тунелю точно під світильніком і точки, що лежить на горизонтальному діаметрі перерізу тунелю.

**Дано:**
$h_1 = 2R$
$h_2 = R$

**Знайти:**
$E_1/E_2$ - ?

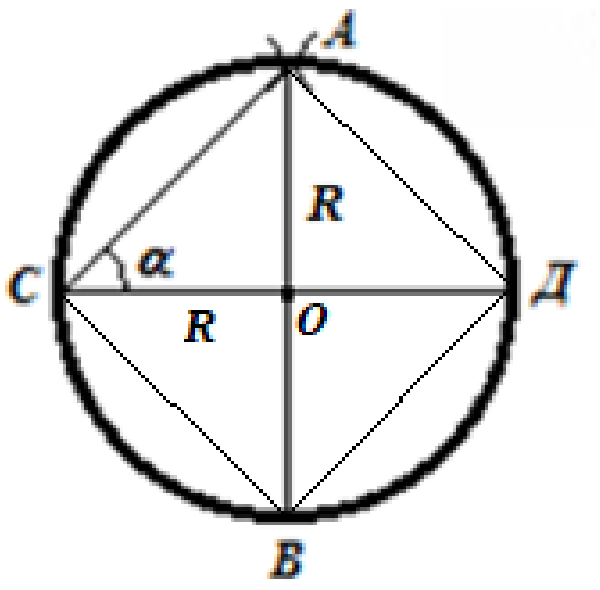

**Розв'язання.**

Якщо з'єднати точки АСВД, то отримаємо квадрат, тобто кут $\alpha = \frac{90^o}{2} = 45^o$.

Освітленість у найнижчої точці тунелю точно під світильником, тобто в точці В знайдемо за формулою:

$$E_B = \frac{I}{(2R)^2} = \frac{I}{4R^2} \qquad \text{тому, що } \cos\alpha = 1, \text{при } \alpha = 0^o$$

Освітленість у точці С знайдемо за формулою:

$$E_C = \frac{I \cdot \cos\alpha}{(AC)^2}; \quad \text{де} \quad AC = \sqrt{R^2 + R^2} = \sqrt{2R^2} = \sqrt{2}R$$

Тоді, при $\cos 45^o = \frac{\sqrt{2}}{2}$:

$$E_C = \frac{I \cdot \frac{\sqrt{2}}{2}}{\left(\sqrt{2}R\right)^2} = \frac{I \cdot \sqrt{2}}{4R^2}$$

$$\frac{E_B}{E_C} = \frac{\frac{I}{4R^2}}{\frac{I \cdot \sqrt{2}}{4R^2}} = \frac{I}{4R^2} \cdot \frac{4R^2}{I \cdot \sqrt{2}} = \frac{1}{\sqrt{2}}$$

**Відповідь:** $\frac{E_B}{E_C} = \frac{1}{\sqrt{2}}$.

**Задача. 5.11.** Над центром квадратної спортивної площадки на висоті 5 м висить лампа. Розрахувати, на якій відстані від центра площадки освітленість поверхні землі в 2 рази менша, ніж в центрі. Вважати, що сила світла лампи по всім напрямкам однакова.

**Дано:**

$h$ =5 м

$\frac{E_0}{E_1} = k = 2.$

**Знайти:**

$d$-?

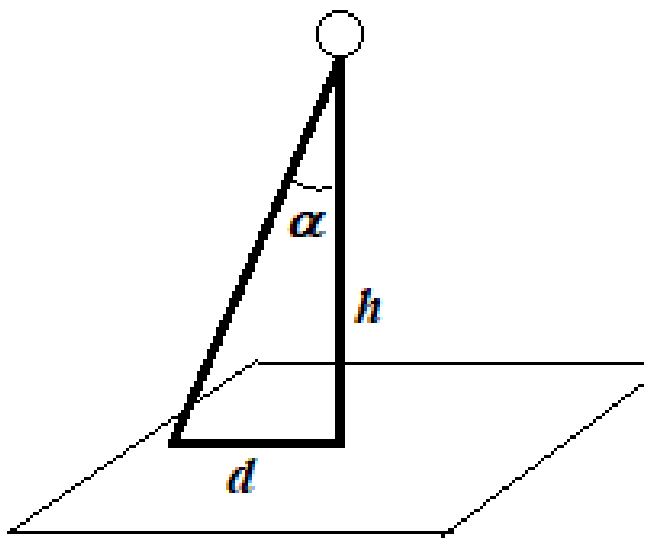

**Розв'язання.**

Освітленість поверхні визначається за формулою:

$$E = \frac{I}{r^2} \cos \alpha$$

Освітленість під лампою, тобто $\cos \alpha = 1$, запишемо як:

$$E_0 = \frac{I}{h^2}$$

Освітленість на відстані $d$:

$$E_1 = \frac{I \cdot \cos \alpha}{h^2 + \mathrm{d}^2}$$

Виразимо $\cos \alpha$ через катет та гіпотенузу прямокутного трикутника:

$$\cos \alpha = \frac{h}{\sqrt{h^2 + \mathrm{d}^2}}$$

Підставимо в $E_1$ і отримаємо:

$$E_1 = \frac{I}{h^2 + \mathrm{d}^2} \cdot \frac{h}{\sqrt{h^2 + \mathrm{d}^2}} = \frac{I \cdot h}{(h^2 + \mathrm{d}^2)^{\frac{3}{2}}}$$

За умовою задачі:

$$k = \frac{E_0}{E_1} = \frac{I}{h^2} \cdot \frac{(h^2 + \mathrm{d}^2)^{\frac{3}{2}}}{I \cdot h} = \frac{(h^2 + \mathrm{d}^2)^{\frac{3}{2}}}{h^3}$$

Представимо $h^3$ як $h^{\frac{2 \cdot 3}{2}}$ тоді:

$$k = \left(1 + \frac{d^2}{h^2}\right)^{\frac{3}{2}}$$

Зведемо обидві частини рівняння у ступінь $\frac{2}{3}$ :

$$k^{\frac{2}{3}} = 1 + \frac{d^2}{h^2} \text{ і виразимо } d$$

$$d = h\sqrt{k^{\frac{2}{3}} - 1} = 5\sqrt{2^{\frac{2}{3}} - 1} = 3{,}8 \text{ м}$$

**Відповідь:** $d = 3{,}8$ м.

## Розділ 6. ЗАКОНИ ГЕОМЕТРИЧНОЇ ОПТИКИ.

**Відбиття світла.** Коли світловий промінь падає на межу поділу двох середовищ, відбувається віддзеркалення світла: промінь змінює напрямок свого ходу і повертається в перше середовище. На рис.5.1. зображені падаюабой промінь AO, відбитий промінь OB, а також перпендикуляр OC, проведений до поверхні KL, що відбиває в точці падіння O.

Кут AOC називається кутом падіння і відраховується від перпендикуляра до поверхні, що відбиває, а не від самої поверхні. Точно так же кут відображення – це кут BOC, утворений відбитим променем і перпендикуляром до поверхні.

**Закон відбиття**:

1) Падаюабой промінь, відбитий промінь і перпендикуляр до поверхні, що відбиває, проведений в точці падіння, лежать в одній площині.

2) Кут відбиття дорівнює куту падіння.

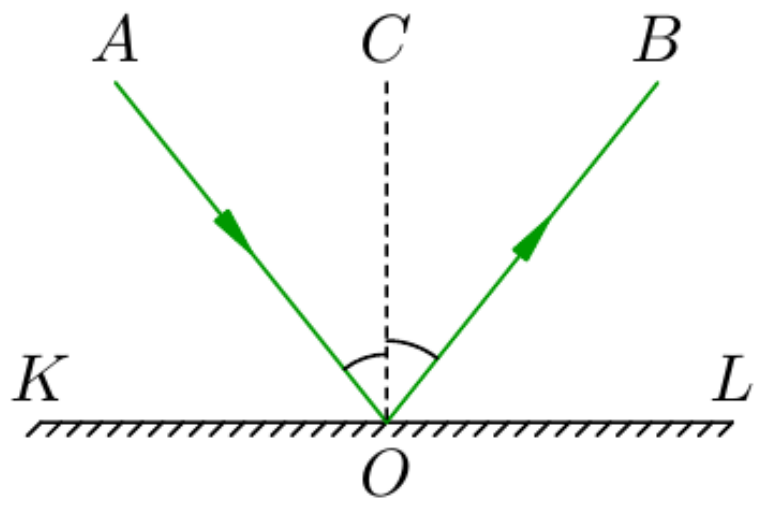

Рис. 5.1. Закон відбиття світла.

**Закон заломлення.**

Нехай світло переходить з середовища 1 з показником заломлення $n_1$ в середу 2 з показником заломлення $n_2$. Середовище з більшим показником заломлення називається оптично щільнішим; відповідно, середовище з меншим показником заломлення називається оптично менш щільним. Переходяабо з оптично менш щільного середовища в оптично щільніше, світловий промінь після заломлення йде ближче до нормалі рис.5.2.а). В цьому випадку кут падіння більше кута заломлення : α> β.

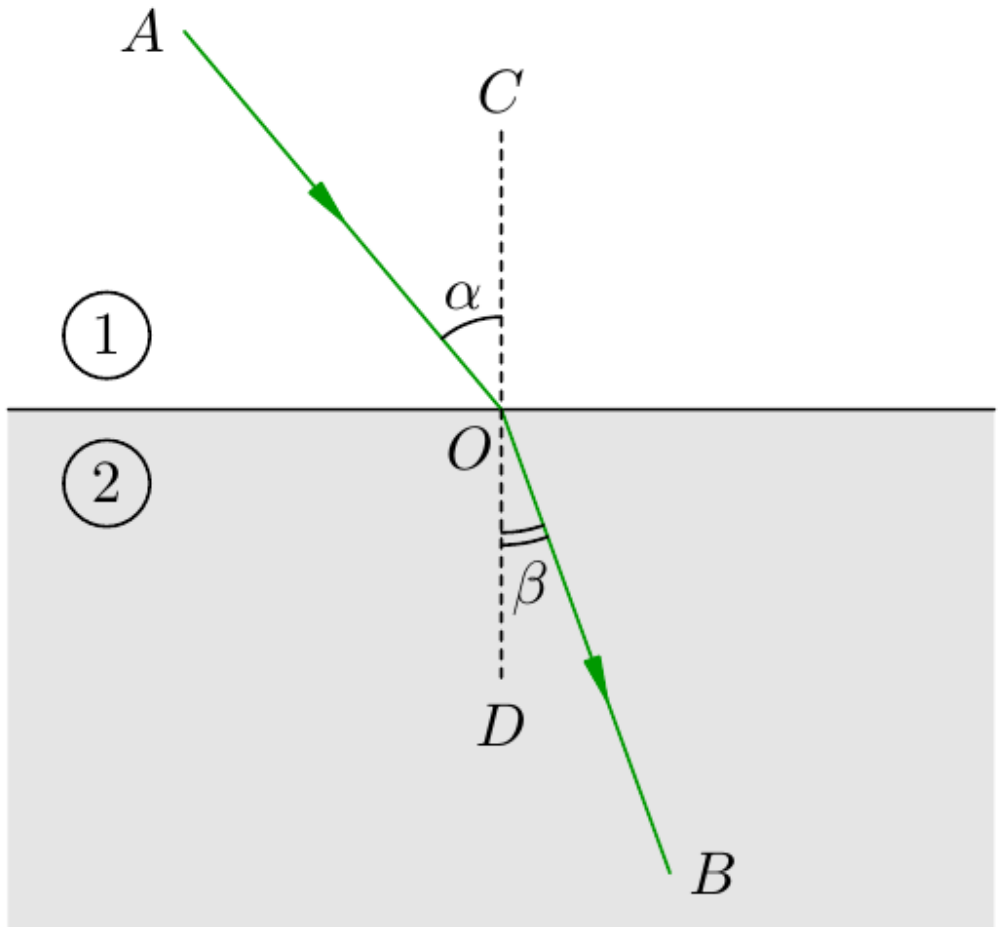

а). $n_1 < n_2 \Rightarrow \alpha > \beta$

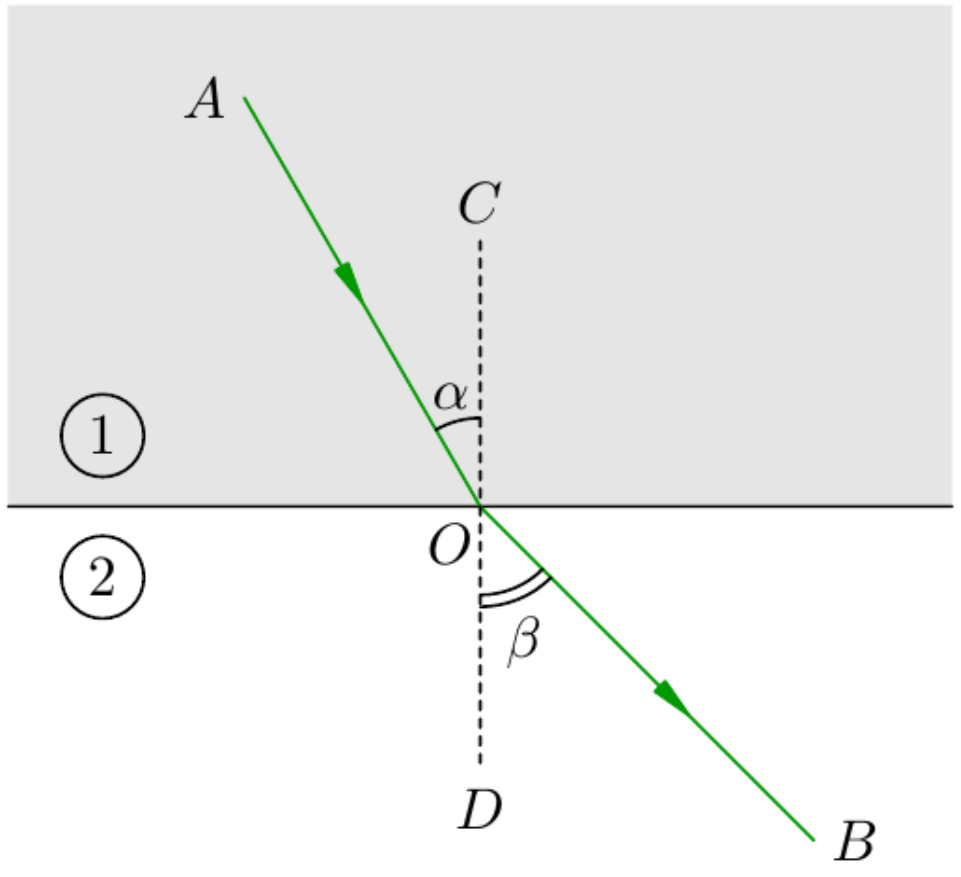

б) $n_1 > n_2 \Rightarrow \alpha < \beta$

Рис. 5.2. Заломлення світла при перетині межі середовищ з різними показниками заломлення: а) перехід променя від менш щільного середовища до більш щільного, наприклад, «повітря-середа»; б) перехід променя від більш щільного середовища до менш щільного.

Навпаки, переходяабо з оптично щільнішого середовища в оптично менш щільне, промінь відхиляється далі від нормалі рис.5.2.б). Тут кут падіння менше кута заломлення : α< β.

**Закон заломлення**:

1) Падаюабой промінь, заломлений промінь і нормаль до поверхні розділу середовищ, проведена в точці падіння, лежать в одній площині.

2) Відношення синуса кута падіння до синуса кута заломлення дорівнює відношенню показника заломлення другого середовища до показника заломлення першого середовища:

$$\frac{\sin\alpha}{\sin\beta} = \frac{n_2}{n_1}$$

Показник заломлення – це відношення швидкості світла у вакуумі до швидкості світла в даному середовищі: $n_1 = c / v_1$, $n_2 = c / v_2$. Підставляюабо це в відношення синусів кутів, отримаємо:

$$\frac{\sin\alpha}{\sin\beta} = \frac{v_1}{v_2}$$

Відстань OF від оптичного центру до фокуса називається фокусною відстанню лінзи і позначається фокусна відстань буквою *F*. Велиабона D, зворотна фокусної відстані, є оптична сила лінзи:

$$D = \frac{1}{F}$$

Оптична сила вимірюється в діоптріях (дптр). Так, якщо фокусна відстань лінзи дорівнює 25 см, то її оптична сила: $D = \frac{1}{0{,}25} = 4$ дптр.

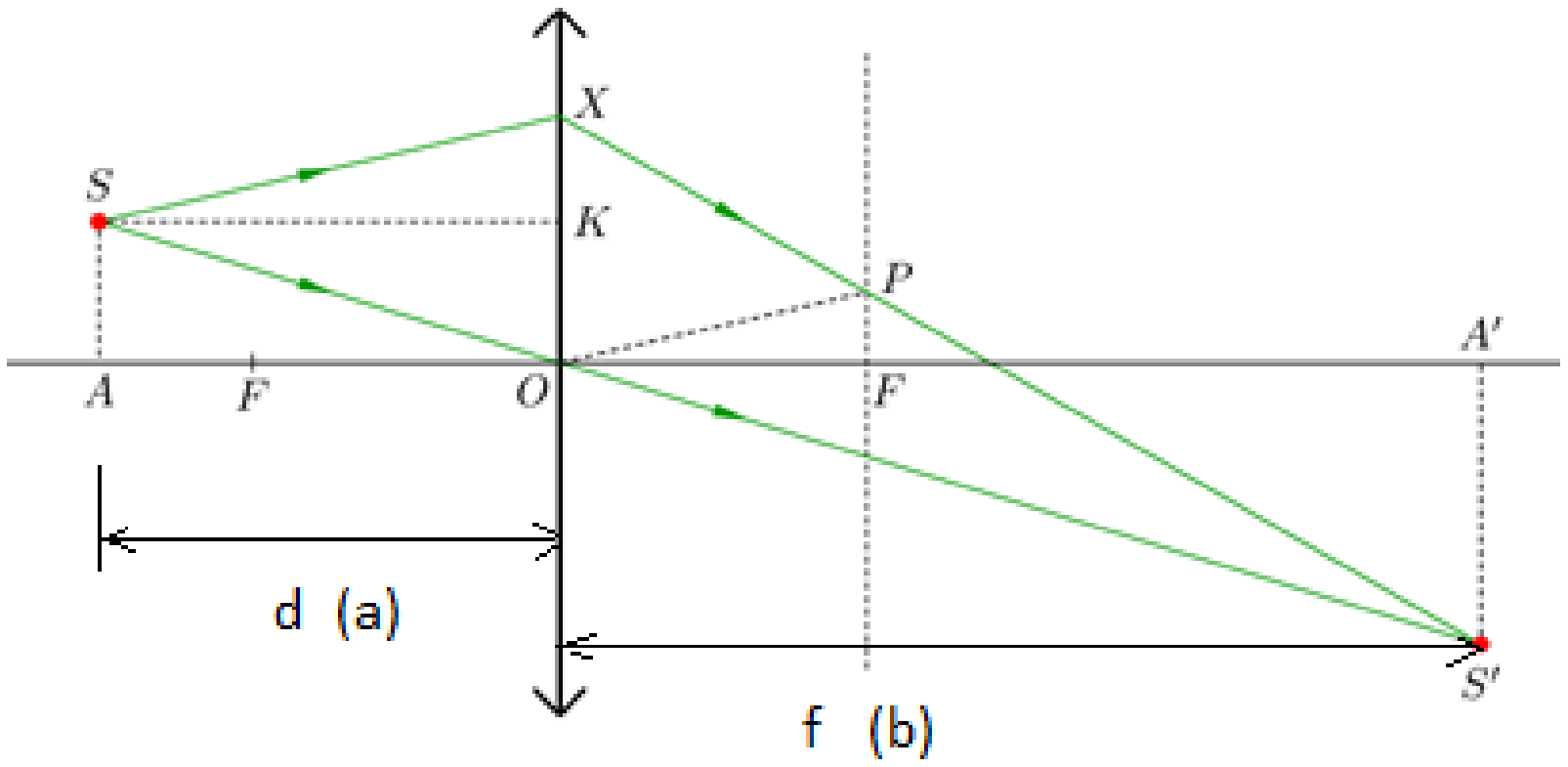

Рис. 5.4. Сема ходу променів через тонку збиральну лінзу, де F – фокус; P – побічний фокус; a – відстань від оптичного центру лінзи до предмета; b – відстань від оптичного центру лінзи до зображення.

Будь-яка пряма, що проходить через оптичний центр лінзи і відмінна від головної оптичної осі, називається побічною оптично віссю. На рис.5.4. зображена побічна оптична вісь – пряма OP, P – побічний фокус. Маємо три формули лінзи – дві для збиральної лінзи і одну для розсіювальної:

$$\frac{1}{a}+\frac{1}{b}=\frac{1}{f}\ ; \quad \frac{1}{a}-\frac{1}{b}=\frac{1}{f}; \quad \frac{1}{a}-\frac{1}{b}=-\frac{1}{f}$$

Ці формули можна записати одноманітно:

$$\frac{1}{a}+\frac{1}{b}=\frac{1}{f},$$

якщо дотримуватись наступних домовленостей о знаках:

• для уявного зображення велиабона b вважається негативною;

• для лінзи, що розсіює велиабона f вважається негативною.

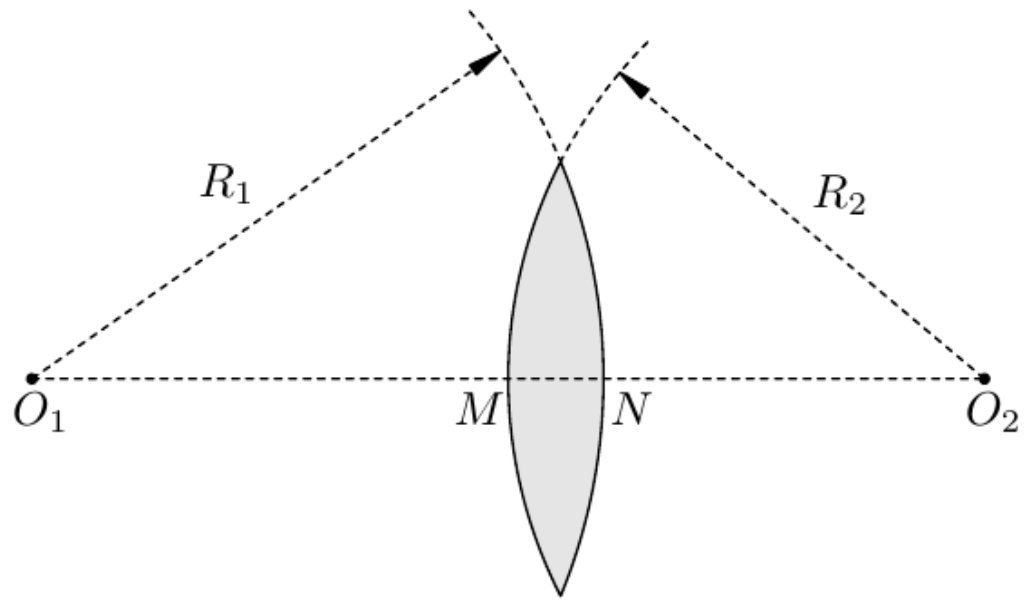

Рис. 5.5. Радіуси кривизни сферичних поверхонь лінзи $R_1$ і $R_2$.

Крім того, формулу тонкої лінзи для двоопуклої тонкої лінзи можна записати через радіуси кривизни сферичних поверхонь лінзи $R_1$ і $R_2$ рис. 5.5.:

$$\frac{1}{a}+\frac{1}{b}=(n-1)\cdot\left(\frac{1}{R_1}+\frac{1}{R_2}\right),$$

де n – показник заломлення матеріалу лінзи; $R_1$ і $R_2$ – радіуси кривизни сферичних поверхонь лінзи. Через ці показники, також можна записати і велиабону, зворотну головній фокусній відстані лінзи, яку називають її оптичної силою:

$$D = \frac{1}{f} = (n-1) \cdot \left( \frac{R_1 + R_2}{R_1 R_2} \right)$$

Кожна з двох заломлюваних поверхонь лінзи має свою оптичну силу:

$$D_1 = \frac{(n-1)}{R_1}; \qquad D_2 = \frac{(n-1)}{R_2}$$

Оптична сила тонкої лінзи дорівнює сумі оптичних сил її обох заломлюваних поверхонь: $D=D_1+D_2$. Оптичну силу вимірюють у діоптріях. Діоптрія – це оптична сила лінзи з фокусною відстанню в 1 метр.

З формули лінзи можна знайти її лінійне збільшення, яке вводиться з формули відношення лінійного розміру зображення h до лінійного розміру предмета H:

$$k = \frac{h}{H} = \frac{b}{a} = \frac{f}{(a-f)}$$

Також для сферичної тонкої лінзи згідно з законом заломлення існує формула для фокусної відстані:

$$F = \frac{1}{\left( \frac{n_{\text{л}}}{n_{\text{ср}}} - 1 \right)} \cdot \frac{1}{\left( \frac{1}{R_1} - \frac{1}{R_2} \right)},$$

де $n_{\text{л}}$ і $n_{\text{ср}}$ – показники заломлення лінзи та середовища, відповідно.

Наведені формули придатні лише для тонких лінз і променів, що утворюють малі кути з головною оптичною віссю лінзи. У практичному застосуванні лінз ці умови зазвичай не виконуються. Тому всі заломлювані лінзою промені вже не збираються в одну точку. Виникає так звана сферична аберація. Для боротьби з нею використовують оптичну систему з кількох лінз з абераціями протилежних знаків так, щоб вони взаємно компенсувались.

**Задача 6.1.** Апарат у воді на глибині $h$=900 м вмикає спрямований догори прожектор. Визнааботь, за який час $t$ світло досягне поверхні води. Швидкість світла в вакуумі $c = 3 \cdot 10^8$м/с, показник заломлення води $n = \frac{4}{3}$ .

| | |
|---|---|
| **Дано:**<br>$h$ =900 м<br>$c = 3 \cdot \frac{10^8 м}{с}$<br>$n = \frac{4}{3}$ | **Розв'язання.** |
| **Знайти:**<br>$t$-? | |

Швидкість світла в воді

$$v = \frac{c}{n},$$

де $n$ – показник заломлення світла.

$$n = \frac{c}{v} \Rightarrow \frac{\sin\alpha}{\sin\beta} = \frac{c}{v}, \quad \text{виразимо } v = \frac{с}{n}$$

Час визначаємо як відношення шляху, в цієї задачі це висота, до швидкості проходження світла до поверхні:

$$t = \frac{h}{v} = \frac{h}{\frac{c}{n}} = \frac{h \cdot n}{c} = \frac{900 \cdot \frac{4}{3}}{3 \cdot 10^8} = 4 \cdot 10^{-6} c$$

**Відповідь:** В= $4 \cdot 10^{-6} c$

**Задача 6.2.** Кут між падаюабом променем і площиною дзеркала в три рази менше кута між падаюабом і відбитим променями. Визнааботь кут падіння променя на дзеркало.

| | |
|---|---|
| **Дано:**<br>$3i = \alpha + \beta$ | |
| **Знайти:**<br>$\alpha$-? | |

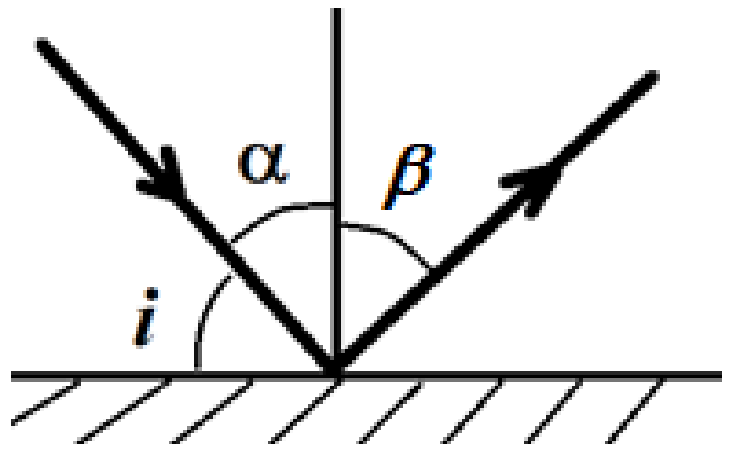

**Розв'язання.**

Маємо кут падіння променя $\alpha$ і кут відбиття $\beta$. За умовою задачі їх сума дорівнюється $3i$, тобто $\alpha + \beta = 3i$, а за законом відбиття «кут відбиття дорівнює куту падіння». Тоді маємо записати, що:

$$2\alpha = 3i, \text{ або } i = \frac{2\alpha}{3}$$

Також можемо записати, що за умовою задачі $i + \alpha = 90^o$ або $i = 90^o - \alpha$. Прирівняємо праві частини отриманих рівнянь.

$$i = \frac{2\alpha}{3} \quad \text{і} \quad i = 90^o - \alpha.$$

Із складеного рівняння виразимо кут падіння світла $\alpha$:

$$90^o - \alpha = \frac{2\alpha}{3} \qquad 3(90^o - \alpha) = 2\alpha$$

$$270 - 3\alpha = 2\alpha \qquad 5\alpha = 270$$

$$\alpha = 54^o \quad \Rightarrow \quad i = 36^o$$

**Відповідь:** $\alpha = 54^o$.

**Задача 6.3.** Промінь світла проходить від повітря в скло з показником заломлення 1,6. Визнааботь косинус кута заломлення, якщо цей кут вдвіче менше кута падіння світла.

**Дано:**
$\alpha = 2\beta$
$n$=1,6

**Знайти:**
$\cos\beta$-?

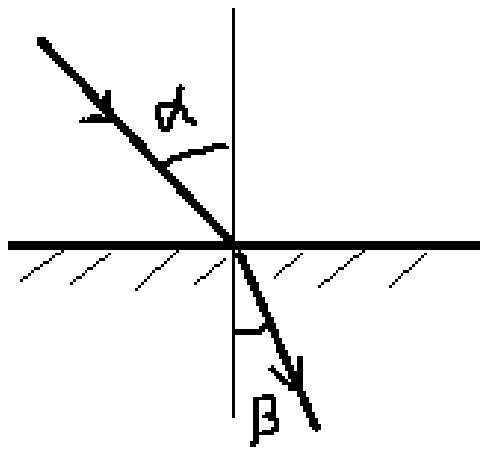

**Розв'язання.**

За законом заломлення світла «відношення синуса кута падіння до синуса кута заломлення дорівнює відношенню показника заломлення другого середовища до показника заломлення першого середовища»:

$$\frac{\sin\alpha}{\sin\beta} = \frac{n_2}{n_1}$$

У задачі перше середовище це повітря у якого $n_1 = 1$, тому:

$$\frac{\sin\alpha}{\sin\beta} = n_2 = 1{,}6$$

За умовою задачі $\alpha = 2\beta$, підставимо замість $\alpha$ кут $2\beta$:

$$\frac{\sin 2\beta}{\sin\beta} = 1{,}6$$

За формулою тригонометричної функції подвійного аргументу маємо:

$$\sin 2\beta = 2\sin\beta \cdot \cos\beta, \quad \text{тоді}$$

$$\frac{2\sin\beta \cdot \cos\beta}{\sin\beta} = 1{,}6 \;\Rightarrow\; 2\cos\beta = 1{,}6$$

$$\cos\beta = \frac{1{,}6}{2} = 0{,}8$$

**Відповідь:** $\cos\beta = 0{,}8$ .

**Задача 6.4.** Промінь світла падає із скла на межу розділу «скло-вода». Визнааботь кут падіння $\alpha$ , при якому відбитий і заломлений промінь перпендикулярні один до одного.

| **Дано:** | **Розв'язання:** |
|---|---|
| $n_1$=1,6<br>$n_2$=1,33 | Для успішного разв'язання зробимо рисунок до умов задачі. |
| **Знайти:**<br>$\alpha$-? | |

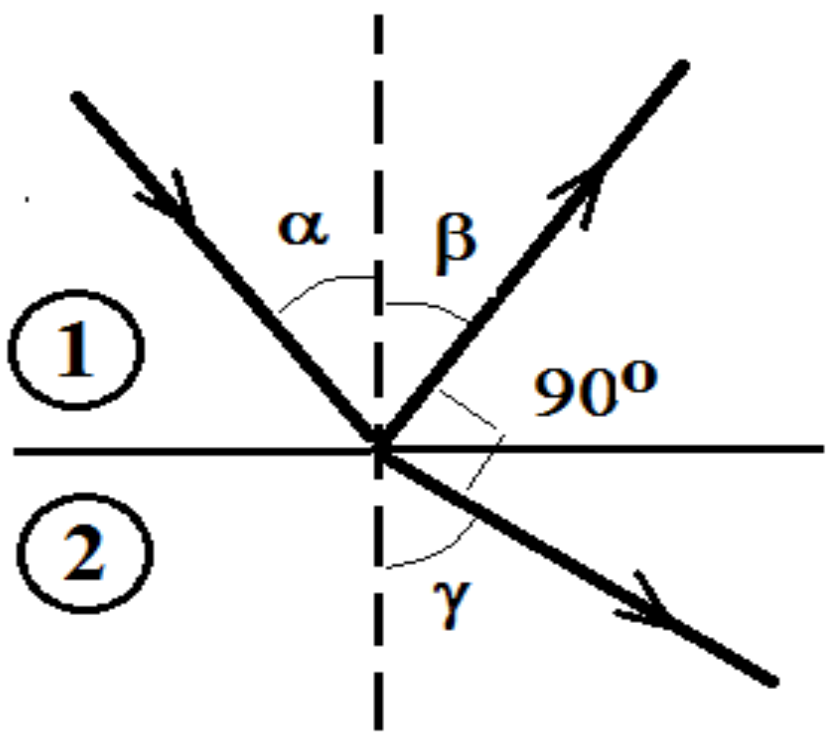

Маємо: середовище 1 з показником заломлення для скла $n_1 = 1{,}6$, та середовище 2 з показником заломлення для води $n_2 = 1{,}33$, $\alpha$ – кут падіння, $\beta$ – відображений кут, $\gamma$ – заломлений кут.

За умовою здачі $\beta \perp \gamma \;\Rightarrow\; \beta + \gamma = 90^{\circ}$, тому що разом створюють розгорнутий кут в 180°, крім цього, за законом відбиття $\alpha = \beta$, а за законом заломлення світла:

$$\frac{\sin\alpha}{\sin\gamma} = \frac{n_2}{n_1}$$

Замінимо в $\beta + \gamma = 90^{\circ}$ $\beta$ на $\alpha$ і отримаємо:

$$\alpha + \gamma = 90^{\circ} \quad \text{або} \quad \gamma = 90^{\circ} - \alpha \qquad \text{підставимо в } \sin\gamma$$

$$\frac{\sin\alpha}{\sin(90^{\circ} - \alpha)} = \frac{n_2}{n_1}$$

$$\sin(90^{\circ} - \alpha) = \cos\alpha, \quad \text{таким абоном:}$$

$$\frac{\sin\alpha}{\cos\alpha} = \frac{n_2}{n_1},$$

$$\text{тобто} \quad \frac{n_2}{n_1} = tg\alpha$$

Підставимо значення показника заломлення для скла $n_1 = 1{,}6$ та для води $n_2 = 1{,}33$.

$$tg\alpha = \frac{1{,}33}{1{,}6} \approx 0{,}83 \qquad \Rightarrow \qquad \alpha = 40^{\circ}$$

**Відповідь:** $\alpha = 40^{\circ}$.

**Задача 6.5.** Промінь світла падає на плоско паралельну пластинку з скла, показник заломлення якого $n = 1{,}73$, під кутом $i = 60^{\circ}$. Визначити товщину пластинки $h$, якщо зміщення променя при виході з пластинки становить $d = 20$ мм.

**Дано:**
$d$ =20 мм
$n$=1,73
i=60°

**Знайти:**
$h$-?

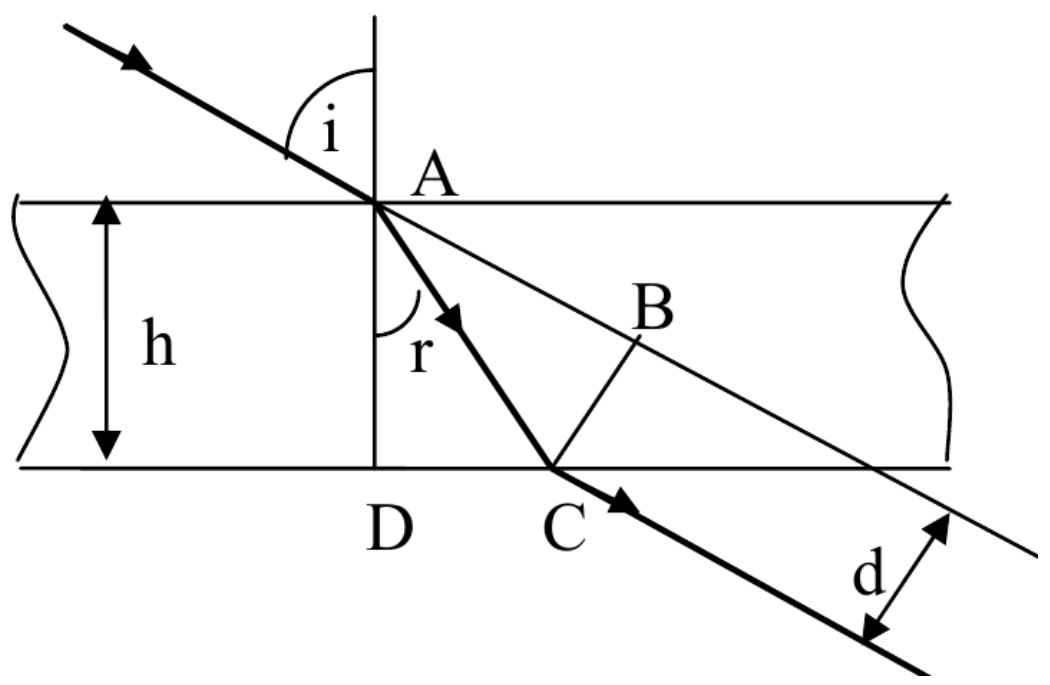

**Розв'язання.**

Маємо промінь, який потрапляє на поверхню скляної пластинки у точці A під кутом $i = 60^{\circ}$, далі заломлюється і йде під кутом заломлення r до точки C. Після точки C промінь знову заломлюється і продовжує хід зовні скляної пластинки паралельно первинному променю, якщо б він йшов не заломлюючись.

Побудований перпендикуляр із точки С до уявного продовження променю без заломлення перетинає у точці В. Саме відрізок СВ і є зміщенням $d$. Розглянемо прямокутний трикутник $\triangle ADC$. Виразимо

$h$ =AD з прямокутного трикутника ADC:

$h$ =AC $\cdot\cos r$;

де $r$ – кут заломлення променя світла.

За законом заломлення світла

$$\frac{\sin i}{\sin r} = n$$

Невідомий кут заломлення $r$ маємо визначити із цього рівняння виразивши $\sin r$:

$$\sin r = \frac{\sin i}{n} = \frac{\sin 60^{o}}{1{,}73} = 0{,}5 \implies$$

$$r = arcsin 0{,}5 = 30^{o}$$

Для знаходження $h$ невідомим залишилась тільки відстань АС, яку можна знайти із прямокутного трикутника $\triangle$ ABC:

$$\text{AC} = \frac{d}{\sin(i-r)}.$$

Отриманий вираз АС підставимо в рівняння $h$ =AC $\cdot\cos r$ і отримаємо:

$$h = \frac{d}{\sin(i-r)}\cos r = \frac{2\cdot 10^{-2}\cdot\cos 30^{o}}{\sin(60^{o}-30^{o})} = \frac{2\cdot 10^{-2}\cdot\frac{\sqrt{3}}{2}}{\sin 30^{o}} = 3{,}46\cdot 10^{-2}\text{м}.$$

**Відповідь:** $h = 3{,}46\cdot 10^{-2}$м.

**Задача 6.6.** Промінь падає на плоско-паралельну пластину товщиною $h = 2$ см. Відомо що $\sin\alpha = 0{,}8$; $\sin\gamma = 0{,}6$ де $\alpha$ та $\gamma$ – відповідно кути падіння та заломлення при вході променю в пластинку. Визначте зміщення b променю після проходження крізь пластинку.

**Дано:**

$h$ =2 см

$\sin\alpha = 0{,}8$

$\sin\gamma = 0{,}6$

**Знайти:**

$b$-?

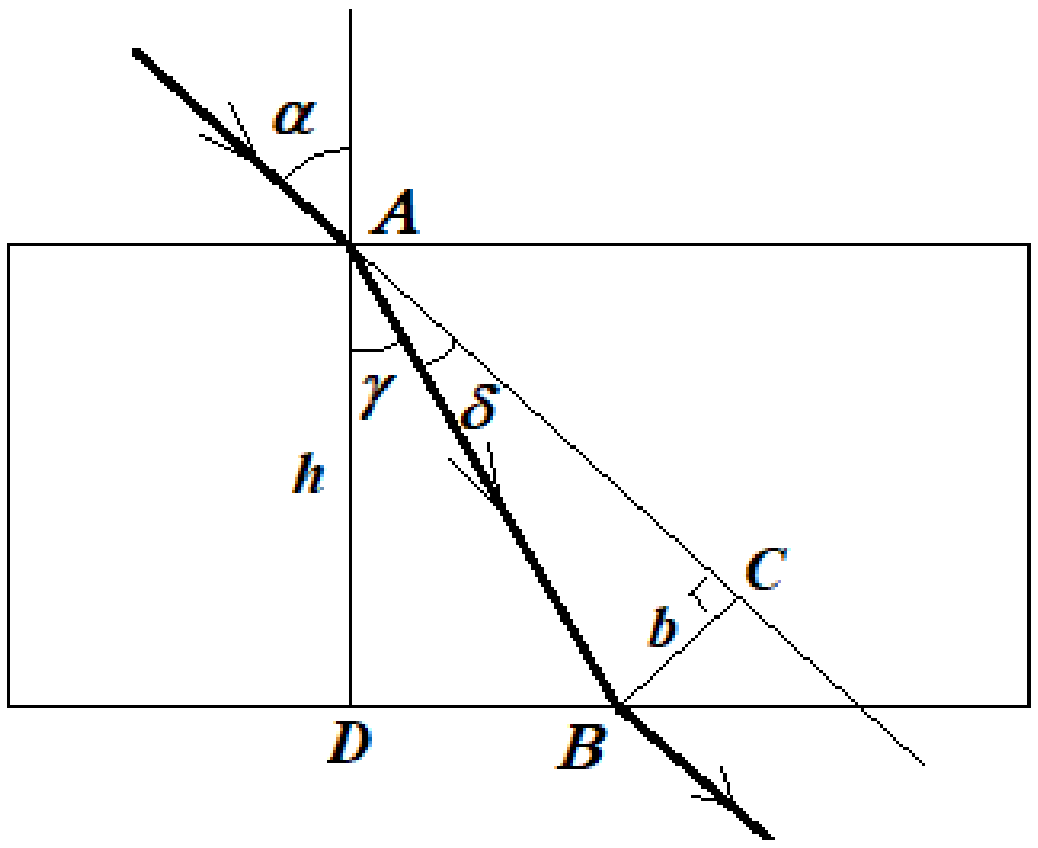

**Розв'язання.**

Задача дуже схожа на попередню. Визначимося з взаємозв'язком між кутами падіння $\alpha$, заломлення $\gamma$ і кутом $\delta$ у вершині А трикутника $\Delta$ ABC:

$$\alpha = \gamma + \delta \;\;\Rightarrow\;\; \delta = \alpha - \gamma$$

Треба визначити діагональ в трикутнику $\Delta$ ABC. Можемо записати, що:

$$\cos\gamma = \frac{h}{\mathrm{AB}}; \quad \mathrm{AB} = \frac{h}{\cos\gamma}$$

Тоді із трикутника ABC

$$\sin\delta = \frac{b}{AB}; \quad b = AB \cdot \sin\delta = \frac{h}{\cos\gamma}\sin(\alpha - \gamma) =$$

Далі існує два способу розв'язання задачі:

1) По таблицям з'ясувати чому дорівнюється кут $\alpha$ із $\sin\alpha = 0{,}8$

і кут $\gamma$ із $\sin\gamma = 0{,}6$.

$\sin 53^o = 0{,}8$ і $\sin 37^o = 0{,}6$ тобто $\alpha = 53^o$ $\gamma = 37^o$ $\alpha - \gamma = 16^o$

Підставимо

$$b = \frac{h}{\cos\gamma}\sin(\alpha-\gamma) = \frac{2\cdot10^{-2}}{\cos 37^o}\sin 16^o = \frac{2\cdot10^{-2}}{0{,}8}0.27 = \frac{0{,}54}{0{,}8}10^{-2} =$$

$$= 0{,}675\cdot10^{-2}\text{ м} = 0{,}00675\text{ м} = 0{,}675\text{ см} = 6{,}75\text{ мм}$$

2) Спосіб:

За тригонометричними формулами додавання:

$$\sin(\alpha-\beta)=\sin\alpha\cdot\cos\beta-\cos\alpha\cdot\sin\beta$$

$$b=AB\cdot\sin\delta=\frac{h}{\cos\gamma}\sin(\alpha-\gamma)=$$

$$=h\frac{\sin\alpha\cos\gamma-\cos\alpha\sin\gamma}{\cos\gamma}=h\left(\sin\alpha-\frac{\cos\alpha\sin\gamma}{\cos\gamma}\right)$$

Дані тільки значення синусів, тому косинуси виразимо через синус із формули

$$\sin^2\alpha+\cos^2\alpha=1$$

отримаємо: $\cos\alpha=\sqrt{1-(\sin\alpha)^2}$ $\cos\gamma=\sqrt{1-(\sin\gamma)^2}$

Підставимо отримані рівняння і значення синусів $\sin\alpha=0{,}8$ ; $\sin\gamma=0{,}6$

в вираз:

$$b=h\left(\sin\alpha-\frac{\sqrt{1-(\sin\alpha)^2}\cdot\sin\gamma}{\sqrt{1-(\sin\gamma)^2}}\right)=2\cdot10^{-2}\left(0{,}8-\frac{\sqrt{1-0{,}8^2}\cdot0{,}6}{\sqrt{1-0{,}6^2}}\right)=$$

$$=2\cdot10^{-2}\left(0{,}8-\frac{\sqrt{0{,}36}\cdot0{,}6}{\sqrt{0{,}64}}\right)=2\cdot10^{-2}\left(0{,}8-\frac{0{,}36}{0{,}8}\right)=2\cdot10^{-2}\cdot0{,}35=$$

$$=0{,}7\cdot10^{-2}\text{ м}=0{,}007\text{ м}=0{,}7\text{ см}=7\text{ мм}$$

Порівняємо з відповіддю, отриманою першим способом:

$$b=0{,}675\cdot10^{-2}\text{ м}=0{,}00675\text{ м}=0{,}675\text{ см}=6{,}75\text{ мм}$$

Розбіжності пов'язані з неточністю при розрахунках.

**Відповідь:** $b=6{,}75$ мм

**Задача 6.7.** На оправі лупи учень прочитав: «$10^x$». Визначте оптичну силу D лупи. Відстань найкращого зору дорівнюється 25 см.

**Дано:**
$d_0$ =25 см
Г=$10^x$

**Знайти:**
*D*-?

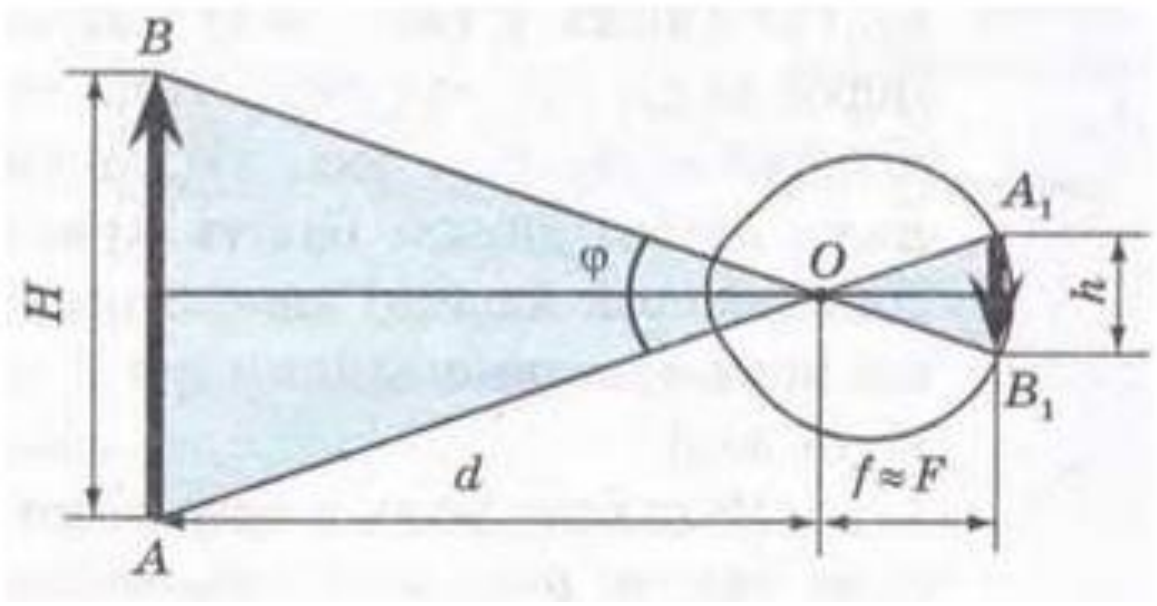

**Розв'язання.**

Напис «$10^x$» означає, що лупа дає збільшення Г=10. Збільшення лупи з фокусною відстанню F складає:

$$\Gamma = \frac{d_0}{F},$$

де $d_0 = 25\,см$ – відстань найкращого зору.

За умовою задачі треба визнааботи оптичну силу D, яка вимірюється в діоптріях (Діоптрія — одиниця виміру оптичної сили лінз і інших оптичних систем, $м^{-1}$, 1 діоптрія дорівнюється оптичній силі лінзи с фокусною відстанню в 1 метр). Тому обов'язково 25 см переводимо в метри:

25 см = 0,25 м.

$$\mathrm{D} = \frac{1}{F}; \qquad F \text{ виразимо із } \quad \Gamma = \frac{d_0}{F}$$

$$F = \frac{d_0}{\Gamma} \quad \text{і} \quad D = \frac{1}{\frac{d_0}{\Gamma}} = \frac{\Gamma}{d_0} = \frac{10}{0{,}25} = 40 \text{ дптр.}$$

**Відповідь:** $D = 40$ дптр.

**Задача 6.8.** Предмет розташований на відстані 50 см від лінзи з оптичною силою –2 дптр. Визначить (в см.) відстань між лінзою і зображенням предмета.

**Дано:**

$d$ =50 см

$D$=-2 дптр

**Знайти:**

$f$-?

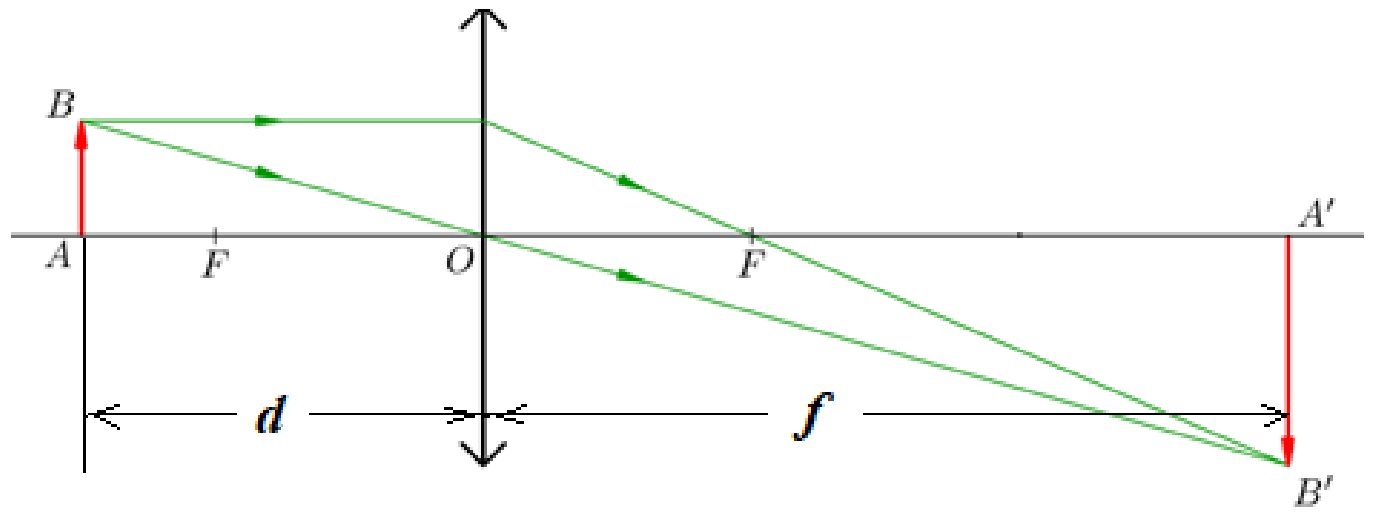

**Розв’язання.**

В умовах задачі надана оптична сила $\mathrm{D} = -2$ дптр, яка пов’язана з фокусом лінзи формулою:

$$\mathrm{D} = \frac{1}{F} \qquad \text{де } F - \text{фокус лінзи}$$

Напишемо формулу для тонкої лінзи:

$$\frac{1}{F} = \frac{1}{d} + \frac{1}{f}$$

За умовою задачі треба знайти $f$ – відстань між лінзою і зображенням:

$$\frac{1}{f} = \frac{1}{F} - \frac{1}{d}, \qquad \text{де } F = \frac{1}{\mathrm{D}} = \frac{1}{-2} = -0{,}5\ \text{м} = -50\ \text{см}$$

$$\text{тоді} \quad \frac{1}{f} = -\frac{1}{50} - \frac{1}{50} = -\frac{2}{50} \quad \Rightarrow \quad f = -\frac{50}{2} = -25\ \text{см}.$$

**Відповідь:** $f = 25$ см.

**Задача 6.9.** Фокусна відстань об’єктива фотоапарата $F = 25$ см. Турист фотографує картину висотою $h = 1$ м з відстані $d = 5$ м. Визначте висоту $H$ зображення, яке утворюється на світлочутливій поверхні.

**Дано:**

$d$ =5 м

$h = 1$ м

$F = 25$ см

**Знайти:**

$H$-?

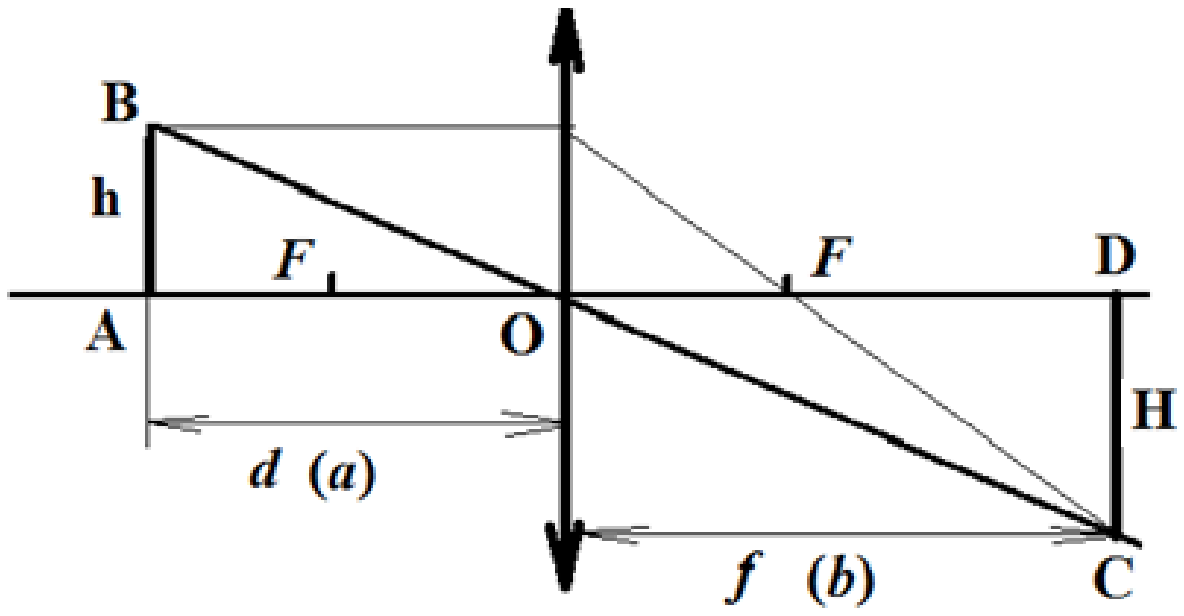

**Розв’язання.**

При вирішенні цієї задачі треба скористатися формулою для лінійного збільшення Г лінзи «Відношення лінійного розміру $H$ зображення предмета до розміру $h$ самого предмету називається лінійним збільшенням Г лінзи».

$$\frac{H}{h} = \Gamma \quad \text{поглянемо на рис.}$$

Бачимо, що трикутники $\triangle$ABO і $\triangle$DCO є подібні, тобто маємо записати:

$$\frac{AB}{CD} = \frac{AO}{OD} = \frac{BO}{OC} = k \quad \text{тоді запишемо:}$$

$$\Gamma = \frac{H}{h} = \frac{|f|}{|d|}.$$

де $d$ – відстань між лінзою та предметом; $f$ – відстань між лінзою і зображенням предмета.

$$\frac{H}{h} = \frac{f}{d}$$

Щоб знайти лінійний розмір зображення $H$ треба ще знайти $f$ - відстань між лінзою та зображенням предмета. Для цього візьмемо формулу для тонкої лінзи:

$$\frac{1}{F} = \frac{1}{d} + \frac{1}{f}$$

$$\frac{1}{f} = \frac{1}{F} - \frac{1}{d} = \frac{1}{0{,}25} - \frac{1}{5} = \frac{19}{5}; \Longrightarrow f = \frac{5}{19} = 0{,}26 \quad \text{підставимо}$$

$$\frac{H}{h} = \frac{f}{d} \qquad H = h\frac{f}{d} = 1\,\frac{0{,}26}{5} = 0{,}052 = 5{,}2 \cdot 10^{-2}\ \text{м} = 5{,}2\ \text{см}$$

**Відповідь:** $H = 5{,}2$ см

**Задача 6.10.** Лінза знаходиться в повітрі, радіуси кривизни поверхонь лінзи $R_1$=10 см і $R_2$ = 15 см. Визначте фокусну відстань лінзи, виготовленої зі скла з показником заломлення $n$ = 1,5, якщо: 1) Лінза двоопукла; 2) Лінза двовгнута; 3) Лінза вгнуто-опукла; 4) Лінза опукло-вгнута;

| **Дано:** | **Розв'язання.** |
|---|---|
| $R_1$=10 см<br>$n$=1,5<br>$R_2$=15 см | Фокусна відстань лінзи визначається речовиною лінзи і середовища, радіусами кривизни поверхонь і формою лінзи. В умовах задачі середовище це повітря, яке має показник заломлення $n_{\text{ср}} = 1$, тому замість відношення показників заломлення можна писати тільки показник заломлення лінзи. |
| **Знайти:**<br>$F$-? | |

Лінза двоопукла.

Обидві сферичні поверхні опуклою стороною дивляться в повітря, оптично менш щільне середовище, отже, радіуси кривизни поверхонь увійдуть в формулу для розрахунку фокуса зі знаком «+».

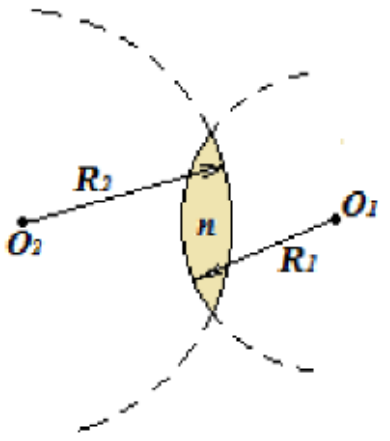

$$\frac{1}{F} = (n-1)\left(\frac{1}{R_1} + \frac{1}{R_2}\right) = (1,5-1)\left(\frac{1}{10} + \frac{1}{15}\right) = \frac{1}{12}\ (\text{см}^{-1})$$

$$F = 12\ \text{см}.$$

Відповідь: Фокусна відстань позитивна - це знааботь, що двоопукла лінза зі скла в повітрі є збиральною з $F = 12$ см.

Лінза двовгнута.

Обидві сферичні поверхні опуклою стороною дивляться в скло, оптично більш щільне середовище, отже, радіуси кривизни поверхонь увійдуть в формулу для розрахунку фокуса зі знаком «-».

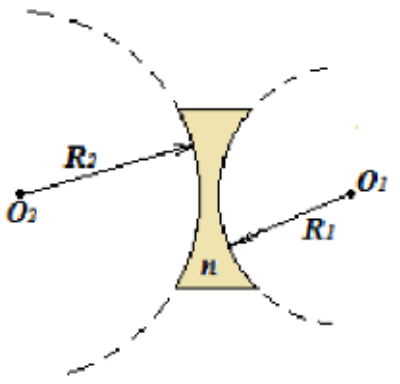

$$\frac{1}{F} = (n-1)\left(-\frac{1}{R_1} - \frac{1}{R_2}\right) = (1,5-1)\left(-\frac{1}{10} - \frac{1}{15}\right) = -\frac{1}{12}\ (\text{см}^{-1})$$

$$F = -12\ \text{см}.$$

Відповідь: Фокусна відстань негативна - це знааботь, що двоопукла лінза зі скла в повітрі є розсіювальною з $F = -12$ см.

Лінза вгнуто-опукла.

Одна сферичня поверхні з $R_1$ опуклою стороною дивиться в повітря, отже увійде в формулу зі знаком «+», але сферичня поверхні з $R_2$ опуклою стороною дивиться в скло, отже увійде в формулу зі знаком «-».

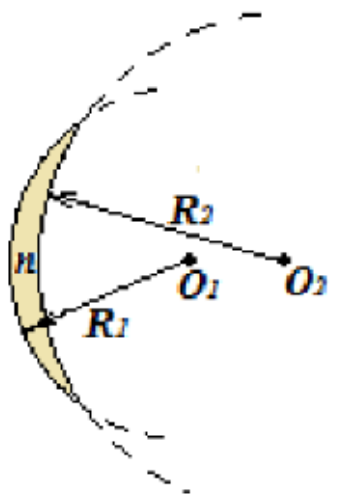

$$\frac{1}{F}=(n-1)\left(\frac{1}{R_1}-\frac{1}{R_2}\right)=(1,5-1)\left(\frac{1}{10}-\frac{1}{15}\right)=\frac{1}{60}\ \left(\text{см}^{-1}\right)$$

$$F=60\ \text{см}.$$

Відповідь: Лінза є збиральною. з $F = 60$ см.

Лінза опукло-вгнута.

Одна сферичня поверхні з $R_1$ опуклою стороною дивиться в скло, отже увійде в формулу зі знаком «-», але сферичня поверхні з $R_2$ опуклою стороною дивиться в повітря, отже увійде в формулу зі знаком «+».

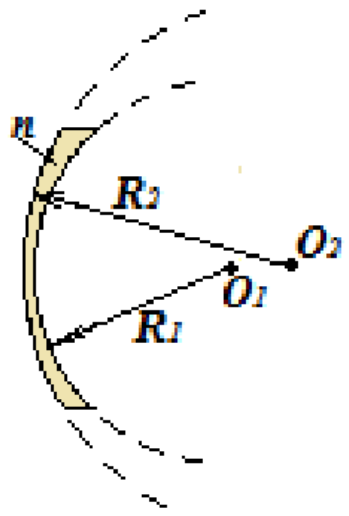

$$\frac{1}{F}=(n-1)\left(-\frac{1}{R_1}+\frac{1}{R_2}\right)=(1,5-1)\left(-\frac{1}{10}+\frac{1}{15}\right)=-\frac{1}{60}\ \left(\text{см}^{-1}\right)$$

$$F=-60\ \text{см}.$$

**Відповідь:** Лінза є розсіювальною з $F = -60$ см.

**Задача 6.11.** Визначте фокусну відстан плоскоопуклої товстої лінзи товщиною $\ell$=5 см, виготовленої зі скла з показником заломлення $n$=1,5. Лінза знаходиться в повітрі, радіус кривизни поверхні лінзи $R$ =2,5 см. Кути заломлення вважати малими і промінь падає на лінзу з плоскої сторони.

**Дано:**
$\ell$=5 см
$n$=1,5
$R$=2,5 см

**Знайти:**
$F$-?

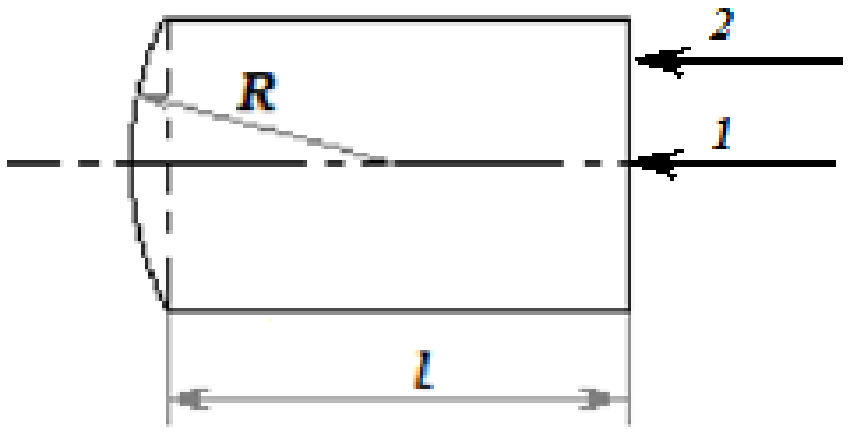

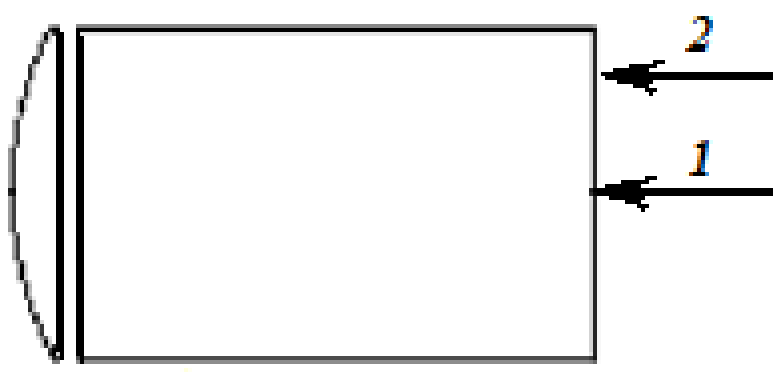

**Розв'язання.**

Уявімо товсту лінзу як сукупність тонкої плоско-опуклої лінзи і плоско-паралельної пластини. Плоско-паралельна пластина не змінює ходу променів 1 і 2, оскільки вони падають перпендикулярно поверхні пластини. Зміна ходу променів відбувається тільки при їхньому проходженні через тонку лінзу. Таким чином, фокусна відстань товстої лінзи збігається з фокусною відстанню тонкої лінзи, яке можна знайти за відомою формулою:

$$\frac{1}{F}=(n-1)\left(\frac{1}{R_1}+\frac{1}{R_2}\right)=(n-1)\left(\frac{1}{R}+\frac{1}{\infty}\right)=\frac{n-1}{R} \qquad F=\frac{R}{n-1}=5 \text{ см.}$$

**Відповідь:** $F=5$ см.

**Задача 6.12.** Промінь падає на грань скляної призми, заломлюючий кут якої . Кут падіння $\alpha=49^{\circ}$ (вважайте, що $\sin\alpha=0{,}75$), показник заломлення скла $n=1{,}5$. Визначте кут $\delta$ відхилення променю від початкового напрямку після проходження скрізь призму.

**Дано:**
$\alpha$ =49°
$n$=1,5
φ=60°

**Знайти:**
$\delta$-?

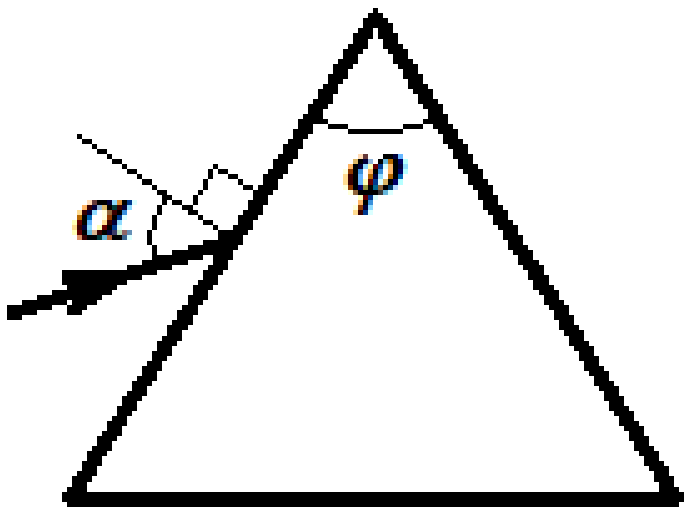

**Розв'язання.**

Щоб визнаботи кут $\delta$, який складається з відхилення променю при входженні в призму, познаабомо як $\delta_1$, та з кута відхилення променю $\delta_2$ при виході з призми, тобто $\delta=\delta_1+\delta_2$.

Позначимо точкою А точку входження променю в призму, через В вихід із призми. Пам'ятаємо, що кут падіння, на рисунку це кут $\alpha_1$, - кут між перпендікуляром до сторони призми і промінем, тоді кут заломлення $\beta_1$ в сумі з кутом відхилення $\delta_1$ будуть дорівнюватись $\alpha_1$ , як кути при прямих, що перетинаються, тобто $\alpha_1 = \beta_1 + \delta_1$ . За умовою задачі при відомому куті $\alpha_1$ треба визначити $\delta_1$, для цього необхідно визначити $\beta_1$. Відома формула:

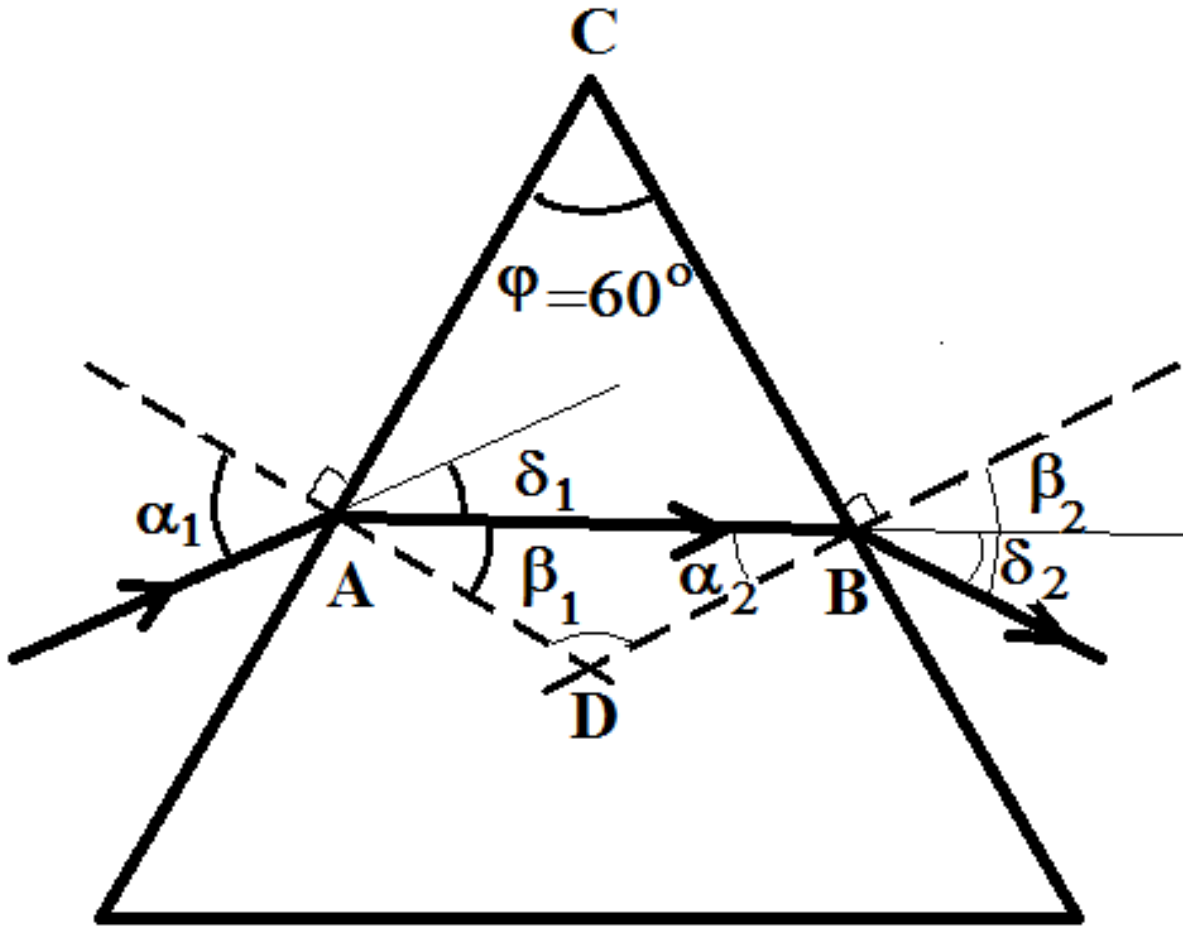

$$\frac{\sin\alpha}{\sin\beta} = \frac{n_2}{n_1}$$

де $n_1 = 1$ – це показник заломлення повітря, а $n_2$ це $n = 1{,}5$.

З цієї формули знаходимо:

$$\sin\beta = \frac{\sin\alpha}{n_2},$$

де $n_2 = 1{,}5$ - показник заломлення скла, який наданий в умовах задачі.

$$\sin\beta = \frac{\sin\alpha}{n_2} = \frac{0{,}75}{1{,}5} = 0{,}5, \qquad \text{тобто} \ \ \beta = 30^\circ$$

Підставимо в $\alpha_1 = \beta_1 + \delta_1$, $\alpha = 49^\circ$ – за умовою задачі і знайдений за $\beta = 30^\circ$ маємо:

$$\delta_1 = \alpha_1 - \beta_1 = 49^\circ - 30^\circ = 19^\circ$$

Для знаходження другого відхилення $\delta_2$ нам не відомий ні кут падіння $\alpha_2$ ні кут заломлення $\delta_2$. Розглянемо трикутник $\Delta$ADB – в ньому нам відомий кут $\beta_1$ і ми можемо визнааботи кут при вершині D.

З геометрії відомо, що сума кутів четирьохкутника = $360^o$. За умовою побудови ліній AD і DB – це є нормалі до сторін призми, тобто кути які вони створюють = $90^o$. Тобто відоми два кута в четирьохкутнику ADBC.

За умовою задачі кут при вершині призми $\varphi = 60^o$, тобто кут при вершині D: $\angle A\text{DB} = 360^o - 2 \cdot 90^o - 60^o = 120^o$,

знаюабо, що сума кутів трикутника = $180^o$ $\Rightarrow$

$$\alpha_2 = 180^o - 120^o - 30^o = 30^o$$

Ще раз застосуємо формулу:

$$\frac{\sin\alpha}{\sin\beta} = \frac{n_2}{n_1}$$

Тільки зараз показник заломлення повітря це вже $n_2 = 1$, тому, що промінь йде зі скла у повітря, тобто:

$$\frac{\sin\alpha_2}{\sin\beta_2} = \frac{1}{n}; \quad \frac{\sin 30^o}{\sin\beta_2} = \frac{1}{1{,}5}$$

$$\sin\beta_2 = 1{,}5 \cdot \sin 30^o = 1{,}5\frac{1}{2} = 0{,}75 \quad \beta_2 = 49^o$$

$$\delta_2 = \beta_2 - \alpha_2 = 49^o - 30^o = 19^o$$

$$\delta = \delta_1 + \delta_2 = 19^o + 19^o = 38^o$$

**Відповідь:** $\delta = 38^o$.

**Задача 6.13.** Фокусна відстань об'єктиву мікроскопа $F_{об} = 4$ мм, а фокусна відстань окуляра $F_{ок} = 20$ мм. Відстань між об'єктивом та окуляром $\ell$ =184 мм. Визначте збільшення мікроскопу $\Gamma$. Відстань найкращого зору $d_0 = 25$ см.

**Дано:**

$\ell$ =184 мм

$F_{об} = 4$ мм

$F_{ок} = 20$ мм

$d_0$=25 см

**Знайти:**

Γ -?

**Розв'язання.**

Збільшення мікроскопу складається зі збільшення об'єктива та окуляра:

$$\Gamma = \Gamma_{об'ектив} \cdot \Gamma_{окуляр}$$

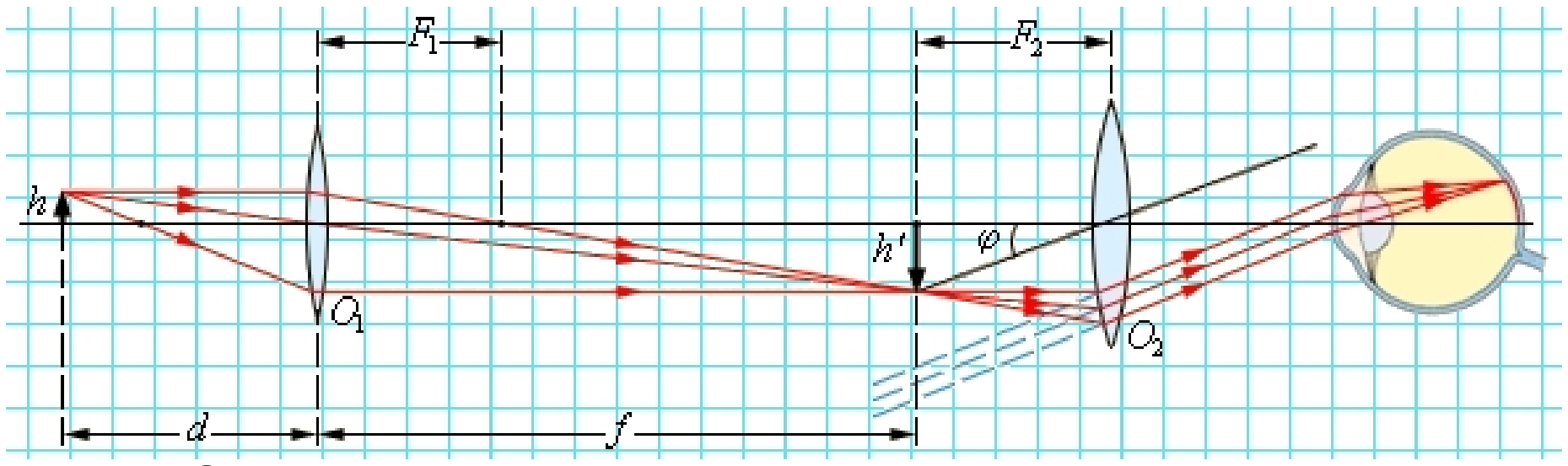

$$\Gamma_{об'ектив} = \frac{f_{об}\ \text{(від об'єктива до зображення)}}{d_{об}\ \text{(від предмета до об'єктива)}} \quad \text{та} \quad \Gamma_{окуляр} = \frac{d_0}{F_{ок}}$$

Для знаходження $\Gamma_{окуляр}$ – збільшення окуляра, всі данні надани в умовах задачі. Тобто треба знайти: $f_{об}$ та $d_{об}$ .

Знайдемо $d_{об}$ за формулою для тонкої лінзи:

$$\frac{1}{d_{об}} + \frac{1}{f_{об}} = \frac{1}{F_{об}}$$

$$\frac{1}{d_{об}} = \frac{1}{F_{об}} - \frac{1}{f_{об}} = \frac{f_{об} - F_{об}}{F_{об} \cdot f_{об}} \quad \text{тобто} \quad d_{об} = \frac{F_{об} \cdot f_{об}}{f_{об} - F_{об}}$$

Поглянемо на рис. $f_{об}$ це є різниця між відстанню від об'єктива до окуляра $\ell$=184 мм. та фокусною відстанню окуляра $F_{окуляр}$.

$$f_{об} = \ell - F_{окуляр}$$

Величини $\ell$ та $F_{окуляр}$ надани і підставивши їх у формулу:

$$d_{об} = \frac{F_{об} \cdot f_{об}}{f_{об} - F_{об}}$$

Тепер можемо знайти збільшення мікроскопу.

Таким чином, збільшення мікроскопу можна розрахувати за формулою:

$$\Gamma = \Gamma_{об'ектив} \cdot \Gamma_{окуляр} = \frac{f_{об'ектив} - F_{об'ектив}}{F_{об'ектив}} \cdot \frac{d_0}{F_{окуляр}};$$

де відстань від зображення до об'єктиву

$$f_{об'ектив} = \ell - F_{окуляр}$$

Підставимо значення і отримаємо:

$$\Gamma = \frac{(\ell - F_{окуляр} - F_{об'ектив})}{F_{об'ектив}} \cdot \frac{d_0}{F_{окуляр}} = \frac{(184 - 20 - 4)250}{4 \cdot 20} = 500 \text{ раз}$$

**Відповідь:** $\Gamma = 500$ раз.

## Задачі для самостійного розв'язання

1. Точкове джерело світла з силою світла $I$=80 кд розташоване в центрі сфери радіусом $R$=2 м. Визначте світловий потік Ф, що проходить через ділянку поверхні сфери площею $S_{діл}$=0.8 м$^2$. [Ф=16 лм].

2. Кругла лампа діаметром $D$=0.1 м має яскравість L=5000 кд/м$^2$ у напрямку, перпендикулярному до її площини. Визначте світловий потік Ф, який випромінюється цією лампою в півсферу, якщо припустити, що вона є ідеальним дифузним випромінювачем. [Ф≈123.37 лм].

3. Якщо настільна лампа дає освітленість $E_1$=300 лк на поверхні столу, розташованій на відстані $r_1$=0.5 м від джерела (промені падають перпендикулярно), то яка освітленість $E_2$ буде на цій же поверхні, якщо лампу підняти на висоту $r_2$=1.0 м? [$E_2$=75 лк].

4. Показник заломлення скла $n$=1,5. Знайти граничний кут повного внутрішнього відбиття β для поверхонь розділу: а) скло-повітря, б) вода-повітря; в) скло-вода. [а) **скло-повітря**: $\alpha_1$ ≈41.8°. б) **вода-повітря**: $\alpha_2$ ≈48.8°. в) **скло-вода**: $\alpha_3$ ≈62.5°].

5. Знайти фокусну відстань $F_2$ лінзи, що занурена у воду, якщо її фокусна відстань в повітрі $F_1$=20 см. Показник заломлення матеріалу лінзи $n$=1,6. [$F_2$≈0.591 м.

## СПИСОК ЛІТЕРАТУРИ